\documentclass{emulateapj}
\usepackage{natbib} 
\usepackage{placeins}
\usepackage{rotate,rotating}
\usepackage{epsfig,graphicx}
\usepackage{float}
\usepackage[pass]{geometry}
\def\lap{\lower.5ex\hbox{$\; \buildrel < \over \sim \;$}}
\def\gap{\lower.5ex\hbox{$\; \buildrel > \over \sim \;$}}

\def\ergcm2s{${\rm erg\ cm^{-2}\ s^{-1}}$}

\def\ergscm2s{${\rm erg\ cm^{-2}\  s^{-1}}$}

\def\cm-2{${\rm cm^{-2}}$}

\begin{document}

\title{Panchromatic Hubble Andromeda Treasury. XIX. The Ancient Star Formation History of the M31 Disk}

\author{Benjamin F. Williams\altaffilmark{1},
Andrew E. Dolphin\altaffilmark{2},
Julianne J. Dalcanton\altaffilmark{1},
Daniel R. Weisz\altaffilmark{1,3},
Eric F. Bell\altaffilmark{4},
Alexia R. Lewis\altaffilmark{1,5,6},
Philip Rosenfield\altaffilmark{7},
Yumi Choi\altaffilmark{1,8},
Evan Skillman\altaffilmark{9},
Antonela Monachesi\altaffilmark{10,11}
}

\altaffiltext{1}{Department of Astronomy, Box 351580, University of Washington, Seattle, WA 98195; ben@astro.washington.edu, jd@astro.washington.edu}
\altaffiltext{2}{Raytheon, 1151 E. Hermans Road, Tucson, AZ 85706; adolphin@raytheon.com}
\altaffiltext{3}{University of California, Berkeley;
  dweisz@berkeley.edu}
\altaffiltext{4}{Department of Astronomy, University of Michigan, 1085 South University Ave., Ann Arbor MI 48109; ericbell@umich.edu}
\altaffiltext{5}{Center for Cosmology and AstroParticle Physics, The Ohio State University, Columbus, OH 43210, USA}
\altaffiltext{6}{Department of Astronomy, The Ohio State University, 140 West 18th Avenue, Columbus, OH 43210, USA; lewis.1590@osu.edu}
\altaffiltext{7}{Harvard-Smithsonian Center for Astrophysics, 60
  Garden St., Cambridge, MA; philip.rosenfield@cfa.harvard.edu}
\altaffiltext{8}{Steward Observatory, University of Arizona, 933 North Cherry Avenue, Tucson, AZ 85721}
\altaffiltext{9}{Department of Astronomy, University of Minnesota, 116 Church St. SE, Minneapolis, MN 55455; skillman@astro.umn.edu}
\altaffiltext{10}{Departamento de F\'isica y Astronom\'ia, Universidad de La Serena, Av. Juan Cisternas 1200 N, La Serena, Chile; antonela@MPA-Garching.MPG.DE}
\altaffiltext{11}{Instituto de Investigaci\'on Multidisciplinar en Ciencia y
Tecnolog\'ia, Universidad de La Serena, Ra\'ul Bitr\'an 1305, La Serena,
Chile}

\begin{abstract}

We map the star formation history across M31 by fitting 
stellar evolution models to color-magnitude diagrams of each
83$''{\times}83''$ (0.3$\times$1.4 kpc, deprojected) region of the PHAT survey
outside of the innermost 6$'{\times}12'$ portion.  We find that most
of the star formation occurred prior to $\sim$8 Gyr ago, followed by a
relatively quiescent period until $\sim$4 Gyr ago, a subsequent star
formation episode about 2 Gyr ago and a return to relative
quiescence. There appears to be little, if any, structure visible for
populations with ages older than 2 Gyr, suggesting significant mixing
since that epoch.  Finally, assuming a Kroupa IMF from 0.1$-$100
M$_{\odot}$, we find that the total amount of star formation over the
past 14 Gyr in the area over which we have fit models is
5${\times}$10$^{10}$~M$_{\odot}$.  Fitting the radial distribution of
this star formation and assuming azimuthal symmetry,
(1.5$\pm$0.2)${\times}$10$^{11}$~M$_{\odot}$ of stars have formed in the
M31 disk as a whole, (9$\pm$2)${\times}$10$^{10}$~M$_{\odot}$ of which
has likely survived to the present after accounting for evolutionary
effects.  This mass is about one fifth of the total dynamical mass of
M31.


\end{abstract}

\section{Introduction}

While large disk galaxies are responsible for the majority of star
formation in the universe \citep{freeman2002,bell2005}, the details of
their formation and evolution are difficult to constrain with observations.
They are complex, having a wide range of stellar populations, multiple
structural components, and complicated dust distributions. Although they are the most common galaxies in large surveys,
integrated light studies of disks from such surveys are generally
limited to their global properties, component structures, and
environments \citep[e.g.,][and many
  others]{tremonti2004,macarthur2004,blanton2009}.  While these measurements are
important for statistical comparisons, only by resolving the
individual stars within a galaxy can we reliably map the stellar
populations that are the products of its entire evolutionary history.
In particular, the resolved stars probe the star formation history
(SFH), which provides the age, mass, and metallicity distributions of
the galaxy's constituent stars.  These fundamental quantities allow
detailed comparisons with numerical simulations and stringent tests of
our ability to infer the masses, metallicities, and star formation
rates of distant galaxies from integrated light measurements.

M31 provides the best opportunity to measure the resolved stellar
populations of a large disk galaxy outside of the Milky Way.  At 770
kpc \citep{mcconnachie2005}, we can resolve M31's individual stars to
below the horizontal branch with HST.  M31 is therefore the only
large, metal-rich disk galaxy whose SFH can be accurately measured and
mapped.  However, M31's large angular size and complex dust distribution have
made it difficult to disentangle its stellar populations. There are
many studies in the literature
\citep[e.g.,][]{williams2002,bellazzini2003,williams2003,brown2003,brown2009,williams2008,bernard2012,bernard2015,bernard2015a,williams2015},
but most have been forced to limit their work to small areas or to the
outskirts of the disk where dust obscuration is minimal. This work has
provided significant insight, including the complexity of the halo
populations \citep{brown2003} and the first clues that the star
formation rate in M31 has decreased significantly over the past
$\sim$Gyr \citep{williams2002}.  Results from these deep drilling
fields, however, have been difficult to incorporate into a global
picture of the complex galaxy.

Over the past few decades, fitting resolved stellar photometry to
measure SFHs has become a powerful tool for constraining galaxy
formation and evolution.  For example, great strides have been made in
understanding the characteristics and evolution of dwarf galaxies
through such measurements \citep[e.g.,][and many
  others]{mateo1998,dohm-palmer2002,gallart2005,dolphin2005,young2007,cole2007,weisz2011,skillman2014,skillman2017}.
These SFHs allow us to test models with detailed observational
constraints.  Because of this, the technique of using resolved stars
to constrain SFHs has revolutionized the study of dwarf galaxies in
the Local Group and beyond.  Unfortunately, due to their greater
complexity, size, and distance to the nearest examples, it has proven difficult to apply this
powerful tool to large disks rather than dwarf galaxies.
Stellar population studies of our own Galaxy are compromised by dust
in the Galactic plane.

Recently, the PHAT survey has begun to shed light on the global SFH of
M31 by fitting resolved stellar photometry.  \citet{lewis2015}
provided spatially-resolved measurements of the recent SFH over the
past 500 Myr using the optical photometry from PHAT.  They found M31's
major star forming ring is a very long-lived structure in the M31
disk.  At older ages in selected low-dust regions,
\citet{williams2015} found that the star formation episode at
$\sim$2-4 Gyr ago, previously detected in the stellar populations of
the outer parts of M31 \citep{bernard2015}, is also seen all the way
in to just a few kpc from the galaxy center, suggesting the intense
event in the history of M31 causing this burst in star formation was
felt all the way to the inner disk.  Intriguingly, this episode
appears relatively weak in the southern disk \citep{bernard2015a},
although their SFHs show somewhat enhanced rates at these ages as
well.

Combining the complete areal coverage of the \citet{lewis2015} recent
SFH maps with the deep analysis of ancient SFHs from
\citet{williams2015} requires a new approach to handle M31's complex
distribution of dust.  Not only must the SFH be measured independently
in each subregion of the disk, the dust must be measured and modeled
properly at each position as well.

In this paper, we produce the first wide-area maps of the full SFH of
the M31 disk.  These maps cover all but the innermost area of the PHAT
footprint (which is too crowded for this work).  Section 2 details the
data analysis techniques used to model the dust and stellar
populations when measuring M31's SFH. Section 3 provides our resulting
SFHs electronic tables, FITS images, and movies to show the
time-resolved build-up of the current stellar mass distribution of the
PHAT footprint of M31.  A companion paper (Williams et al., in
preparation) will discuss the resulting map of stellar mass, along
with the resulting spatially-resolved mass-to-light ratio relations
for M31.  Throughout the paper, we assume a distance modulus for M31
of 24.47 \citep{mcconnachie2005}.  The nominal foreground extinction
is $A_{\rm VFG}=0.17$ \citep{schlafly2011}.

\section{Data Analysis}

The data for this study come from the Panchromatic Hubble Andromeda
Treasury \citep[PHAT;][]{dalcanton2012,williams2014}.  Briefly, PHAT is a
multiwavelength HST survey mapping 414 contiguous HST fields of the
northern M31 disk and bulge in 6 broad wavelength bands from the
near-ultraviolet to the near-infrared.  The broad wavelength coverage
required that the region be covered with 3 HST detectors.  The survey
obtained data in the F275W and F336W bands with the UVIS detectors of
the WFC3 camera, the F475W and F814W bands in the WFC detectors of the
ACS camera, and the F110W and F160W bands in the IR detectors of the
WFC3 camera.

For this work, we use the photometry and artificial star tests
provided in \citet{williams2014}, as well as the dust extinction maps
from \citet{dalcanton2015}, to fit for the SFH of $83''{\times}83''$
subregions of the survey area.  The methods for producing these
catalogs and maps are described in detail in \citet{williams2014} and
\citet{dalcanton2015}, respectively.  The data are homogeneous in
exposure time in each band over the survey, but they vary
significantly in photometric depth and precision due to crowding
effects.   In Figure~\ref{stellar_density_map}, we provide a map
  of the stellar density of bright stars with
  18.5$<$m$_{F160W}{<}$19.5, for which the PHAT survey is complete at
  all stellar densities.  This metric results in a smooth distribution
  of stellar densities over the PHAT survey area.  Figures~\ref{cmds},
  \ref{ircmds}, and \ref{rms} show example color-magnitude diagrams (CMDs),
  completeness limits, and photometric uncertainties for a range of
  stellar densities show in the map in
  Figure~\ref{stellar_density_map}.  A more detailed discussion of how
  these measurements were made is provided in \citet{williams2014}.
  We show examples here to demonstrate the wide range of photometric
  quality over the survey area.

As the photometric analysis is detailed in \citet{williams2014},
  we focus here on the additional analysis necessary to measure
  spatially-resolved ancient star formation histories from these
  data, in particular, fitting models to the PHAT photometry.  We use
the software package MATCH 2.6
\citep{dolphin2002,dolphin2012,dolphin2013} to find the combination of
model stellar ages and metallicities that best fit the CMD of each
sample of stars brighter than the 50\% completeness limit.  This
software package has been well tested and proven to provide reliable
measurements \citep[e.g.,][and many
  others]{dolphin2005,williams2008,mcquinn2010,williams2011,weisz2011,cole2014,williams2015}.

We use this package in a similar way to these studies, comparing
results when fitting with the Padova \citep{marigo2008,girardi2010}
models to those from fitting with the PARSEC \citep{bressan2012},
BaSTI \citep{basti,basti2,basti3}, and MIST \citep{mist} models.
In addition, testing fits with different models allows us to assess the
uncertainties (and ultimately the reliability) associated with these
late stages of evolution upon which most of our measurements
depend. In Figure~\ref{model_cmds}, we show comparison CMDs of these
models in F475W-F814W, shifted to the distance of M31, at 2 ages (2
and 10 Gyr) and 2 metallicities ([Fe/H]=$-$1.0 and 0.0).  For
  comparisons with other relevant ages, we plot isochrones of 1, 2, 4,
  and 10 Gyr in the figure as well.  There are significant
differences in feature color and shape, especially in the red clump,
sub-giant branch and RGB.   One well known difference between
  these model sets is that the BaSTI set has an extended horizontal
  branch (HB) and the other sets do not.  There is very little
  evidence for an extended HB in the PHAT photometry outside of the
  bulge and minor-axis fields \citep{rosenfield2012,williams2012}, and
  the feature is so weak compared to the red clump and red-giant
  branch that our results depend very little on it. 

The shape and position of the red clump is very sensitive to age and
metallicity \citep[e.g.,][]{rejkuba2005,williams2008}.  With the
age-metallicity relation fixed, the age distribution plays a strong
role in the ability to match this feature.  Slight differences in the
color and brightness of the red clump as a function of age in
different model sets therefore can significantly affect the age
distribution.  The Padova and PARSEC models come from the
same research group and have similar red clumps, whereas the others are
different.  Thus, our uncertainty at these ages in any one set of fits
is large, and there could be systematic differences between the
results of the innermost regions ($<$5 kpc), where the red clump stars cannot be
measured, and the rest of the PHAT footprint (see
\S~\ref{systematics}).  Adopting long time
  bins at old ages, as we have done, helps to account for these uncertainties and improves the consistency of the results across these different model sets. 

Our data are more complicated than those typically fit using MATCH.  We
have more than 2 observed bands, covering a wide range of depth and
crowding, and a very complex dust distribution.  Below we describe the
details of the CMD modeling, which takes advantage of many powerful
tools that MATCH provides, including sophisticated dust extinction
models, simultaneous fitting of multiple CMDs, and the ability to
force a chemical enrichment history.

\subsection{Artificial Star Tests}

The PHAT survey covers over 800 HST fields of M31, which presents a
daunting computational problem when trying to characterize the
photometric biases by analyzing the recovery of artificial stars.  As
a compromise,
\citet{williams2014} only performed artificial star tests at 6
representative positions along the major axis.  These tests were enough to
characterize the photometric quality as a function of stellar density,
but were not comprehensive as they did not sample every
location in the survey.

To fit the CMDs over the entire survey area with MATCH as described in
detail in \citet{williams2014}, it was necessary to insert and recover
artificial stars to measure the bias, uncertainty, and completeness of
the photometry as a function of color and magnitude at every location
in the survey.  As shown in \citet{williams2014}, the stellar density
varies dramatically from the dense inner bulge to the diffuse outer
disk in spite of the equal exposure times. Given this density spread,
spatial variation in photometric quality, biases and depth are
completely dominated by variations in the density of stars (i.e.,
crowding). We therefore use stellar density as an indicator of the
expected crowding in each region.  We therefore use stellar
  density (i.e., Figure~\ref{stellar_density_map}) as an indicator of
  the expected crowding in each region.

For each stellar density sampled in \citet{williams2014}, we measured
the bias, skew, variance, and covariance in the differences between
the input and output magnitudes of the artificial stars in
\citet{williams2014} as well as the probability that a star is
recovered (completeness).  We then characterize these quantities in
0.05 mag $\times$ 0.05 mag color-magnitude bins.  These parameters all
follow smooth relationships when plotted against the density of bright
stars with 18.5$<$m$_{F160W}{<}$19.5.  We therefore fit curves to
these quantities, and used the fitted values to derive bias, skew,
variance, covariance and completeness values for each color,
magnitude, and stellar density in the PHAT survey.

Using the fits to the artificial star results, we can simulate
the properties of artificial star tests in every survey region.  We
generated simulated artificial star catalogs for each of the
$83''{\times}83''$ regions in our study based on the density of stars,
where the simulated fake star catalogs used as inputs to MATCH are created to
have the bias, skew, variance, covariance, and completeness matched to
the stellar density of the region.

We tested the efficacy of our manufactured fake star catalogs by
running SFH fits with MATCH for both the actual fake star tests, as
well as with our simulated fake star tests in one of the 6 regions
from \citet{williams2015} where we had both.  The resulting SFHs are
shown in Figure~\ref{compare_sfhs}.  Here, the gray shaded areas
  mark the $\sim$1$\sigma$ random uncertainties from each fit (see
  Section~\ref{uncertainties} for uncertainty determination).  The
  results agree to well within these random uncertainties.

\subsection{Fitting}

Although MATCH has been extensively tested and used for SFH
measurements in the literature, there were aspects of the PHAT
photometry that required us to use relatively new capabilities of the
software.  In what follows, we describe our methods for applying these
capabilities for extinction, chemical enrichment, multi-band
photometry, and uncertainty determination.

\subsubsection{Extinction Modeling}

An extremely challenging aspect to measuring the ancient SFH of the
M31 disk is proper modeling of its complex dust extinction.  Across
the PHAT survey, the range of crowding effects and differential
extinction effects are noticeable.  Crowding causes fewer faint stars
to be detected at high stellar densities, and it causes the CMD
features to be broader \citep{williams2014}.  Differential extinction
also broadens the features, but only along the reddening vector.  As
can be seen clearly at lower stellar densities where the features are
well-defined, the red giant branch and red clump are split into a
reddened component behind the dust and a foreground component in front
of the dust.  This difference in the effects of crowding and reddening
allowed the IR data to be used for the extinction mapping in
\citet{dalcanton2015}, and it also allows us to fit the CMD
features to obtain reliable estimates of the age and metallicity
distribution of the stars, as described below.

\citet{dalcanton2015} made a major innovation in our ability to model
the differential extinction in the M31 disk.  They found that the
spread in reddening of red giant branch stars in the PHAT survey, if
taken over small spatial areas, had two components: an unreddened
component and a reddened component.  Further, the reddened component
was well-represented by a log-normal distribution with the following
parameterization: the median $A_{\rm V}$ ($\mu$), the dimensionless
width of the log-normal $A_{\rm V}$ distribution ({$\sigma$), and the
  fraction of stars being affected by the dust (f$_{\rm red}$).  They
  used these parameters to produce the most detailed and comprehensive
  map of the dust distribution of the M31 disk to date.

In order to use these maps for our fitting, we needed to degrade the
very fine resolution of the \citealp{dalcanton2015} maps.  Due to
internal dynamics and interactions, the old populations ($\gap$1 Gyr)
should be well-mixed on spatial scales much larger than the
\citet{dalcanton2015} pixel scale.  Furthermore, the number of stars
in these small regions is too low to provide good statistics for SFH
fitting.  Fortunately, we were able to develop a way to combine pixels
in the \citet{dalcanton2015} maps to allow larger samples of stars to
be fitted simultaneously using sums of log-normal distributions, which
we describe below.

We began with the \citet{dalcanton2015} extinction maps, which are on
a 3.3$''$ (13 pc) pixel$^{-1}$ scale. We took stellar photometry from
each of these $3.3''{\times}3.3''$ regions separately.  For each of
these samples, we know the correct differential extinction
distribution from the \citet{dalcanton2015} measurements.  Next, we
combined the $3.3''{\times}3.3''$ regions into $83''{\times}83''$
regions ($25{\times}25$ extinction map pixels) across the survey.
However, each of these areas contained 625 extinction distributions in
the extinction maps, and the wide range in $\mu$ and $\sigma$ values
present results in a combined reddening distribution that no longer
resembles our log-normal model (see
Figure~\ref{combined_log_normal}). Therefore, some subdivision of the
sample in each $83''{\times}83''$ region was required.

In order to effectively apply our differential extinction model, we
divided each set of 625 extinction distributions into quartiles of
A$_{\rm V}$.  Once this division was
done, the reddening distribution within each quartile resembled a
single log-normal (see Figure~\ref{combined_log_normal}).  We then fit
that log-normal to determine the parameters of the reddening
distribution for that quartile of that region.  Examples of fits to an
unranked 625 pixel region and to a single quartile of the same
sub-region are shown in Figure~\ref{combined_log_normal}.  Within each
$83''{\times}83''$ region, we assigned each star to an extinction
quartile based on its location in the \citet{dalcanton2015} maps.
This technique resulted in 826 regions, each with 4 photometric
subsamples (one for each quartile in extinction space), each of which
could be analyzed independently using the reddening distribution
appropriate for that quartile.  The individual quartiles typically
contained 10000 to 50000 stars.

Another major milestone in our modeling ability was including this
log-normal extinction distribution parameterization into MATCH.  As of
MATCH 2.5, in addition to applying foreground reddening to model CMDs,
one can apply the $diskav$ option to convolve model CMDs with an
additional log-normal extinction distribution identical to those
measured in the \citet{dalcanton2015} maps.  The log-normal is applied
to the fraction of RGB stars behind the M31 disk, which is also
measured in the \citet{dalcanton2015} maps.

The $diskav$ dust model in MATCH has 7 parameters.  The three
parameters that control the redistribution of stars in the RGB are the
parameters of the log-normal reddening distribution ($\mu$, $\sigma$,
and f$_{\rm red}$).  These three we take directly from the processing
of the \citet{dalcanton2015} dust maps described above.  The other 4
parameters include: 1) a parameter that allows for differential
foreground extinction that affects all stars, but since our regions
are very small on the sky, this parameter is zero for our study; 2)
three parameters that allow the young stars (less than the transition
age) to have a larger fraction of stars affected by dust than old
stars.  One of the three is the ratio of the scale height of the young
stars to the dust, another is the ratio of the scale height of the old
stars to the dust, and the last is the age at which this transition
occurs.  We ran many hundreds of tests covering a large grid of
  values for the scale height and transition age parameters, and we
  compared the resulting fit values for these tests to determine which
  parameters provided the best fits. We found that 1) within a
relatively broad range, the resulting ancient SFH was essentially
unaffected by our choice of these parameters (the differences in fit
quality were very small) and 2) the best fits were typically those
with a low transition age parameter ($\sim$0.1 Gyr) and high values
for the scale heights of the old stars to the dust (10-20).  We thus
fixed these parameters to 0.1 and 20 for all locations.

The only free reddening parameter in the fitting was the
  foreground extinction, which we will call $A_{\rm VFG}$ to
  distinguish it from $A_{\rm V}$ that we obtain from the
  \citet{dalcanton2015} maps, applied to the entire CMD.  This value
  was not constrained by \citet{dalcanton2015}, as all of their
  measurements were done relative to the ``unreddened'' component of
  the RGB at each location.  However, these ``unreddened'' stars are
  likely to still be reddened by foreground dust in the Galaxy and/or
  by dust in M31 that was not fully accounted for in the dust model.
  We compensate for this unknown component in the extinction in MATCH
  by applying a extinction value $A_{\rm VFG}$ to all stars in the
  CMD.  When fitting each sample, we limited the range to
  $0.1{<}A_{\rm VFG}{<}0.7$, which is reasonable given the nominal
  foreground value of $A_{\rm VFG}=0.17$ \citep{schlafly2011}.
  Including this single free extinction parameter in our fitting
  improved the fits tremendously, making the relative likelihoods (as
  determined from the Poisson $-ln(p)$ value)  hundreds of orders of
  magnitude larger than fits that did not contain this parameter or
  fixed it to the same value for the entire survey.

\subsubsection{Enrichment Modeling}

When fitting each $A_{\rm V}$ quartile of each $83''{\times}83''$
region of the PHAT photometry, we fixed the chemical evolution model
used by the fitting software. For data of the depth in PHAT, jointly
constraining the chemical enrichment history and star formation rate
history is particularly challenging and must be treated carefully.

Given that our data quality required us to limit the freedom of the
fitting routine, we decided on limiting the fitting to follow a sensible
chemical evolution model. To determine an appropriate model, we
started by fitting the data assuming several different potential
chemical enrichment histories. Most SFH fitting work to date has
focused on dwarf galaxies, and because such dwarfs are relatively
simple systems, forcing simple monotonically increasing
age-metallicity relations during the fitting of shallower data sets
has been very successful \citep[e.g.,][]{weisz2011}.  We tried these
relations (constant metallicity, constant enrichment rate, rapid early
enrichment); however, they had little flexibility, making them poor
approximations of the populations in the PHAT data which are likely
much more complex than those of most dwarf galaxies.  As a large
spiral, M31's enrichment history is likely to be at least as complex
as that of the Galaxy \citep[e.g.,][and references
  therein]{chiappini2001,freeman2002,minchev2014,bland-hawthorn2016}
making it more difficult to apply these previously-adopted simple
enrichment laws across the entire galaxy.

To provide more flexibility in the available age-metallicity
relations, we updated the MATCH package to allow the user to set
exponentially decreasing enrichment rates with various exponential
decay rates (\.{Z} $\propto e^{-{\rm P\rm X}}$ , where X is zero at
the oldest time and one at present).  There is evidence for the early
enrichment in both observations \citep[e.g.,][]{molla1997} and simple
models \citep[e.g.,][]{naab2006} of large disk galaxies.  Our
exponentially-decreasing function approximates such early enrichment.

Using this function, we found the parameters that provided the best
fits to our data.  We fitted with assumed P values ranging from 0 to
10 for all of our model sets.  We found that all model sets preferred
lower values for P at larger radii (i.e., earlier enrichment at
smaller radii).  We therefore found it reasonable to adopt a P-value
of 0.6 outside of 12 kpc (outer disk), a value of 2.4 from 5 kpc to 12
kpc (inner disk), and a P-value of 4.8 inside of 5 kpc (bulge).  Our
model enrichment histories are shown in
Figure~\ref{enrichment_histories}. For the BaSTI model set, the fits
were better with higher P-values across the entire galaxy, so we
adopted 2.4 (outer disk), 4.8 (inner disk), and 7.2 (bulge) for the
BaSTI fits.  All of the models also preferred a relatively large
Gaussian metallicity spread with a full-width half-maximum (FWHM) value set
by the user.  The FWHM that most often provided the best fit values
was 0.25.  We therefore fixed this value to 0.25 for all of the final fits.

Overall, our adopted enrichment model appears reasonable.  While we
found the best chemical enrichment model by comparing the quality of
the fits for many different possibilities, adopting more gradual
enrichment at larger radii turned out to be generally consistent with
the possibility that the enrichment history of M31 is more like that
of the Galaxy \citep[e.g.,][]{minchev2014} than like that of a typical
dwarf galaxy.  Furthermore, the regions with the earliest enrichment
(the inner regions) tend to have the highest star formation rates at
early times.  Thus, our adopted enrichment history is generally
self-consistent even though it was not forced to be.  In addition, our
adopted enrichment history naturally reproduces the known metallicity
gradient in M31 (see Section~\ref{metal_enrichment}), even though this
was also not forced.

 In the Appendix, we provide the results of fitting the data
  without applying any assumptions about the enrichment history. We
note that the overall metallicity distribution is similar to our fixed
enrichment model result. In addition, the adopted enrichment model
turned out to be generally similar to the overall trends seen in the
free metallicity case and without allowing the relatively chaotic
short timescale changes seen in those results (see Appendix), which
provides additional evidence that our adopted model provides a
reasonable prior for constraining the fitting.

The general pattern of the SFH in the Appendix is the same as the
results from our adopted chemical evolution model, with the highest
rate prior to 8 Gyr ago and at 2-4 Gyr ago; however, the epoch from
4-8 Gyr ago has more star formation.  Such a difference is expected
because it is more likely to find stars of every age when every
metallicity is allowed at every age.  With this much freedom, the fit
is often improved by adding small outlier populations to fit any
artifacts or contamination present in the CMD. In essence, by adding
populations at unlikely metallicities at those ages, the fit can be
improved, but those additions are more likely to be compensating for
deficiencies in the models and photometry than they are to be
revealing populations that are truly present in M31. These small
outlier populations would likely be scattered rather randomly in age,
pushing the SFH to look more constant.  Much deeper resolved
photometry would be necessary to reliably measure such populations
with no forced enrichment model.

Leaving the software free to use any and all metallicities at every
age to fit the data is prone to introduce erroneous populations to
compensate for deficiencies in the data such as foreground stars,
imperfect reddening models, imperfect fake star statistics, and/or
variable stars.  These erroneous populations then corrupt the solution
where they intersect with real CMD features.  Even in the presence of
very deep photometry, solutions with constrained enrichment are
generally the most reliable because they force the fitting routine to
ignore data points in the CMD that are unlikely to be associated with
a true population of stars.

Furthermore, PHAT photometry contains little information that could be
used to resolve age-reddening-metallicity degeneracies at ages over a
few Gyr ago.  When resolved stellar photometry does not include the
ancient main sequence, breaking degeneracies between age and
metallicity depends on the details of features in the CMD where
current stellar evolution models tend to be less well-constrained,
particularly for older populations.

While the fits in the Appendix may provide an interesting comparison,
they are less reliable than those with a fixed enrichment history
presented in the main text and should not be used for modeling choices
or conclusions. 

\subsubsection{Simultaneous Multi-band Fitting}

The PHAT survey measured stars in six bands: F275W, F336W, F475W,
F814W, F110W, and F160W.  The near ultraviolet images (F275W and
F336W) did not detect the red giant branch, and thus were only
sensitive to the youngest stars in the survey.  For the purposes of
this study, we are interested in constraining the age distributions of
the stars older than a few hundred Myr, and thus limited our fitting
to the optical and near infrared bands.

Setting MATCH to fit the optical and IR data simultaneously in
  principle helps to break degeneracies in the fits.  When fitting
only the optical CMD, the position of the red clump with respect to
the red giant branch provides strong constraints on age and
metallicity, but with some degeneracy.  However, when the IR data are
included, the model RGB must also have the correct IR color, slope,
and TRGB, adding significantly to the constraints. We therefore use
joint optical and NIR CMD fitting for our final SFHs.  Examples of the
SFHs of a region as measured with optical data only, IR data only, and
simultaneously fitting both are shown in
Figure~\ref{compare_multiple_band_sfhs}.  The results are
  consistent within uncertainties, suggesting only minor improvement
  is gained by fitting both, but we found the fits across quartiles to
  be more consistent when both optical and NIR CMDs were fit
  simultaneously.  Therefore, since including both CMDs provides
  additional information to the fitting, and the results appear to be
  consistent with both of the individual CMD fits, we include both
  CMDs in our analysis.

Once we had determined the reddening model, enrichment model, optical
(F475W-F814W) and NIR (F110W-F160W) sample from each $A_{\rm V}$
quartile of each $83''{\times}83''$ region, we fit for the SFHs with
the MATCH utility $calcsfh$ four times (once for each set of stellar
evolution models) using 0.1 dex resolution in log age and 0.1 dex
resolution log metallicity.  The CMDs of each sample were binned to
0.05 mag in color and 0.1 mag in luminosity. During the fitting,
  the total fit value is optimized, which is a combination of the
  residuals across all color-magnitude bins in both CMDs.  This means
  that the optical CMD carries more overall statistical weight because
  it has more color-magnitude bins due to its greater depth and wider
  color baseline.

The initial results of the fitting were
13216 SFH measurements.  We then processed these initial results to
determine the uncertainties in the measurements, as described below.

\subsection{Random Uncertainty Estimates}\label{uncertainties}

\citet{dolphin2012} and \citet{dolphin2013} describe the most reliable
methods for determining both systematic and random uncertainties when
fitting stellar evolution models to CMDs.  Here we describe our method
for determining the random uncertainties, which is done through tests
on the data themselves.  The systematic uncertainties require analysis
of the results from multiple model sets, and is described in
Section~\ref{systematics}.

Random uncertainties are generally due to a combination of the number
of stars in the CMD and how well those stars define the CMD features.
When looking for relative changes in populations across a dataset, the
most relevant uncertainties are the random uncertainties because
differences between populations change the distribution of stars
within the CMD.  These differing distributions will sensitively change
the resulting fit when applying the same models to the different
regions.  Thus, when fitting neighboring regions with the same
  models, the random uncertainties provide a robust test for
  statistically significant differences between the resident
  populations.  

\citet{dolphin2013} describes a technique that
  provides reliable random uncertainty estimates.  This technique
  applies a Hybrid Monte Carlo algorithm \citep{duane1987}.  Briefly,
  this algorithm propagates points within the probability space using
  Hamiltonian dynamics, allowing large motions while ensuring very
  high acceptance rates.  Such an approach is ideal for the large
  dimensional space used in SFH fitting.  We implemented this
  technique with the $hybridMC$ tool within the MATCH package.  We
  applied this tool to all of the $calcsfh$ output, including
  uncertainty data.  We then combined the SFHs and uncertainties for
the different quartiles into a single SFH for each region using the
MATCH tasks $outcombine$ (to combine the multiple extinction
quartiles), $zcombine$ and $zcmerge$ (to process the output into total
rates and uncertainties in each age bin).

Table~\ref{z4popbox_table}, which gives only the best-fit rates for
every age and metallicity in our model, no uncertainties are included.
In Table~\ref{sfh}, we give the random uncertainties, as these are
most useful for comparing different regions in the disk.  The
  large covariance between adjacent ages often causes highly
  asymmetric uncertainties.  Statistically-significant differences in
the CMDs between one region and another will cause differences in SFH
larger than these uncertainties.  However, the uncertainty on the
absolute SFH of any region is larger, and should take into account the
differences between model sets.

Once we had completed all of the steps above, we organized and
combined the SFH measurements to study the evolution and stellar mass
of the M31 disk and to estimate our systematic uncertainties, as
described below.

\section{Results and Discussion}\label{results}

In Table~\ref{z4popbox_table}, we provide the star formation rate in each time
bin at each metallicity for each model suite, in each
83$''{\times}83''$ region. In Table~\ref{sfh}, we provide the star
formation rate in each time bin, along with the random
uncertainties, for each model suite, in each of our 83$''{\times}83''$
regions. Table~\ref{total_rates} provides the total SFRs for each age
bin for our entire survey area, and Table~\ref{total_mass} provides
the cumulative stellar mass of our entire survey area at the end of
each age bin.  Uncertainties are the linear sum of the uncertainties
in each subregion.  In Figure~\ref{compare_model_sfhs}, we show the
cumulative SFHs for all regions, color-coded by stellar density, for
all model sets.  In Figure~\ref{total_sfh}, we show the sum of the
SFHs covering the entire PHAT footprint resulting from fits using each
of the 4 model sets.  In Figure~\ref{radial_sfhs}, we show the sums of
the SFHs in 4 radial bins for the Padova model fits.  Due to
systematic uncertainties described in Section~\ref{systematics}, we
present our results with 8-14 Gyr as a single epoch. Furthermore, we
present all ages $<$300 Myr as a single epoch; see \citet{lewis2015}
for a detailed analysis of the recent SFH of M31 from the PHAT
photometry.  The SFH of M31 at these young ages changes significantly
on smaller spatial scales than we have used here.  In addition, our
dust model was optimized for fitting the old stellar population, as it
was based on the color distribution of the RGB stars in the IR data.

In Figure~\ref{z4hist}, we show the total metallicity distribution for
the survey from all model sets, as well as for three radial bins, for
the Padova model set.  These metallicity distributions are provided in
Tables~\ref{total_metallicity} and \ref{radial_metallicity}, and were
calculated from the best-fit full populations in
Table~\ref{z4popbox_table}.  All of our results are described in more
detail below.

\subsection{M31 Global Star Formation History}

The total SFH of the PHAT survey region (Figure~\ref{total_sfh}) shows
three strong features.  These features are associated with ages
spanning 1 to 14 Gyr.  These ages are mostly constrained by the
detailed ratios and colors of stars on in the red clump, AGB bump, and
red giant branch (see Figure~\ref{isochrones} and \citealp[e.g.,][]{williams2008,williams2015}). Because our fitting includes complex dust effects,
it is not simple to point to a single feature in the CMD that is
responsible for these ages; however, the position of an unreddened red
clump moves in relation to the red giant branch for different ages and
metallicities.  Thus, these SFH features best reproduced the observed
red clump, AGB bump, red giant branch morphology in all model fits,
demonstrating the power of full CMD fitting.

The first distinct feature of the SFH is that most of the stellar mass was
formed prior to 8 Gyr ago.  We cannot resolve any large bursts that
may have occurred from 8 to 14 Gyr ago, but our results show that the
mean rate over that long epoch was as high or higher than any rate we
measure more recently, even over much shorter epochs.  The second
feature is that there was relatively little star formation going on in
M31 from $\sim$4-8 Gyr ago.  Perhaps the gas supply was depleted
during this epoch; however, the third feature shows that $\sim$2-4 Gyr
ago there was a high star formation rate again.  This feature suggests
that something likely occurred that either replenished the gas supply,
triggered star formation, or both.  Whatever the cause, since this
episode, the rate has dropped back to relative quiescence at the
present day.  This evolutionary history is broadly consistent with our
results when the metallicity is left as a free parameter (see
Appendix).

We note that while \citet{bernard2015} found a significant global
burst of star formation at $\sim$2 Gyr ago, \citet{bernard2015a} did
not detect a similar burst in each and every one of their HST fields
in the southern disk, and suggested that perhaps the episode was
confined to the outer disk.  However, our measurements suggest that
the episode may only be less prominent in the more heavily populated
inner disk.  With large samples, a burst appears consistently, but
because the burst is less significant in the high-density inner disk,
it may not be easily detected in every field measured.  For example, 2
of the 3 fields in \citet{bernard2015a} showed a clear burst above the
mean star formation rate between 2 and 4 Gyr ago.  The third field
does not show such a statistically significant enhancement, but does
show lower rates immediately preceding and following this epoch.
Resolving the lower main-sequence of the inner disk over a large area
of M31 will be required to provide conclusive proof of this result.

\subsection{Growth of Mass and Metallicity}\label{metal_enrichment}

In Figure~\ref{total_sfh}, we show the total stellar mass
formed in the PHAT footprint over the past 14 Gyr.  We obtain a precise measurement of the total stellar mass formed of 5${\times}10^{10}$~M$_{\odot}$; however, this is only an
upper-limit on the stellar mass {\it present}, as most of the
populations are quite old so that most of the stars of more than a few
solar masses will have put a large fraction of their mass back into
the interstellar gas.  If we neglect stellar remnants and assume
  all stars more massive than 1 solar mass have returned their mass
  to the ISM, then the correction factor for a Kroupa IMF is 0.6, which would make the
present day stellar mass inside the PHAT footprint closer to
3${\times}10^{10}$~M$_{\odot}$.  

Even though we fixed the chemical enrichment model, our results still
allow some interpretation about the metallicity of M31. In
Figure~\ref{z4hist}, we show the total metallicity distribution for
the survey from all model fits.  We note that the high SFRs occur mainly in the earliest epoch when the enrichment rate was
the highest.  Thus, our chemical enrichment model appears reasonable
in that the high enrichment rate occurs during the high SFR epoch.
Furthermore, the vast majority of the
stellar mass of M31 has [Fe/H]${\geq}{-}$0.5, with the largest number
of stars at or above solar metallicity.  Thus, consistent with previous measurements, we find the M31 disk to be dominated by metal-rich stars.  Furthermore, the SFH suggests that most of these metal-rich stars are older than 8~Gyr.

\subsection{Radial Distribution of Populations }

We would like to probe the growth of the stellar disk.  However, while
our data provide much information about the radial distributions of
populations of various ages at the present day, it provides no
information about where those populations actually formed.  In a
massive spiral such as M31, the stars are likely to have undergone
significant radial migration on Gyr timescales
\citep[e.g.][]{roskar2008}.  Furthermore, our results suggest a major
event in the evolution of M31 2-4 Gyr ago, which could have
significantly redistributed the stars throughout the disk.  Thus,
while we investigate the detailed radial distribution of the stellar
mass associated with various ages below, it is important to note that
the distributions are likely strongly affected by mixing and migration
since those populations formed. 

We first examine the radial distribution of the metallicity of the
total population.  In Figure~\ref{z4hist}, we show the metallicity
distribution for three radial bins from the Padova fits.  These
histograms were calculated by weighting each metallicity by the amount
of stellar mass produced at that metallicity in every region measured.
They are then corrected for the fraction of the full elliptical region
covered by the survey data (which is $\sim$0.3 for all 3 radial bins),
so that the absolute values reflect the total M31 disk assuming the
rest of the disk out to 20 kpc is similar to the PHAT footprint.  The
radial metallicity distribution for all model fits peaks at or above
solar, except in the outer disk, where it peaks at slightly lower
metallicity.  The peak and median metallicities move 0.2-0.3 dex in
$\sim$15 kpc, for a rough metallicity gradient of ${\sim}{-}$0.02 dex
kpc$^{-1}$, which is consistent with the gradient found directly from
RGB fitting \citep{gregersen2015}.

Next, we examine the radial distribution of populations as a function
of age.  To show the distribution of stellar ages with position, we
produced maps of the SFR radial profiles of the cumulative stellar mass
surface density with age in Figure~\ref{density_vs_radius}.  There appears to be very
little change in the age distribution with radius. If we assume
  the old stars formed in arm structures and/or in a more
  centrally-concentrated radial profile than the present-day profile,
  they now appear to be well-mixed as they have no structure beyond the smooth exponential profile of the present-day disk.  If we fit all of
these profiles with exponentials, we can estimate the size and mass of
the disk as a function of lookback time.  These are shown in
Figure~\ref{scale_norm}.  We note that Figure~\ref{sfr_maps} suggests
that the ring structure is present in all populations younger than 1
Gyr, making it unclear if the appearance of the ring is related
  to the last major star forming event.

Looking at Figure~\ref{scale_norm}, as we would expect from the total
SFH, most of the stellar mass in the disk was formed by 8 Gyr ago.  Then the disk went through a
relatively quiescent period, followed by a possible star forming event
$\sim$4~Gyr ago, although that even is limited to the Padova and
PARSEC model fits.  Finally, the disk made another significant amount
of stellar mass $\sim$2 Gyr ago, common to all model fits. The scale
length appears to change little with lookback time (stellar age), and all model fits appear
consistent with the currently-measured scale-length (gray band;
\citealp{courteau2011}).  

This result is consistent with the little
change in age with radius and shows more quantitatively that the old populations appear well-mixed, which is consistent with the measurements along the southwest major axis \citep{bernard2015a} which used much deeper data over a much smaller region in the other half of the disk.  The consistency of our measurements with theirs is encouraging that we considering our data are much shallower.  While a direct comparison would be an ideal test of consistency, none of their fields overlap with the PHAT footprint.

Integrating the present-day values for the normalization and
scale-length consistent with all of the model fits (800-900 and
5.0-5.5, respectively) gives the total stellar mass formed over the
lifetime of the M31 disk, which is
1.5($\pm$0.2)$\times$10$^{11}$~M$_{\odot}$.  As with the mass formed
inside of the PHAT footprint, most of it is old and the massive
component has largely gone back into the interstellar gas.  As in that
case, if 60\% of the stellar mass formed is still in the form of
stars, which would make the present day stellar mass of the disk
closer to 9${\times}10^{10}$~M$_{\odot}$. Thus, our survey contains
about one third of the total stellar mass in the disk.  We note that
this result is consistent with the recent results from integrated
light spectral energy distribution fitting, such as that of
\citet{sick2015}, who inferred a stellar mass of
1.0$\pm$0.2$\times$10$^{11}$~M$_{\odot}$. This is consistent with
stellar mass being responsible for $\sim$20\% of the dynamical mass
\citep[4.7 $\times$10$^{11}$~M$_{\odot}$;][]{chemin2009} out to 38
kpc.

\subsection{Systematic Uncertainties}\label{systematics}

While random uncertainties are generally due to the depth and size of
the photometric sample, and were able to be reliably estimated using
techniques from \citet{dolphin2013}, systematic uncertainties are
generally due to uncertainties in the assumed underlying physics in
the stellar evolution models and the assumed extinction model.  When
constraining the absolute age of a stellar population, the systematic
uncertainties are most relevant, as they quantify the possible
differences in ages and metallicities when different model sets are
fitted to the data.  We therefore compare the results from fitting
four different model sets to take these systematic uncertainties into
account when making any interpretations based on our SFH measurements.
Furthermore, we assessed the reliability of our extinction model by
examining the best fit foreground reddening and comparing results of
different extinction quartiles.

\subsubsection{Stellar Evolution Systematics}

Even though the details of the age distribution within each spatial
region varied depending on the model set used during the fitting, the
total amount of stellar mass produced was consistent across model sets
to a precision of $\sim$20\%.  In Figure~\ref{mass_maps}, we show the
total stellar mass as a function of position across the PHAT survey
from our measurements.  While the BaSTI masses (right panel) are
systematically higher by $\sim$20\% at 8 Gyr, the structure is
consistent across all sets.  In Figure~\ref{total_sfh}, we show the
total stellar mass formed in the PHAT footprint over the past 14 Gyr,
as calculated from the fits to all 4 model sets.  The masses are
consistent across all model sets to within 20\% at all ages with the
exception of the most crowded innermost portion of the survey.  In
fact, the agreement is within 10\% for the total mass produced over
the entire 14 Gyr period, showing that our measurements of the total
stellar mass formed as a function of position in the PHAT footprint
are robust.  Furthermore, the similarity remains whether or not we
force a specific enrichment history (see Appendix).

There is also general agreement on the overall metallicity
distribution of the stellar disk (see Figure~\ref{z4hist}).  We note
that all of the model fits result in very similar metallicity
distributions, dominated by metal-rich stars.  As with the overall
disk mass, this consistency also remains whether or not we force a
specific enrichment history (see Appendix).

With our adopted time binning, all of the models give results
that are consistent with one another within the uncertainties in the
cumulative SFH for ages younger than 8 Gyr.  Thus, the variance
observed in the differential SFH is due to covariance between age bins
that is not represented with the error bars.  Furthermore, if we break the total into 4
radial bins, as shown in Figure~\ref{radial_sfhs}, we see how depth
affects the systematic uncertainties because our photometry was deeper
at larger radii due to the effects of crowding.  The largest
discrepancies actually appear at intermediate ages ($\sim$4 Gyr),
which is expected because the red clump drives much of
the systematic uncertainties in our fitting.

The Padova and PARSEC model fits are
improved by the inclusion of a 4 Gyr old population at radii that
include the red clump, and the other 2 model sets do not favor such a
population.  These intermediate age systematics are not apparent in the innermost radii,
likely because the red clump is below the crowding limit in this
region.   If the PARSEC and Padova models are correct in their fits at
the outer radii, this intermediate-age population may be present in
the inner disk, but may be undetected without photometry that included
the red clump.  In short, these differences appear consistent with a
factor of 2 age uncertainty due to model uncertainties.  Thus, it
would be reasonable to add a systematic uncertainty of a factor of 2
to all of the ages in Table~\ref{sfh} if attempting to constrain
absolute ages using our SFHs.

\subsubsection{Extinction Systematics}

Another potential source of systematic uncertainty is deficiencies in
our extinction model, which we assessed both by looking at the
best-fitting foreground extinction and by comparing results across
different $A_{\rm V}$ quartiles.  

One way to assess systematics in the reddening model is by checking
other free parameters in the fitting.  One such parameter is the
foreground reddening, which adjusts the colors of all of the models.
This reddening is due to the Milky Way, and, in principle, should be
uncorrelated with M31 structure (though it still may have structure on
scales smaller than the PHAT survey footprint).  We did not see any
trends with stellar density.  The quartiles of the same regions could
often have quite different internal extinction properties, but their
foreground extinctions agreed very well.  Comparisons of the
foreground extinctions for different quartiles of all regions (e.g.,
all first quartile extinctions vs all third quartile extinctions)
showed a median offset of 0 and standard deviation of 0.04 to 0.06
depending on the quartiles being compared.  These results suggest that
the foreground $A_{\rm VFG}$ is not being driven by photometric depth effects
or our technique of splitting the data in quartiles of differential
extinction.

To further investigate the foreground extinction in our fitting, in
Figure~\ref{av_maps}, we plot maps of the fitted $A_{\rm VFG}$ values.
These maps clearly show structure associated with M31, suggesting that
the foreground extinction, which should only be associated with the
Milky Way, is compensating for other deficiencies in the model.  To check the impact of $A_{\rm VFG}$ on our fitting, we also fit the
  data with a fixed foreground extinction across the survey.  We found
  the fits to be substantially worse, with relative probabilities
  hundreds of orders of magnitude lower than the fits with the
  foreground reddening free to vary by about half of a V band
  magnitude.  The fact that the best-fitting foreground reddening
  turned out to be strongly correlated with M31 features clearly
  indicates it allowed the fitting to compensate for deficiencies in
  the fixed log-normal reddening parameters.  Therefore, requiring it
  to be constant across the survey would introduce spatially-dependent
  systematic errors into the solution.  We apparently are largely mitigating against this systematic with this additional and highly-restricted free
  parameter.

While the IR-measured differential extinction parameters have greatly
improved the fitting, they do not provide precision measurements for
extinctions of $A_V{\lap}0.5$ \citep{dalcanton2015}.  Furthermore,
regions with and without young upper main sequence stars, whose
intrinsic colors are well-defined, will have different features with
which to constrain the foreground extinction.  Therefore it is not
surprising to see structures at this level in the best-fit foreground
values.  In addition, in the BaSTI fits, the best-fit foreground
reddening values appear to be systematically higher than the nominal
value \citep[$A_{\rm VFG}=0.17$;][]{schlafly2011}, suggesting that the
BaSTI models may be systematically bluer than the others, which is
consistent with the systematically older fitted ages.

Finally, we were able to assess possible systematics due to the
extinction model by comparing the SFH results of different quartiles
for the same region.  We found that these results typically agreed
within the uncertainties, as shown in Figure~\ref{compare_quartiles},
despite the considerable differences in their amount of dust present.  These results show that the $diskav$
dust model reliably compensates for the effects of differential
extinction within M31 itself, at least when informed by the extinction
maps of \citet{dalcanton2015}. Thus, the systematic uncertainties due to uncertainties in the extinction model appear to be small compared to those due to uncertainties in the stellar evolution models.

\section{Conclusions}

We have fit four sets of stellar evolution models to the full optical
and NIR photometry from the PHAT survey, outside of the inner bulge,
where crowding makes the resolved stellar photometry highly biased and
difficult to model.  We have applied independent constraints on the
dust distribution to our model fits, and we have adopted an
exponentially-decreasing chemical enrichment rate during our fitting.

A few features in the SFHs persist independent of the models used to
fit the data or the adopted enrichment history, strongly suggesting
they are reliable.  All of our measurements support an evolutionary
story where most of M31 was built prior to 8 Gyr ago, followed by an
extended, relatively quiescent period, giving way to a widespread
episode of star formation at $\sim$2 Gyr ago. Finally, since this burst, the rate of
star formation has decreased significantly and the global structure of
the disk has remained unchanged.  The star forming ring is then visible at all
ages $\lap$1 Gyr.  This overall picture is consistent with what has
been seen in previous studies \citep{bernard2015,lewis2015}, but the
consistency across the face of the disk suggests significant radial
mixing of the populations older than $\sim$1 Gyr.

Furthermore, our most robust constraints are on the total stellar mass
formed in the PHAT footprint (outside of the inner bulge), which we
measure to be 5.0-5.5$\times$10$^{10}$~M$_{\odot}$, consistent with
all of our fits. We also find a total stellar mass formed in the disk,
based on an exponential fit to the present day stellar density
profile, of 1.5$\pm$0.2$\times$10$^{11}$~M$_{\odot}$, consistent with
all of our fits.  After accounting for the evolution of the massive
stars, the total stellar mass still present is
$\sim$3$\times$10$^{10}$~M$_{\odot}$ and
$\sim$9$\times$10$^{10}$~M$_{\odot}$ in the PHAT footprint and the
entire disk, respectively. This mass accounts for 20\% of the
dynamical mass of M31 inside of 38 kpc.

By fitting the data with a variety of model sets and enrichment
assumptions, we have shown that we can only constrain the ages of the
old population to an absolute precision of a factor of $\sim$2. The
cumulative stellar masses derived from our fits are consistent across
model sets to within 20\% at early times and 10\% at more recent times, and the total masses are
consistent within 20\% for each region.  Thus, we have produced spatially-resolved
maps of the total stellar mass in M31 with 83$''$ resolution for a
complete range of stellar ages.


\clearpage
\begin{turnpage}
\tabletypesize{\tiny}

\begin{deluxetable}{cccccccccccccccccccc}\tablewidth{9.0in}
\tablecaption{M31 Star Formation Rates by Metallicity (M$_{\odot}$~yr$^{-1}$) for each survey region for our fixed enrichment model fits; Columns are age range in Gyr; full table available in electronic format only}
\tablehead{
\colhead{Model Set} &
\colhead{RA} &
\colhead{Dec} &
\colhead{[Fe/H]} &
\colhead{0.0-0.3} & 
\colhead{0.3-0.4} & 
\colhead{0.4-0.5} & 
\colhead{0.5-0.6} & 
\colhead{0.6-0.8} & 
\colhead{0.8-1.0} & 
\colhead{1.0-1.3} & 
\colhead{1.3-1.6} & 
\colhead{1.6-2.0} & 
\colhead{2.0-2.5} & 
\colhead{2.5-3.2} & 
\colhead{3.2-4.0} & 
\colhead{4.0-5.0} & 
\colhead{5.0-6.3} & 
\colhead{6.3-7.9} & 
\colhead{7.9-14.1}
}
\startdata
Padova & 0:42:45.03 & 41:22:17.0 & -2.25 & 6.0e-24 & 7.1e-23 & 8.5e-23 & 3.2e-23 & 0.0e+00 & 0.0e+00 & 0.0e+00 & 1.6e-23 & 5.0e-22 & 1.2e-21 & 3.0e-22 & 7.2e-24 & 0.0e+00 & 0.0e+00 & 0.0e+00 & 1.0e-04\\
Padova & 0:42:45.03 & 41:22:17.0 & -2.15 & 2.2e-22 & 2.6e-21 & 3.2e-21 & 1.2e-21 & 0.0e+00 & 0.0e+00 & 0.0e+00 & 6.0e-22 & 1.9e-20 & 4.5e-20 & 1.1e-20 & 2.6e-22 & 0.0e+00 & 0.0e+00 & 0.0e+00 & 5.8e-05\\
Padova & 0:42:45.03 & 41:22:17.0 & -2.05 & 7.3e-21 & 8.6e-20 & 1.0e-19 & 3.9e-20 & 0.0e+00 & 0.0e+00 & 0.0e+00 & 2.0e-20 & 6.0e-19 & 1.5e-18 & 3.5e-19 & 8.4e-21 & 0.0e+00 & 0.0e+00 & 0.0e+00 & 7.8e-05\\
Padova & 0:42:45.03 & 41:22:17.0 & -1.95 & 2.0e-19 & 2.4e-18 & 2.8e-18 & 1.1e-18 & 0.0e+00 & 0.0e+00 & 0.0e+00 & 5.4e-19 & 1.7e-17 & 4.0e-17 & 9.7e-18 & 2.3e-19 & 0.0e+00 & 0.0e+00 & 0.0e+00 & 1.0e-04\\
Padova & 0:42:45.03 & 41:22:17.0 & -1.85 & 4.8e-18 & 5.6e-17 & 6.7e-17 & 2.6e-17 & 0.0e+00 & 0.0e+00 & 0.0e+00 & 1.3e-17 & 3.9e-16 & 9.4e-16 & 2.3e-16 & 5.4e-18 & 0.0e+00 & 0.0e+00 & 0.0e+00 & 1.3e-04\\
Padova & 0:42:45.03 & 41:22:17.0 & -1.75 & 9.6e-17 & 1.1e-15 & 1.4e-15 & 5.1e-16 & 0.0e+00 & 0.0e+00 & 0.0e+00 & 2.6e-16 & 7.9e-15 & 1.9e-14 & 4.6e-15 & 1.1e-16 & 0.0e+00 & 0.0e+00 & 0.0e+00 & 1.7e-04\\
Padova & 0:42:45.03 & 41:22:17.0 & -1.65 & 1.7e-15 & 2.0e-14 & 2.3e-14 & 8.9e-15 & 0.0e+00 & 0.0e+00 & 0.0e+00 & 4.4e-15 & 1.3e-13 & 3.2e-13 & 7.8e-14 & 1.8e-15 & 0.0e+00 & 0.0e+00 & 0.0e+00 & 2.1e-04\\
Padova & 0:42:45.03 & 41:22:17.0 & -1.55 & 2.4e-14 & 2.9e-13 & 3.4e-13 & 1.3e-13 & 0.0e+00 & 0.0e+00 & 0.0e+00 & 6.4e-14 & 2.0e-12 & 4.7e-12 & 1.1e-12 & 2.6e-14 & 0.0e+00 & 0.0e+00 & 0.0e+00 & 2.7e-04\\
Padova & 0:42:45.03 & 41:22:17.0 & -1.45 & 3.0e-13 & 3.6e-12 & 4.3e-12 & 1.6e-12 & 0.0e+00 & 0.0e+00 & 0.0e+00 & 8.0e-13 & 2.5e-11 & 5.9e-11 & 1.4e-11 & 3.3e-13 & 0.0e+00 & 0.0e+00 & 0.0e+00 & 3.4e-04\\
Padova & 0:42:45.03 & 41:22:17.0 & -1.35 & 3.3e-12 & 3.8e-11 & 4.6e-11 & 1.7e-11 & 0.0e+00 & 0.0e+00 & 0.0e+00 & 8.6e-12 & 2.6e-10 & 6.3e-10 & 1.5e-10 & 3.4e-12 & 0.0e+00 & 0.0e+00 & 0.0e+00 & 4.4e-04\\
Padova & 0:42:45.03 & 41:22:17.0 & -1.25 & 3.0e-11 & 3.5e-10 & 4.2e-10 & 1.6e-10 & 0.0e+00 & 0.0e+00 & 0.0e+00 & 7.8e-11 & 2.4e-09 & 5.7e-09 & 1.4e-09 & 3.1e-11 & 0.0e+00 & 0.0e+00 & 0.0e+00 & 5.6e-04\\
Padova & 0:42:45.03 & 41:22:17.0 & -1.15 & 2.3e-10 & 2.7e-09 & 3.3e-09 & 1.2e-09 & 0.0e+00 & 0.0e+00 & 0.0e+00 & 6.1e-10 & 1.9e-08 & 4.4e-08 & 1.0e-08 & 2.4e-10 & 0.0e+00 & 0.0e+00 & 0.0e+00 & 7.2e-04\\
Padova & 0:42:45.03 & 41:22:17.0 & -1.05 & 1.5e-09 & 1.8e-08 & 2.2e-08 & 8.2e-09 & 0.0e+00 & 0.0e+00 & 0.0e+00 & 4.0e-09 & 1.2e-07 & 2.9e-07 & 6.9e-08 & 1.6e-09 & 0.0e+00 & 0.0e+00 & 0.0e+00 & 9.3e-04\\
Padova & 0:42:45.03 & 41:22:17.0 & -0.95 & 8.8e-09 & 1.0e-07 & 1.2e-07 & 4.7e-08 & 0.0e+00 & 0.0e+00 & 0.0e+00 & 2.3e-08 & 6.9e-07 & 1.6e-06 & 3.9e-07 & 8.8e-09 & 0.0e+00 & 0.0e+00 & 0.0e+00 & 1.2e-03\\
Padova & 0:42:45.03 & 41:22:17.0 & -0.85 & 4.3e-08 & 5.0e-07 & 6.0e-07 & 2.3e-07 & 0.0e+00 & 0.0e+00 & 0.0e+00 & 1.1e-07 & 3.4e-06 & 7.9e-06 & 1.9e-06 & 4.2e-08 & 0.0e+00 & 0.0e+00 & 0.0e+00 & 1.5e-03\\
Padova & 0:42:45.03 & 41:22:17.0 & -0.75 & 1.8e-07 & 2.1e-06 & 2.5e-06 & 9.4e-07 & 0.0e+00 & 0.0e+00 & 0.0e+00 & 4.5e-07 & 1.4e-05 & 3.3e-05 & 7.7e-06 & 1.7e-07 & 0.0e+00 & 0.0e+00 & 0.0e+00 & 1.9e-03\\
Padova & 0:42:45.03 & 41:22:17.0 & -0.65 & 6.3e-07 & 7.3e-06 & 8.8e-06 & 3.3e-06 & 0.0e+00 & 0.0e+00 & 0.0e+00 & 1.6e-06 & 4.9e-05 & 1.1e-04 & 2.7e-05 & 6.0e-07 & 0.0e+00 & 0.0e+00 & 0.0e+00 & 2.4e-03\\
Padova & 0:42:45.03 & 41:22:17.0 & -0.55 & 1.9e-06 & 2.2e-05 & 2.6e-05 & 1.0e-05 & 0.0e+00 & 0.0e+00 & 0.0e+00 & 4.8e-06 & 1.5e-04 & 3.4e-04 & 8.0e-05 & 1.8e-06 & 0.0e+00 & 0.0e+00 & 0.0e+00 & 2.9e-03\\
Padova & 0:42:45.03 & 41:22:17.0 & -0.45 & 4.9e-06 & 5.7e-05 & 6.8e-05 & 2.6e-05 & 0.0e+00 & 0.0e+00 & 0.0e+00 & 1.2e-05 & 3.8e-04 & 8.8e-04 & 2.1e-04 & 4.6e-06 & 0.0e+00 & 0.0e+00 & 0.0e+00 & 3.4e-03\\
Padova & 0:42:45.03 & 41:22:17.0 & -0.35 & 1.1e-05 & 1.3e-04 & 1.5e-04 & 5.7e-05 & 0.0e+00 & 0.0e+00 & 0.0e+00 & 2.7e-05 & 8.2e-04 & 1.9e-03 & 4.5e-04 & 9.9e-06 & 0.0e+00 & 0.0e+00 & 0.0e+00 & 3.8e-03\\
Padova & 0:42:45.03 & 41:22:17.0 & -0.25 & 2.0e-05 & 2.4e-04 & 2.8e-04 & 1.1e-04 & 0.0e+00 & 0.0e+00 & 0.0e+00 & 5.1e-05 & 1.5e-03 & 3.6e-03 & 8.3e-04 & 1.8e-05 & 0.0e+00 & 0.0e+00 & 0.0e+00 & 4.0e-03\\
Padova & 0:42:45.03 & 41:22:17.0 & -0.15 & 3.2e-05 & 3.8e-04 & 4.5e-04 & 1.7e-04 & 0.0e+00 & 0.0e+00 & 0.0e+00 & 8.2e-05 & 2.5e-03 & 5.8e-03 & 1.3e-03 & 2.9e-05 & 0.0e+00 & 0.0e+00 & 0.0e+00 & 4.0e-03\\
Padova & 0:42:45.03 & 41:22:17.0 & -0.05 & 4.4e-05 & 5.2e-04 & 6.2e-04 & 2.3e-04 & 0.0e+00 & 0.0e+00 & 0.0e+00 & 1.1e-04 & 3.4e-03 & 7.8e-03 & 1.8e-03 & 3.9e-05 & 0.0e+00 & 0.0e+00 & 0.0e+00 & 3.7e-03\\
Padova & 0:42:45.03 & 41:22:17.0 &  0.05 & 2.2e-04 & 2.6e-03 & 3.0e-03 & 1.2e-03 & 0.0e+00 & 0.0e+00 & 0.0e+00 & 5.5e-04 & 1.6e-02 & 3.8e-02 & 8.7e-03 & 1.9e-04 & 0.0e+00 & 0.0e+00 & 0.0e+00 & 8.9e-03\\
\enddata
\label{z4popbox_table}
\end{deluxetable}

\begin{deluxetable*}{cccp{0.8cm}p{0.8cm}p{0.8cm}p{0.8cm}p{0.8cm}p{0.8cm}p{0.8cm}p{0.8cm}p{0.8cm}p{0.8cm}p{0.8cm}p{0.8cm}p{0.8cm}p{0.8cm}p{0.8cm}p{0.8cm}p{0.8cm}}
\tablecaption{M31 Star Formation Rates (10$^{-4}$~M$_{\odot}$~yr$^{-1}$~arcmin$^{-2}$) for fixed model enrichment rates; Columns are age range in Gyr; full table available in electronic format only}
\tablehead{
\colhead{Model Set} &
\colhead{RA} &
\colhead{Dec} &
\colhead{0.0-0.3} & 
\colhead{0.3-0.4} & 
\colhead{0.4-0.5} & 
\colhead{0.5-0.6} & 
\colhead{0.6-0.8} & 
\colhead{0.8-1.0} & 
\colhead{1.0-1.3} & 
\colhead{1.3-1.6} & 
\colhead{1.6-2.0} & 
\colhead{2.0-2.5} & 
\colhead{2.5-3.2} & 
\colhead{3.2-4.0} & 
\colhead{4.0-5.0} & 
\colhead{5.0-6.3} & 
\colhead{6.3-7.9} & 
\colhead{7.9-14.1}
}
\startdata
Padova & 0:43:36.26 & 41:25:06.2 &  0.3$^{+0.4}_{-0.2}$ &  0.6$^{+0.2}_{-0.3}$ &  0.9$^{+0.7}_{-0.5}$ & 22.1$^{+2.4}_{-1.5}$ &  7.4$^{+0.9}_{-3.7}$ &  0.0$^{+1.2}_{-0.0}$ &  0.0$^{+0.7}_{-0.0}$ &  0.5$^{+2.7}_{-0.0}$ & 120.3$^{+13.8}_{-9.9}$ & 232.0$^{+11.0}_{-14.6}$ &  1.2$^{+2.1}_{-0.6}$ &  3.0$^{+3.4}_{-1.5}$ &  3.1$^{+2.2}_{-1.9}$ &  8.2$^{+0.6}_{-6.7}$ &  3.9$^{+2.6}_{-2.8}$ & 159.4$^{+4.1}_{-3.0}$ \\
Padova & 0:43:36.29 & 41:23:43.2 &  0.6$^{+0.5}_{-0.3}$ &  1.6$^{+0.0}_{-0.9}$ &  0.3$^{+1.5}_{-0.0}$ & 39.8$^{+2.9}_{-2.3}$ & 12.8$^{+0.6}_{-5.4}$ &  0.0$^{+1.7}_{-0.0}$ &  0.0$^{+0.9}_{-0.0}$ &  9.3$^{+2.9}_{-3.0}$ & 58.9$^{+9.5}_{-12.8}$ & 295.1$^{+16.8}_{-9.2}$ &  0.0$^{+3.9}_{-0.0}$ & 19.5$^{+7.5}_{-7.8}$ &  1.4$^{+10.2}_{-0.0}$ & 37.8$^{+0.0}_{-21.4}$ &  8.0$^{+18.8}_{-2.2}$ & 149.8$^{+5.1}_{-3.8}$ \\
Padova & 0:43:36.32 & 41:22:20.1 &  1.6$^{+1.0}_{-0.7}$ &  3.2$^{+0.4}_{-1.2}$ &  3.6$^{+1.9}_{-1.1}$ & 22.9$^{+3.1}_{-2.2}$ & 15.6$^{+2.5}_{-3.0}$ &  0.0$^{+0.8}_{-0.0}$ &  0.0$^{+0.5}_{-0.0}$ &  0.1$^{+2.1}_{-0.0}$ & 113.4$^{+13.6}_{-9.1}$ & 226.1$^{+7.1}_{-18.9}$ &  0.0$^{+3.6}_{-0.0}$ & 10.6$^{+1.7}_{-7.4}$ &  2.5$^{+3.9}_{-1.4}$ &  3.6$^{+4.0}_{-2.1}$ &  2.4$^{+6.4}_{-1.3}$ & 155.1$^{+1.5}_{-4.0}$ \\
Padova & 0:43:36.36 & 41:20:57.0 &  1.1$^{+0.8}_{-0.6}$ &  3.7$^{+0.2}_{-1.7}$ &  3.2$^{+2.7}_{-1.0}$ & 48.3$^{+3.2}_{-2.6}$ &  4.3$^{+2.4}_{-2.4}$ &  0.0$^{+0.7}_{-0.0}$ &  0.0$^{+0.4}_{-0.0}$ & 10.0$^{+1.2}_{-5.5}$ & 185.9$^{+21.9}_{-6.0}$ & 144.9$^{+8.9}_{-21.5}$ &  1.7$^{+4.6}_{-1.1}$ &  8.3$^{+3.2}_{-6.3}$ &  4.1$^{+4.6}_{-2.9}$ &  4.7$^{+2.9}_{-4.0}$ &  4.2$^{+2.1}_{-3.5}$ & 163.0$^{+4.3}_{-3.2}$ \\
Padova & 0:43:36.39 & 41:19:34.0 &  0.8$^{+0.5}_{-0.4}$ &  3.0$^{+0.9}_{-1.1}$ &  4.3$^{+1.8}_{-2.2}$ & 27.7$^{+4.0}_{-0.9}$ &  4.3$^{+0.9}_{-3.7}$ &  0.0$^{+1.0}_{-0.0}$ &  0.0$^{+0.7}_{-0.0}$ &  2.6$^{+0.8}_{-2.5}$ & 118.4$^{+22.2}_{-10.4}$ & 192.7$^{+12.1}_{-22.7}$ &  0.2$^{+3.7}_{-0.1}$ &  2.3$^{+6.7}_{-1.7}$ &  7.4$^{+0.7}_{-6.7}$ &  4.9$^{+4.9}_{-4.4}$ &  3.2$^{+6.5}_{-2.9}$ & 143.1$^{+4.6}_{-4.3}$ \\
Padova & 0:43:36.42 & 41:18:10.9 &  0.3$^{+0.3}_{-0.2}$ &  2.1$^{+0.4}_{-0.9}$ &  8.3$^{+0.8}_{-2.1}$ & 18.3$^{+3.4}_{-0.1}$ &  4.6$^{+0.0}_{-2.4}$ &  0.0$^{+0.4}_{-0.0}$ &  0.0$^{+0.2}_{-0.0}$ &  1.5$^{+0.0}_{-1.2}$ & 109.6$^{+14.9}_{-0.2}$ & 177.5$^{+4.9}_{-12.1}$ &  2.3$^{+0.8}_{-1.8}$ &  7.9$^{+3.4}_{-4.7}$ &  8.1$^{+5.4}_{-4.2}$ & 20.1$^{+0.5}_{-10.9}$ &  3.0$^{+2.3}_{-2.2}$ & 120.9$^{+4.6}_{-1.2}$ \\
Padova & 0:43:36.45 & 41:16:47.8 &  0.7$^{+0.8}_{-0.5}$ &  7.8$^{+0.4}_{-2.7}$ &  5.0$^{+3.6}_{-1.5}$ & 43.6$^{+4.0}_{-4.8}$ & 37.8$^{+5.9}_{-4.3}$ &  0.0$^{+1.7}_{-0.0}$ & 20.4$^{+4.1}_{-7.8}$ & 52.2$^{+9.5}_{-5.3}$ & 118.4$^{+9.9}_{-13.0}$ & 135.2$^{+8.5}_{-13.9}$ &  1.3$^{+3.3}_{-0.7}$ &  7.7$^{+4.0}_{-3.6}$ &  3.4$^{+5.6}_{-2.0}$ &  7.6$^{+2.6}_{-5.7}$ &  0.9$^{+7.8}_{-0.1}$ & 190.6$^{+8.2}_{-9.5}$ \\
Padova & 0:43:36.48 & 41:15:24.8 &  0.4$^{+0.3}_{-0.2}$ &  2.8$^{+0.0}_{-1.0}$ &  1.4$^{+1.5}_{-0.2}$ & 18.8$^{+1.1}_{-1.4}$ &  9.1$^{+1.5}_{-1.3}$ &  3.4$^{+0.7}_{-1.0}$ &  0.0$^{+0.8}_{-0.0}$ &  3.8$^{+1.2}_{-1.7}$ & 124.2$^{+4.8}_{-2.5}$ &  7.6$^{+1.1}_{-4.9}$ &  0.0$^{+1.7}_{-0.0}$ &  0.0$^{+4.2}_{-0.0}$ & 16.4$^{+0.1}_{-7.7}$ &  0.6$^{+5.0}_{-0.0}$ &  0.0$^{+3.3}_{-0.0}$ & 101.2$^{+3.6}_{-4.9}$ \\
Padova & 0:43:36.51 & 41:14:01.7 &  0.6$^{+0.4}_{-0.3}$ &  1.1$^{+0.6}_{-0.3}$ &  3.9$^{+0.4}_{-1.2}$ & 18.1$^{+1.6}_{-0.7}$ &  8.0$^{+0.7}_{-1.3}$ &  0.0$^{+0.6}_{-0.0}$ & 16.5$^{+0.4}_{-1.5}$ &  0.0$^{+1.2}_{-0.0}$ & 82.1$^{+3.3}_{-2.8}$ &  5.7$^{+2.7}_{-2.8}$ &  0.0$^{+3.7}_{-0.0}$ & 22.9$^{+1.8}_{-7.9}$ & 13.2$^{+8.6}_{-2.6}$ & 10.6$^{+2.2}_{-5.6}$ &  0.0$^{+4.0}_{-0.0}$ & 79.4$^{+4.4}_{-4.6}$ \\
Padova & 0:43:36.54 & 41:12:38.7 &  1.2$^{+0.4}_{-0.4}$ &  3.7$^{+0.9}_{-0.4}$ &  2.7$^{+0.6}_{-1.1}$ & 18.6$^{+1.5}_{-0.7}$ & 14.7$^{+0.7}_{-1.1}$ &  0.0$^{+0.2}_{-0.0}$ & 12.2$^{+0.2}_{-1.2}$ &  0.0$^{+1.8}_{-0.0}$ & 49.1$^{+1.5}_{-4.7}$ &  7.4$^{+3.3}_{-3.8}$ &  0.0$^{+3.5}_{-0.0}$ & 33.3$^{+4.8}_{-5.6}$ & 14.8$^{+3.5}_{-7.0}$ &  4.7$^{+5.1}_{-1.5}$ &  1.8$^{+6.6}_{-0.1}$ & 61.9$^{+2.2}_{-4.2}$ \\
Padova & 0:43:36.57 & 41:11:15.6 &  1.0$^{+0.6}_{-0.4}$ &  4.5$^{+0.3}_{-0.9}$ &  1.7$^{+0.9}_{-0.5}$ & 14.3$^{+0.9}_{-0.8}$ & 12.3$^{+0.5}_{-0.9}$ &  0.0$^{+0.5}_{-0.0}$ & 16.6$^{+0.0}_{-1.3}$ &  0.0$^{+1.3}_{-0.0}$ & 62.6$^{+1.8}_{-2.4}$ &  5.3$^{+1.3}_{-2.6}$ &  0.0$^{+1.9}_{-0.0}$ &  5.3$^{+2.3}_{-2.4}$ &  8.0$^{+1.6}_{-4.0}$ &  2.8$^{+4.5}_{-0.8}$ &  0.0$^{+3.9}_{-0.0}$ & 49.7$^{+1.5}_{-3.4}$ \\
Padova & 0:43:36.60 & 41:09:52.5 &  1.6$^{+0.5}_{-0.4}$ &  0.8$^{+0.3}_{-0.4}$ &  1.9$^{+0.3}_{-0.7}$ & 10.0$^{+0.8}_{-0.6}$ &  7.6$^{+0.4}_{-0.7}$ &  0.0$^{+0.3}_{-0.0}$ & 13.2$^{+0.5}_{-0.8}$ &  2.2$^{+1.5}_{-0.5}$ & 56.3$^{+1.6}_{-2.0}$ &  2.9$^{+1.1}_{-1.4}$ &  0.0$^{+1.2}_{-0.0}$ &  4.3$^{+0.7}_{-1.9}$ &  1.3$^{+1.7}_{-0.5}$ &  0.0$^{+1.9}_{-0.0}$ &  0.0$^{+2.3}_{-0.0}$ & 47.3$^{+1.5}_{-3.0}$ \\
Padova & 0:43:43.37 & 41:36:11.1 &  0.9$^{+0.5}_{-0.4}$ &  3.2$^{+0.5}_{-1.0}$ &  4.0$^{+1.3}_{-1.1}$ & 10.4$^{+1.6}_{-1.2}$ &  6.6$^{+1.9}_{-1.1}$ &  8.8$^{+2.0}_{-2.3}$ & 12.3$^{+2.0}_{-1.7}$ &  0.0$^{+0.3}_{-0.0}$ & 85.7$^{+1.0}_{-4.0}$ &  0.0$^{+2.7}_{-0.0}$ &  0.0$^{+1.5}_{-0.0}$ & 50.8$^{+0.0}_{-9.8}$ &  0.0$^{+5.2}_{-0.0}$ &  1.1$^{+8.7}_{-0.1}$ & 18.1$^{+2.7}_{-11.9}$ & 38.5$^{+2.6}_{-1.3}$ \\
\enddata
\label{sfh}
\end{deluxetable*}

\begin{deluxetable*}{cccccccccccccc}
\tablewidth{9.0in}
\tablecaption{Total Star Formation Rates \tablenotemark{a}}
\tablehead{
\colhead{Lookback Start (Years)} &
\colhead{Lookback End (Years)} &
\colhead{Padova SFR (M$_{\odot}$~yr$^{-1}$)} & 
\colhead{$+$error} &
\colhead{$-$error} &
\colhead{BaSTI SFR} &
\colhead{$+$error} &
\colhead{$-$error} &
\colhead{PARSEC SFR} & 
\colhead{$+$error} &
\colhead{$-$error} &
\colhead{MIST SFR} & 
\colhead{$+$error} &
\colhead{$-$error} 
}
\startdata
3.2e+08 & 4.0e+06 & 1.0e-01 & 3.7e-02 & 4.2e-02 &  1.6e-01 &  6.3e-02 &  5.5e-02 &  1.3e-01 &  4.7e-02 & 5.2e-02 & 1.8e-01 & 6.2e-02 &  5.9e-02\\
4.0e+08 & 3.2e+08 & 2.9e-01 & 5.6e-02 & 8.6e-02 &  7.2e-02 &  6.7e-02 &  2.1e-02 &  1.3e-01 &  4.9e-02 & 4.5e-02 & 5.9e-01 & 6.4e-02 &  1.4e-01\\
5.0e+08 & 4.0e+08 & 4.7e-01 & 1.2e-01 & 9.9e-02 &  4.9e-01 &  5.4e-02 &  1.2e-01 &  2.7e-01 &  5.4e-02 & 8.8e-02 & 5.3e-01 & 1.4e-01 &  8.7e-02\\
6.3e+08 & 5.0e+08 & 1.6e+00 & 1.4e-01 & 1.6e-01 &  9.8e-01 &  1.0e-01 &  1.5e-01 &  7.2e-01 &  1.2e-01 & 9.4e-02 & 8.4e-01 & 9.5e-02 &  1.1e-01\\
7.9e+08 & 6.3e+08 & 1.1e+00 & 1.3e-01 & 1.4e-01 &  1.0e+00 &  1.9e-01 &  6.8e-02 &  9.8e-01 &  1.2e-01 & 1.3e-01 & 4.7e-01 & 9.5e-02 &  8.5e-02\\
1.0e+09 & 7.9e+08 & 5.3e-01 & 1.2e-01 & 7.4e-02 &  1.8e+00 &  1.5e-01 &  2.1e-01 &  1.6e+00 &  1.7e-01 & 1.5e-01 & 5.3e-01 & 9.9e-02 &  7.3e-02\\
1.3e+09 & 1.0e+09 & 1.4e+00 & 9.6e-02 & 1.0e-01 &  2.3e+00 &  2.0e-01 &  1.9e-01 &  1.8e+00 &  2.2e-01 & 2.0e-01 & 1.6e+00 & 1.2e-01 &  2.0e-01\\
1.6e+09 & 1.3e+09 & 4.5e-01 & 9.1e-02 & 8.3e-02 &  2.0e+00 &  2.3e-01 &  2.4e-01 &  3.0e+00 &  1.9e-01 & 2.6e-01 & 1.6e+00 & 2.2e-01 &  2.7e-01\\
2.0e+09 & 1.6e+09 & 6.3e+00 & 3.0e-01 & 3.6e-01 &  5.7e+00 &  3.2e-01 &  3.9e-01 &  1.6e+00 &  2.5e-01 & 2.1e-01 & 5.1e+00 & 5.8e-01 &  5.3e-01\\
2.5e+09 & 2.0e+09 & 4.0e+00 & 3.6e-01 & 3.9e-01 &  1.6e+00 &  2.0e-01 &  3.5e-01 &  6.2e+00 &  2.9e-01 & 4.5e-01 & 6.0e+00 & 6.2e-01 &  7.3e-01\\
3.2e+09 & 2.5e+09 & 1.7e-01 & 3.5e-01 & 6.2e-02 &  2.0e-01 &  2.2e-01 &  9.4e-02 &  5.5e-01 &  2.5e-01 & 1.5e-01 & 1.5e+00 & 3.4e-01 &  5.6e-01\\
4.0e+09 & 3.2e+09 & 2.8e+00 & 2.4e-01 & 6.7e-01 &  1.9e-01 &  1.6e-01 &  1.1e-01 &  1.9e+00 &  3.6e-01 & 3.5e-01 & 2.1e-01 & 3.1e-01 &  1.1e-01\\
5.0e+09 & 4.0e+09 & 8.8e-01 & 4.8e-01 & 3.2e-01 &  8.5e-02 &  1.8e-01 &  4.2e-02 &  1.6e+00 &  4.0e-01 & 4.1e-01 & 3.6e-01 & 2.1e-01 &  2.0e-01\\
6.3e+09 & 5.0e+09 & 5.1e-01 & 3.9e-01 & 2.3e-01 &  1.2e-01 &  1.7e-01 &  6.8e-02 &  1.2e-01 &  2.4e-01 & 4.3e-02 & 1.6e-01 & 2.8e-01 &  8.5e-02\\
7.9e+09 & 6.3e+09 & 6.1e-01 & 4.7e-01 & 2.1e-01 &  1.3e-01 &  2.3e-01 &  6.6e-02 &  1.3e+00 &  4.3e-01 & 3.0e-01 & 6.6e-01 & 3.0e-01 &  2.4e-01\\
1.4e+10 & 7.9e+09 & 6.3e+00 & 2.8e-01 & 3.3e-01 &  8.0e+00 &  2.9e-01 &  3.4e-01 &  6.0e+00 &  3.1e-01 & 3.4e-01 & 7.1e+00 & 3.9e-01 &  3.9e-01\\
\enddata
\tablenotetext{a}{In area analyzed.  To scale these to total M31 rates, multiply these rates by 3.}
\label{total_rates}
\end{deluxetable*}

\begin{deluxetable*}{ccccccccccccc}
\tablecaption{Total Cumulative Mass \tablenotemark{a}}
\tablehead{
\colhead{Lookback Time (Years)} &
\colhead{Padova Mass (M$_{\odot}$)} & 
\colhead{$+$error} &
\colhead{$-$error} & 
\colhead{BaSTI Mass (M$_{\odot}$)} & 
\colhead{$+$error} &
\colhead{$-$error} & 
\colhead{PARSEC Mass (M$_{\odot}$)} & 
\colhead{$+$error} &
\colhead{$-$error} & 
\colhead{MIST Mass (M$_{\odot}$)} & 
\colhead{$+$error} &
\colhead{$-$error}  
}
\startdata
7.9e+09 & 3.9e+10 & 1.8e+09 & 2.1e+09 & 4.9e+10 & 1.8e+09 & 2.1e+09 & 3.7e+10 & 1.9e+09 & 2.1e+09 & 4.4e+10 & 2.4e+09 & 2.4e+09\\
6.3e+09 & 4.0e+10 & 2.5e+09 & 2.4e+09 & 4.9e+10 & 2.2e+09 & 2.2e+09 & 3.9e+10 & 2.6e+09 & 2.6e+09 & 4.5e+10 & 2.9e+09 & 2.8e+09\\
5.0e+09 & 4.1e+10 & 3.0e+09 & 2.7e+09 & 5.0e+10 & 2.4e+09 & 2.3e+09 & 3.9e+10 & 3.0e+09 & 2.6e+09 & 4.5e+10 & 3.3e+09 & 2.9e+09\\
4.0e+09 & 4.2e+10 & 3.5e+09 & 3.0e+09 & 5.0e+10 & 2.5e+09 & 2.4e+09 & 4.1e+10 & 3.4e+09 & 3.0e+09 & 4.6e+10 & 3.5e+09 & 3.1e+09\\
3.2e+09 & 4.4e+10 & 3.7e+09 & 3.6e+09 & 5.0e+10 & 2.7e+09 & 2.4e+09 & 4.3e+10 & 3.7e+09 & 3.3e+09 & 4.6e+10 & 3.8e+09 & 3.2e+09\\
2.5e+09 & 4.4e+10 & 4.0e+09 & 3.6e+09 & 5.0e+10 & 2.8e+09 & 2.5e+09 & 4.3e+10 & 3.8e+09 & 3.4e+09 & 4.7e+10 & 4.0e+09 & 3.6e+09\\
2.0e+09 & 4.6e+10 & 4.1e+09 & 3.8e+09 & 5.1e+10 & 2.9e+09 & 2.7e+09 & 4.6e+10 & 4.0e+09 & 3.7e+09 & 5.0e+10 & 4.3e+09 & 4.0e+09\\
1.6e+09 & 4.9e+10 & 4.3e+09 & 4.0e+09 & 5.3e+10 & 3.1e+09 & 2.8e+09 & 4.7e+10 & 4.1e+09 & 3.7e+09 & 5.2e+10 & 4.6e+09 & 4.2e+09\\
1.3e+09 & 4.9e+10 & 4.3e+09 & 4.0e+09 & 5.4e+10 & 3.1e+09 & 2.9e+09 & 4.8e+10 & 4.2e+09 & 3.8e+09 & 5.3e+10 & 4.6e+09 & 4.3e+09\\
1.0e+09 & 4.9e+10 & 4.3e+09 & 4.0e+09 & 5.4e+10 & 3.2e+09 & 3.0e+09 & 4.8e+10 & 4.2e+09 & 3.9e+09 & 5.3e+10 & 4.7e+09 & 4.3e+09\\
7.9e+08 & 4.9e+10 & 4.3e+09 & 4.0e+09 & 5.5e+10 & 3.2e+09 & 3.0e+09 & 4.9e+10 & 4.2e+09 & 3.9e+09 & 5.3e+10 & 4.7e+09 & 4.4e+09\\
6.3e+08 & 5.0e+10 & 4.4e+09 & 4.0e+09 & 5.5e+10 & 3.2e+09 & 3.0e+09 & 4.9e+10 & 4.3e+09 & 3.9e+09 & 5.3e+10 & 4.7e+09 & 4.4e+09\\
5.0e+08 & 5.0e+10 & 4.4e+09 & 4.1e+09 & 5.5e+10 & 3.3e+09 & 3.0e+09 & 4.9e+10 & 4.3e+09 & 4.0e+09 & 5.3e+10 & 4.7e+09 & 4.4e+09\\
4.0e+08 & 5.0e+10 & 4.4e+09 & 4.1e+09 & 5.5e+10 & 3.3e+09 & 3.1e+09 & 4.9e+10 & 4.3e+09 & 4.0e+09 & 5.3e+10 & 4.7e+09 & 4.4e+09\\
3.2e+08 & 5.0e+10 & 4.4e+09 & 4.1e+09 & 5.5e+10 & 3.3e+09 & 3.1e+09 & 4.9e+10 & 4.3e+09 & 4.0e+09 & 5.3e+10 & 4.7e+09 & 4.4e+09\\
4.0e+06 & 5.0e+10 & 4.4e+09 & 4.1e+09 & 5.5e+10 & 3.3e+09 & 3.1e+09 & 4.9e+10 & 4.3e+09 & 4.0e+09 & 5.3e+10 & 4.7e+09 & 4.4e+09\\
\enddata
\tablenotetext{a}{In area analyzed.  To scale to total M31 disk stellar mass, multiply these masses by 3.}
\label{total_mass}
\end{deluxetable*}

\begin{deluxetable*}{cccccccccc}
\tablecaption{Metallicity Distribution for the M31 Disk}
\tablehead{
\colhead{[Fe/H]$_{low}$} &
\colhead{[Fe/H]$_{high}$} &
\colhead{Padova Mass (M$_{\odot}$)} & 
\colhead{Padova Scaled (M$_{\odot}$)} &
\colhead{BaSTI Mass (M$_{\odot}$)} & 
\colhead{BaSTI Scaled (M$_{\odot}$)} &
\colhead{PARSEC Mass (M$_{\odot}$)} & 
\colhead{PARSEC Scaled (M$_{\odot}$)} &
\colhead{MIST Mass (M$_{\odot}$)} & 
\colhead{MIST Scaled (M$_{\odot}$)} 
}
\startdata
-2.4 & -2.2 & 8.5e+07 & 2.5e+08 & 4.8e+07 & 1.5e+08 & 0.0 & 0.0 & 1.4e+08 & 4.3e+08\\
-2.2 & -2.0 & 1.2e+08 & 3.5e+08 & 6.5e+07 & 2.0e+08 & 1.5e+08 & 4.6e+08 & 2.0e+08 & 5.9e+08\\
-2.0 & -1.8 & 2.0e+08 & 5.9e+08 & 1.1e+08 & 3.4e+08 & 1.8e+08 & 5.3e+08 & 3.4e+08 & 1.0e+09\\
-1.8 & -1.6 & 3.2e+08 & 9.6e+08 & 1.8e+08 & 5.5e+08 & 2.9e+08 & 8.7e+08 & 5.5e+08 & 1.6e+09\\
-1.6 & -1.4 & 5.2e+08 & 1.5e+09 & 2.9e+08 & 8.8e+08 & 4.7e+08 & 1.4e+09 & 8.8e+08 & 2.6e+09\\
-1.4 & -1.2 & 8.3e+08 & 2.5e+09 & 4.8e+08 & 1.4e+09 & 7.6e+08 & 2.3e+09 & 1.4e+09 & 4.3e+09\\
-1.2 & -1.0 & 1.3e+09 & 4.0e+09 & 8.1e+08 & 2.4e+09 & 1.2e+09 & 3.7e+09 & 2.3e+09 & 6.9e+09\\
-1.0 & -0.8 & 2.1e+09 & 6.4e+09 & 1.5e+09 & 4.4e+09 & 2.0e+09 & 6.1e+09 & 3.7e+09 & 1.1e+10\\
-0.8 & -0.6 & 3.4e+09 & 1.0e+10 & 3.0e+09 & 9.0e+09 & 3.3e+09 & 9.8e+09 & 5.4e+09 & 1.6e+10\\
-0.6 & -0.4 & 5.5e+09 & 1.7e+10 & 6.2e+09 & 1.9e+10 & 5.0e+09 & 1.5e+10 & 7.3e+09 & 2.2e+10\\
-0.4 & -0.2 & 8.7e+09 & 2.6e+10 & 1.1e+10 & 3.3e+10 & 7.1e+09 & 2.1e+10 & 8.9e+09 & 2.7e+10\\
-0.2 & 0.0 & 1.1e+10 & 3.3e+10 & 1.3e+10 & 4.0e+10 & 9.2e+09 & 2.8e+10 & 9.4e+09 & 2.8e+10\\
0.0 & 0.2 & 1.6e+10 & 4.8e+10 & 1.8e+10 & 5.5e+10 & 1.9e+10 & 5.8e+10 & 1.3e+10 & 3.9e+10\\
\enddata
\label{total_metallicity}
\end{deluxetable*}

\begin{deluxetable}{ccccccccccc}
\tablewidth{10.0in}
\tablecaption{Metallicity Distribution in Radial Bins.}
\tablehead{
\colhead{Radial Range (kpc)} &
\colhead{[Fe/H]$_{low}$} &
\colhead{[Fe/H]$_{high}$} &
\colhead{Padova Mass (M$_{\odot}$)} & 
\colhead{Padova Scaled (M$_{\odot}$)} &
\colhead{BaSTI Mass (M$_{\odot}$)} & 
\colhead{BaSTI Scaled (M$_{\odot}$)} &
\colhead{PARSEC Mass (M$_{\odot}$)} & 
\colhead{PARSEC Scaled (M$_{\odot}$)} &
\colhead{MIST Mass (M$_{\odot}$)} & 
\colhead{MIST Scaled (M$_{\odot}$)} 
}
\startdata
$<$5 & -2.4 & -2.2 & 4.2e+07 & 6.0e+07 & 3.1e+07 & 1.1e+08 & 1.2e+07 & 7.3e+07 & 2.3e+07 & 3.5e+07\\
$<$5 & -2.2 & -2.0 & 5.7e+07 & 8.2e+07 & 4.2e+07 & 1.5e+08 & 1.6e+07 & 1.0e+08 & 3.0e+07 & 4.7e+07\\
$<$5 & -2.0 & -1.8 & 9.7e+07 & 1.4e+08 & 7.2e+07 & 2.6e+08 & 2.8e+07 & 1.7e+08 & 5.2e+07 & 8.1e+07\\
$<$5 & -1.8 & -1.6 & 1.6e+08 & 2.3e+08 & 1.2e+08 & 4.3e+08 & 4.5e+07 & 2.8e+08 & 8.5e+07 & 1.3e+08\\
$<$5 & -1.6 & -1.4 & 2.6e+08 & 3.7e+08 & 1.9e+08 & 6.9e+08 & 7.2e+07 & 4.4e+08 & 1.4e+08 & 2.1e+08\\
$<$5 & -1.4 & -1.2 & 4.2e+08 & 6.0e+08 & 3.0e+08 & 1.1e+09 & 1.1e+08 & 6.8e+08 & 2.2e+08 & 3.5e+08\\
$<$5 & -1.2 & -1.0 & 6.9e+08 & 9.9e+08 & 4.9e+08 & 1.8e+09 & 1.6e+08 & 9.9e+08 & 3.7e+08 & 5.8e+08\\
$<$5 & -1.0 & -0.8 & 1.2e+09 & 1.7e+09 & 7.4e+08 & 2.7e+09 & 2.4e+08 & 1.4e+09 & 6.5e+08 & 1.0e+09\\
$<$5 & -0.8 & -0.6 & 1.9e+09 & 2.8e+09 & 1.1e+09 & 3.9e+09 & 4.1e+08 & 2.5e+09 & 1.2e+09 & 1.9e+09\\
$<$5 & -0.6 & -0.4 & 3.1e+09 & 4.4e+09 & 1.7e+09 & 6.1e+09 & 8.0e+08 & 4.8e+09 & 2.5e+09 & 3.8e+09\\
$<$5 & -0.4 & -0.2 & 4.6e+09 & 6.6e+09 & 2.8e+09 & 1.0e+10 & 1.2e+09 & 7.4e+09 & 4.6e+09 & 7.1e+09\\
$<$5 & -0.2 & 0.0 & 5.7e+09 & 8.3e+09 & 3.8e+09 & 1.4e+10 & 1.3e+09 & 7.7e+09 & 6.3e+09 & 9.7e+09\\
$<$5 & 0.0 & 0.2 & 8.8e+09 & 1.3e+10 & 5.7e+09 & 2.1e+10 & 1.4e+09 & 8.6e+09 & 9.6e+09 & 1.5e+10\\
5$-$12 & -2.4 & -2.2 & 3.1e+07 & 1.1e+08 & 1.2e+07 & 7.3e+07 & 2.3e+07 & 3.5e+07 & 1.8e+07 & 5.8e+07\\
5$-$12 & -2.2 & -2.0 & 4.2e+07 & 1.5e+08 & 1.6e+07 & 1.0e+08 & 3.0e+07 & 4.7e+07 & 2.5e+07 & 7.9e+07\\
5$-$12 & -2.0 & -1.8 & 7.2e+07 & 2.6e+08 & 2.8e+07 & 1.7e+08 & 5.2e+07 & 8.1e+07 & 4.3e+07 & 1.4e+08\\
5$-$12 & -1.8 & -1.6 & 1.2e+08 & 4.3e+08 & 4.5e+07 & 2.8e+08 & 8.5e+07 & 1.3e+08 & 7.0e+07 & 2.2e+08\\
5$-$12 & -1.6 & -1.4 & 1.9e+08 & 6.9e+08 & 7.2e+07 & 4.4e+08 & 1.4e+08 & 2.1e+08 & 1.1e+08 & 3.5e+08\\
5$-$12 & -1.4 & -1.2 & 3.0e+08 & 1.1e+09 & 1.1e+08 & 6.8e+08 & 2.2e+08 & 3.5e+08 & 1.8e+08 & 5.8e+08\\
5$-$12 & -1.2 & -1.0 & 4.9e+08 & 1.8e+09 & 1.6e+08 & 9.9e+08 & 3.7e+08 & 5.8e+08 & 3.1e+08 & 9.7e+08\\
5$-$12 & -1.0 & -0.8 & 7.4e+08 & 2.7e+09 & 2.4e+08 & 1.4e+09 & 6.5e+08 & 1.0e+09 & 5.6e+08 & 1.8e+09\\
5$-$12 & -0.8 & -0.6 & 1.1e+09 & 3.9e+09 & 4.1e+08 & 2.5e+09 & 1.2e+09 & 1.9e+09 & 1.2e+09 & 3.7e+09\\
5$-$12 & -0.6 & -0.4 & 1.7e+09 & 6.1e+09 & 8.0e+08 & 4.8e+09 & 2.5e+09 & 3.8e+09 & 2.5e+09 & 8.0e+09\\
5$-$12 & -0.4 & -0.2 & 2.8e+09 & 1.0e+10 & 1.2e+09 & 7.4e+09 & 4.6e+09 & 7.1e+09 & 4.5e+09 & 1.4e+10\\
5$-$12 & -0.2 & 0.0 & 3.8e+09 & 1.4e+10 & 1.3e+09 & 7.7e+09 & 6.3e+09 & 9.7e+09 & 5.6e+09 & 1.8e+10\\
5$-$12 & 0.0 & 0.2 & 5.7e+09 & 2.1e+10 & 1.4e+09 & 8.6e+09 & 9.6e+09 & 1.5e+10 & 7.0e+09 & 2.2e+10\\
$>$12 & -2.4 & -2.2 & 1.2e+07 & 7.3e+07 & 2.3e+07 & 3.5e+07 & 1.8e+07 & 5.8e+07 & 7.2e+06 & 3.9e+07\\
$>$12 & -2.2 & -2.0 & 1.6e+07 & 1.0e+08 & 3.0e+07 & 4.7e+07 & 2.5e+07 & 7.9e+07 & 9.7e+06 & 5.3e+07\\
$>$12 & -2.0 & -1.8 & 2.8e+07 & 1.7e+08 & 5.2e+07 & 8.1e+07 & 4.3e+07 & 1.4e+08 & 1.7e+07 & 9.0e+07\\
$>$12 & -1.8 & -1.6 & 4.5e+07 & 2.8e+08 & 8.5e+07 & 1.3e+08 & 7.0e+07 & 2.2e+08 & 2.7e+07 & 1.5e+08\\
$>$12 & -1.6 & -1.4 & 7.2e+07 & 4.4e+08 & 1.4e+08 & 2.1e+08 & 1.1e+08 & 3.5e+08 & 4.4e+07 & 2.4e+08\\
$>$12 & -1.4 & -1.2 & 1.1e+08 & 6.8e+08 & 2.2e+08 & 3.5e+08 & 1.8e+08 & 5.8e+08 & 7.2e+07 & 3.9e+08\\
$>$12 & -1.2 & -1.0 & 1.6e+08 & 9.9e+08 & 3.7e+08 & 5.8e+08 & 3.1e+08 & 9.7e+08 & 1.3e+08 & 7.0e+08\\
$>$12 & -1.0 & -0.8 & 2.4e+08 & 1.4e+09 & 6.5e+08 & 1.0e+09 & 5.6e+08 & 1.8e+09 & 2.7e+08 & 1.5e+09\\
$>$12 & -0.8 & -0.6 & 4.1e+08 & 2.5e+09 & 1.2e+09 & 1.9e+09 & 1.2e+09 & 3.7e+09 & 6.4e+08 & 3.4e+09\\
$>$12 & -0.6 & -0.4 & 8.0e+08 & 4.8e+09 & 2.5e+09 & 3.8e+09 & 2.5e+09 & 8.0e+09 & 1.2e+09 & 6.8e+09\\
$>$12 & -0.4 & -0.2 & 1.2e+09 & 7.4e+09 & 4.6e+09 & 7.1e+09 & 4.5e+09 & 1.4e+10 & 1.7e+09 & 9.4e+09\\
$>$12 & -0.2 & 0.0 & 1.3e+09 & 7.7e+09 & 6.3e+09 & 9.7e+09 & 5.6e+09 & 1.8e+10 & 1.6e+09 & 8.8e+09\\
$>$12 & 0.0 & 0.2 & 1.4e+09 & 8.6e+09 & 9.6e+09 & 1.5e+10 & 7.0e+09 & 2.2e+10 & 1.6e+09 & 8.4e+09\\
\enddata
\label{radial_metallicity}
\end{deluxetable}

\end{turnpage}


\begin{figure*}
\begin{center}
\includegraphics[height=3.1in]{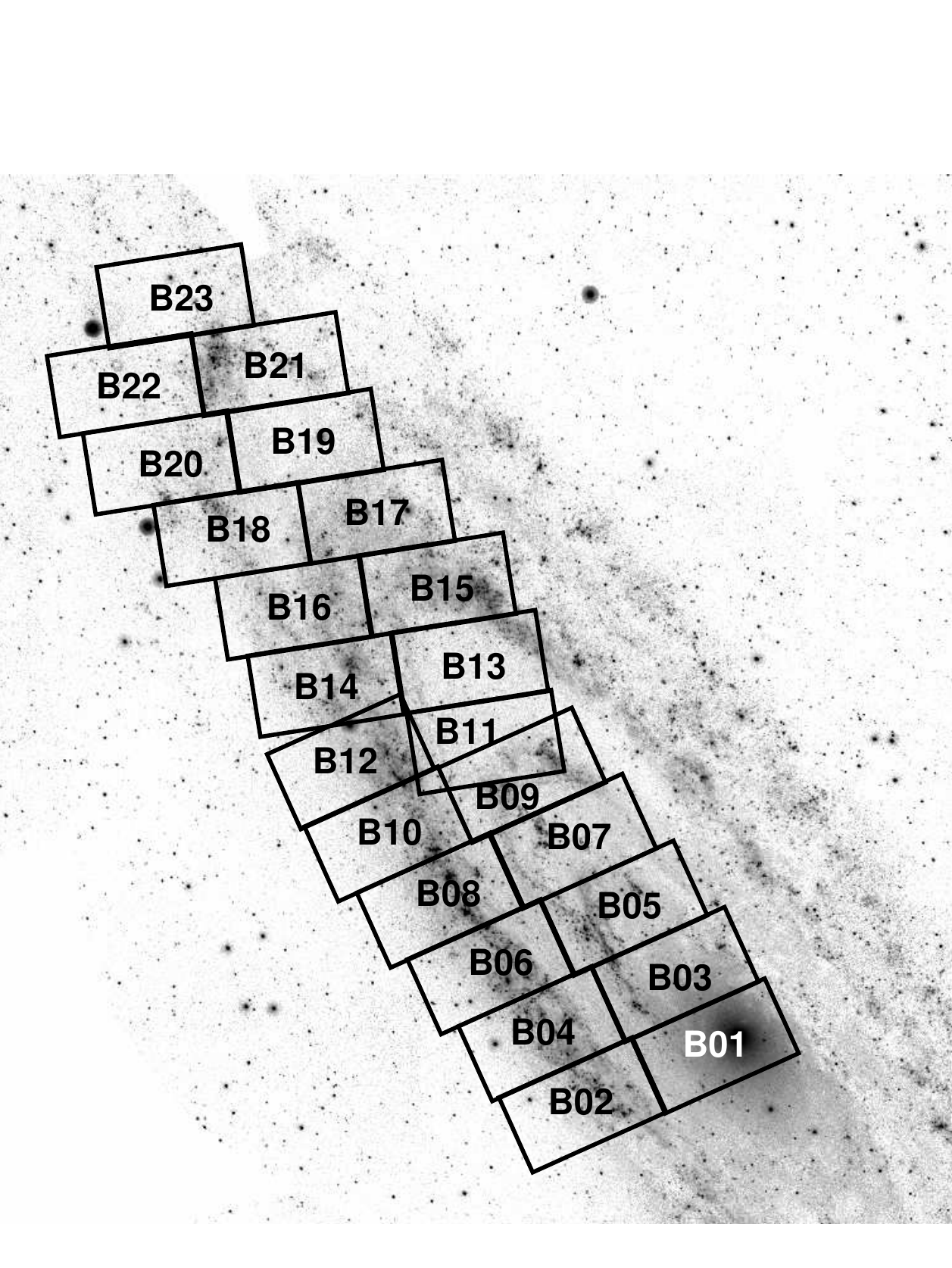}
\includegraphics[width=3.1in]{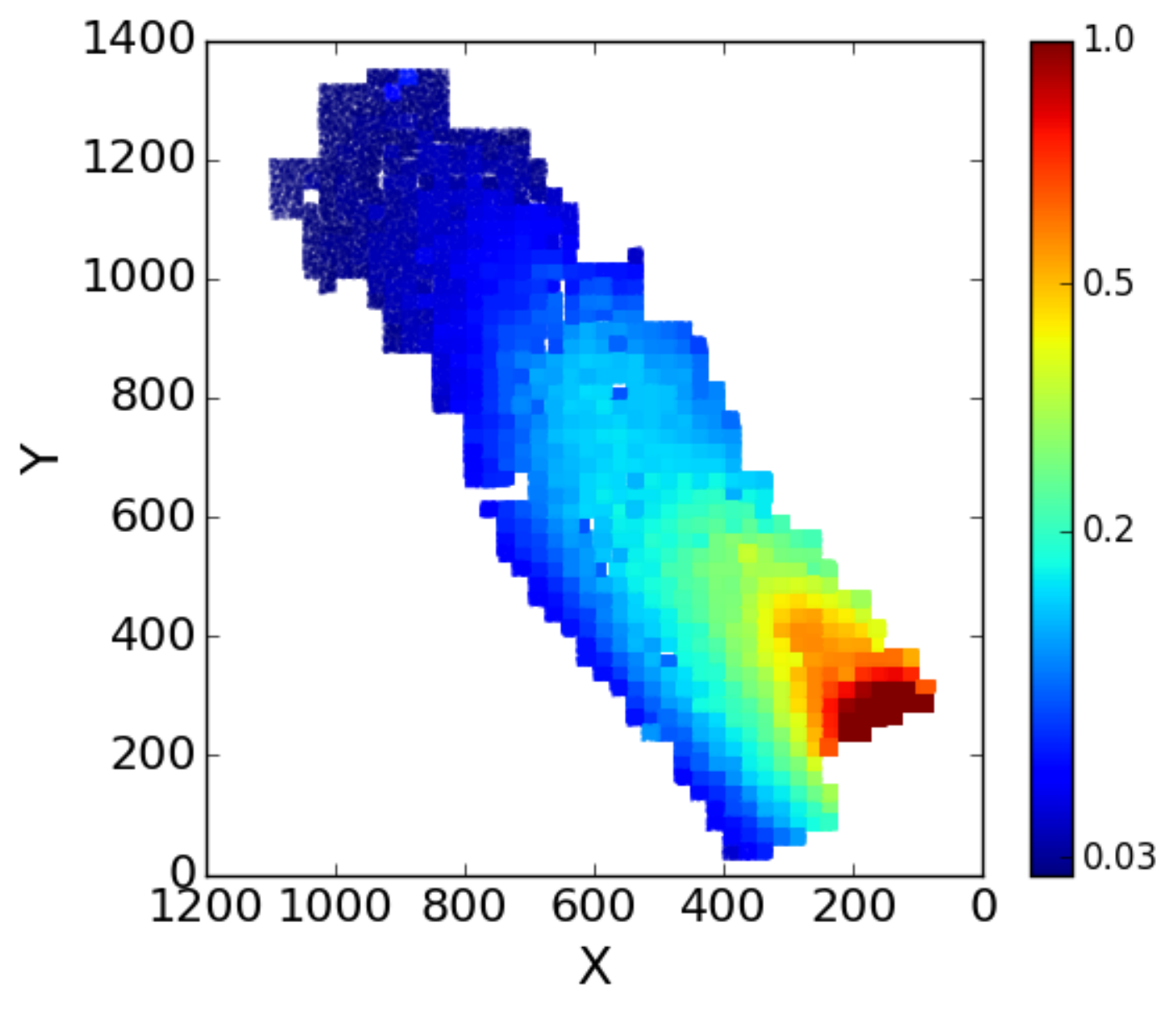}
\end{center}
\caption{Map of the portions of the PHAT survey for which we measured SFHs. {\it Left:} PHAT region from \citet{williams2014} is shown on a GALEX NUV image of M31. {\it Right} Stellar density map of the PHAT region outlined in {\it right.} The areas measured are color-coded by stellar density for 18.5$<m_{F160W}<$19.5 (stars per arcsec$^2$).} 
\label{stellar_density_map}
\end{figure*}

\begin{figure*}
\begin{center}
\includegraphics[width=3.1in]{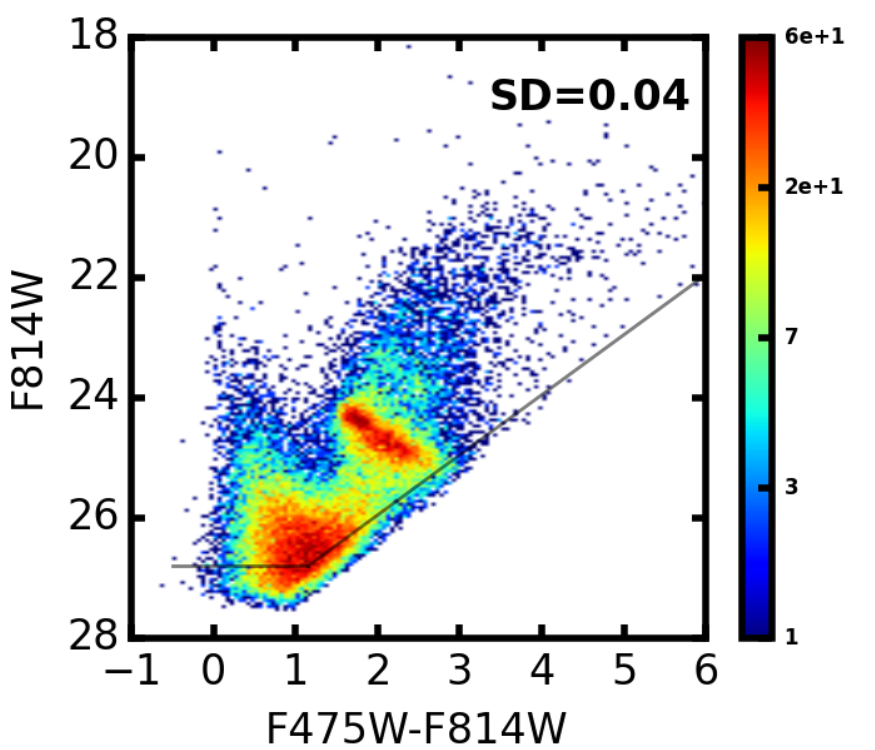}
\includegraphics[width=3.1in]{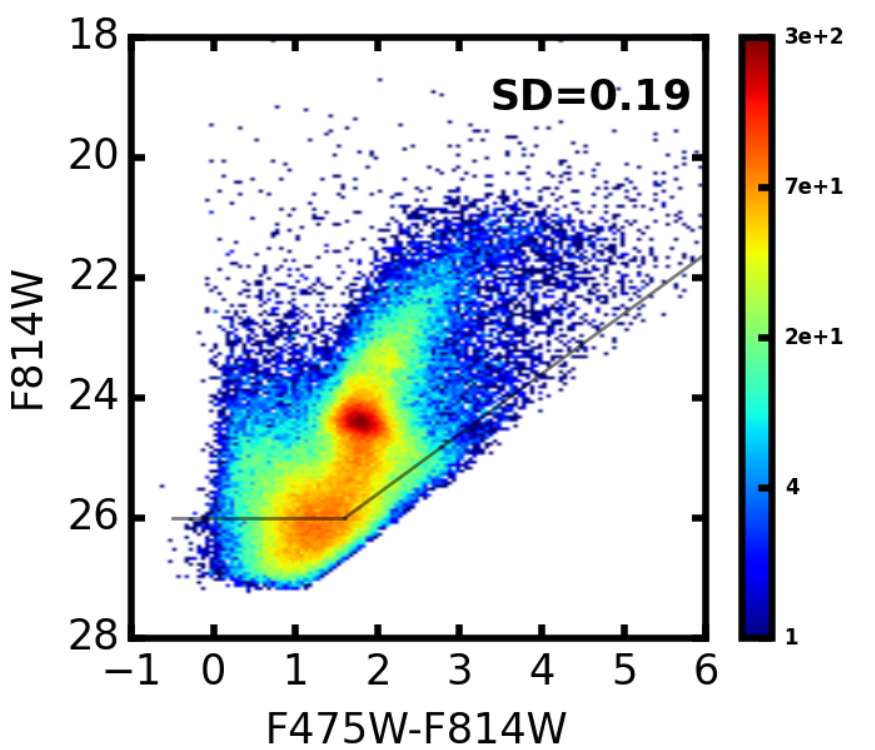}
\includegraphics[width=3.1in]{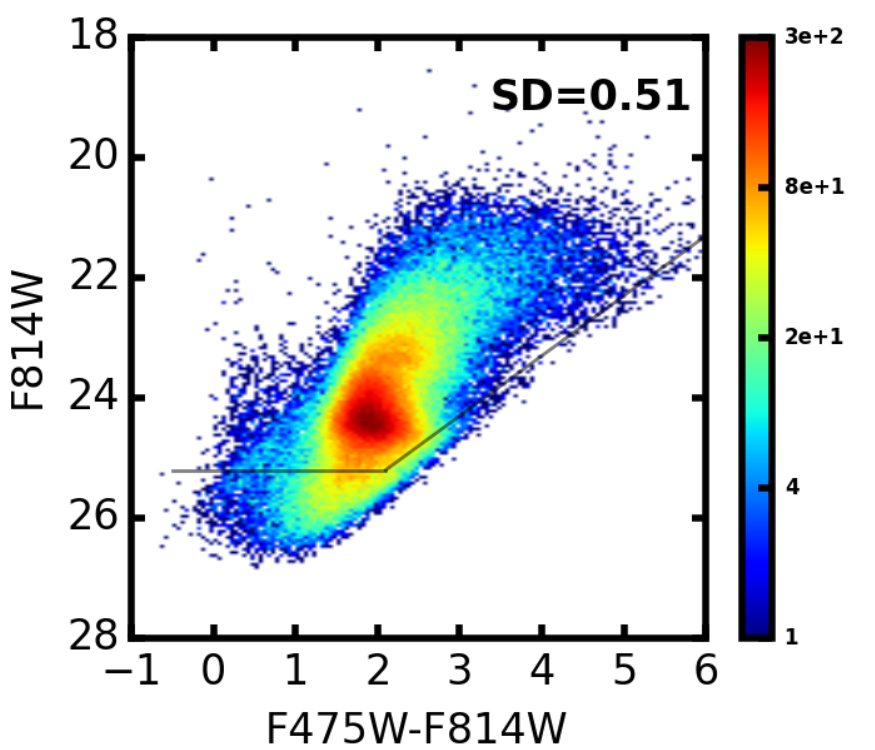}
\includegraphics[width=3.1in]{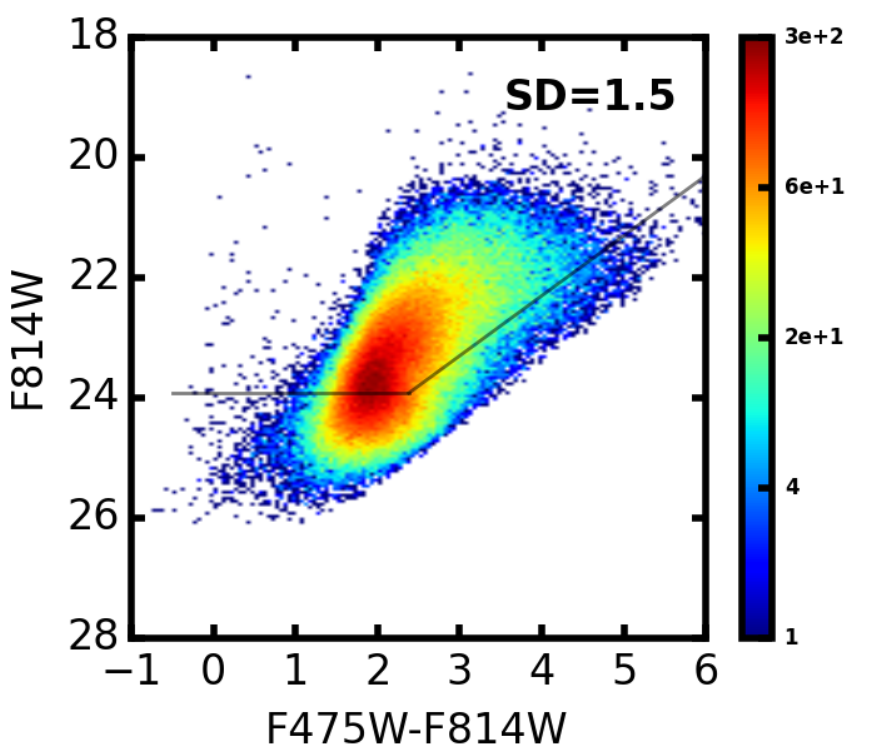}
\end{center}
\caption{Optical CMDs from 4 of our 83$''{\times}83''$ regions covering the full range of stellar densities. Color indicates the number of data points in each place in the CMD. Black lines show the 50\% completeness limits. {\it Upper Left:} Surface density (SD) = 0.04. {\it Upper Right:} SD=0.19. {\it Lower Left:} SD=0.51. {\it Lower Right:} SD=1.5.} 
\label{cmds}
\end{figure*}

\begin{figure*}
\begin{center}
\includegraphics[width=3.1in]{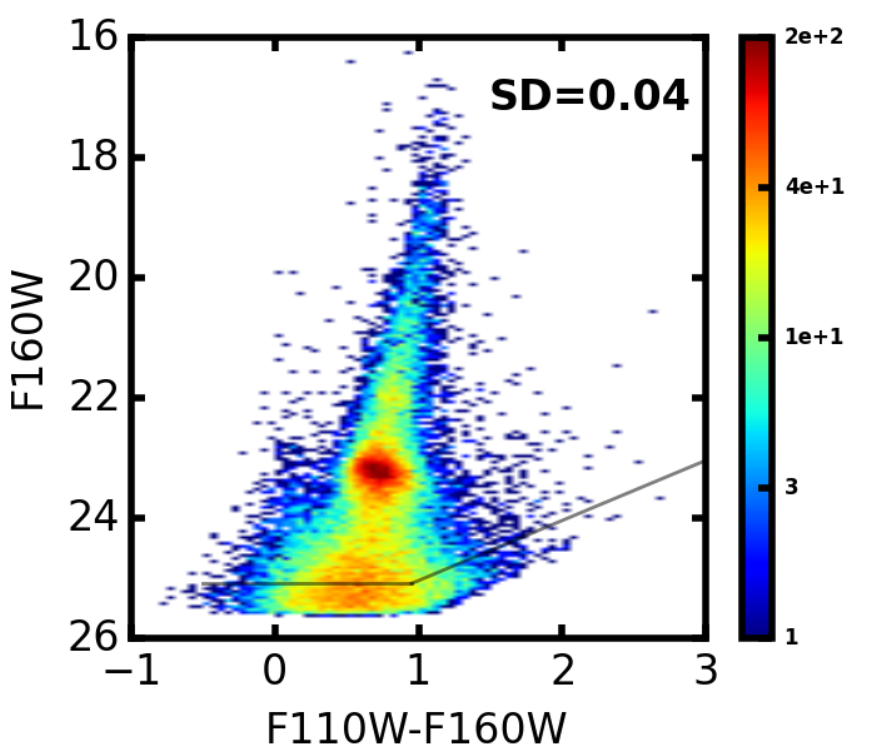}
\includegraphics[width=3.1in]{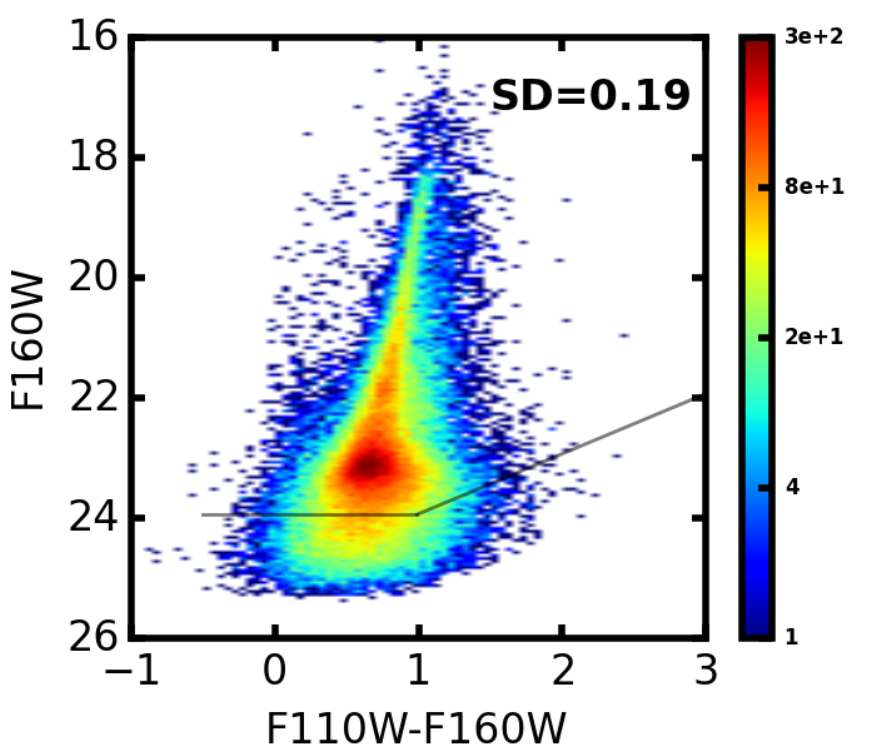}
\includegraphics[width=3.1in]{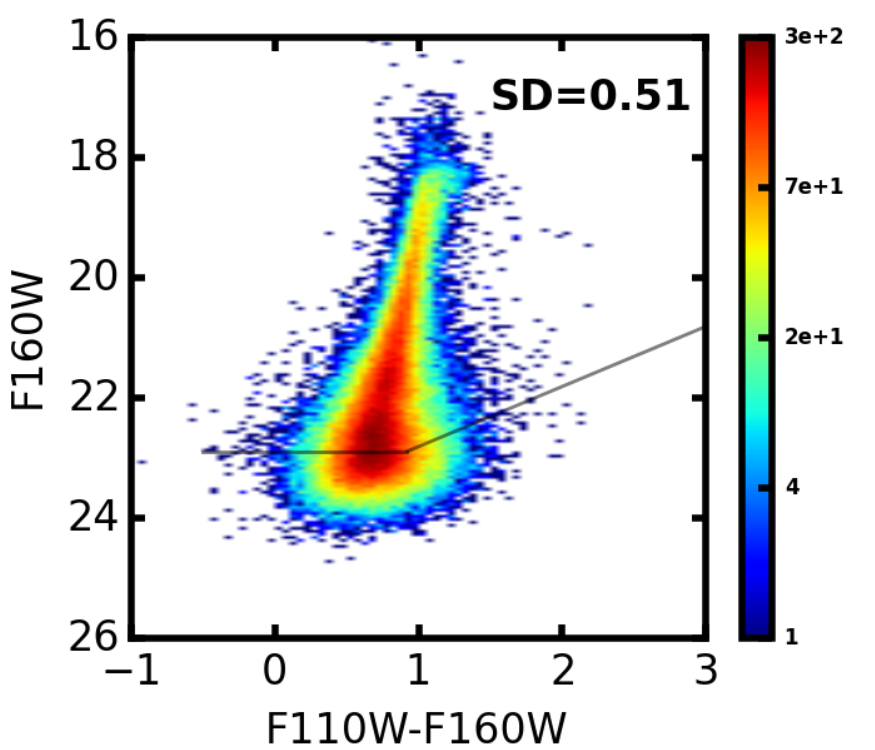}
\includegraphics[width=3.1in]{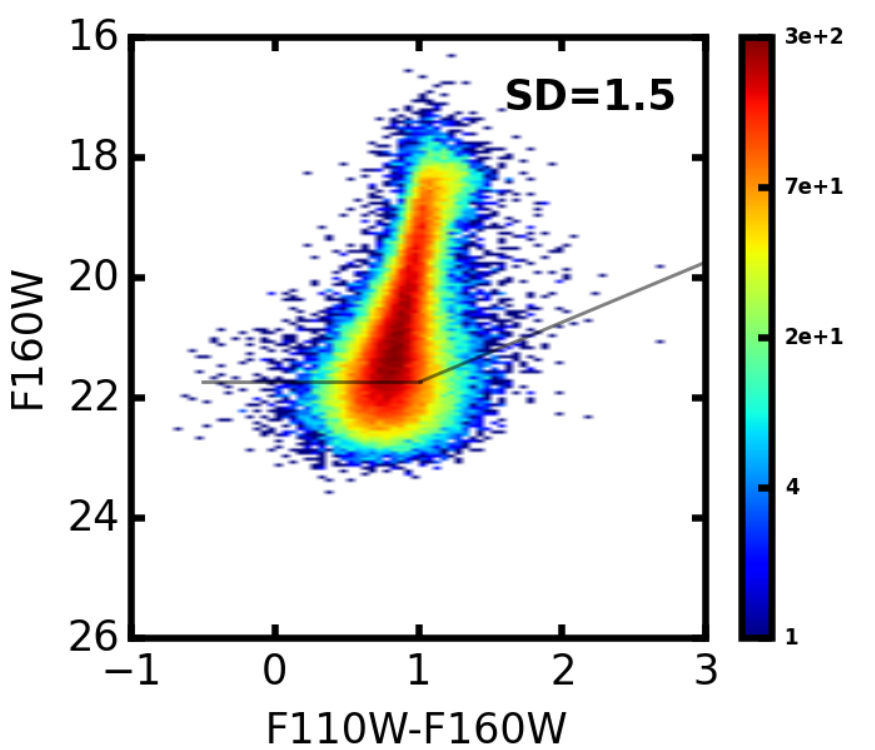}
\end{center}
\caption{Near IR CMDs from 4 of our regions covering the full range of stellar densities. Color indicates the number of data points in each place in the CMD. Black lines show the 50\% completeness limits. {\it Upper Left:} 0.04. {\it Upper Right:} 0.19. {\it Lower Left:} 0.51. {\it Lower Right:} 1.5.} 
\label{ircmds}
\end{figure*}

\begin{figure*}
\begin{center}
\includegraphics[width=5.1in]{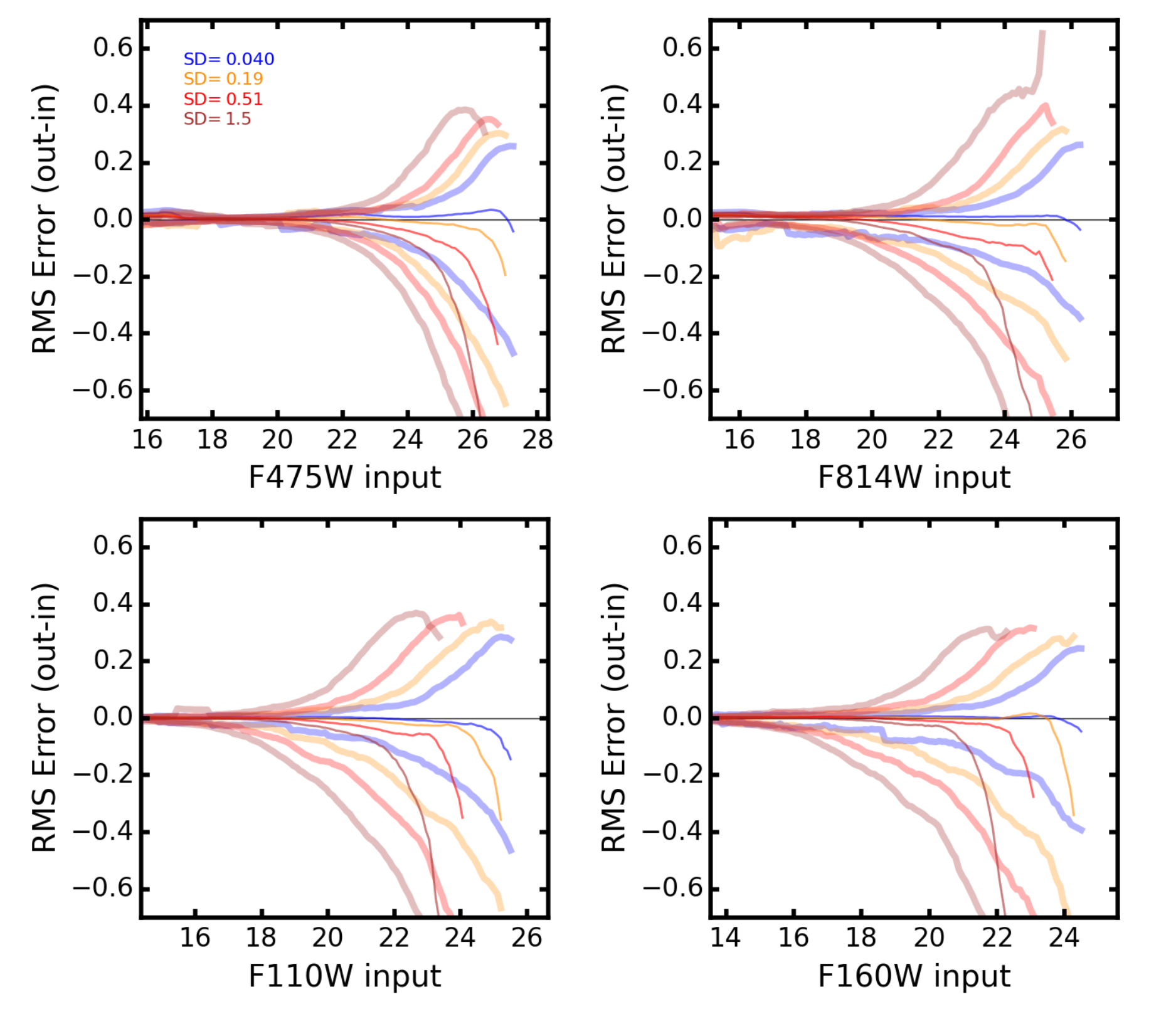}
\end{center}
\caption{Uncertainties as a function of magnitude at several stellar densities for the four bands used in our analysis of the PHAT photometry measured with the output magnitude minus the input magnitude from our artificial star tests.  Color-coded broad lines mark the root-mean-square scatter in the positive and negative directions. Narrow lines mark the median difference (photometric bias).  Redder lines denote higher stellar densities.  Each panel shows the relation for a different band. \label{rms}}
\end{figure*}


\begin{turnpage}

\begin{figure*}
\begin{center}
\includegraphics[width=9.0in]{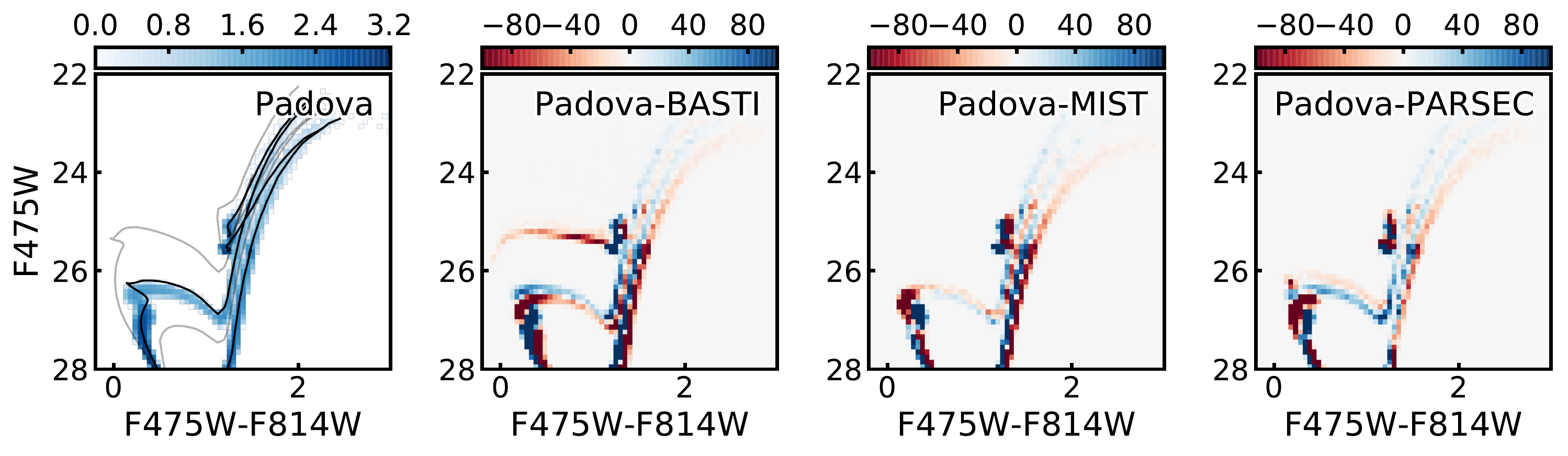}
\includegraphics[width=9.0in]{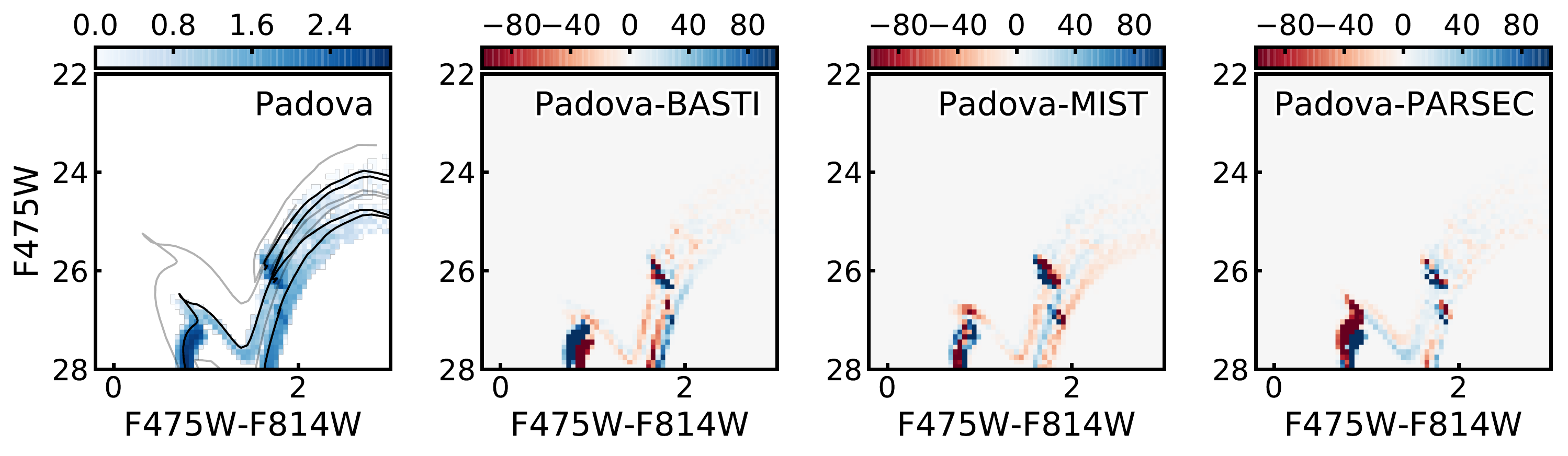}
\end{center}
\caption{{\it Top}: Left panel shows an example CMD of the Padova models at with ages  of 2 and 10 Gyr at metallicity Z=0.0018.  Dispersion is due to the use of finite age and metallicity ranges (0.05 dex) to generate the models.  Black lines mark the corresponding isochrones for ages of 2 and 10 Gyr.  Gray lines mark isochrones for 1 Gyr and 4 Gyr.  Other panels show the
  differences between this CMD and those with the same ages and metallicity for
  the other three models sets used in our analysis.  {\it
    Bottom}: Same as {\it Top}, but for Z=0.018. }
\label{model_cmds}
\end{figure*}

\end{turnpage}

\begin{figure*}
\includegraphics[width=3.45in]{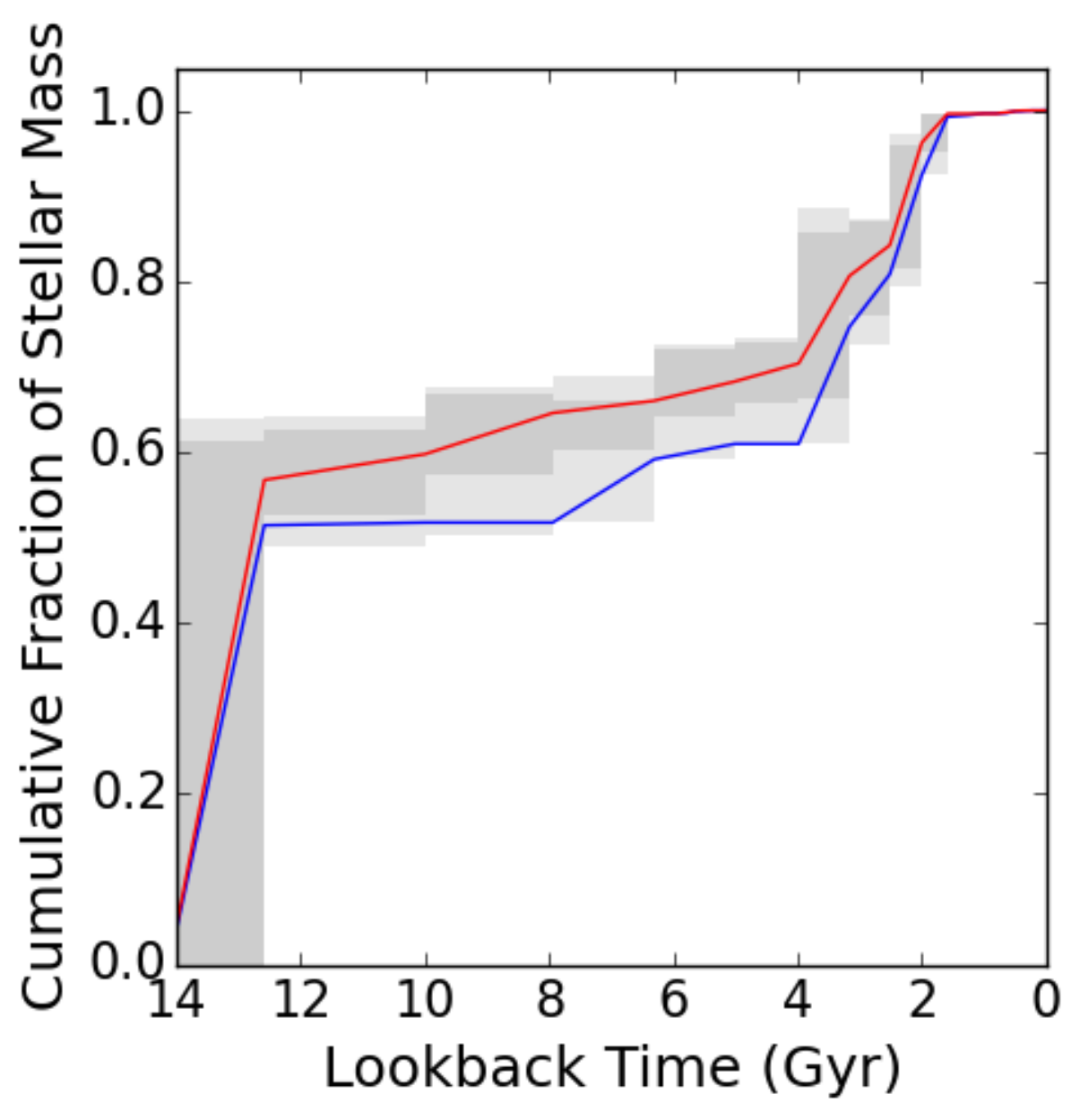}
\includegraphics[width=3.5in]{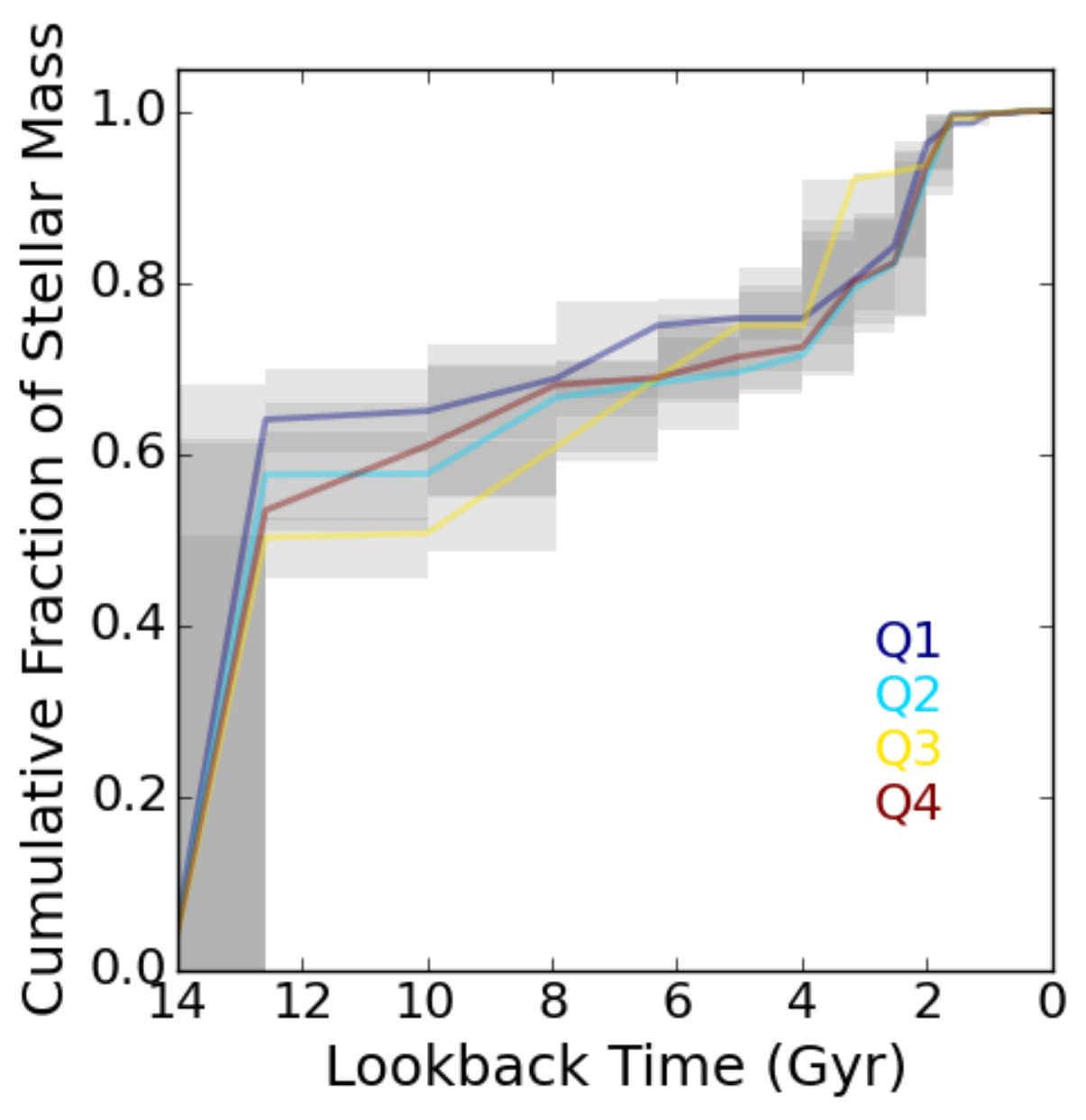}
\caption{{\it Left:} Cumulative SFH for a region as measured with ASTs
  performed directly on the data (blue) are overplotted with the SFH
  of for the same region as measured with ASTs calculated using our
  stellar density function fitting technique to infer the photometric
  quality as a function of color and magnitude at the stellar density
  of the region (red). Gray area shows the random uncertainties from the measurement, areas where the uncertainties overlap are darker gray.  {\it Right:} Cumulative SFH for a region as measured
  by the samples taken from the 4 quartiles of A$_{\rm V}$ ($\mu$)
  values.  Colors indicate the different quartiles.  There is no trend
  with quartiles, and the measurements agree within their
  uncertainties. \label{compare_sfhs}}

\end{figure*}



\begin{figure*}
\includegraphics[width=2.2in]{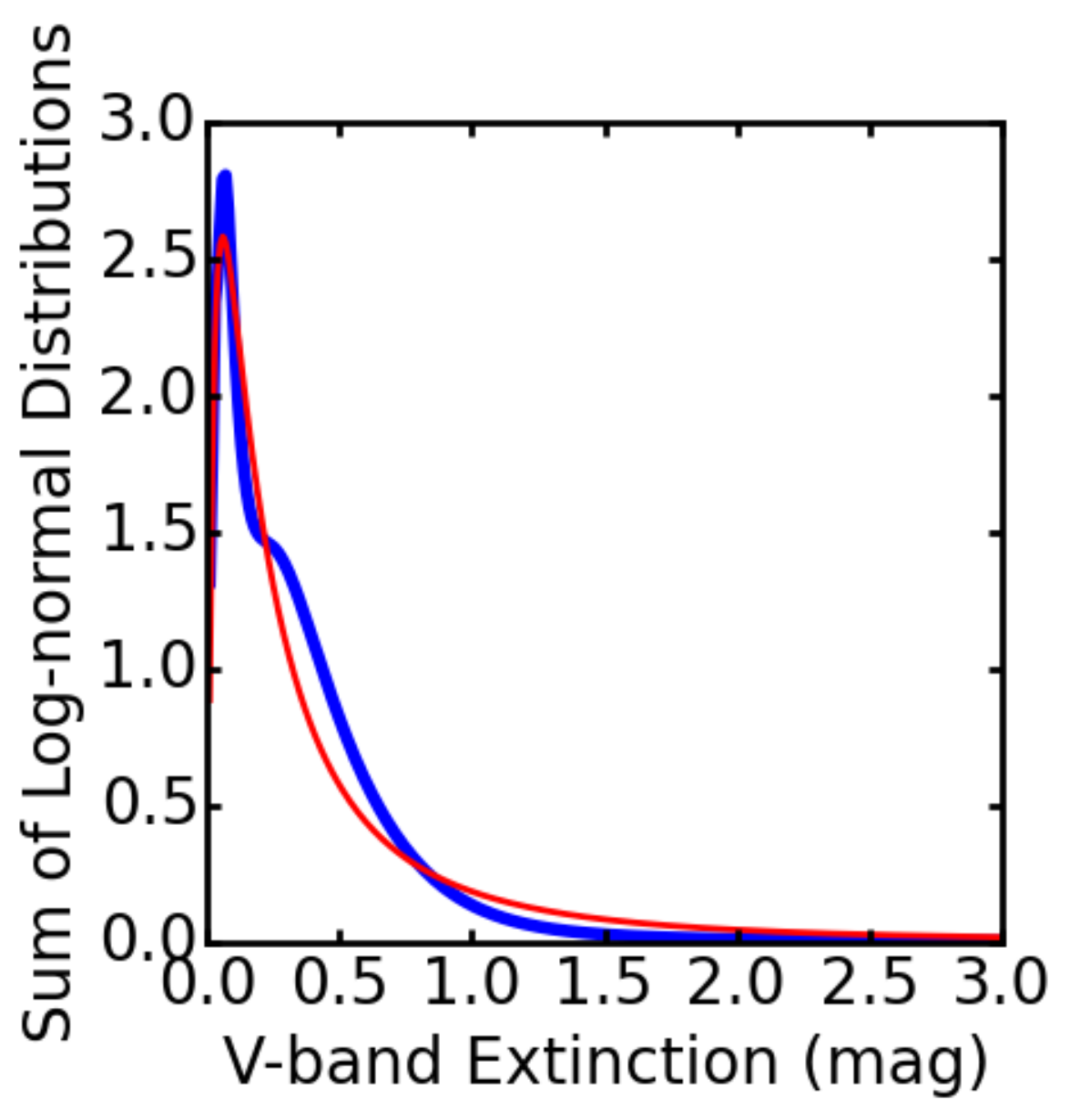}
\includegraphics[width=2.1in]{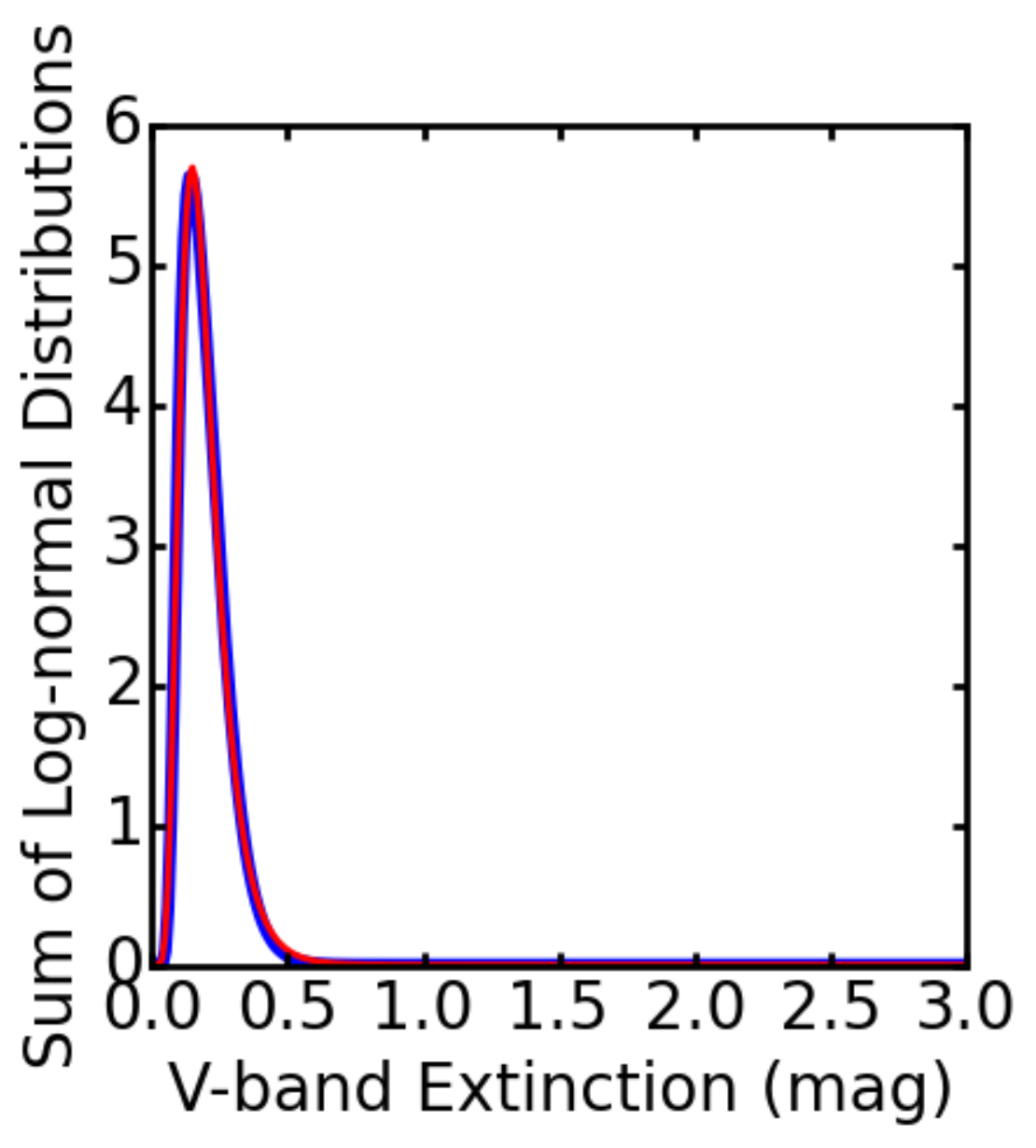}
\includegraphics[width=2.2in]{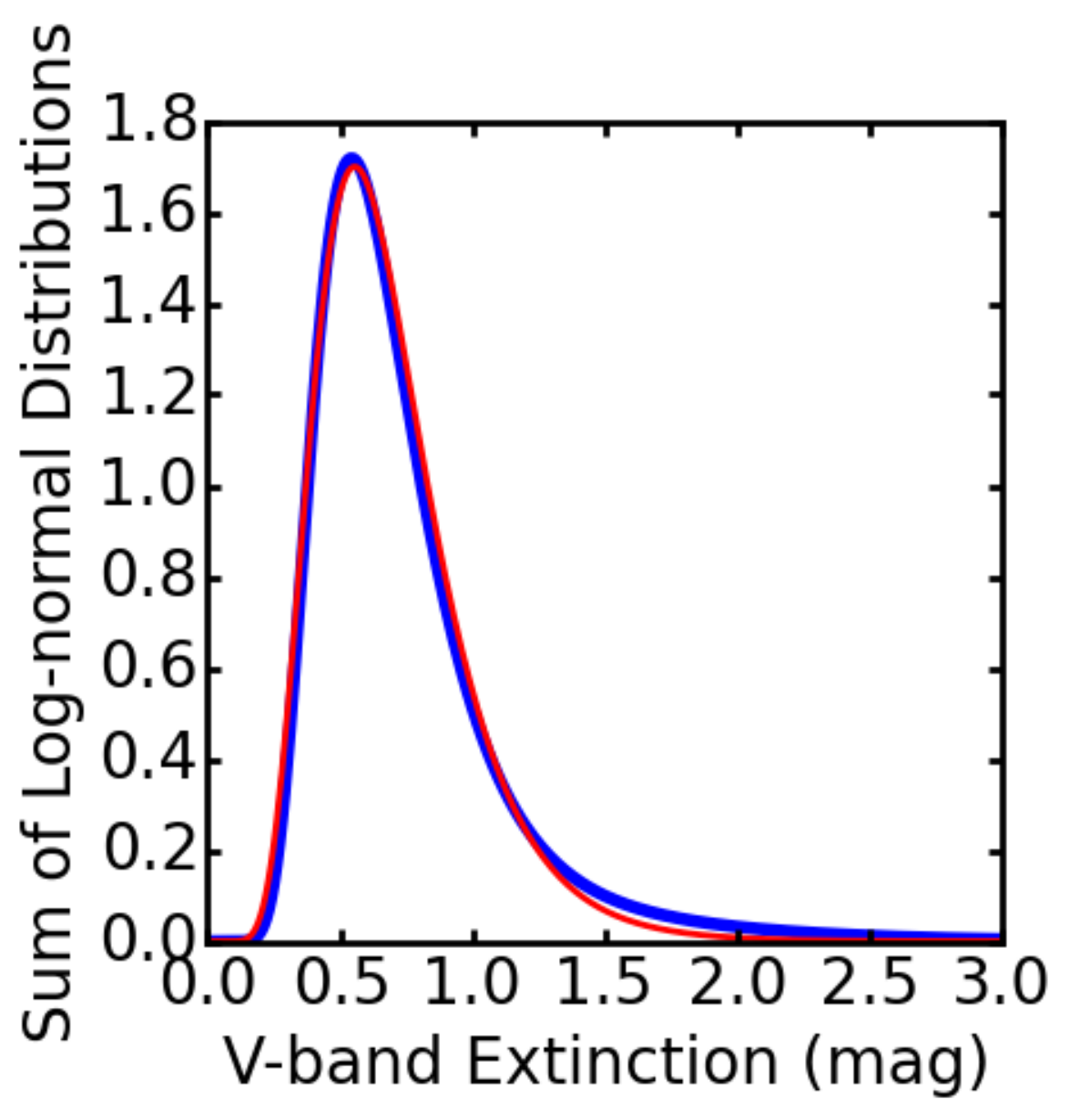}
\caption{{\it Left:} Log-normal fit to a sum of the log-normal functions of all of the pixels in the \citet{dalcanton2015} maps of one of our sub-regions.  {\it Center} Log-normal fit to a sum of the log-normal functions of pixels in the second quartile of $\mu$ values in \citet{dalcanton2015} maps of one of our sub-regions ($\mu=0.18, \sigma=0.43$). {\it Right} Log-normal fit to a sum of the log-normal functions of pixels in the forth quartile of $\mu$ values in \citet{dalcanton2015} maps of one of our sub-regions ($\mu=0.65, \sigma=0.39$).\label{combined_log_normal}}
\end{figure*}


\begin{figure*}
\begin{center}
\includegraphics[width=5.1in]{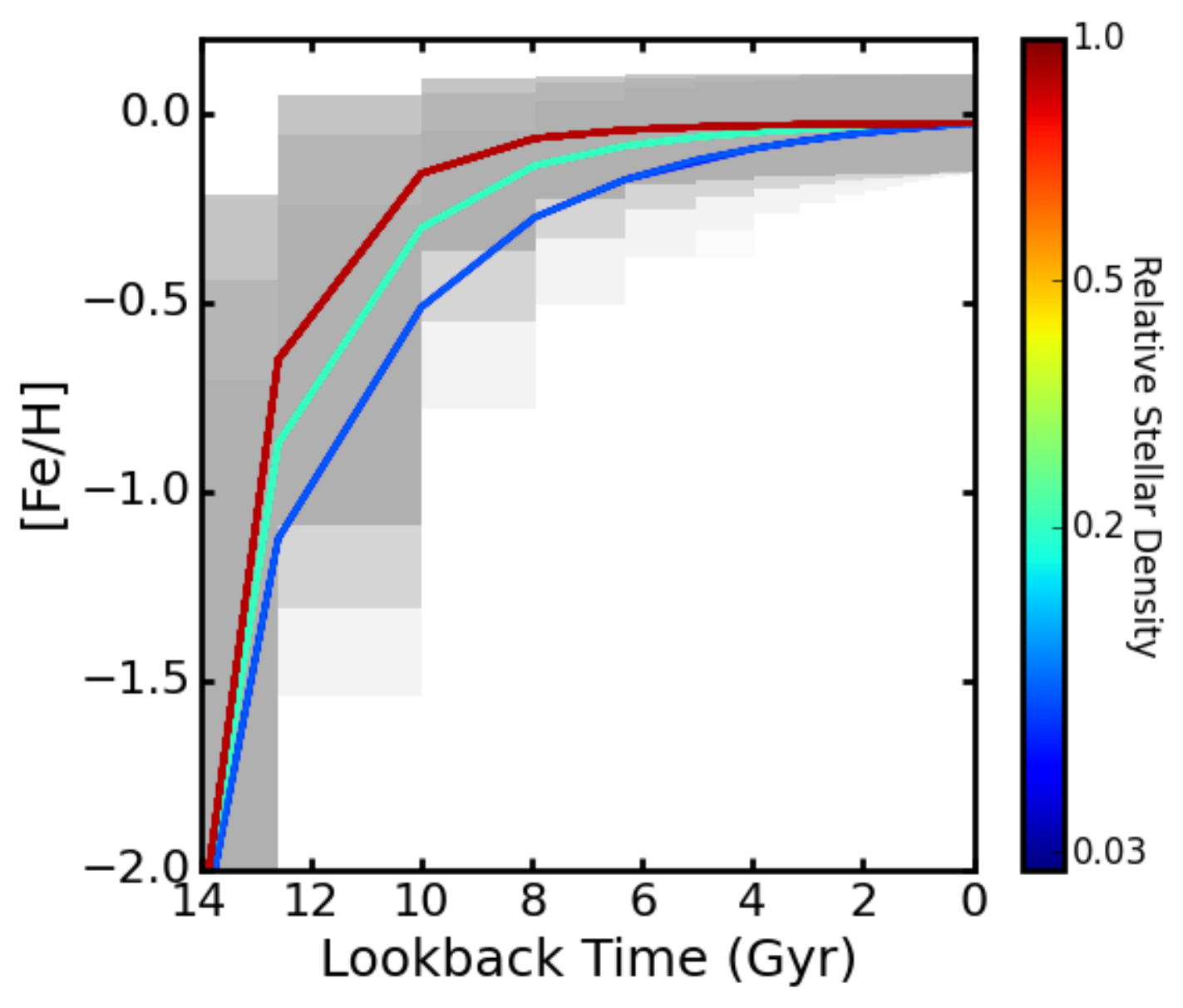}
\end{center}
\caption{Enrichment models adopted for three radial divisions, using the same color-coding as in Figure~\ref{stellar_density_map}.  Gray areas show the spread in metallicity allowed in each epoch.  The inner regions enrich faster earlier, while the outer regions have more constant enrichment.}
\label{enrichment_histories}
\end{figure*}

\begin{figure*}
\begin{center}
\includegraphics[width=5.1in]{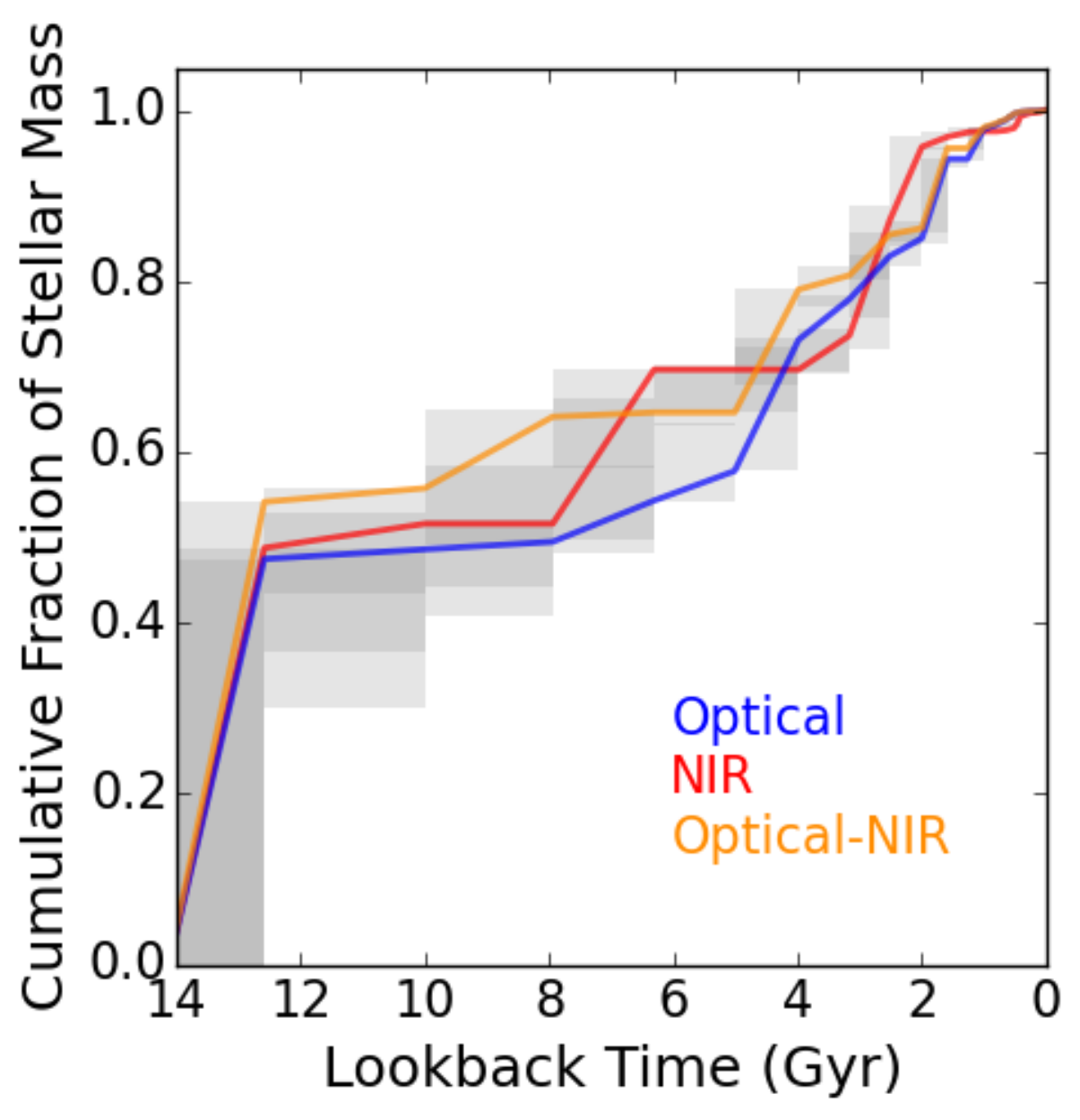}
\end{center}
\caption{Cumulative stellar mass with age measured for one of our
  regions by fitting the near-IR data only (red), the optical data
  only (blue), and simultaneously fitting both (orange). The gray
  shaded areas show the random uncertainties for each fit. Darker
  areas show where the uncertainties from the different fits
  overlap.  \label{compare_multiple_band_sfhs}}

\end{figure*}

\begin{figure*}
\begin{center}
\includegraphics[width=3.1in]{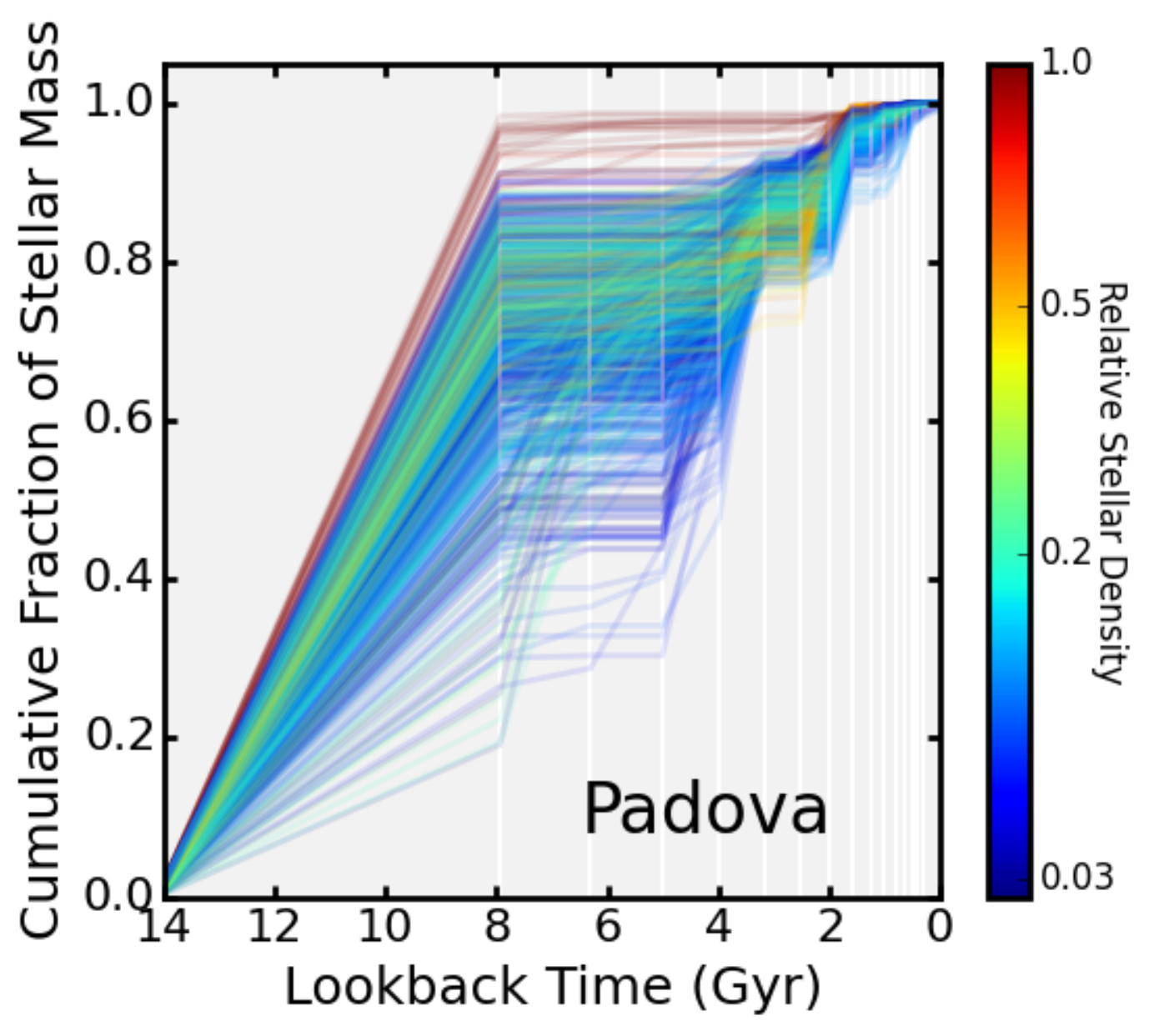}
\includegraphics[width=3.1in]{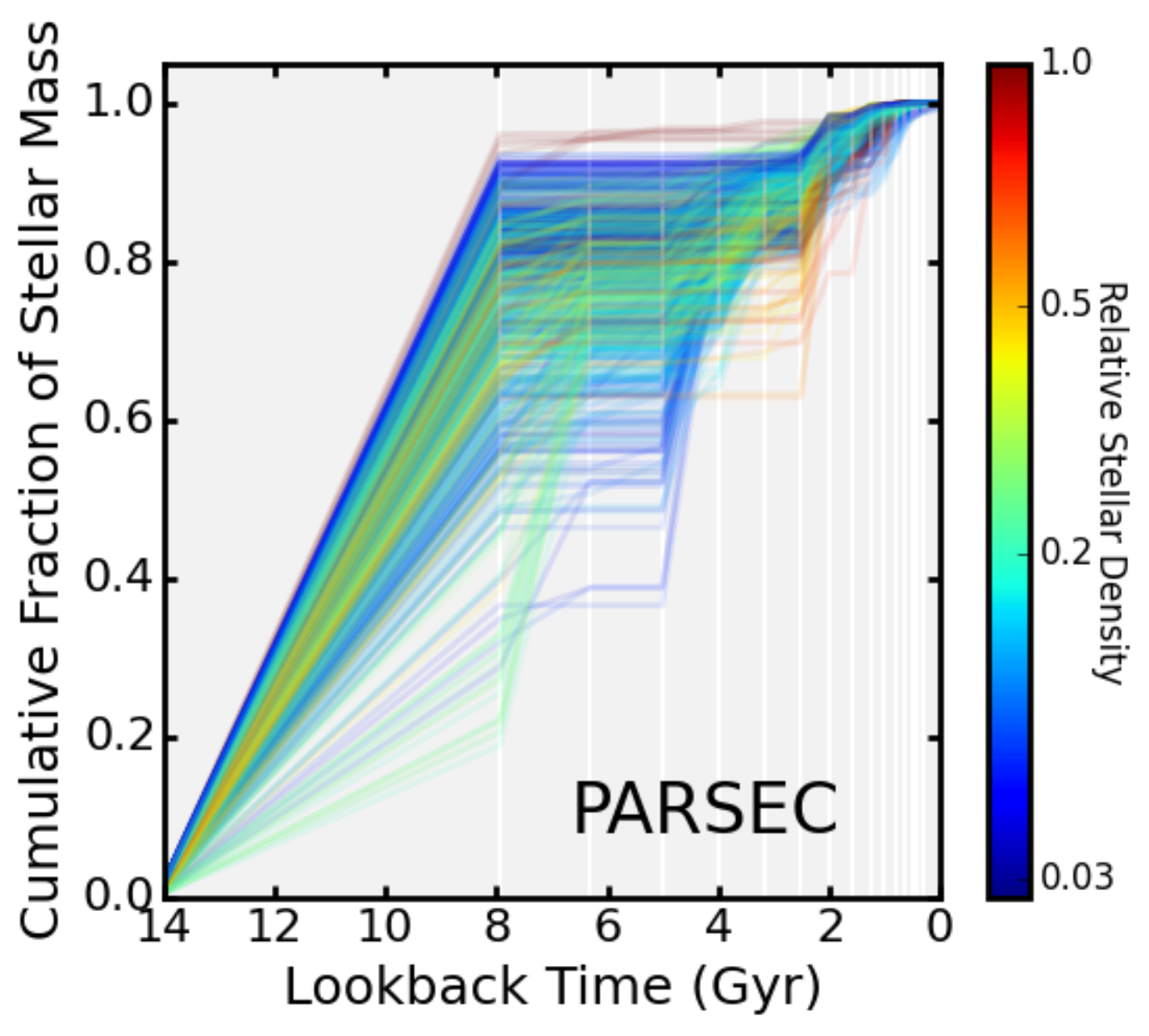}
\includegraphics[width=3.1in]{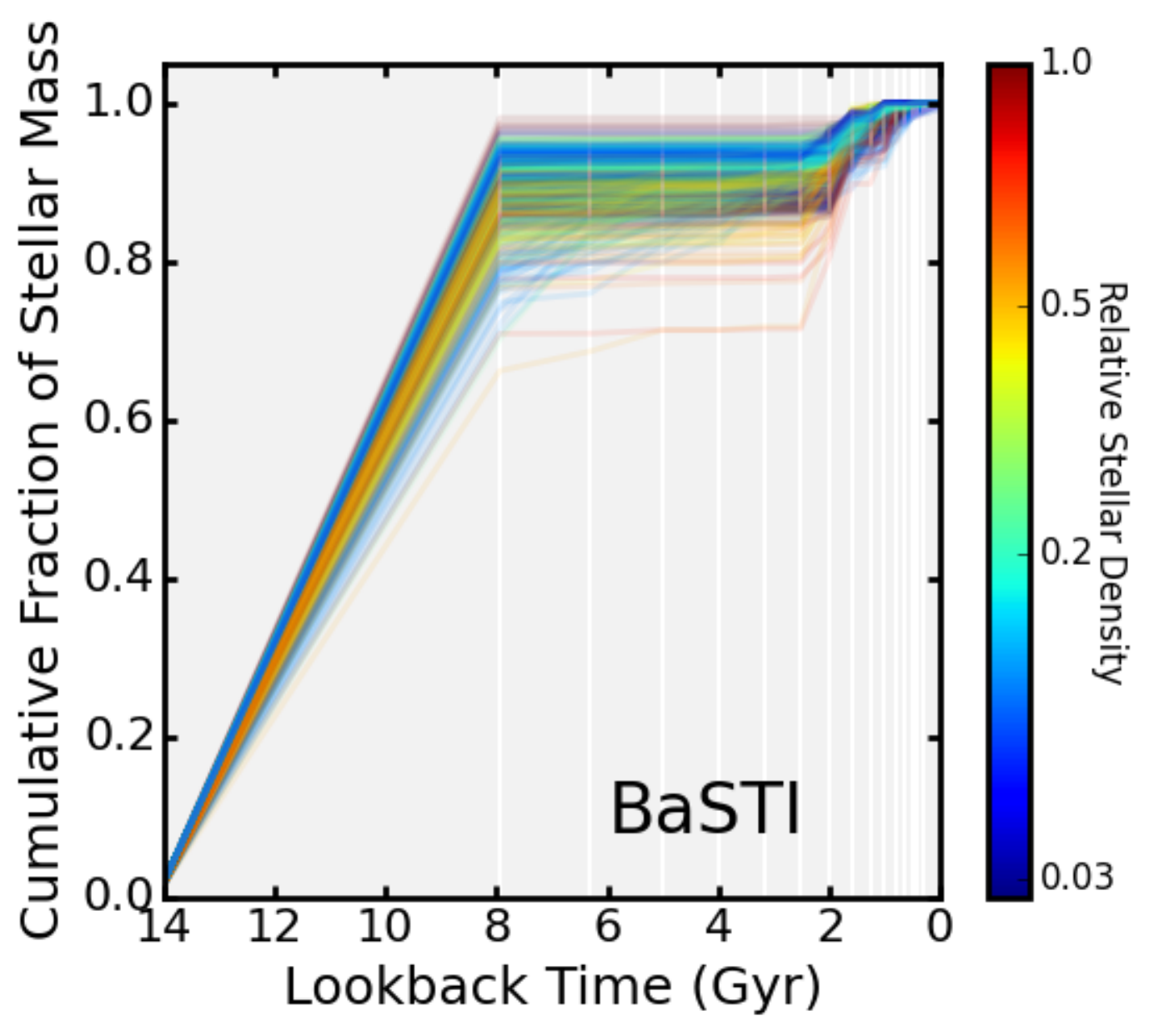}
\includegraphics[width=3.1in]{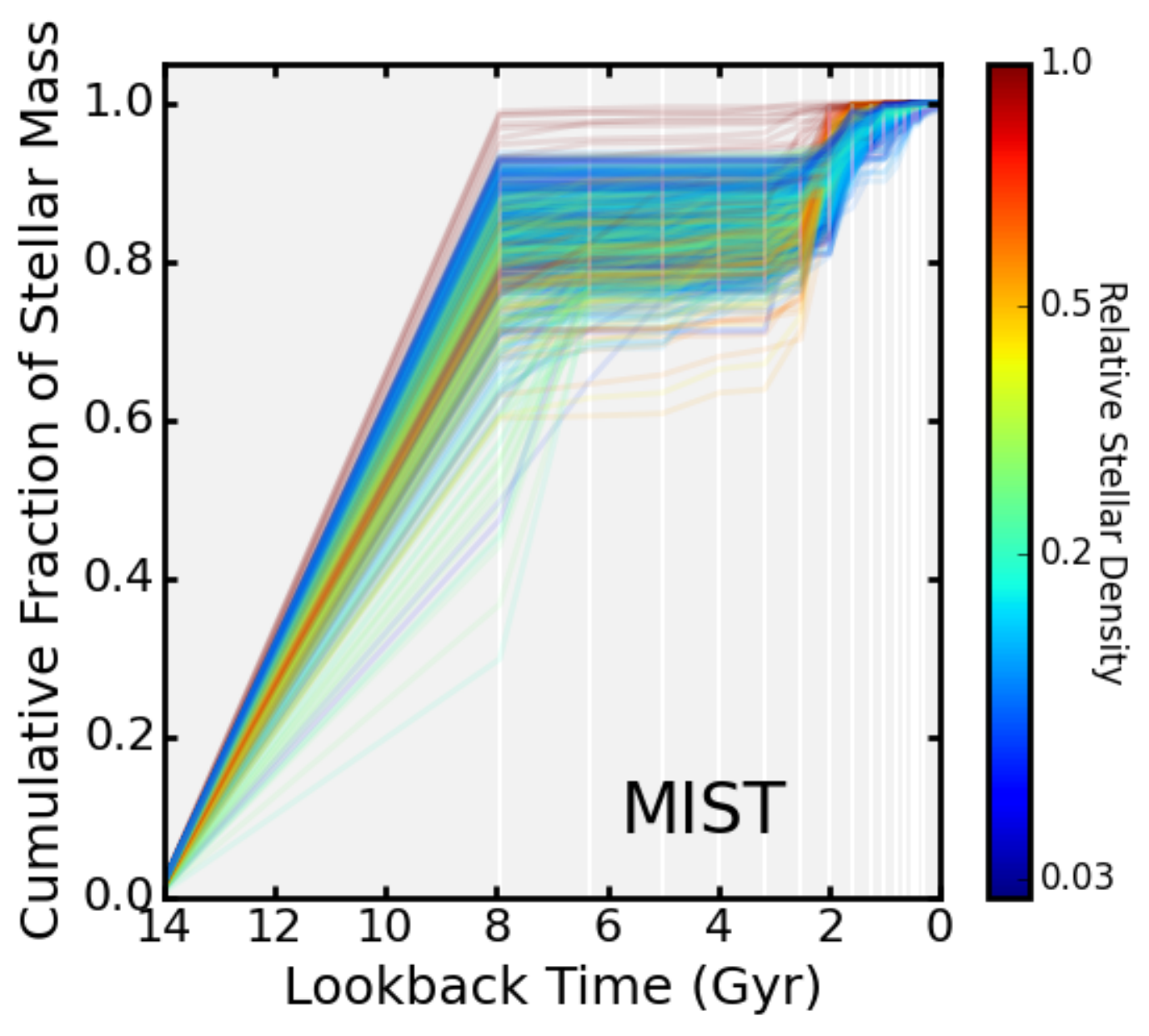}
\end{center}
\caption{Cumulative SFHs for the survey with different stellar
  evolution models for a forced enrichment scenario. Lines are
  color-coded by their values in Figure~\ref{stellar_density_map}, and
  panels are labeled with their corresponding isochrones set. \label{compare_model_sfhs} }

\end{figure*}

\clearpage

\begin{turnpage}
\begin{figure*}
\begin{center}
\includegraphics[width=3.0in]{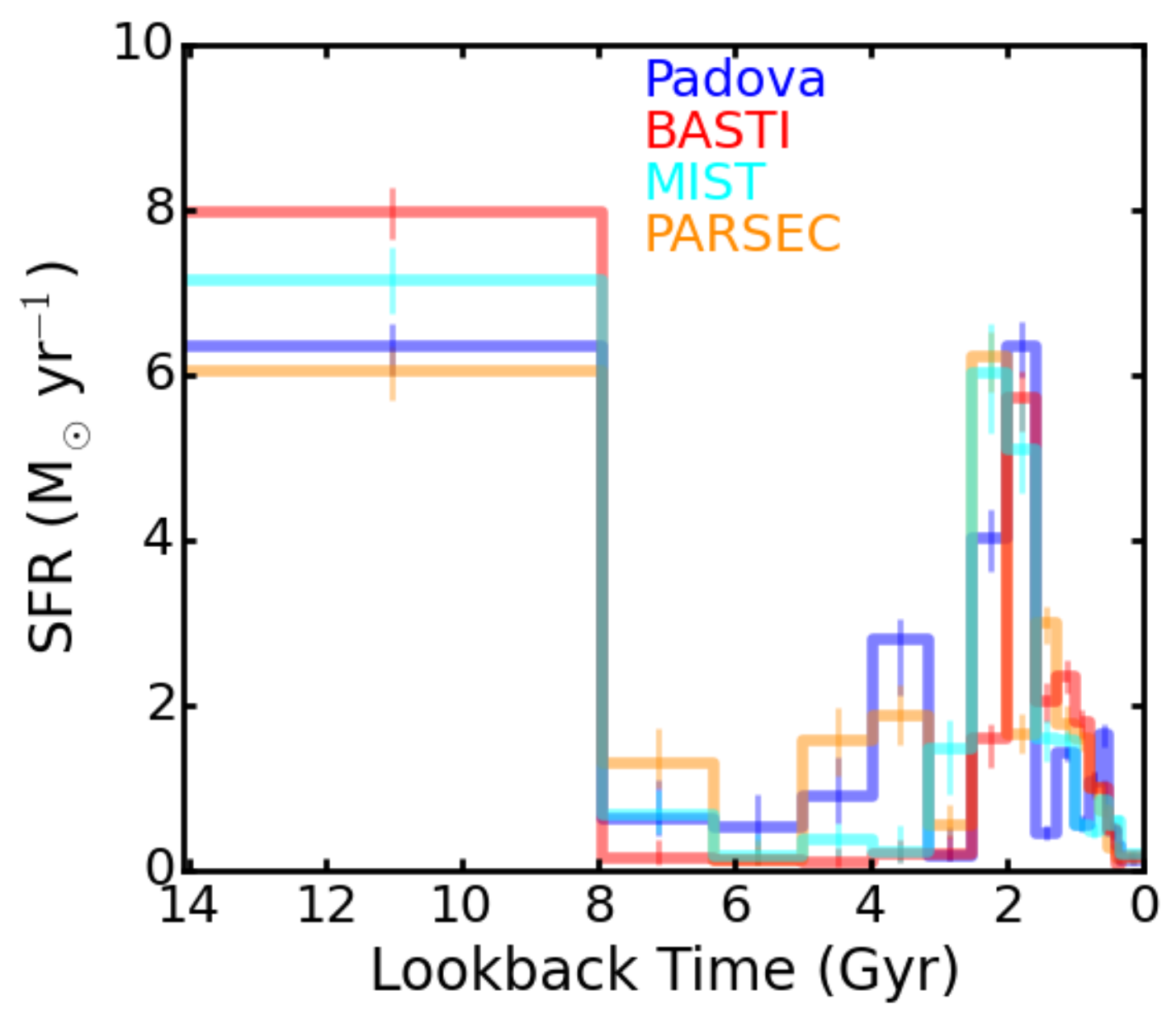}
\includegraphics[width=3.0in]{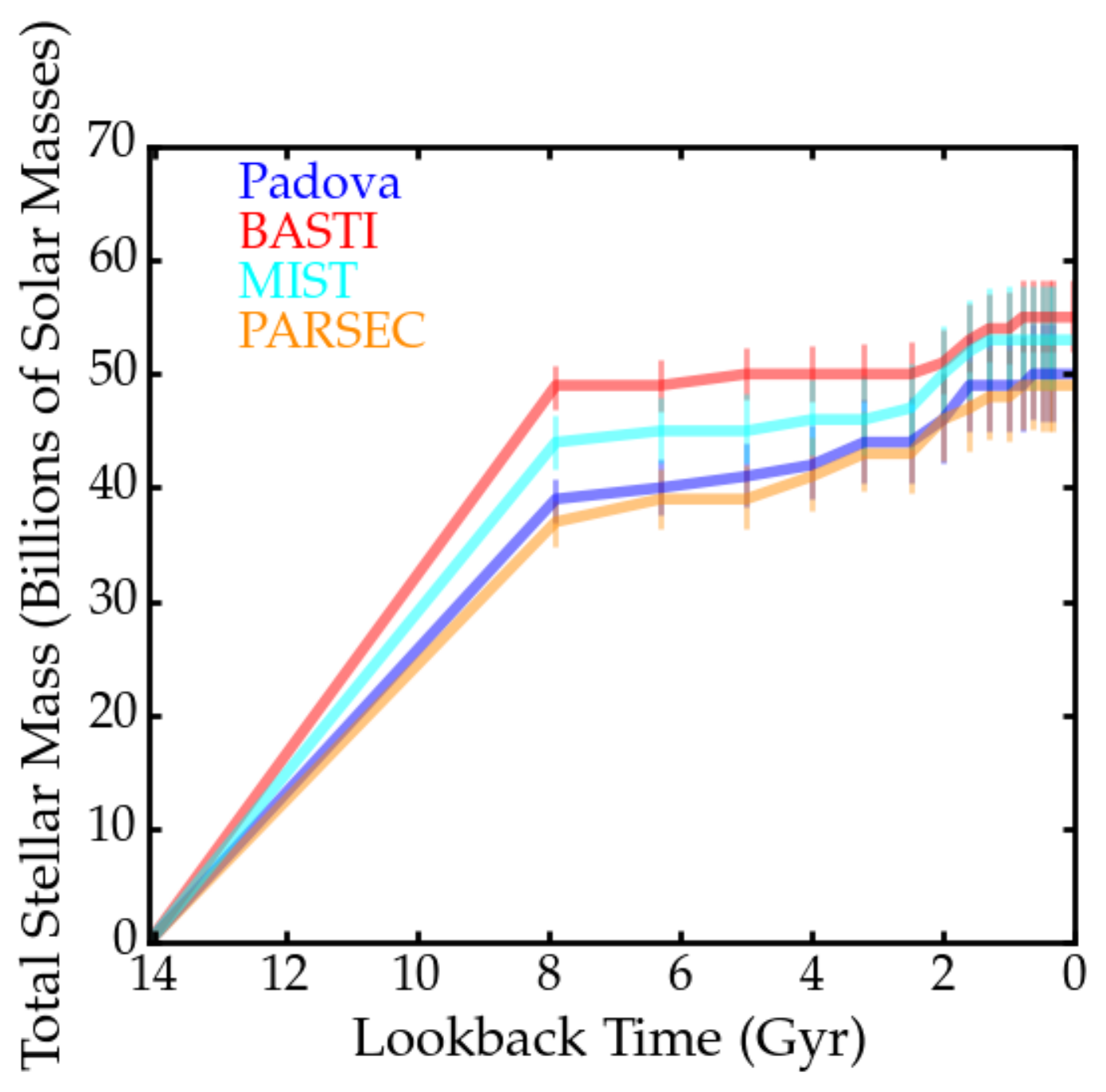}
\includegraphics[width=3.0in]{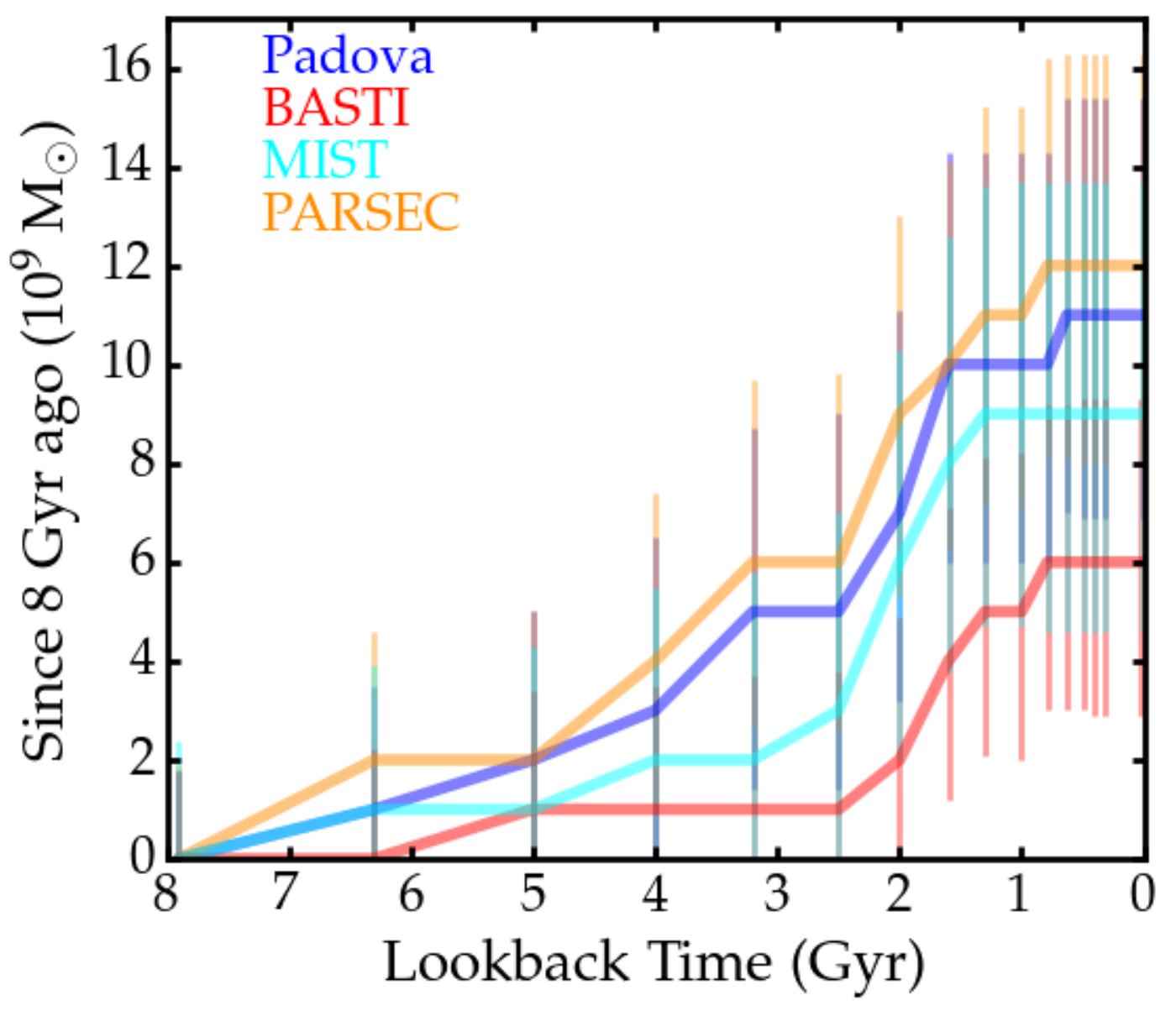}
\end{center}
\caption{{\it Left}: Total star formation rate vs. age for the entire
  PHAT footprint, as measured using 4 different model sets.  {\it
    Middle}: The cumulative stellar mass formed as a function of time. {\it Right}: Same as {\it Middle}, but starting removing the oldest bin of star formation to more clearly compare the results at younger ages.  \label{total_sfh}}
\end{figure*}

\end{turnpage}

\begin{figure*}
\begin{center}
\includegraphics[width=5.1in]{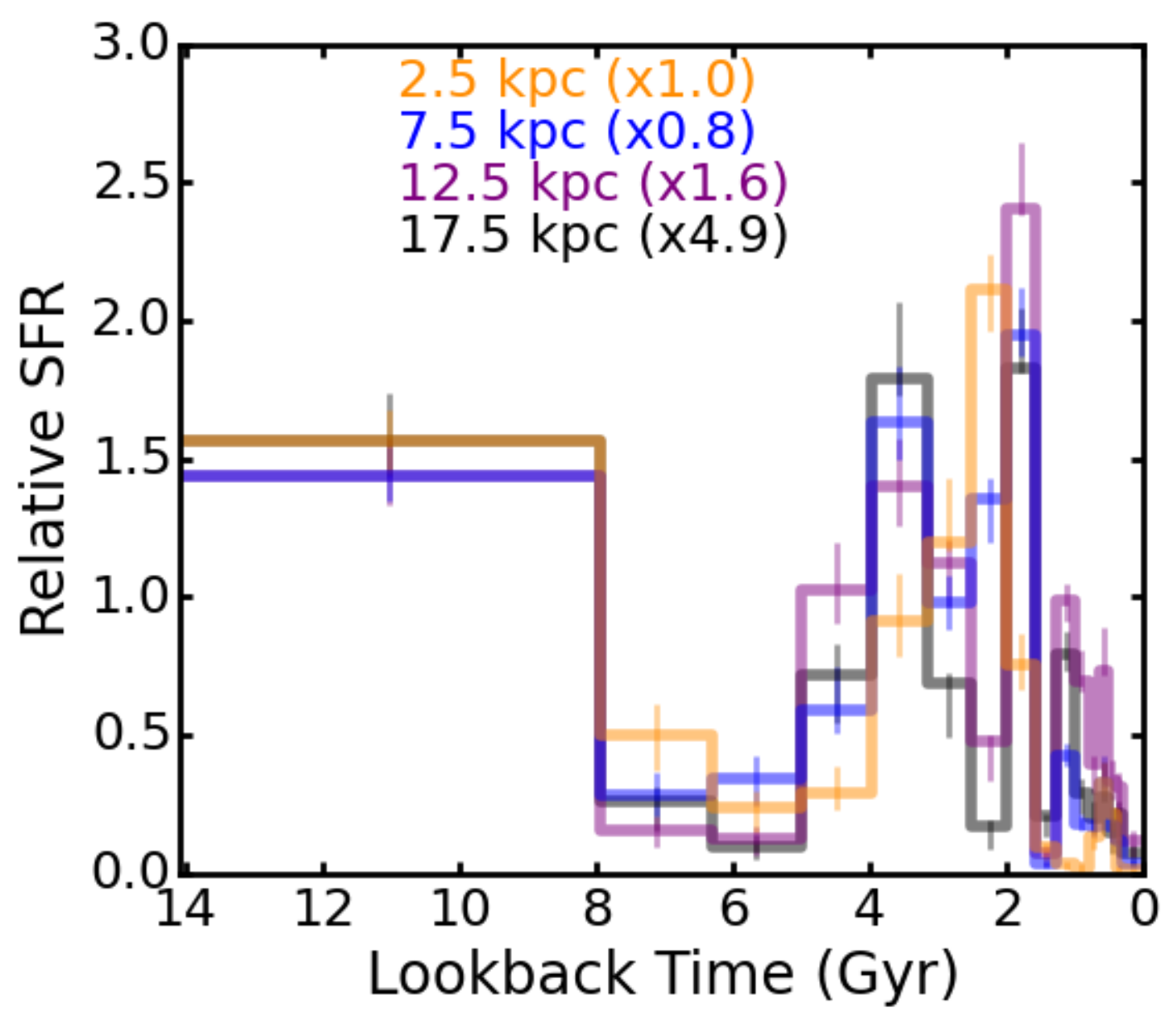}
\end{center}
\caption{Total relative star formation rate vs. age from the Padova
  model fits for 4 radial bins, color-coded by the radius used for the
  measurements, normalized to have the same mean rate of 1
    M$_{\odot}$~yr$^{-1}$. \label{radial_sfhs}}

\end{figure*}


\begin{figure*}
\begin{center}
\includegraphics[width=3.1in]{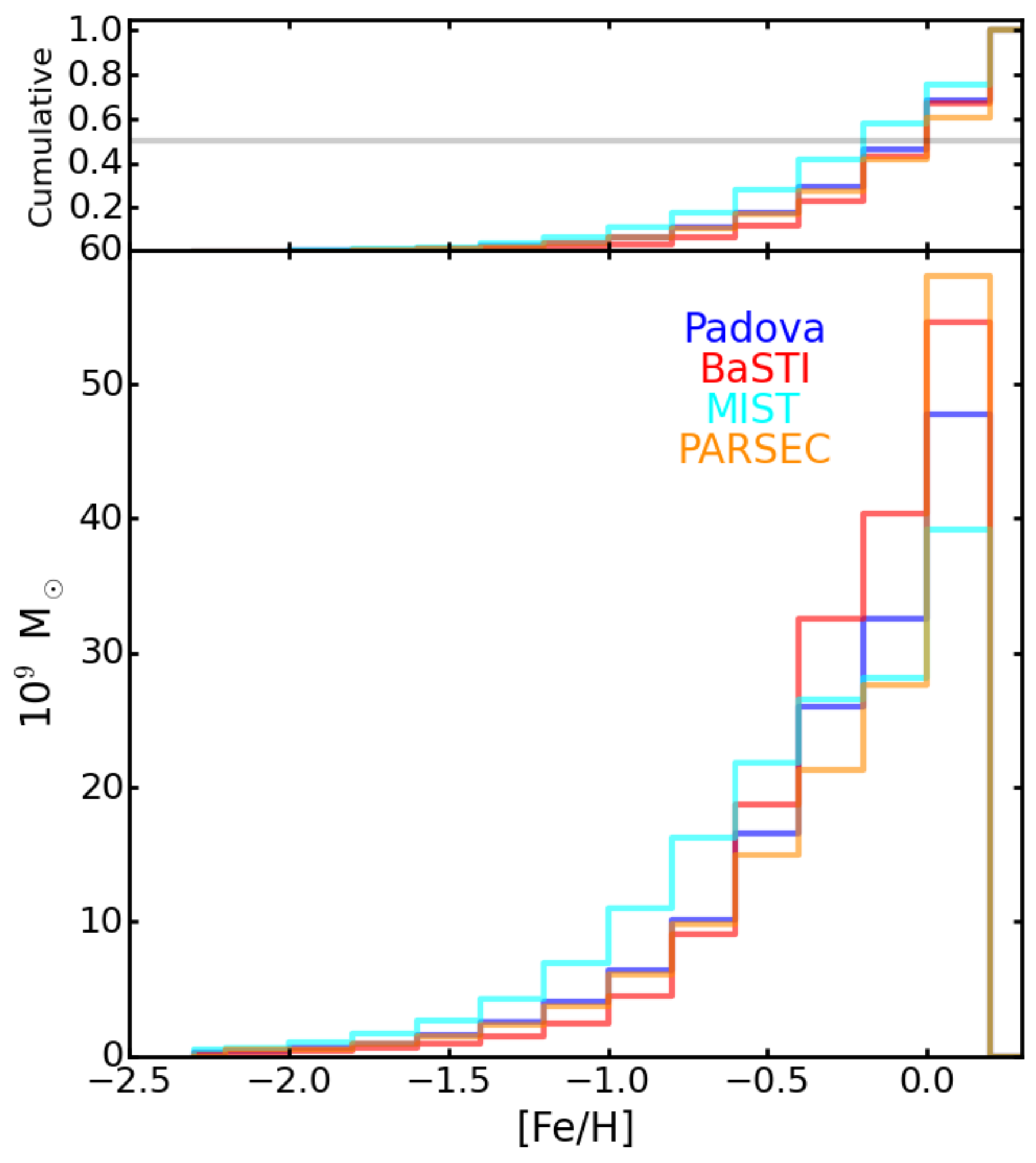}
\includegraphics[width=3.1in]{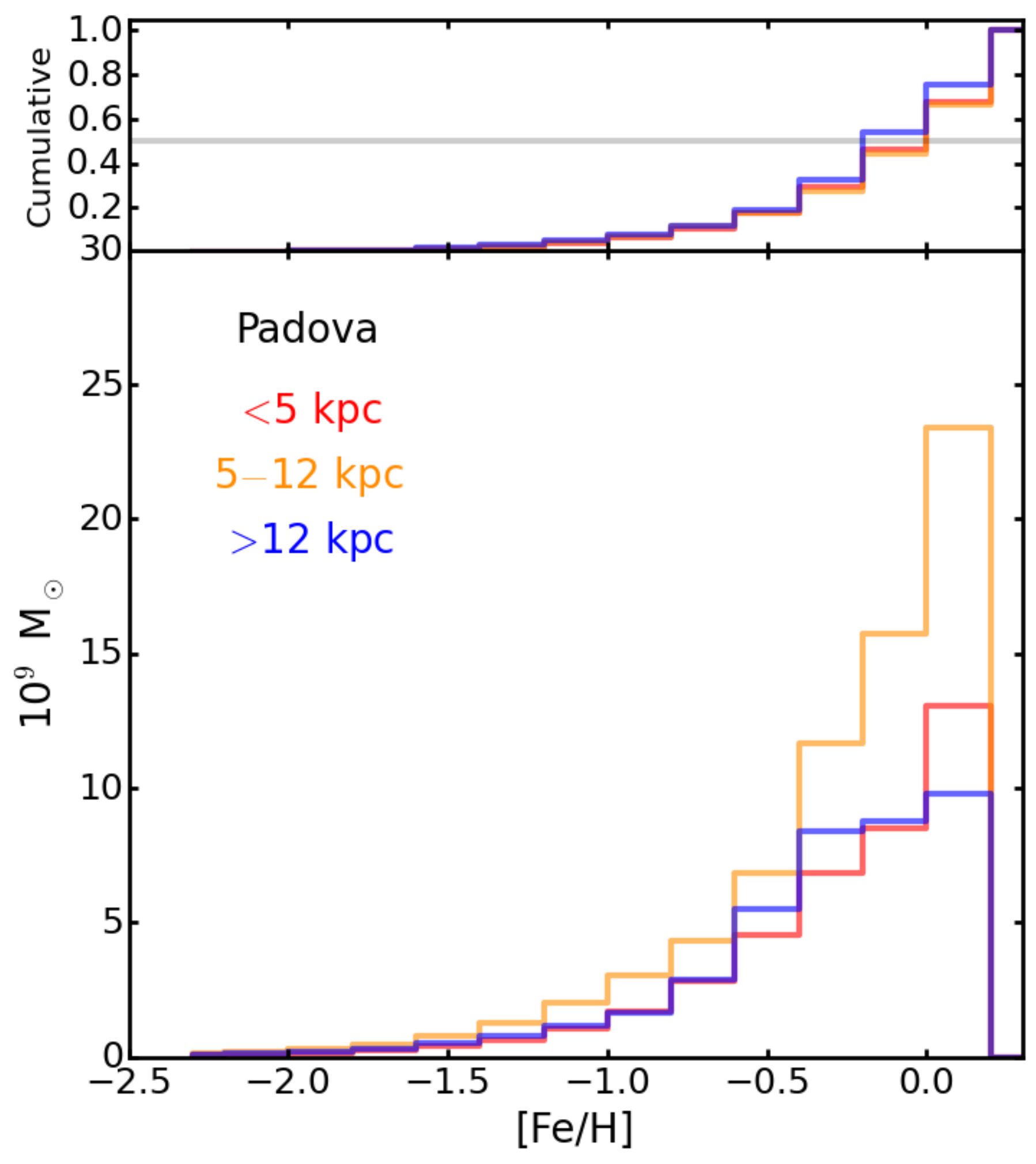}
\end{center}
\caption{{\it Left:} Metallicity histograms for the entire M31 disk when using our
  fixed enrichment model, extrapolated from the entire area measured.
  The cumulative fraction of stars as a function on metallicity is
  included in the top panel.  
  The distribution for all 4 model sets is shown. {\it Right:}
  Metallicity histograms for 3 radial bins when using our fixed
  enrichment model and fitting with the Padova stellar evolution
  models, correcting for area coverage. The cumulative fraction of stars as a function on metallicity is
  included in the top panel.   }
\label{z4hist}
\end{figure*}

\begin{figure*}
\begin{center}
\includegraphics[width=3.1in]{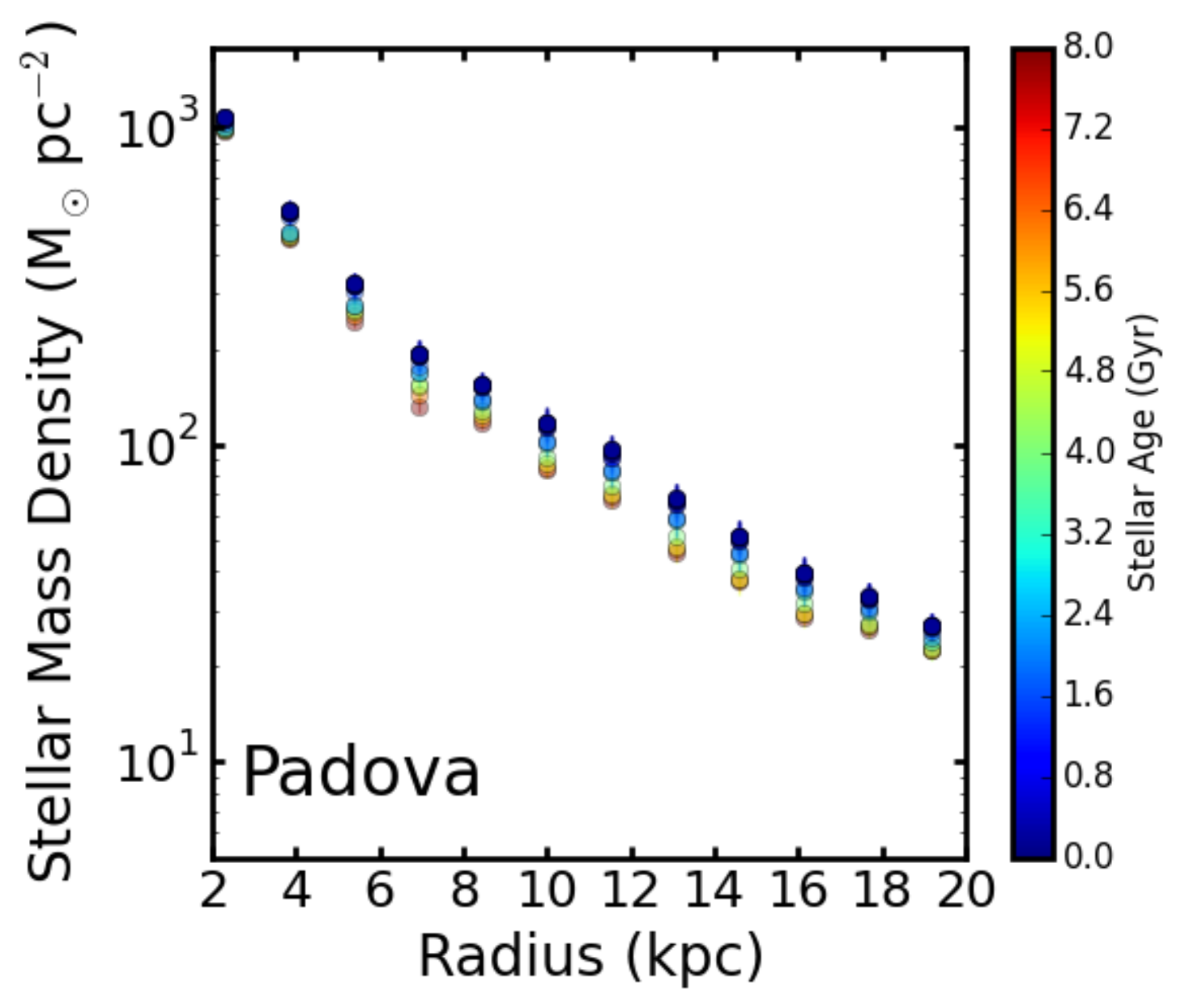}
\includegraphics[width=3.1in]{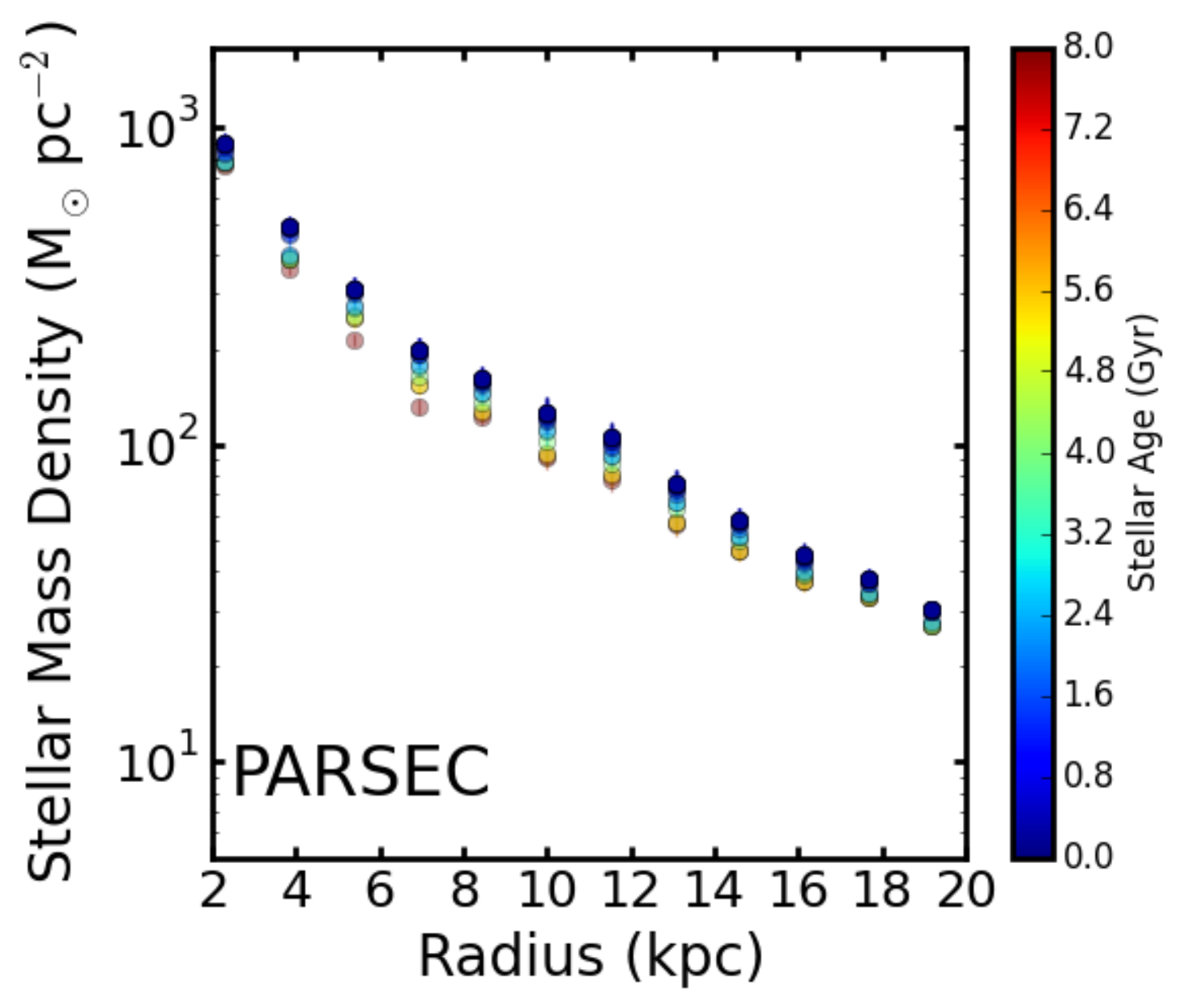}
\includegraphics[width=3.1in]{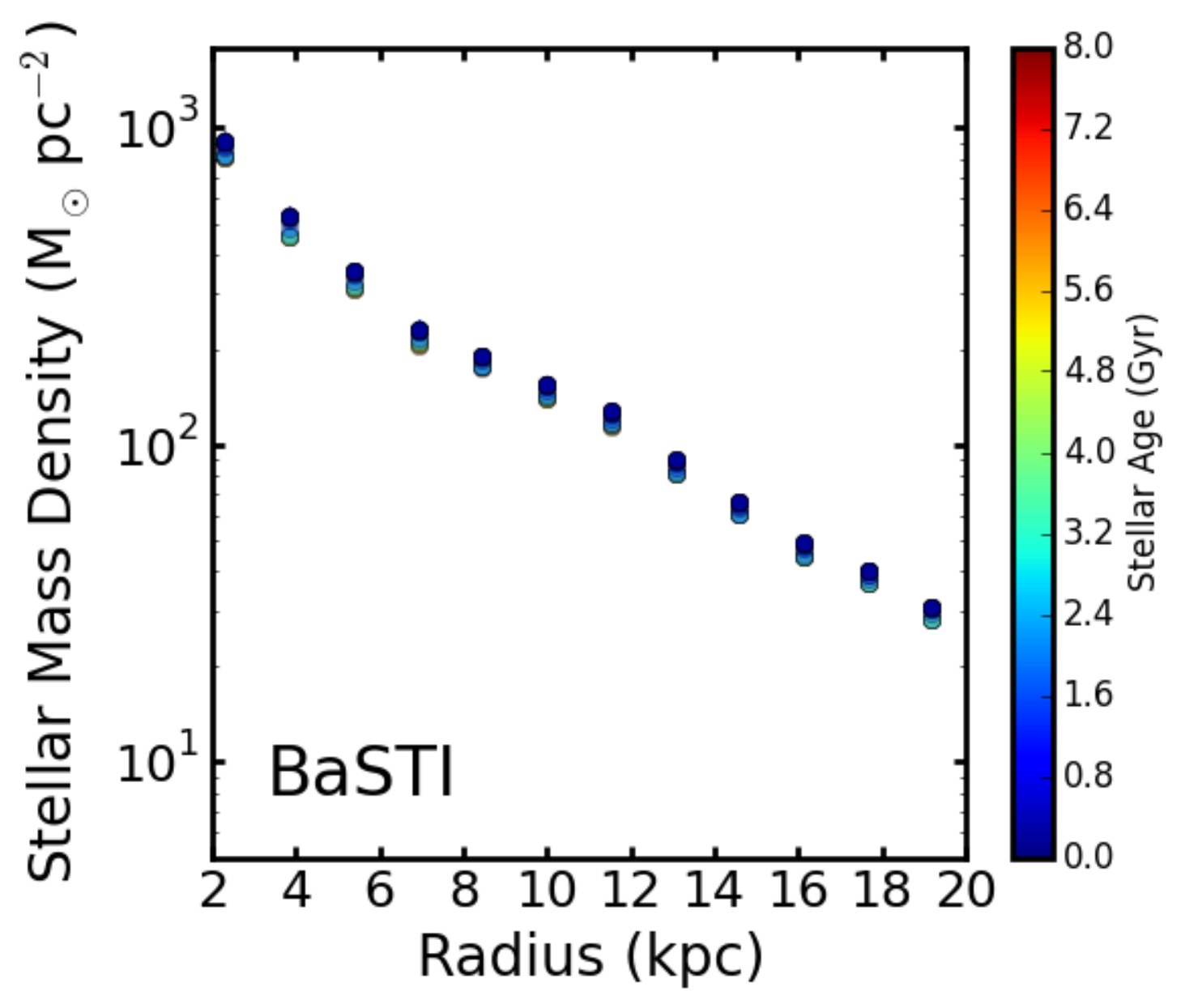}
\includegraphics[width=3.1in]{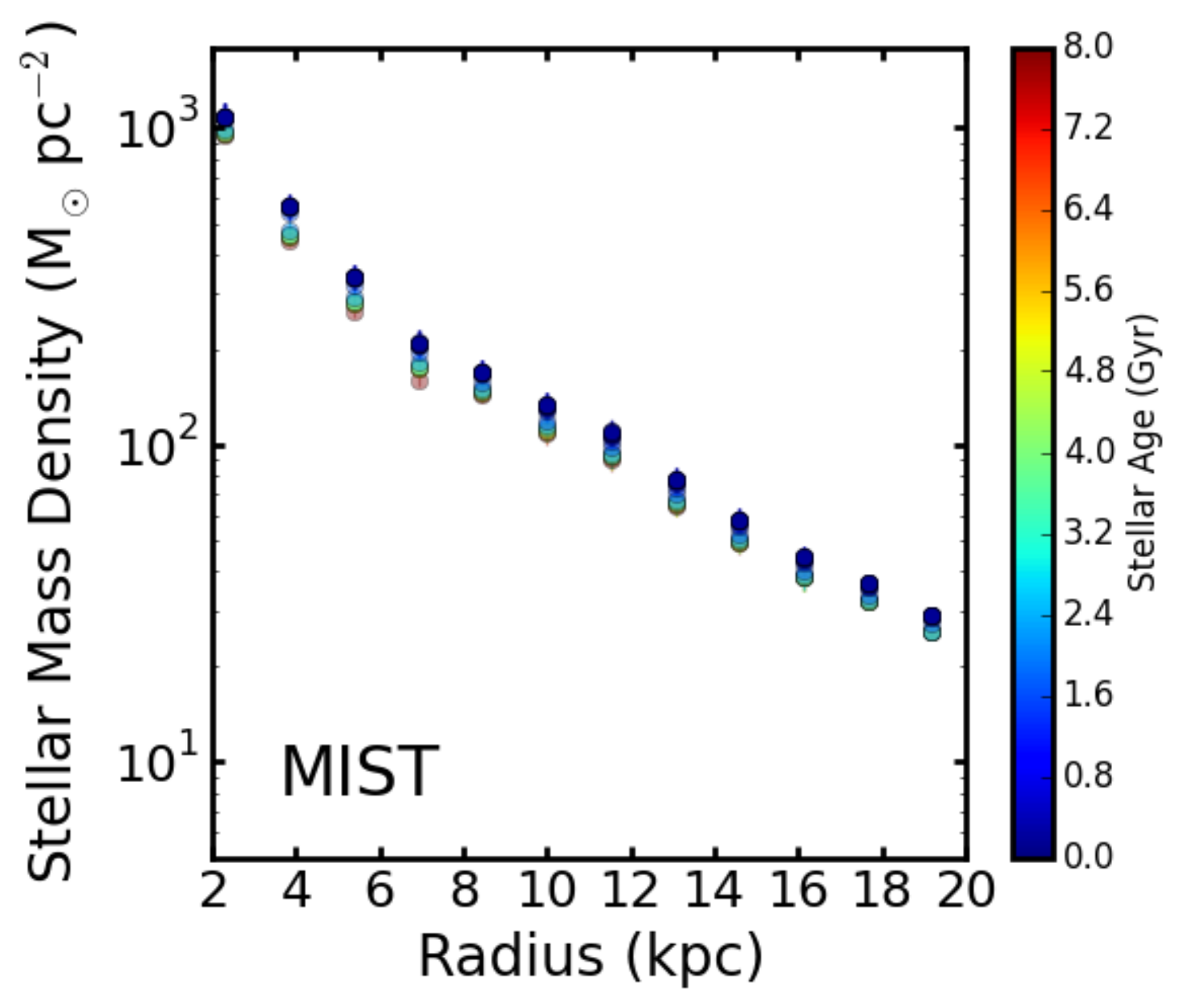}
\end{center}
\caption{Cumulative stellar mass surface density formed as a function
  of radius at several ages (lookback times), as calculated from the
  fits to 4 model sets. The error bars are the linear sum of the random
uncertainties on the total star formation in each spatial bin in the
annulus at each age.  {\it Upper-left} The mass profile obtained for
each age from the fits to the Padova models.  {\it Other panels:} Same
as upper-left from the other model fits.  Each panel is labeled with the appropriate model set name. \label{density_vs_radius}}

\end{figure*}


\begin{figure*}
\begin{center}
\includegraphics[width=3.1in]{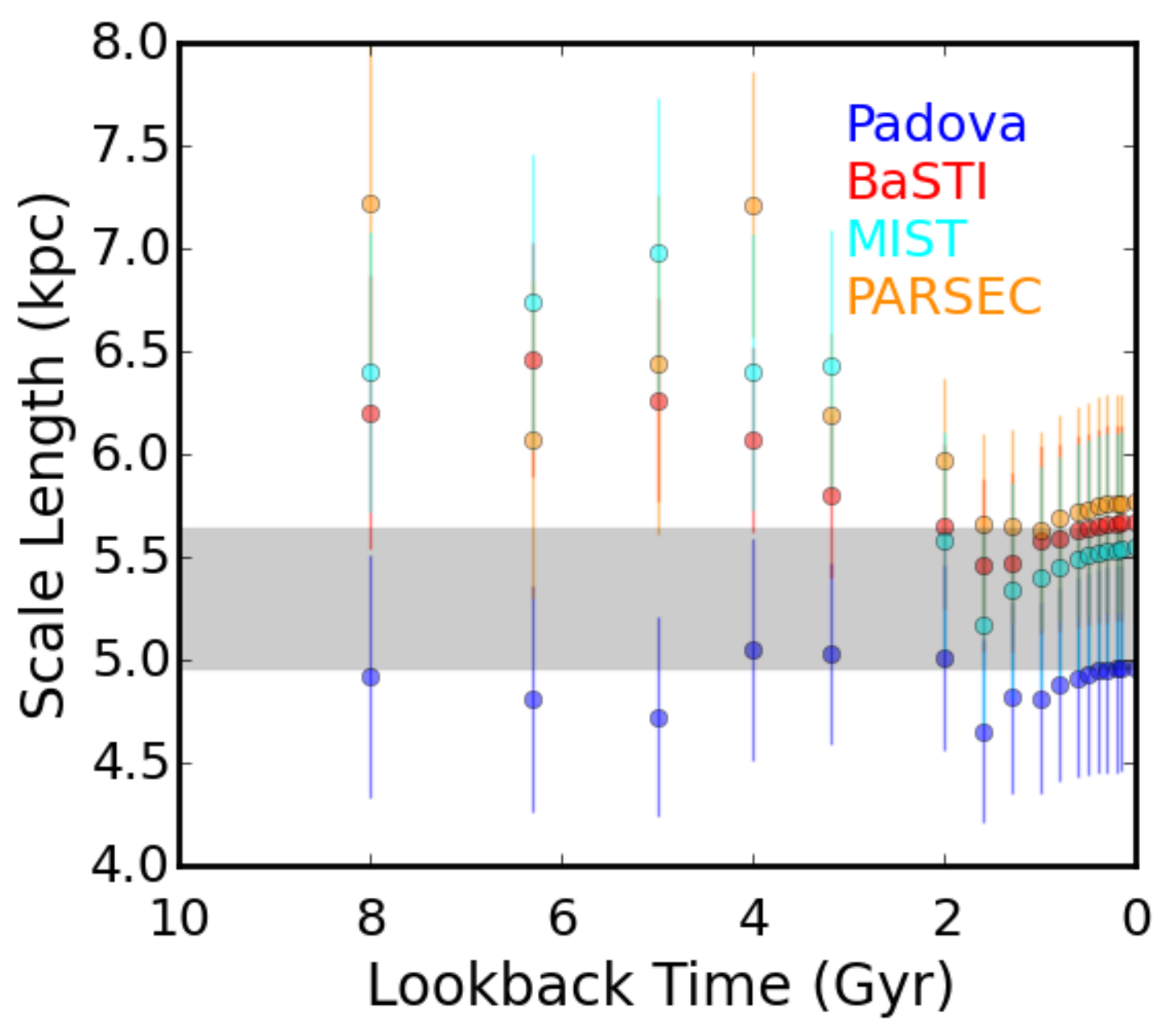}
\includegraphics[width=3.1in]{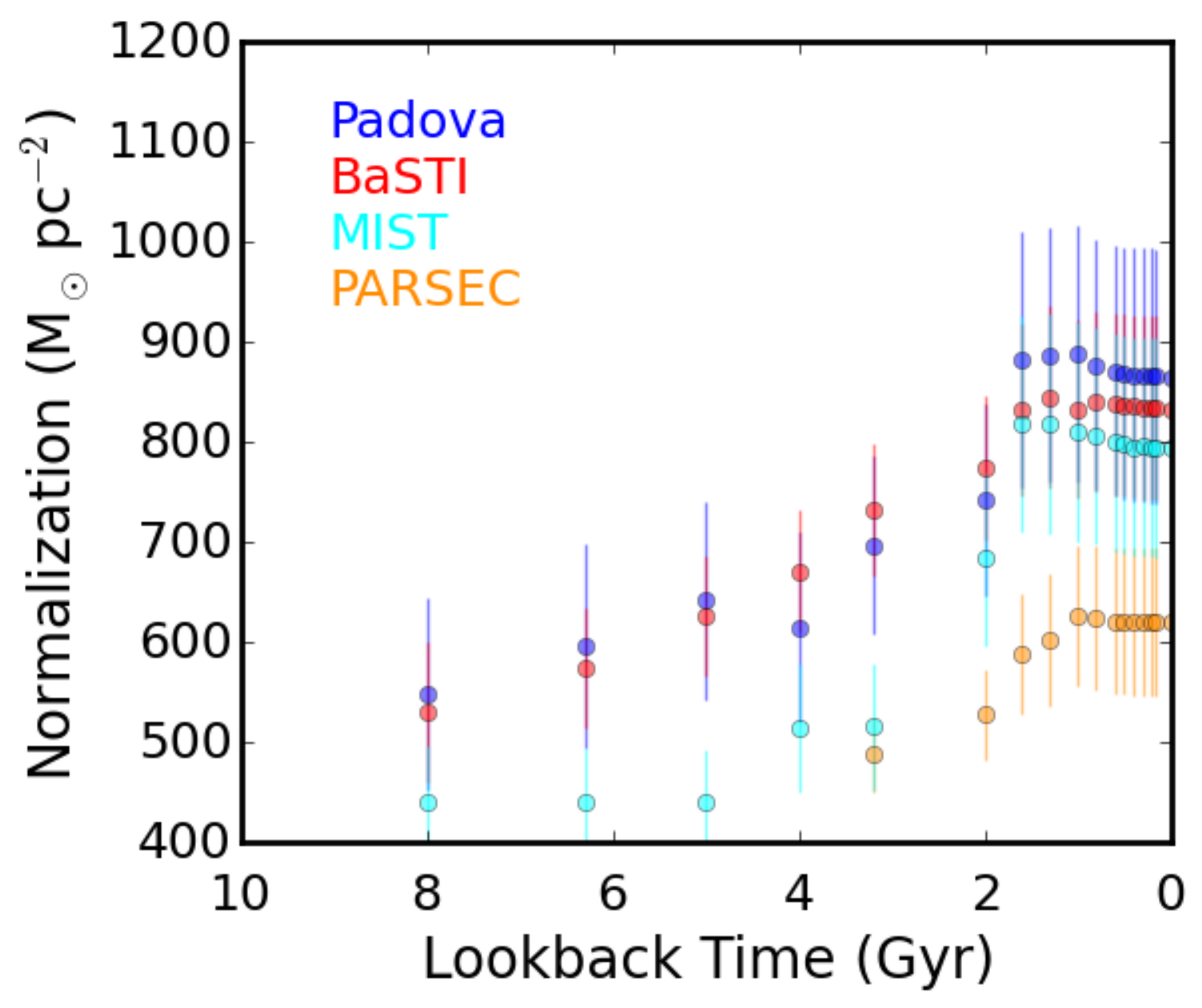}
\end{center}
\caption{Exponential fits to the disk profile of M31 beyond 4~kpc as a
  function of lookback time as calculated from our fits using the
  scipy function {\tt curve\_fit}.  {\it Left:}
  Disk scale length as a function of lookback time. {\it Right:}
  Normalization of the exponential disk as a function of lookback
  time.}
\label{scale_norm}
\end{figure*}

\begin{figure*}
\begin{center}
\includegraphics[height=2.2in]{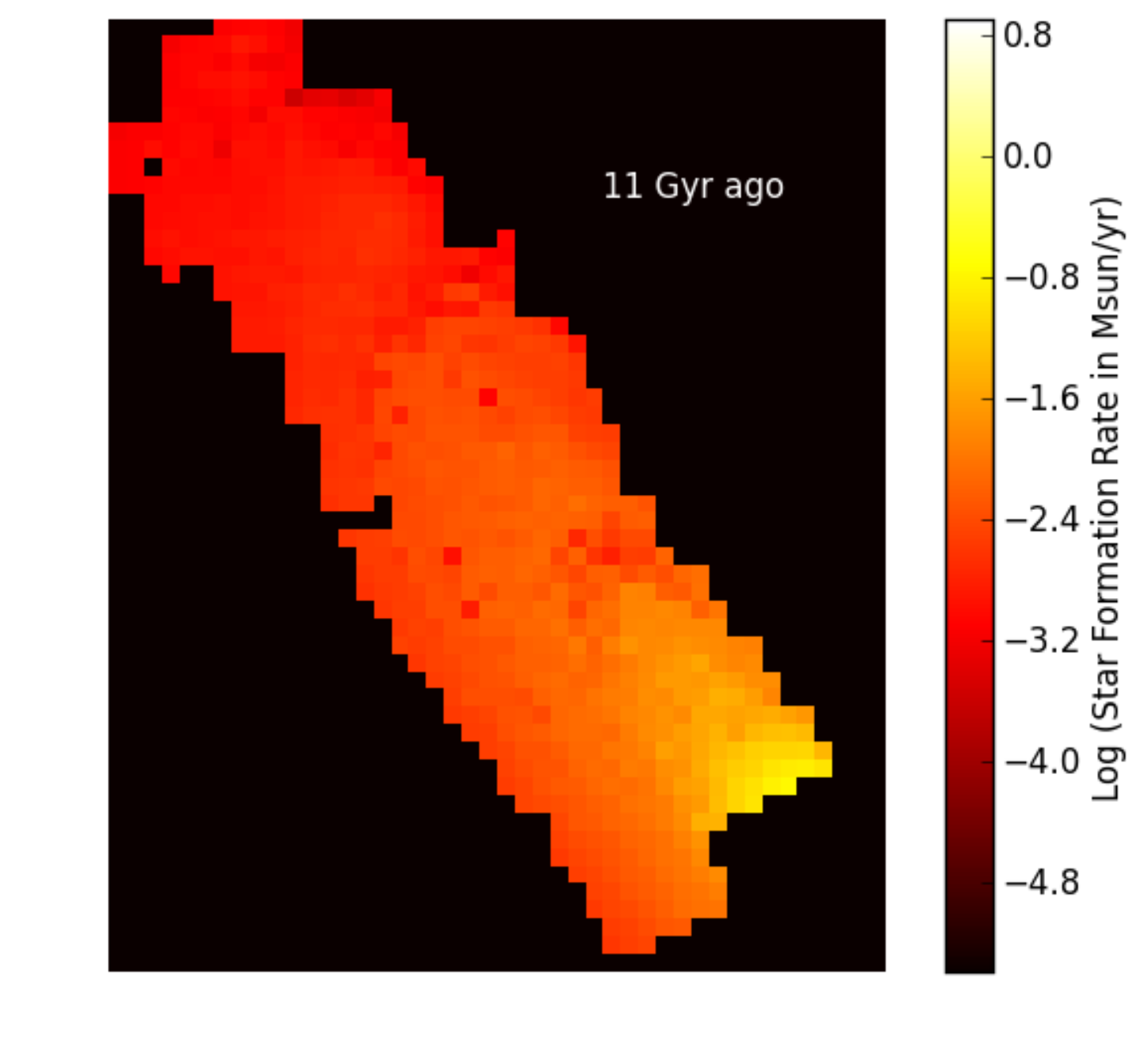}
\includegraphics[height=2.2in]{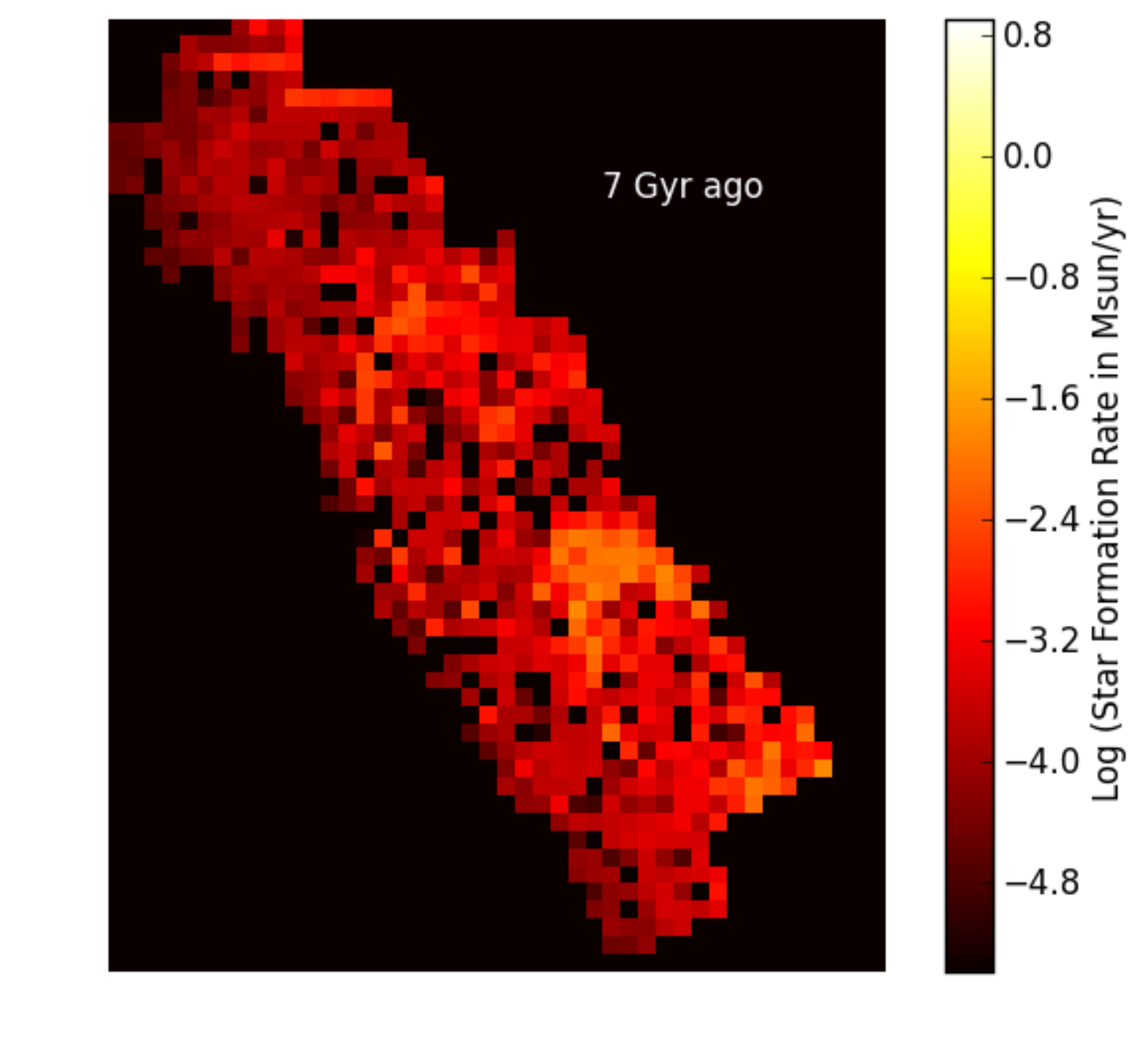}
\includegraphics[height=2.2in]{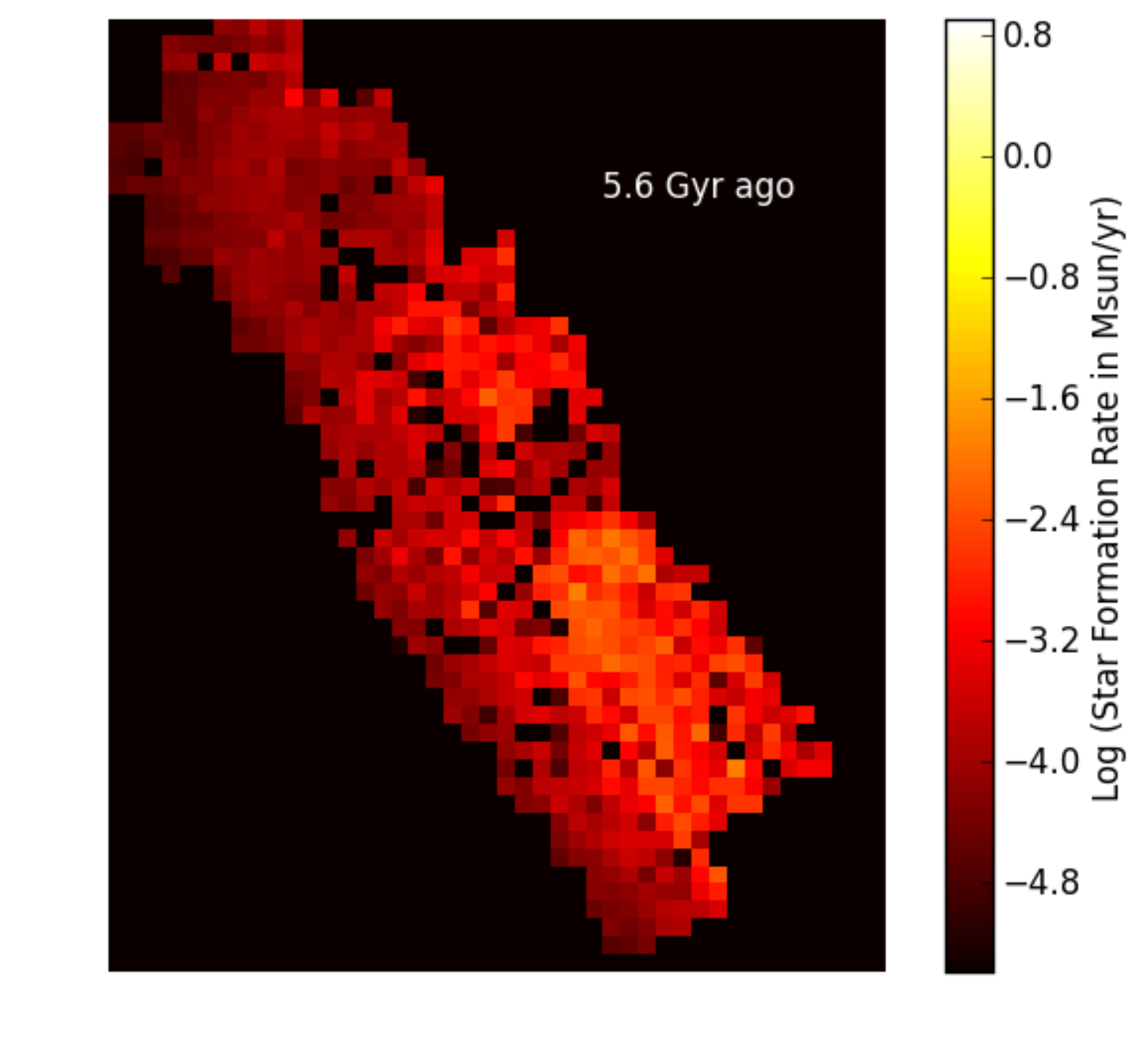}
\includegraphics[height=2.2in]{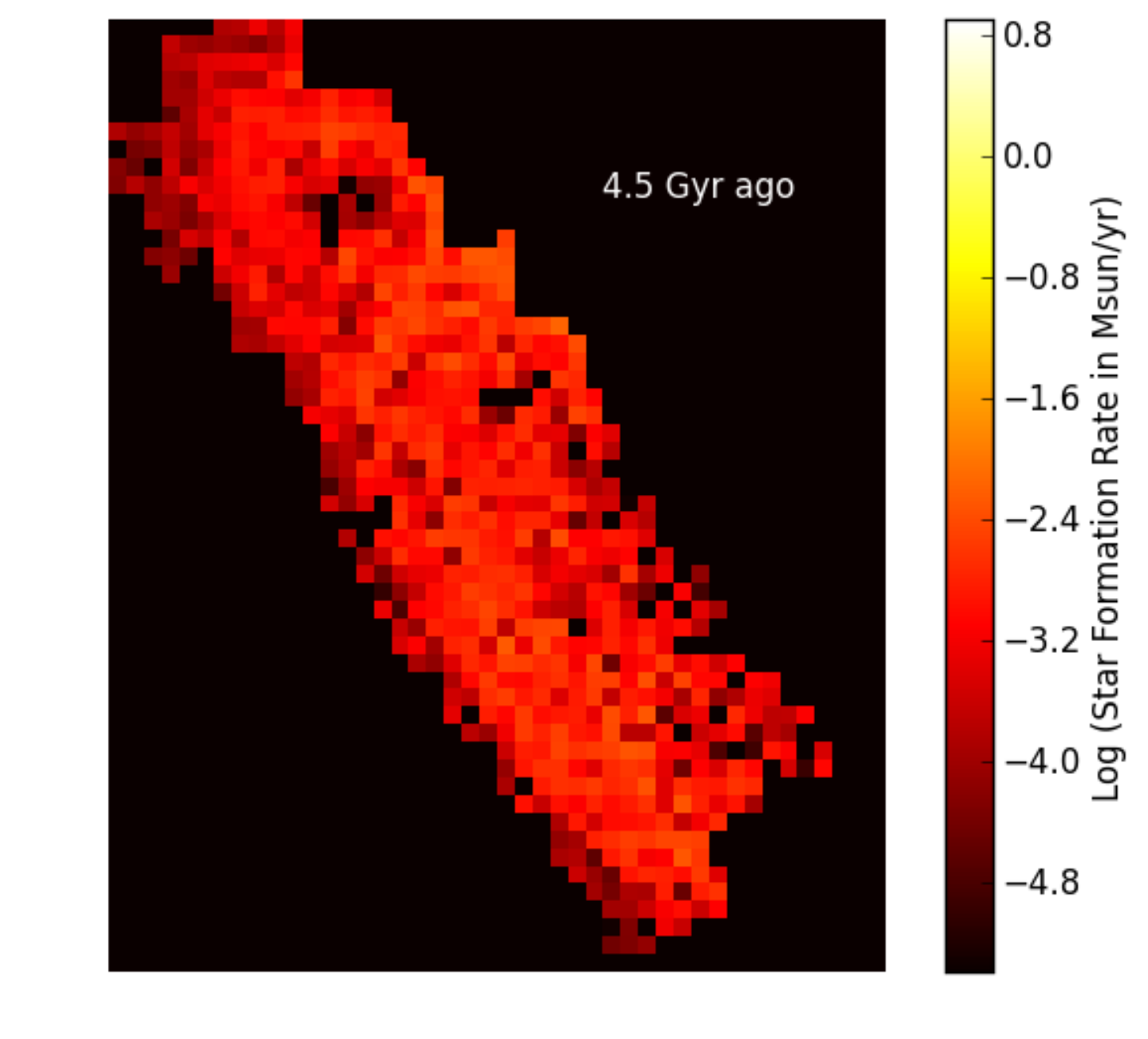}
\includegraphics[height=2.2in]{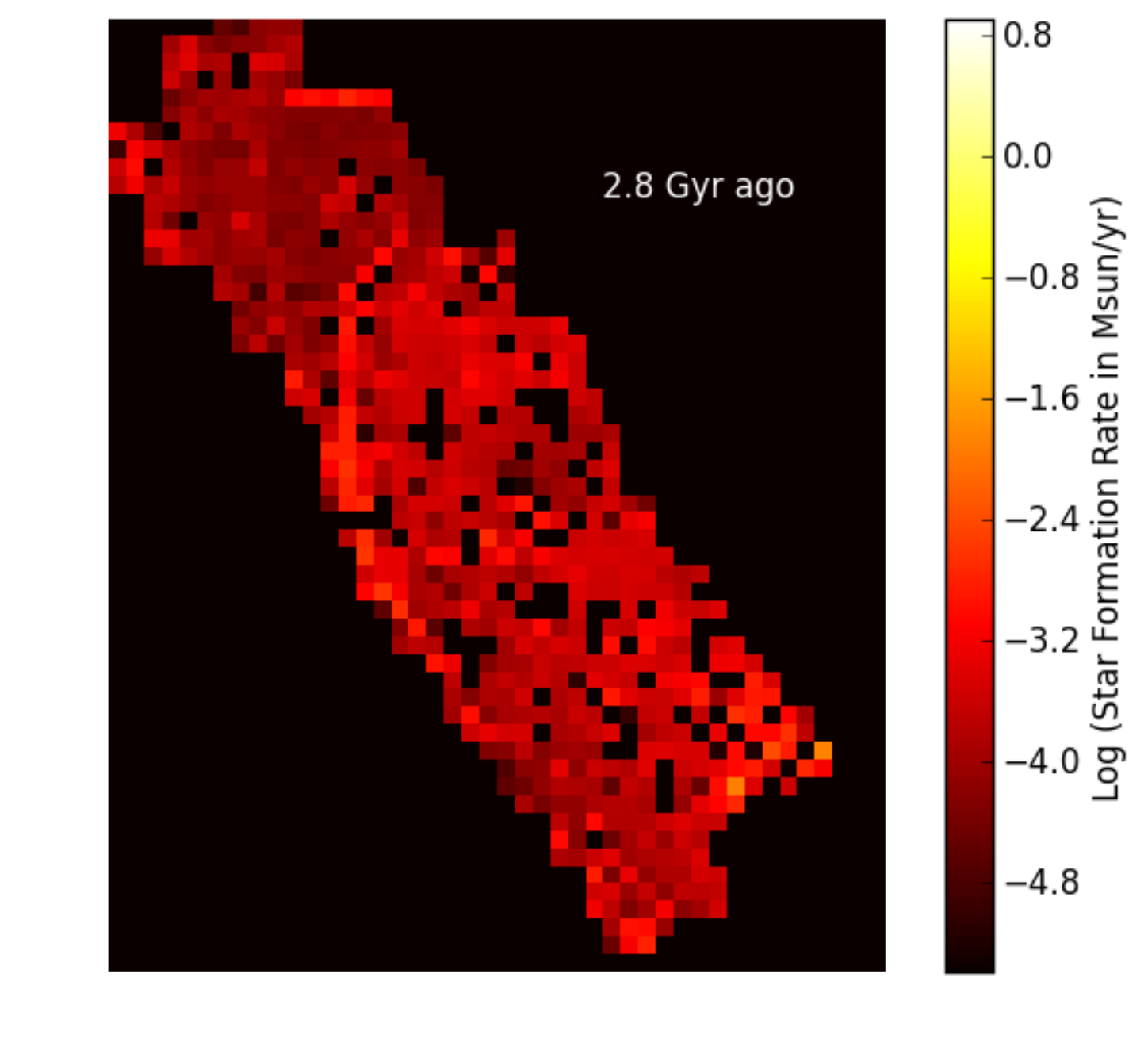}
\includegraphics[height=2.2in]{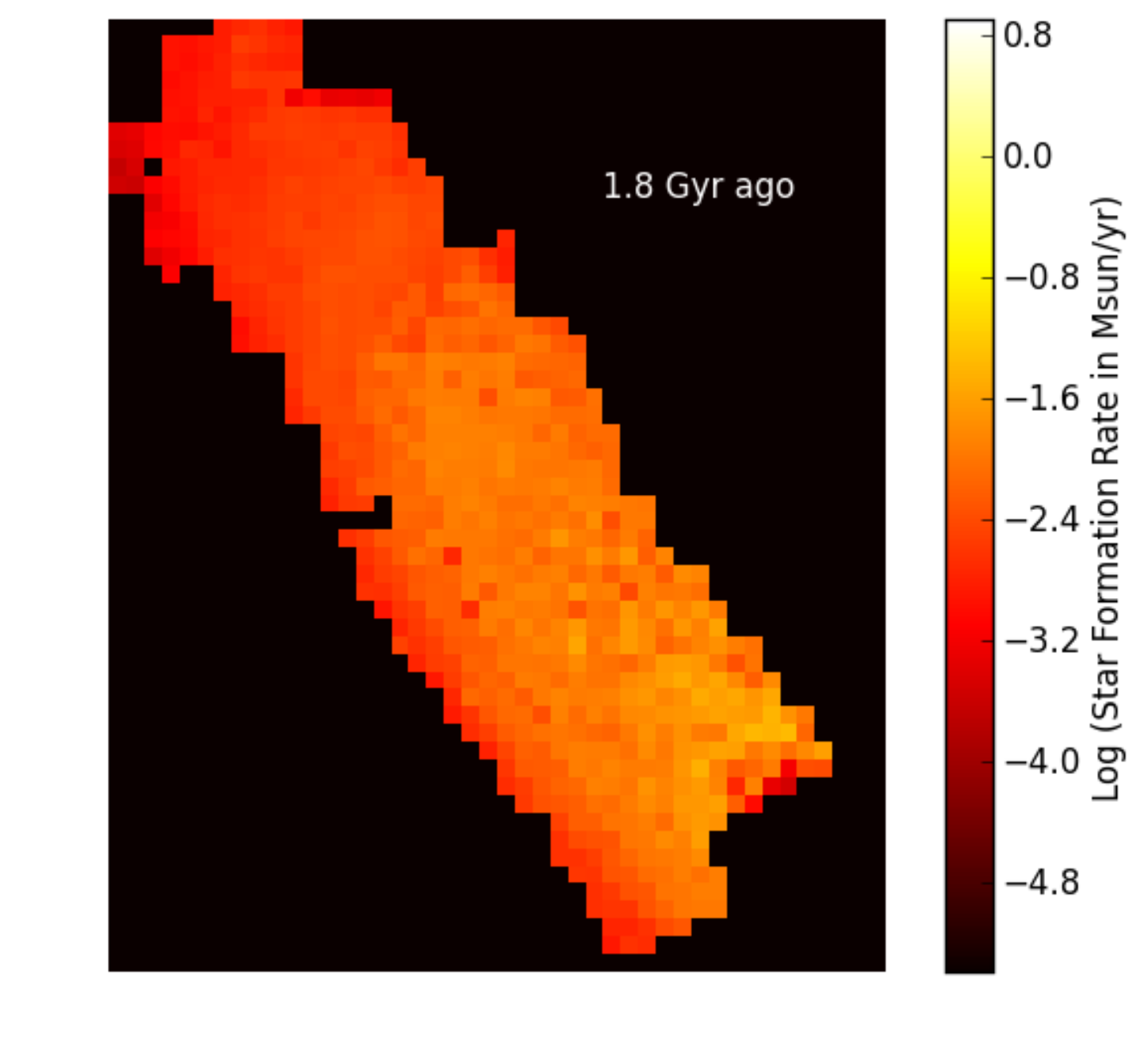}
\includegraphics[height=2.2in]{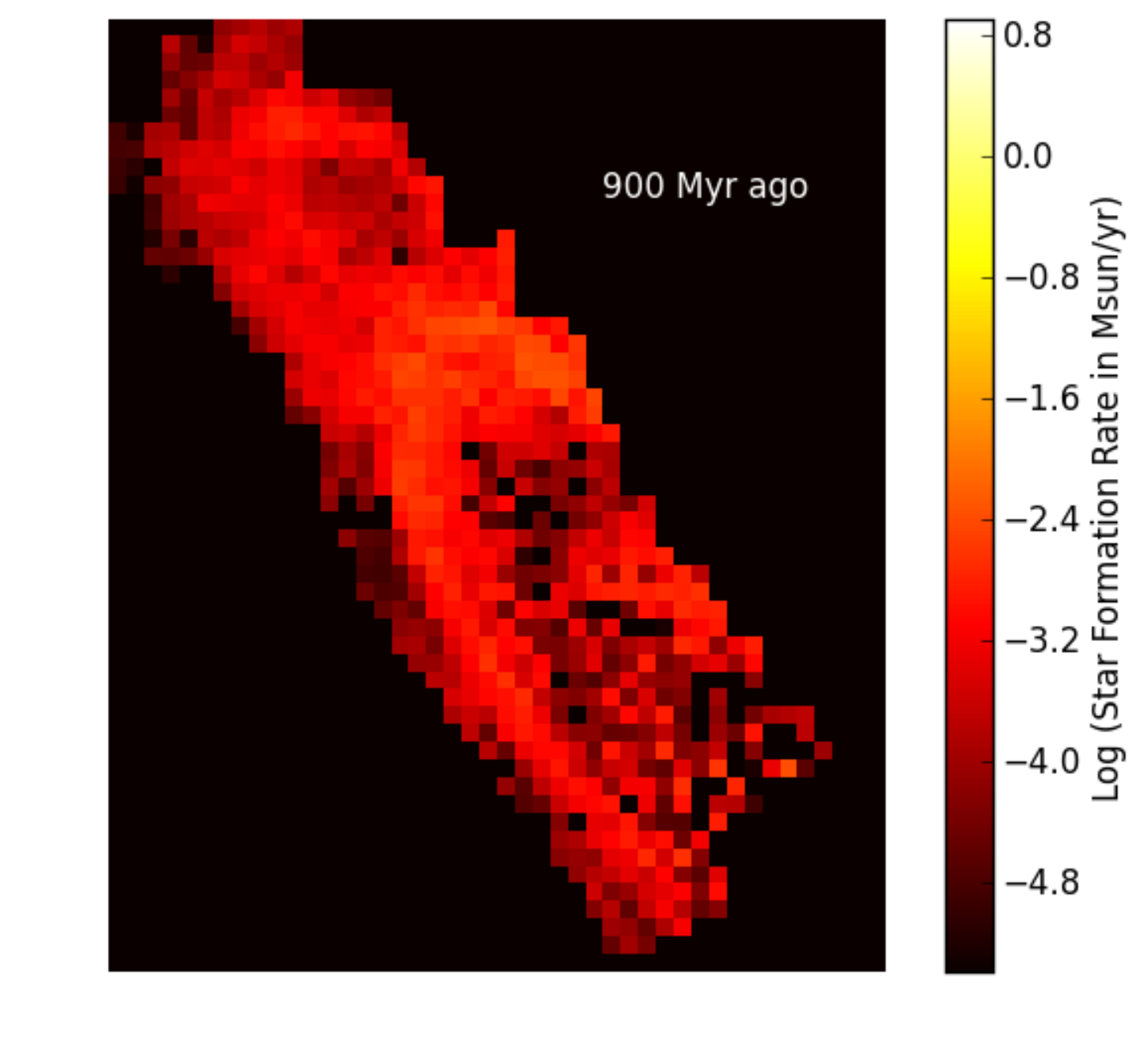}
\includegraphics[height=2.2in]{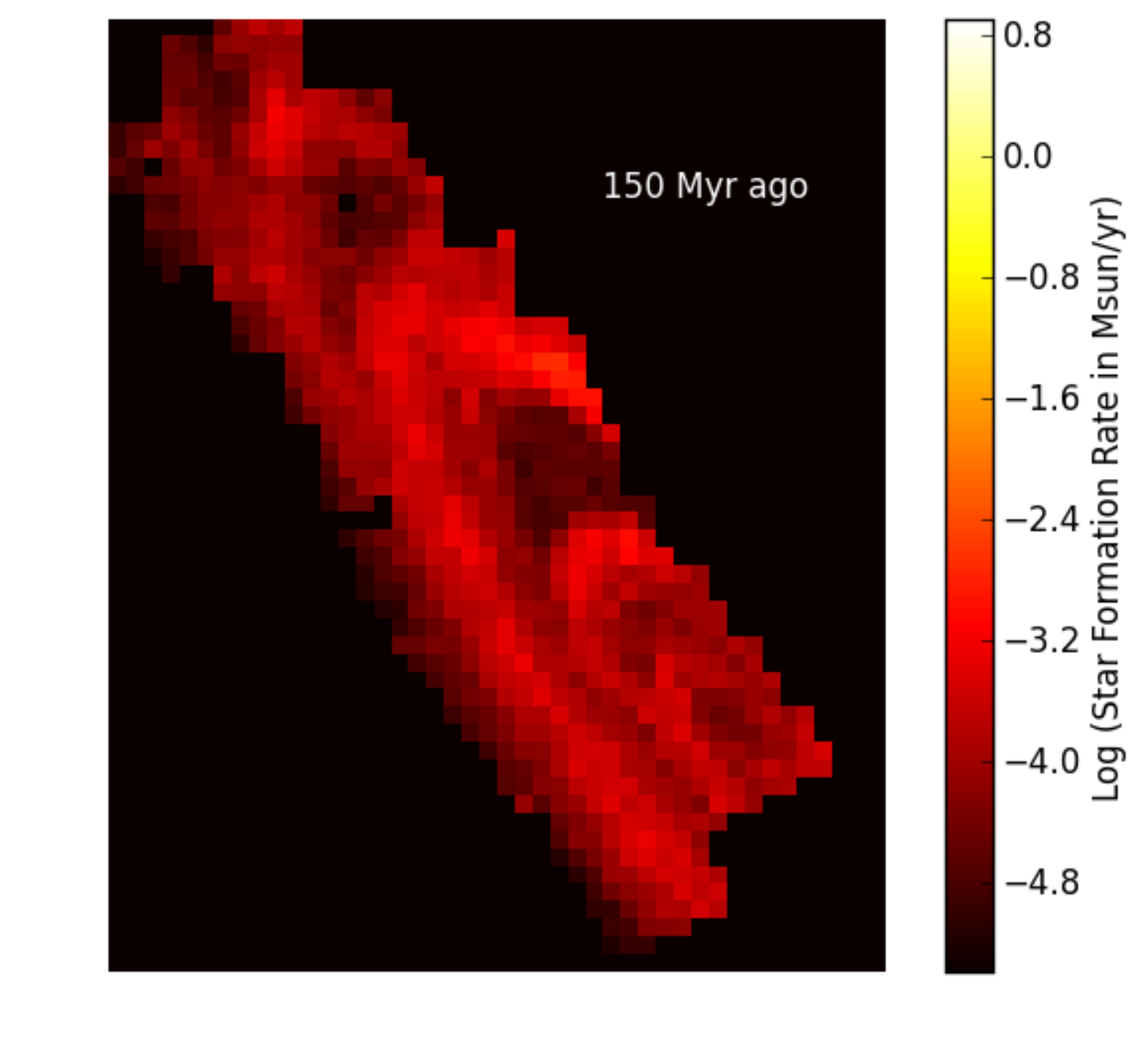}
\end{center}
\caption{Subsample of SFH maps from the star formation movie made from the Padova model fits to the PHAT survey.}
\label{sfr_maps}
\end{figure*}

\begin{figure*}
\begin{center}
\includegraphics[width=6.2in]{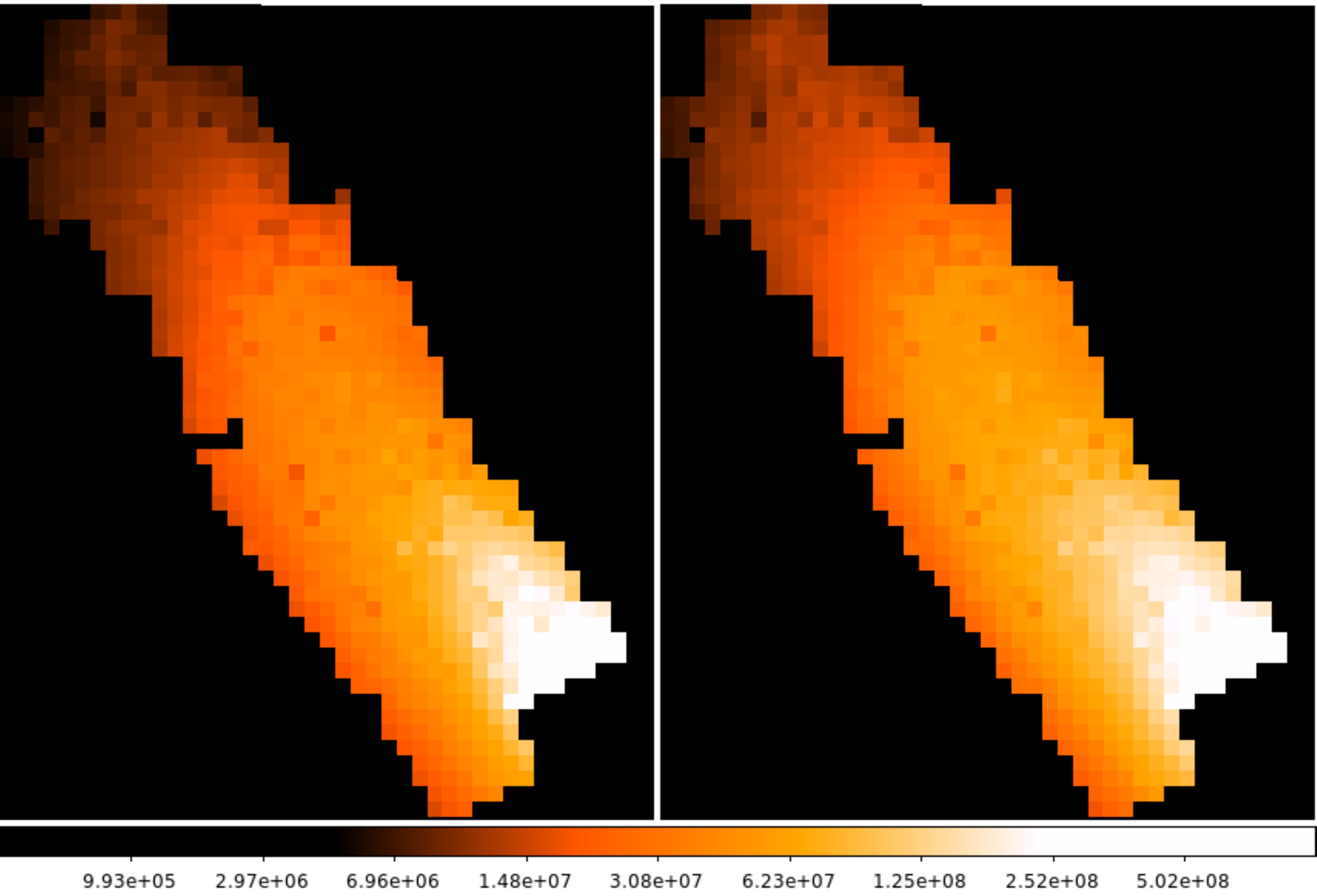}
\includegraphics[width=6.2in]{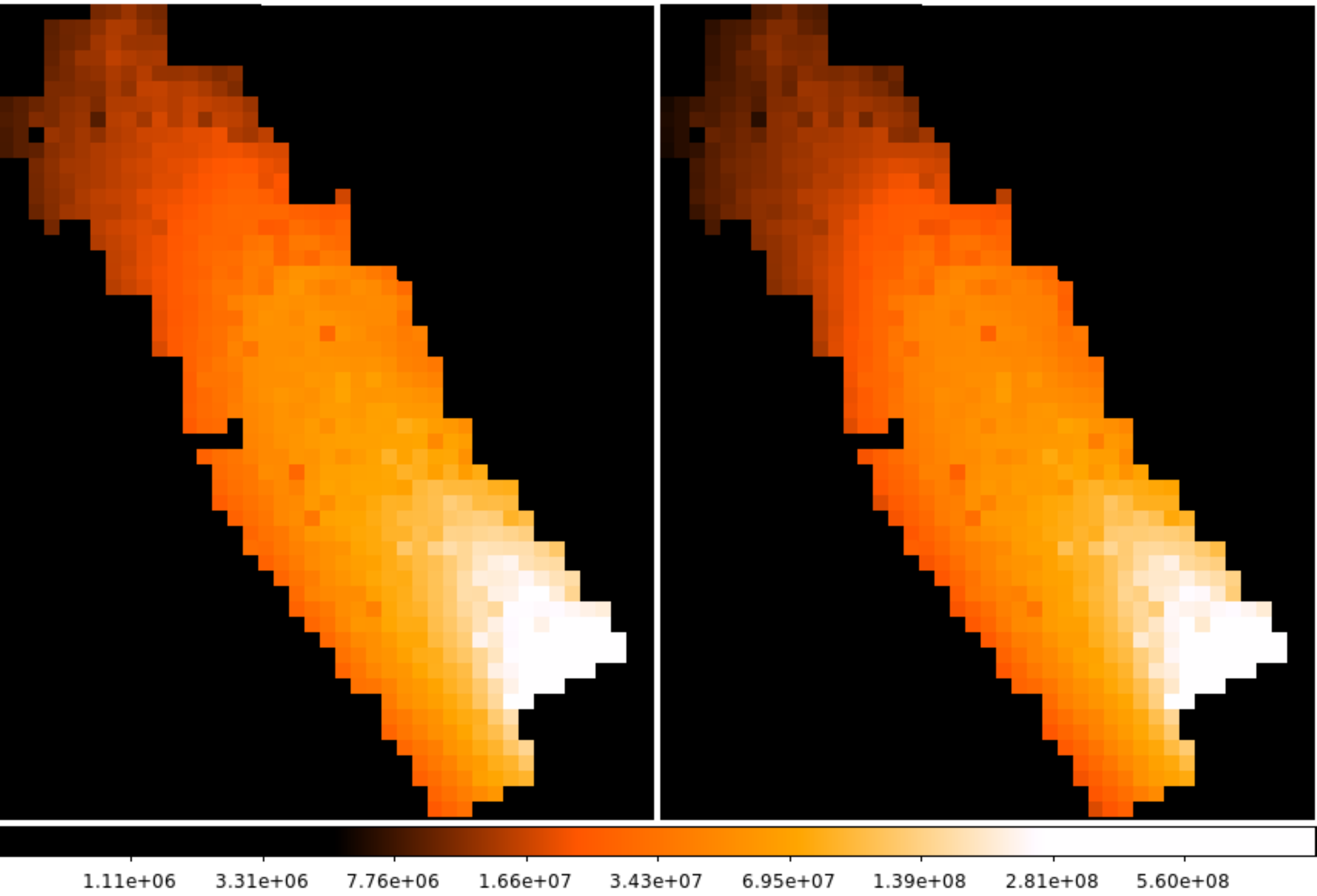}
\end{center}
\caption{Total stellar mass in each 83$''{\times}83''$ region of the PHAT survey from the 4 different model sets.  {\it Upper Left:}  Padova fits. {\it Upper Right:} BaSTI fits. {\it Lower Left:}  PARSEC fits. {\it Lower Right:} MIST fits.    \label{mass_maps}}
\end{figure*}

\begin{figure*}
\begin{center}
\includegraphics[width=6.2in]{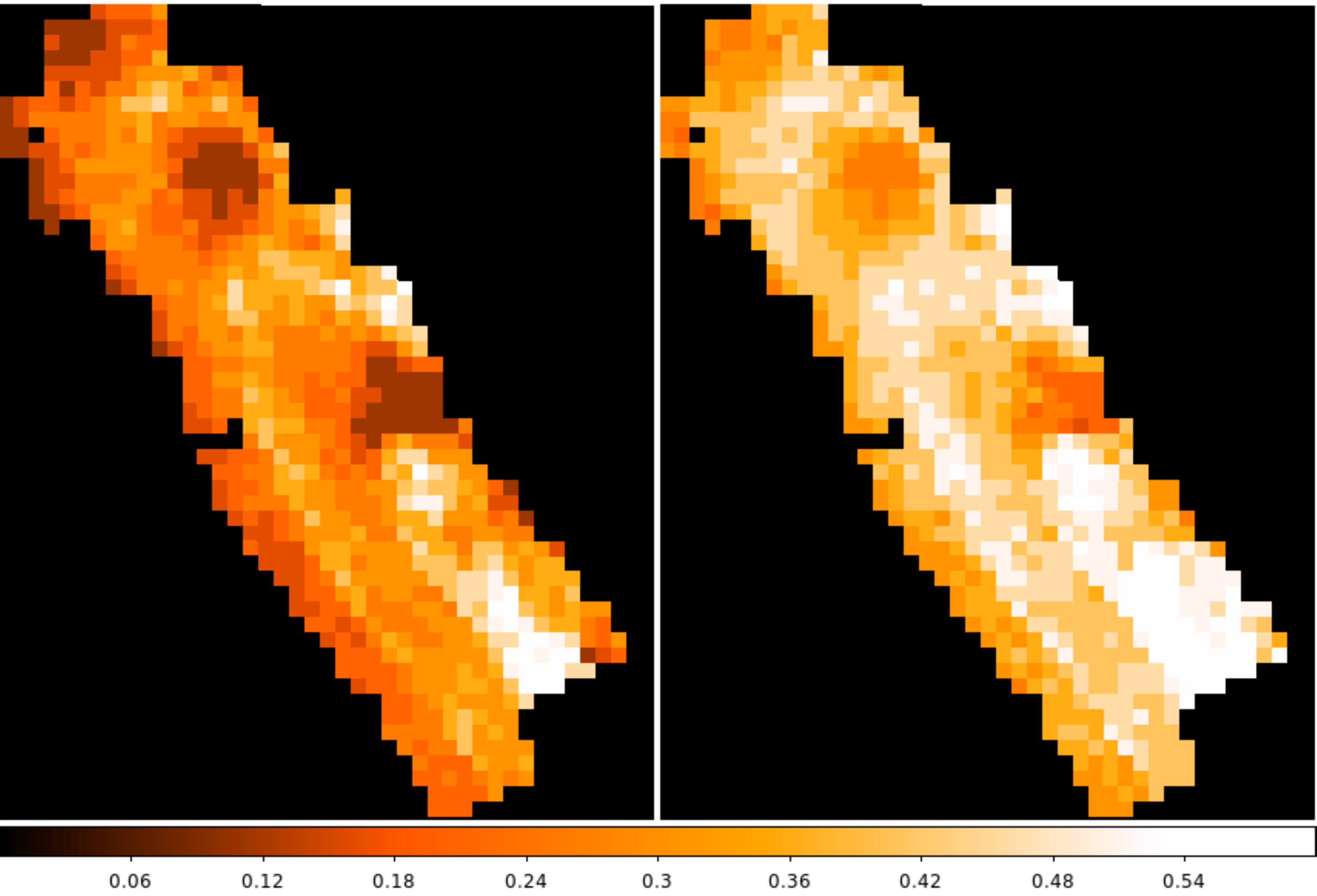}
\includegraphics[width=6.2in]{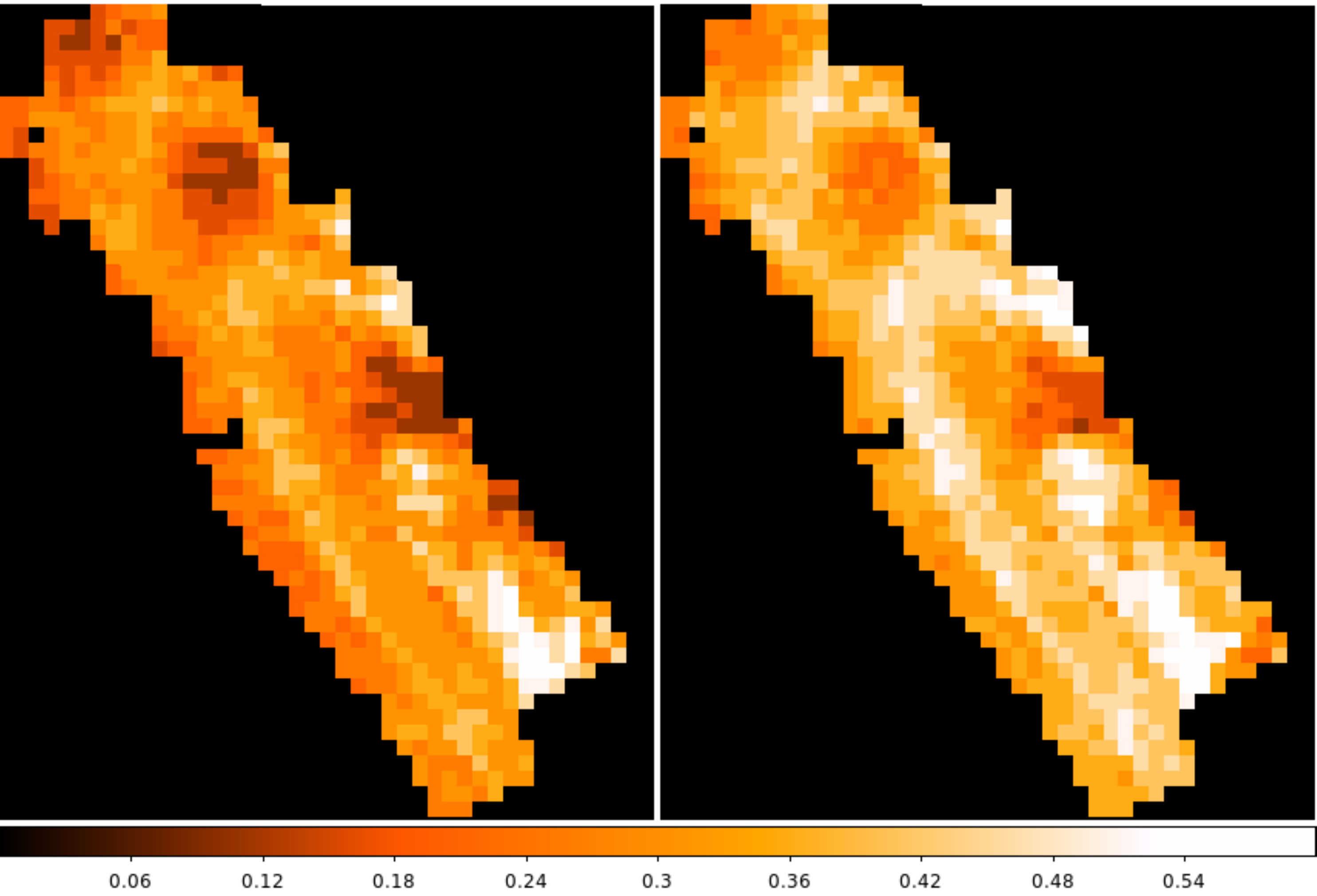}
\end{center}
\caption{Maps of the best-fitting foreground $A_{\rm VFG}$ values in each 83$''{\times}83''$ region of the PHAT survey from the 4 different model sets.  {\it Upper Left:}  Padova fits. {\it Upper Right:} BaSTI fits. {\it Lower Left:}  PARSEC fits. {\it Lower Right:} MIST fits.    \label{av_maps}}
\end{figure*}

\begin{figure*}
\begin{center}
\includegraphics[width=3.1in]{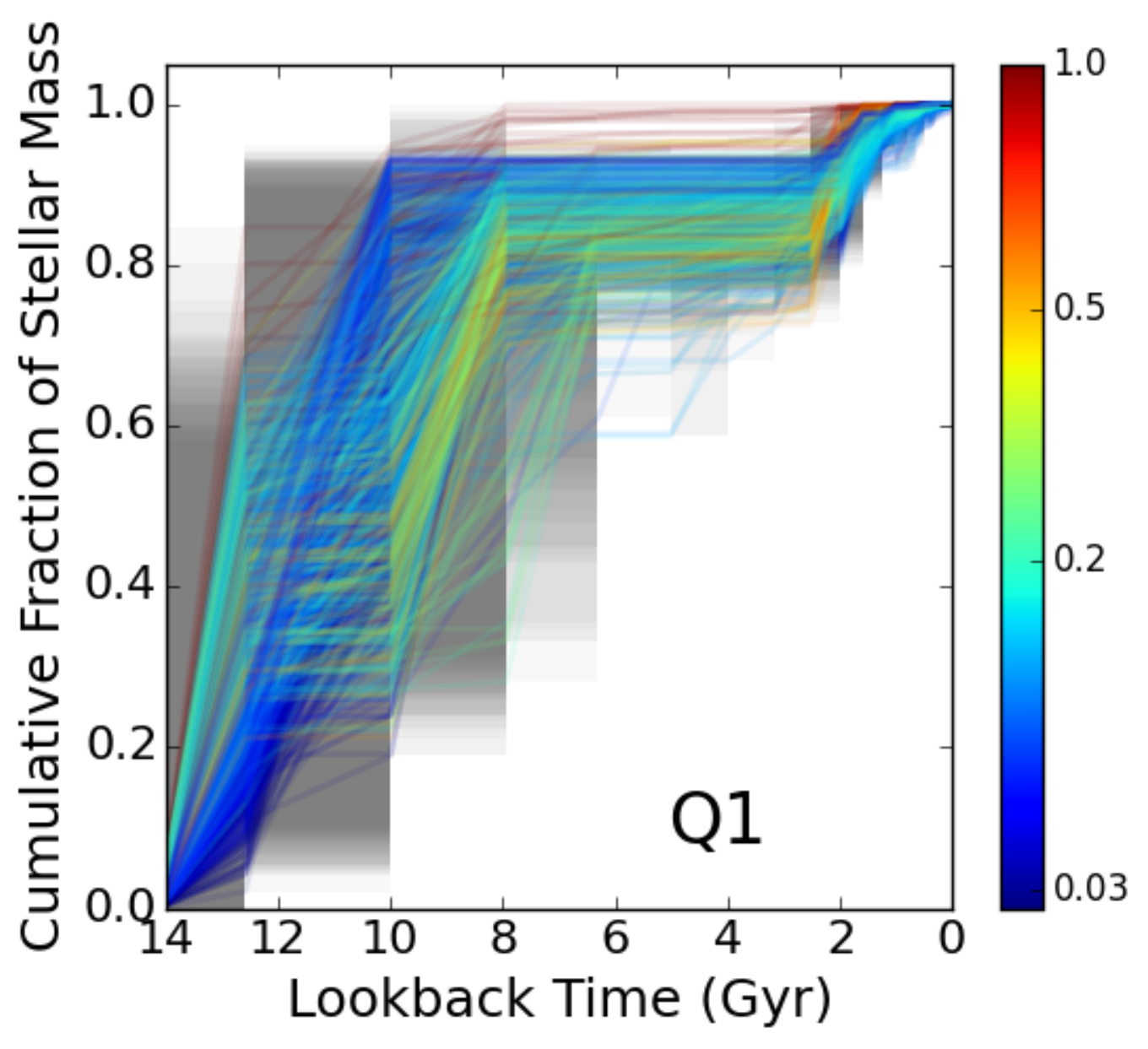}
\includegraphics[width=3.1in]{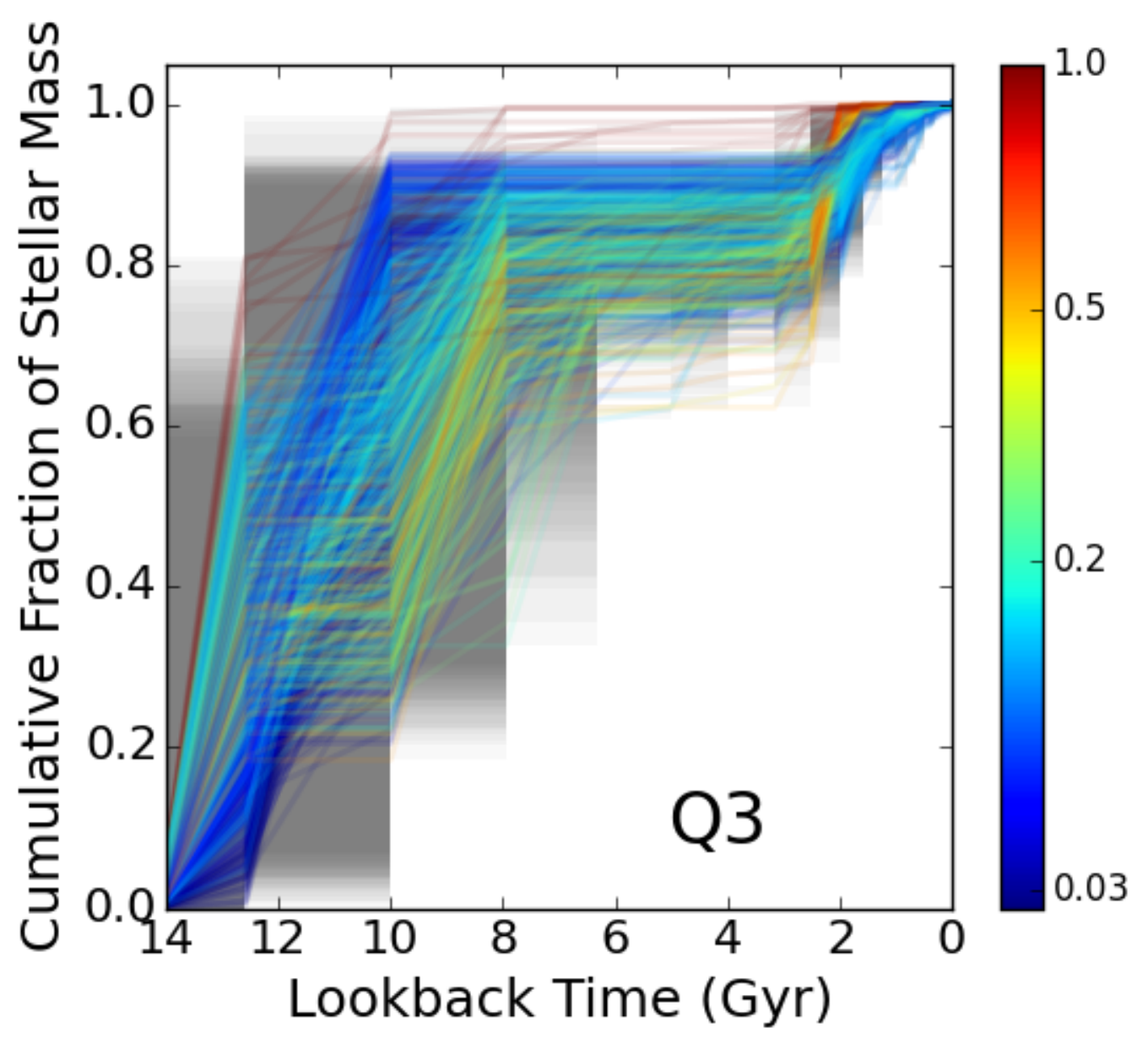}
\end{center}
\caption{{\it Left:} The SFHs measured for the first quartile of $\mu$ values in each region, color coded by stellar density using the same colors as Figure~\ref{stellar_density_map}. {\it Right:} Same as {\it center}, but for the third quartile in all fields. These examples are from the MIST model fits. \label{compare_quartiles}}
\end{figure*}

\clearpage


\section{Appendix: Metallicity as a Free Parameter}

In this Appendix, we show the SFH results when fitting the data with
the metallicity as a free parameter for each age.  The resulting
enrichment histories are shown in Figure~\ref{free_metallicity}. SFR
in each metallicity and time are provided in Table~\ref{popbox_table};
total rates and uncertainties for each time bin are provided in
Table~\ref{free_sfh}.  Total rates and masses are given in
Tables~\ref{free_total_rates} and \ref{free_total_mass}.  In
Figure~\ref{free_sfhs}, we show the SFH results when the metallicity
is fitted as a free parameter at all ages.  This figure is analogous
to Figure~\ref{compare_model_sfhs}. In Figure~\ref{free_total_sfh}, we
show the total SFH of the PHAT footprint.  This figure is analogous to
Figure~\ref{total_sfh}. In Figure~\ref{free_radial_sfhs}, we break the
total into 4 radial bins, as in
Figure~\ref{radial_sfhs}. Figure~\ref{free_metallicity_hist} shows
histograms of the resulting total metallicity distribution of the M31
disk for these SFHs, as well as the variations with radius in the
Padova fits, comparable to Figure~\ref{z4hist} in the main text.  In
Figure~\ref{free_density_vs_radius}, we show the resulting build-up of
the stellar mass in the disk analogous to
Figure~\ref{density_vs_radius}. In
Figure~\ref{free_exp_fits}, we show the exponential parameters of the
stellar mass profile as a function of lookback time.  The Padova and
BaSTI results are shown; analogous to Figure~\ref{scale_norm}.  In
Figure~\ref{free_sfr_maps}, we show maps of the stellar mass formed at
a variety of ages, analogous to Figure~\ref{sfr_maps} in the main
text.  Below, we compare these to the results with our adopted
chemical enrichment model. Figure~\ref{free_mass_maps} shows the total
stellar mass maps, analogous to Figure~\ref{mass_maps}.
Figure~\ref{free_av_maps} shows a map of the best-fit foreground
extinction values for these SFHs, analogous to Figure~\ref{av_maps}.

The largest systematic offset between models in the free metallicity
fits is seen in the PARSEC models with the metallicity fitted as a
free parameter.  Those fits result in systematically lower stellar
masses, and younger ages, at smaller radii, which is not consistent
with previous work \citep[e.g.][]{saglia2010,dong2015}.  This
difference is not apparent in the fits presented in the main text,
suggesting that fixing the enrichment model was most important when
applying this model set.

Another interesting feature is seen in the enrichment histories
(Figures~\ref{free_metallicity}) is the dip in metallicity seen at 2-3
Gyr, which is clearest in the Padova fits, but also present in the
other model fits.  This dip in metallicity is coincident in age with
the last global star formation episode.  This episode is seen in all
of the SFHs, regardless of the enrichment history, and it is clearly
visible in the 1.8 Gyr panel of the SFR maps we provide in
Figures~\ref{sfr_maps} and \ref{free_sfr_maps}.  However, since the
outburst appears with or without a forced enrichment history, the data
appear relatively insensitive to the metallicity at this age.  Thus,
while it may appear that this feature of the free metallicity fits
provides some tantalizing evidence that this powerful star forming
episode may have been accompanied by the accretion of low metallicity
gas, deeper data would be necessary to make such a claim.  For now, we
can only detect the large stellar mass present at this age range.

As a final note on the effects of fixing the enrichment history, in
Figure~\ref{isochrones} we show the NIR CMD of some dust-free regions
of the PHAT survey (as measured by \citealp{dalcanton2015}) overlaid
with Padova isochrones. The narrow RGBs in the NIR show that there is
a dominant metallicity, which is [Fe/H]${\sim}-$0.5. Because the
adopted enrichment history only allows specific metallicities to be
fitted at specific ages, the fits will necessarily increase the rates
at ages where [Fe/H]${\sim}-$0.5.  In our best-fitting enrichment
model, presented in the main text, the metallicity reaches this range
at $>$8 Gyr ago.  Thus both fitting techniques agree that the bulk of
the stellar mass was formed prior to 8 Gyr ago.

\clearpage

\begin{turnpage}
\tabletypesize{\tiny}

\begin{deluxetable}{cccccccccccccccccccc}\tablewidth{9.0in}
\tablecaption{M31 Star Formation Rates by Metallicity (M$_{\odot}$~yr$^{-1}$) for the free metallicity fits; Columns are age range in Gyr; full table available in electronic format only}
\tablehead{
\colhead{Model Set} &
\colhead{RA} &
\colhead{Dec} &
\colhead{[Fe/H]} &
\colhead{0.0-0.3} & 
\colhead{0.3-0.4} & 
\colhead{0.4-0.5} & 
\colhead{0.5-0.6} & 
\colhead{0.6-0.8} & 
\colhead{0.8-1.0} & 
\colhead{1.0-1.3} & 
\colhead{1.3-1.6} & 
\colhead{1.6-2.0} & 
\colhead{2.0-2.5} & 
\colhead{2.5-3.2} & 
\colhead{3.2-4.0} & 
\colhead{4.0-5.0} & 
\colhead{5.0-6.3} & 
\colhead{6.3-7.9} & 
\colhead{7.9-14.1}
}
\startdata
Padova & 0:42:45.03 & 41:22:17.0 & -2.25 & 2.4e-04 & 0.0e+00 & 0.0e+00 & 0.0e+00 & 0.0e+00 & 0.0e+00 & 0.0e+00 & 0.0e+00 & 0.0e+00 & 0.0e+00 & 0.0e+00 & 0.0e+00 & 0.0e+00 & 0.0e+00 & 1.1e-04 & 4.6e-03\\
Padova & 0:42:45.03 & 41:22:17.0 & -2.15 & 1.8e-04 & 0.0e+00 & 0.0e+00 & 0.0e+00 & 0.0e+00 & 0.0e+00 & 0.0e+00 & 0.0e+00 & 0.0e+00 & 0.0e+00 & 0.0e+00 & 0.0e+00 & 0.0e+00 & 0.0e+00 & 0.0e+00 & 0.0e+00\\
Padova & 0:42:45.03 & 41:22:17.0 & -2.05 & 1.6e-06 & 0.0e+00 & 0.0e+00 & 0.0e+00 & 0.0e+00 & 0.0e+00 & 0.0e+00 & 0.0e+00 & 0.0e+00 & 0.0e+00 & 0.0e+00 & 0.0e+00 & 0.0e+00 & 0.0e+00 & 0.0e+00 & 0.0e+00\\
Padova & 0:42:45.03 & 41:22:17.0 & -1.95 & 8.1e-05 & 0.0e+00 & 0.0e+00 & 0.0e+00 & 0.0e+00 & 0.0e+00 & 0.0e+00 & 0.0e+00 & 0.0e+00 & 0.0e+00 & 0.0e+00 & 0.0e+00 & 0.0e+00 & 0.0e+00 & 0.0e+00 & 0.0e+00\\
Padova & 0:42:45.03 & 41:22:17.0 & -1.85 & 4.3e-06 & 0.0e+00 & 0.0e+00 & 0.0e+00 & 0.0e+00 & 0.0e+00 & 0.0e+00 & 6.4e-04 & 0.0e+00 & 0.0e+00 & 0.0e+00 & 0.0e+00 & 0.0e+00 & 0.0e+00 & 0.0e+00 & 0.0e+00\\
Padova & 0:42:45.03 & 41:22:17.0 & -1.75 & 2.8e-05 & 0.0e+00 & 0.0e+00 & 0.0e+00 & 0.0e+00 & 0.0e+00 & 0.0e+00 & 0.0e+00 & 0.0e+00 & 0.0e+00 & 0.0e+00 & 0.0e+00 & 0.0e+00 & 0.0e+00 & 0.0e+00 & 0.0e+00\\
Padova & 0:42:45.03 & 41:22:17.0 & -1.65 & 8.2e-05 & 0.0e+00 & 0.0e+00 & 0.0e+00 & 0.0e+00 & 0.0e+00 & 0.0e+00 & 0.0e+00 & 0.0e+00 & 0.0e+00 & 0.0e+00 & 0.0e+00 & 0.0e+00 & 0.0e+00 & 0.0e+00 & 6.0e-05\\
Padova & 0:42:45.03 & 41:22:17.0 & -1.55 & 0.0e+00 & 0.0e+00 & 0.0e+00 & 0.0e+00 & 0.0e+00 & 0.0e+00 & 0.0e+00 & 0.0e+00 & 0.0e+00 & 0.0e+00 & 0.0e+00 & 0.0e+00 & 0.0e+00 & 0.0e+00 & 0.0e+00 & 0.0e+00\\
Padova & 0:42:45.03 & 41:22:17.0 & -1.45 & 1.7e-04 & 0.0e+00 & 0.0e+00 & 0.0e+00 & 0.0e+00 & 0.0e+00 & 0.0e+00 & 0.0e+00 & 0.0e+00 & 0.0e+00 & 0.0e+00 & 0.0e+00 & 0.0e+00 & 0.0e+00 & 0.0e+00 & 6.5e-05\\
Padova & 0:42:45.03 & 41:22:17.0 & -1.35 & 7.2e-05 & 1.3e-05 & 0.0e+00 & 0.0e+00 & 0.0e+00 & 0.0e+00 & 0.0e+00 & 0.0e+00 & 0.0e+00 & 0.0e+00 & 0.0e+00 & 0.0e+00 & 0.0e+00 & 0.0e+00 & 0.0e+00 & 7.6e-04\\
Padova & 0:42:45.03 & 41:22:17.0 & -1.25 & 6.3e-05 & 3.2e-06 & 0.0e+00 & 0.0e+00 & 0.0e+00 & 0.0e+00 & 0.0e+00 & 0.0e+00 & 0.0e+00 & 0.0e+00 & 0.0e+00 & 0.0e+00 & 0.0e+00 & 0.0e+00 & 0.0e+00 & 7.3e-04\\
Padova & 0:42:45.03 & 41:22:17.0 & -1.15 & 7.2e-05 & 0.0e+00 & 0.0e+00 & 0.0e+00 & 0.0e+00 & 0.0e+00 & 0.0e+00 & 0.0e+00 & 0.0e+00 & 0.0e+00 & 0.0e+00 & 0.0e+00 & 0.0e+00 & 0.0e+00 & 0.0e+00 & 0.0e+00\\
Padova & 0:42:45.03 & 41:22:17.0 & -1.05 & 0.0e+00 & 0.0e+00 & 0.0e+00 & 0.0e+00 & 4.0e-05 & 0.0e+00 & 0.0e+00 & 0.0e+00 & 0.0e+00 & 0.0e+00 & 0.0e+00 & 0.0e+00 & 0.0e+00 & 0.0e+00 & 0.0e+00 & 1.7e-04\\
Padova & 0:42:45.03 & 41:22:17.0 & -0.95 & 3.2e-06 & 0.0e+00 & 0.0e+00 & 0.0e+00 & 6.3e-04 & 0.0e+00 & 0.0e+00 & 0.0e+00 & 5.5e-03 & 0.0e+00 & 0.0e+00 & 0.0e+00 & 0.0e+00 & 0.0e+00 & 0.0e+00 & 4.7e-04\\
Padova & 0:42:45.03 & 41:22:17.0 & -0.85 & 0.0e+00 & 0.0e+00 & 0.0e+00 & 1.2e-03 & 0.0e+00 & 7.9e-05 & 0.0e+00 & 0.0e+00 & 9.2e-03 & 0.0e+00 & 0.0e+00 & 0.0e+00 & 0.0e+00 & 0.0e+00 & 0.0e+00 & 4.6e-04\\
Padova & 0:42:45.03 & 41:22:17.0 & -0.75 & 0.0e+00 & 4.4e-06 & 3.3e-03 & 2.8e-03 & 1.2e-03 & 4.7e-04 & 0.0e+00 & 0.0e+00 & 9.1e-05 & 1.8e-02 & 0.0e+00 & 0.0e+00 & 0.0e+00 & 0.0e+00 & 0.0e+00 & 0.0e+00\\
Padova & 0:42:45.03 & 41:22:17.0 & -0.65 & 1.2e-05 & 3.1e-03 & 1.1e-02 & 1.8e-03 & 0.0e+00 & 0.0e+00 & 0.0e+00 & 0.0e+00 & 0.0e+00 & 2.2e-02 & 0.0e+00 & 0.0e+00 & 0.0e+00 & 0.0e+00 & 0.0e+00 & 0.0e+00\\
Padova & 0:42:45.03 & 41:22:17.0 & -0.55 & 1.9e-06 & 0.0e+00 & 4.5e-03 & 0.0e+00 & 0.0e+00 & 0.0e+00 & 0.0e+00 & 0.0e+00 & 3.9e-04 & 0.0e+00 & 2.4e-02 & 0.0e+00 & 0.0e+00 & 0.0e+00 & 0.0e+00 & 0.0e+00\\
Padova & 0:42:45.03 & 41:22:17.0 & -0.45 & 8.0e-06 & 0.0e+00 & 0.0e+00 & 0.0e+00 & 0.0e+00 & 0.0e+00 & 0.0e+00 & 0.0e+00 & 9.8e-04 & 0.0e+00 & 4.3e-02 & 1.7e-03 & 0.0e+00 & 0.0e+00 & 1.5e-03 & 1.5e-03\\
Padova & 0:42:45.03 & 41:22:17.0 & -0.35 & 0.0e+00 & 0.0e+00 & 0.0e+00 & 0.0e+00 & 0.0e+00 & 0.0e+00 & 0.0e+00 & 0.0e+00 & 0.0e+00 & 0.0e+00 & 0.0e+00 & 0.0e+00 & 0.0e+00 & 0.0e+00 & 1.0e-04 & 1.1e-02\\
Padova & 0:42:45.03 & 41:22:17.0 & -0.25 & 7.2e-06 & 0.0e+00 & 3.7e-04 & 0.0e+00 & 0.0e+00 & 0.0e+00 & 0.0e+00 & 0.0e+00 & 3.2e-04 & 0.0e+00 & 0.0e+00 & 5.1e-04 & 0.0e+00 & 0.0e+00 & 0.0e+00 & 0.0e+00\\
Padova & 0:42:45.03 & 41:22:17.0 & -0.15 & 7.0e-06 & 0.0e+00 & 0.0e+00 & 0.0e+00 & 0.0e+00 & 0.0e+00 & 0.0e+00 & 0.0e+00 & 0.0e+00 & 0.0e+00 & 7.5e-04 & 0.0e+00 & 0.0e+00 & 0.0e+00 & 2.8e-04 & 0.0e+00\\
Padova & 0:42:45.03 & 41:22:17.0 & -0.05 & 4.6e-06 & 0.0e+00 & 0.0e+00 & 0.0e+00 & 0.0e+00 & 0.0e+00 & 0.0e+00 & 0.0e+00 & 0.0e+00 & 3.6e-04 & 0.0e+00 & 1.2e-03 & 0.0e+00 & 3.4e-04 & 3.5e-04 & 0.0e+00\\
Padova & 0:42:45.03 & 41:22:17.0 &  0.05 & 0.0e+00 & 0.0e+00 & 0.0e+00 & 5.4e-04 & 0.0e+00 & 0.0e+00 & 0.0e+00 & 0.0e+00 & 3.8e-04 & 3.2e-04 & 1.1e-04 & 0.0e+00 & 0.0e+00 & 0.0e+00 & 4.1e-03 & 1.0e-02\\
Padova & 0:42:45.08 & 41:20:53.9 & -2.25 & 4.7e-06 & 0.0e+00 & 0.0e+00 & 0.0e+00 & 0.0e+00 & 0.0e+00 & 0.0e+00 & 0.0e+00 & 0.0e+00 & 0.0e+00 & 0.0e+00 & 0.0e+00 & 0.0e+00 & 0.0e+00 & 0.0e+00 & 8.4e-05\\
\enddata
\label{popbox_table}
\end{deluxetable}

\begin{deluxetable*}{cccp{0.8cm}p{0.8cm}p{0.8cm}p{0.8cm}p{0.8cm}p{0.8cm}p{0.8cm}p{0.8cm}p{0.8cm}p{0.8cm}p{0.8cm}p{0.8cm}p{0.8cm}p{0.8cm}p{0.8cm}p{0.8cm}p{0.8cm}}
\tablewidth{9.0in}
\tablecaption{M31 Star Formation Rates (10$^{-4}$~M$_{\odot}$~yr$^{-1}$~arcmin$^{-2}$ for free metallicity fits; Columns are age range in Gyr; full table available in electronic format only)}
\tablehead{
\colhead{Model Set} &
\colhead{RA} &
\colhead{Dec} &
\colhead{0.0-0.3} & 
\colhead{0.3-0.4} & 
\colhead{0.4-0.5} & 
\colhead{0.5-0.6} & 
\colhead{0.6-0.8} & 
\colhead{0.8-1.0} & 
\colhead{1.0-1.3} & 
\colhead{1.3-1.6} & 
\colhead{1.6-2.0} & 
\colhead{2.0-2.5} & 
\colhead{2.5-3.2} & 
\colhead{3.2-4.0} & 
\colhead{4.0-5.0} & 
\colhead{5.0-6.3} & 
\colhead{6.3-7.9} & 
\colhead{7.9-14.1}
}
\startdata
Padova & 0:43:36.26 & 41:25:06.2 &  0.5$^{+0.6}_{-0.5}$ &  0.6$^{+0.3}_{-0.6}$ &  2.8$^{+0.0}_{-2.4}$ & 17.6$^{+3.5}_{-1.5}$ &  0.2$^{+1.5}_{-0.2}$ &  1.2$^{+0.6}_{-1.2}$ &  2.6$^{+0.3}_{-2.4}$ &  3.8$^{+0.0}_{-3.3}$ & 86.9$^{+20.6}_{-8.3}$ & 273.1$^{+8.2}_{-22.7}$ & 82.4$^{+17.7}_{-2.5}$ & 115.8$^{+32.5}_{-16.4}$ & 26.1$^{+35.9}_{-13.8}$ &  5.6$^{+1.9}_{-5.6}$ & 59.2$^{+16.5}_{-6.3}$ & 102.2$^{+4.6}_{-9.9}$ \\
Padova & 0:43:36.29 & 41:23:43.2 &  0.9$^{+0.9}_{-0.8}$ &  0.6$^{+0.7}_{-0.5}$ &  6.3$^{+0.0}_{-5.0}$ & 17.9$^{+9.7}_{-1.2}$ &  8.8$^{+6.1}_{-4.9}$ &  1.7$^{+1.5}_{-1.7}$ &  4.9$^{+3.4}_{-3.7}$ &  5.5$^{+2.0}_{-4.7}$ & 66.5$^{+27.9}_{-0.0}$ & 319.5$^{+1.9}_{-35.5}$ & 97.0$^{+22.5}_{-0.2}$ & 158.0$^{+4.7}_{-20.8}$ &  2.4$^{+6.8}_{-2.0}$ & 28.6$^{+6.7}_{-9.1}$ & 56.3$^{+15.4}_{-3.4}$ & 79.1$^{+5.5}_{-7.6}$ \\
Padova & 0:43:36.32 & 41:22:20.1 &  2.2$^{+1.6}_{-1.6}$ &  0.6$^{+0.7}_{-0.4}$ &  6.3$^{+0.0}_{-3.9}$ & 21.1$^{+5.9}_{-0.0}$ &  3.5$^{+2.7}_{-2.2}$ &  0.8$^{+0.8}_{-0.8}$ &  3.1$^{+0.8}_{-2.2}$ &  5.7$^{+0.7}_{-3.4}$ & 88.6$^{+19.0}_{-2.5}$ & 263.0$^{+7.5}_{-19.5}$ & 73.7$^{+18.4}_{-3.3}$ & 137.2$^{+15.2}_{-19.0}$ & 35.6$^{+21.3}_{-10.3}$ & 27.0$^{+7.7}_{-16.4}$ & 65.1$^{+22.1}_{-1.6}$ & 74.9$^{+4.0}_{-9.6}$ \\
Padova & 0:43:36.36 & 41:20:57.0 &  1.5$^{+1.0}_{-1.1}$ &  0.8$^{+0.8}_{-0.6}$ &  7.0$^{+1.5}_{-3.8}$ & 29.0$^{+6.1}_{-3.3}$ &  5.0$^{+2.4}_{-2.7}$ &  0.0$^{+1.2}_{-0.0}$ &  1.0$^{+0.6}_{-0.9}$ &  4.6$^{+1.1}_{-2.6}$ & 128.4$^{+22.5}_{-0.0}$ & 159.4$^{+12.8}_{-13.5}$ & 207.8$^{+16.1}_{-4.1}$ & 53.7$^{+0.0}_{-24.1}$ &  8.6$^{+8.6}_{-2.7}$ &  7.0$^{+7.8}_{-5.0}$ & 88.3$^{+15.3}_{-3.8}$ & 88.1$^{+9.8}_{-9.2}$ \\
Padova & 0:43:36.39 & 41:19:34.0 &  1.0$^{+0.7}_{-0.7}$ &  1.2$^{+0.7}_{-0.8}$ &  6.0$^{+0.0}_{-3.7}$ & 21.3$^{+5.6}_{-0.0}$ &  3.0$^{+1.5}_{-1.5}$ &  1.0$^{+3.0}_{-0.4}$ &  2.5$^{+2.0}_{-1.4}$ &  2.4$^{+0.0}_{-2.2}$ & 79.8$^{+18.6}_{-1.3}$ & 231.1$^{+0.0}_{-21.4}$ & 63.9$^{+18.6}_{-0.0}$ & 91.4$^{+7.9}_{-16.0}$ & 26.1$^{+11.1}_{-0.2}$ & 58.1$^{+3.7}_{-13.3}$ & 27.8$^{+10.3}_{-6.8}$ & 83.0$^{+5.8}_{-7.2}$ \\
Padova & 0:43:36.42 & 41:18:10.9 &  0.4$^{+0.4}_{-0.2}$ &  2.8$^{+0.0}_{-2.0}$ &  5.8$^{+0.6}_{-2.5}$ & 14.3$^{+4.6}_{-0.0}$ &  2.5$^{+0.8}_{-1.3}$ &  0.1$^{+1.1}_{-0.0}$ &  2.6$^{+0.0}_{-1.5}$ &  2.0$^{+0.0}_{-1.5}$ & 124.9$^{+7.1}_{-5.8}$ & 201.3$^{+11.5}_{-1.5}$ & 39.3$^{+8.3}_{-1.7}$ & 60.3$^{+0.6}_{-15.9}$ & 23.5$^{+11.9}_{-1.3}$ & 38.7$^{+5.2}_{-6.3}$ & 18.8$^{+2.4}_{-9.5}$ & 83.8$^{+7.0}_{-6.0}$ \\
Padova & 0:43:36.45 & 41:16:47.8 &  0.9$^{+1.1}_{-0.5}$ &  7.1$^{+0.0}_{-5.9}$ &  5.3$^{+8.1}_{-0.1}$ & 38.3$^{+5.6}_{-4.5}$ &  5.3$^{+1.6}_{-2.8}$ &  2.8$^{+8.2}_{-0.8}$ & 89.0$^{+0.1}_{-11.7}$ &  2.2$^{+0.1}_{-1.7}$ & 68.6$^{+10.7}_{-1.7}$ & 165.5$^{+10.1}_{-6.5}$ & 58.8$^{+10.1}_{-10.6}$ & 123.2$^{+0.2}_{-23.5}$ & 32.5$^{+25.4}_{-0.2}$ & 129.6$^{+15.7}_{-10.0}$ & 24.9$^{+6.0}_{-11.8}$ & 98.4$^{+7.9}_{-4.6}$ \\
Padova & 0:43:36.48 & 41:15:24.8 &  0.5$^{+0.4}_{-0.3}$ &  2.8$^{+0.0}_{-1.7}$ &  4.1$^{+0.3}_{-1.8}$ & 13.9$^{+4.2}_{-0.0}$ &  8.8$^{+0.0}_{-3.4}$ &  3.2$^{+2.2}_{-0.6}$ &  3.5$^{+0.0}_{-2.2}$ &  3.3$^{+0.0}_{-2.1}$ & 107.9$^{+14.5}_{-0.0}$ & 56.4$^{+0.1}_{-7.8}$ & 14.2$^{+0.5}_{-6.1}$ & 24.9$^{+5.3}_{-4.5}$ & 34.2$^{+6.0}_{-3.2}$ & 26.9$^{+4.5}_{-5.8}$ & 31.2$^{+0.3}_{-9.1}$ & 68.3$^{+5.1}_{-2.3}$ \\
Padova & 0:43:36.51 & 41:14:01.7 &  0.7$^{+0.5}_{-0.3}$ &  3.0$^{+0.0}_{-2.4}$ &  8.1$^{+1.4}_{-1.1}$ & 17.0$^{+3.2}_{-0.0}$ &  1.0$^{+0.5}_{-0.6}$ &  3.3$^{+1.5}_{-0.6}$ & 28.2$^{+2.0}_{-0.8}$ &  0.3$^{+0.2}_{-0.2}$ & 33.9$^{+4.2}_{-3.6}$ & 52.0$^{+1.1}_{-5.8}$ & 39.5$^{+2.5}_{-1.9}$ & 35.2$^{+7.2}_{-2.5}$ & 23.9$^{+4.4}_{-5.4}$ & 13.0$^{+4.6}_{-4.0}$ & 14.2$^{+1.2}_{-4.1}$ & 60.8$^{+3.5}_{-2.0}$ \\
Padova & 0:43:36.54 & 41:12:38.7 &  0.8$^{+0.7}_{-0.3}$ &  4.9$^{+0.0}_{-3.1}$ &  5.8$^{+1.6}_{-1.1}$ & 19.8$^{+3.9}_{-0.0}$ &  7.8$^{+0.0}_{-2.4}$ &  3.0$^{+2.0}_{-0.0}$ & 19.9$^{+1.4}_{-0.0}$ &  1.0$^{+0.0}_{-0.8}$ &  6.5$^{+0.0}_{-4.0}$ & 14.9$^{+0.1}_{-6.5}$ & 29.4$^{+3.6}_{-1.9}$ & 51.6$^{+4.0}_{-1.9}$ & 44.4$^{+8.3}_{-0.0}$ & 12.7$^{+1.5}_{-4.8}$ & 10.4$^{+1.9}_{-3.4}$ & 45.3$^{+2.3}_{-2.3}$ \\
Padova & 0:43:36.57 & 41:11:15.6 &  0.9$^{+0.5}_{-0.5}$ &  6.8$^{+0.4}_{-1.3}$ &  3.6$^{+0.6}_{-1.0}$ & 20.5$^{+1.9}_{-0.0}$ &  0.8$^{+0.0}_{-0.7}$ &  7.1$^{+1.7}_{-0.0}$ & 21.9$^{+1.1}_{-0.6}$ &  0.8$^{+0.0}_{-0.6}$ & 47.5$^{+4.6}_{-0.7}$ & 10.5$^{+0.0}_{-3.3}$ & 38.0$^{+3.9}_{-1.0}$ & 12.6$^{+0.0}_{-4.1}$ & 11.1$^{+0.8}_{-3.8}$ &  6.4$^{+7.0}_{-0.0}$ &  4.0$^{+0.9}_{-2.7}$ & 39.9$^{+2.0}_{-2.3}$ \\
Padova & 0:43:36.60 & 41:09:52.5 &  1.4$^{+0.7}_{-0.3}$ &  1.6$^{+0.0}_{-1.0}$ &  4.4$^{+0.4}_{-1.0}$ & 12.0$^{+1.9}_{-0.0}$ &  0.7$^{+0.0}_{-0.6}$ &  4.9$^{+0.4}_{-0.6}$ & 16.9$^{+1.6}_{-0.0}$ &  2.0$^{+0.0}_{-1.5}$ & 47.4$^{+3.8}_{-0.0}$ &  4.4$^{+0.0}_{-3.1}$ & 20.5$^{+3.8}_{-0.0}$ & 23.5$^{+0.0}_{-5.5}$ &  2.2$^{+2.5}_{-0.3}$ &  6.3$^{+3.2}_{-0.1}$ &  2.6$^{+0.1}_{-2.2}$ & 38.4$^{+1.9}_{-1.5}$ \\
Padova & 0:43:43.37 & 41:36:11.1 &  0.7$^{+0.7}_{-0.4}$ &  6.3$^{+0.0}_{-2.6}$ &  2.5$^{+0.4}_{-0.9}$ &  6.8$^{+3.6}_{-0.0}$ & 10.4$^{+0.2}_{-2.5}$ &  2.3$^{+0.0}_{-1.8}$ & 18.5$^{+3.8}_{-0.0}$ &  0.0$^{+0.2}_{-0.0}$ & 59.7$^{+1.3}_{-3.4}$ &  1.9$^{+0.0}_{-1.4}$ & 18.3$^{+2.4}_{-0.9}$ & 83.6$^{+9.5}_{-0.1}$ & 15.2$^{+0.1}_{-5.4}$ &  0.0$^{+0.6}_{-0.0}$ &  1.9$^{+0.0}_{-1.6}$ & 36.2$^{+2.2}_{-2.3}$ \\
\enddata
\label{free_sfh}
\end{deluxetable*}

\begin{deluxetable*}{cccccccccccccc}
\tablewidth{9.0in}
\tablecaption{Total Star Formation Rates \tablenotemark{a}}
\tablehead{
\colhead{Lookback Start (Years)} &
\colhead{Lookback End (Years)} &
\colhead{Padova SFR (M$_{\odot}$~yr$^{-1}$)} & 
\colhead{$+$error} &
\colhead{$-$error} &
\colhead{BaSTI SFR} &
\colhead{$+$error} &
\colhead{$-$error} &
\colhead{PARSEC SFR} & 
\colhead{$+$error} &
\colhead{$-$error} &
\colhead{MIST SFR} & 
\colhead{$+$error} &
\colhead{$-$error} 
}
\startdata
3.2e+08 & 4.0e+06 & 1.5e-01 & 6.3e-02 & 8.1e-02 &  2.1e-01 &  3.2e-02 &  2.9e-02 &  1.6e-01 &  7.1e-02 & 8.5e-02 & 2.1e-01 & 8.3e-02 &  8.0e-02\\
4.0e+08 & 3.2e+08 & 4.2e-01 & 7.0e-02 & 1.7e-01 &  2.1e-01 &  2.7e-02 &  3.1e-02 &  1.6e-01 &  4.8e-02 & 7.9e-02 & 5.3e-01 & 1.1e-01 &  1.8e-01\\
5.0e+08 & 4.0e+08 & 7.5e-01 & 1.3e-01 & 2.1e-01 &  6.1e-01 &  5.6e-02 &  5.7e-02 &  3.9e-01 &  6.1e-02 & 1.5e-01 & 7.5e-01 & 1.9e-01 &  1.4e-01\\
6.3e+08 & 5.0e+08 & 1.3e+00 & 3.4e-01 & 8.9e-02 &  9.6e-01 &  8.3e-02 &  7.7e-02 &  1.1e+00 &  2.5e-01 & 1.2e-01 & 5.0e-01 & 1.8e-01 &  6.9e-02\\
7.9e+08 & 6.3e+08 & 6.7e-01 & 1.1e-01 & 1.8e-01 &  3.4e-01 &  8.9e-02 &  4.7e-02 &  8.5e-01 &  2.2e-01 & 1.4e-01 & 2.7e-01 & 5.7e-02 &  8.6e-02\\
1.0e+09 & 7.9e+08 & 7.8e-01 & 1.4e-01 & 9.8e-02 &  1.3e+00 &  1.0e-01 &  8.5e-02 &  6.5e-01 &  1.2e-01 & 2.0e-01 & 7.3e-01 & 9.7e-02 &  9.0e-02\\
1.3e+09 & 1.0e+09 & 1.4e+00 & 1.2e-01 & 1.3e-01 &  2.7e+00 &  2.3e-01 &  1.2e-01 &  2.3e+00 &  2.7e-01 & 2.7e-01 & 1.9e+00 & 2.4e-01 &  1.4e-01\\
1.6e+09 & 1.3e+09 & 2.4e-01 & 4.0e-02 & 1.1e-01 &  2.4e+00 &  1.6e-01 &  1.9e-01 &  5.4e+00 &  4.5e-01 & 1.9e-01 & 1.1e+00 & 1.5e-01 &  2.4e-01\\
2.0e+09 & 1.6e+09 & 5.3e+00 & 5.5e-01 & 2.3e-01 &  2.7e+00 &  2.9e-01 &  1.9e-01 &  2.3e+00 &  2.7e-01 & 3.0e-01 & 4.6e+00 & 6.1e-01 &  2.5e-01\\
2.5e+09 & 2.0e+09 & 4.4e+00 & 2.7e-01 & 4.8e-01 &  2.6e+00 &  2.2e-01 &  1.8e-01 &  3.5e+00 &  3.5e-01 & 3.7e-01 & 6.2e+00 & 4.5e-01 &  5.0e-01\\
3.2e+09 & 2.5e+09 & 3.5e+00 & 4.7e-01 & 3.0e-01 &  1.2e+00 &  1.5e-01 &  1.6e-01 &  2.2e+00 &  2.5e-01 & 4.3e-01 & 5.9e+00 & 5.4e-01 &  4.5e-01\\
4.0e+09 & 3.2e+09 & 4.4e+00 & 6.2e-01 & 4.3e-01 &  1.2e+00 &  1.3e-01 &  1.5e-01 &  4.5e+00 &  5.6e-01 & 3.3e-01 & 5.9e-01 & 2.3e-01 &  2.4e-01\\
5.0e+09 & 4.0e+09 & 1.9e+00 & 4.3e-01 & 2.9e-01 &  9.7e-01 &  1.4e-01 &  1.7e-01 &  3.8e+00 &  6.1e-01 & 2.8e-01 & 1.7e+00 & 3.1e-01 &  3.4e-01\\
6.3e+09 & 5.0e+09 & 8.1e-01 & 2.0e-01 & 2.9e-01 &  1.2e+00 &  1.4e-01 &  2.1e-01 &  1.6e+00 &  2.7e-01 & 2.4e-01 & 1.0e+00 & 2.7e-01 &  3.1e-01\\
7.9e+09 & 6.3e+09 & 1.2e+00 & 3.1e-01 & 3.0e-01 &  2.4e+00 &  2.5e-01 &  1.9e-01 &  2.7e+00 &  3.4e-01 & 4.7e-01 & 1.1e+00 & 2.2e-01 &  3.1e-01\\
1.4e+10 & 7.9e+09 & 5.2e+00 & 4.0e-01 & 3.6e-01 &  6.6e+00 &  3.8e-01 &  3.0e-01 &  3.3e+00 &  3.0e-01 & 2.7e-01 & 5.8e+00 & 4.2e-01 &  3.2e-01\\
\enddata
\tablenotetext{a}{In area analyzed.  To scale these to total M31 rates, multiply these rates by 3.}
\label{free_total_rates}
\end{deluxetable*}

\begin{deluxetable*}{ccccccccccccc}
\tablecaption{Total Cumulative Mass \tablenotemark{a}}
\tablehead{
\colhead{Lookback Time (Years)} &
\colhead{Padova Mass (M$_{\odot}$)} & 
\colhead{$+$error} &
\colhead{$-$error} & 
\colhead{BaSTI Mass (M$_{\odot}$)} & 
\colhead{$+$error} &
\colhead{$-$error} & 
\colhead{PARSEC Mass (M$_{\odot}$)} & 
\colhead{$+$error} &
\colhead{$-$error} & 
\colhead{MIST Mass (M$_{\odot}$)} & 
\colhead{$+$error} &
\colhead{$-$error}  
}
\startdata
7.9e+09 & 3.2e+10 & 2.5e+09 & 2.2e+09 & 4.1e+10 & 2.3e+09 & 1.8e+09 & 2.0e+10 & 1.9e+09 & 1.7e+09 & 3.6e+10 & 2.6e+09 & 2.0e+09\\
6.3e+09 & 3.4e+10 & 3.0e+09 & 2.7e+09 & 4.4e+10 & 2.7e+09 & 2.1e+09 & 2.5e+10 & 2.4e+09 & 2.4e+09 & 3.8e+10 & 3.0e+09 & 2.5e+09\\
5.0e+09 & 3.5e+10 & 3.2e+09 & 3.1e+09 & 4.6e+10 & 2.9e+09 & 2.4e+09 & 2.7e+10 & 2.8e+09 & 2.7e+09 & 3.9e+10 & 3.3e+09 & 2.9e+09\\
4.0e+09 & 3.7e+10 & 3.7e+09 & 3.4e+09 & 4.7e+10 & 3.1e+09 & 2.6e+09 & 3.1e+10 & 3.4e+09 & 3.0e+09 & 4.1e+10 & 3.7e+09 & 3.2e+09\\
3.2e+09 & 4.1e+10 & 4.2e+09 & 3.8e+09 & 4.8e+10 & 3.2e+09 & 2.7e+09 & 3.4e+10 & 3.8e+09 & 3.3e+09 & 4.1e+10 & 3.8e+09 & 3.4e+09\\
2.5e+09 & 4.3e+10 & 4.5e+09 & 4.0e+09 & 4.9e+10 & 3.3e+09 & 2.8e+09 & 3.6e+10 & 4.0e+09 & 3.6e+09 & 4.5e+10 & 4.2e+09 & 3.7e+09\\
2.0e+09 & 4.5e+10 & 4.6e+09 & 4.2e+09 & 5.0e+10 & 3.4e+09 & 2.9e+09 & 3.7e+10 & 4.2e+09 & 3.8e+09 & 4.8e+10 & 4.4e+09 & 4.0e+09\\
1.6e+09 & 4.7e+10 & 4.8e+09 & 4.3e+09 & 5.1e+10 & 3.5e+09 & 3.0e+09 & 3.8e+10 & 4.3e+09 & 3.9e+09 & 5.0e+10 & 4.7e+09 & 4.1e+09\\
1.3e+09 & 4.7e+10 & 4.9e+09 & 4.3e+09 & 5.2e+10 & 3.6e+09 & 3.0e+09 & 4.0e+10 & 4.4e+09 & 3.9e+09 & 5.1e+10 & 4.7e+09 & 4.2e+09\\
1.0e+09 & 4.8e+10 & 4.9e+09 & 4.4e+09 & 5.3e+10 & 3.6e+09 & 3.1e+09 & 4.1e+10 & 4.5e+09 & 4.0e+09 & 5.1e+10 & 4.8e+09 & 4.2e+09\\
7.9e+08 & 4.8e+10 & 4.9e+09 & 4.4e+09 & 5.3e+10 & 3.6e+09 & 3.1e+09 & 4.1e+10 & 4.5e+09 & 4.1e+09 & 5.1e+10 & 4.8e+09 & 4.2e+09\\
6.3e+08 & 4.8e+10 & 4.9e+09 & 4.4e+09 & 5.3e+10 & 3.6e+09 & 3.1e+09 & 4.1e+10 & 4.6e+09 & 4.1e+09 & 5.1e+10 & 4.8e+09 & 4.2e+09\\
5.0e+08 & 4.8e+10 & 5.0e+09 & 4.4e+09 & 5.3e+10 & 3.7e+09 & 3.1e+09 & 4.1e+10 & 4.6e+09 & 4.1e+09 & 5.1e+10 & 4.8e+09 & 4.2e+09\\
4.0e+08 & 4.8e+10 & 5.0e+09 & 4.5e+09 & 5.3e+10 & 3.7e+09 & 3.1e+09 & 4.1e+10 & 4.6e+09 & 4.1e+09 & 5.2e+10 & 4.9e+09 & 4.2e+09\\
3.2e+08 & 4.8e+10 & 5.0e+09 & 4.5e+09 & 5.3e+10 & 3.7e+09 & 3.1e+09 & 4.1e+10 & 4.6e+09 & 4.1e+09 & 5.2e+10 & 4.9e+09 & 4.3e+09\\
4.0e+06 & 4.8e+10 & 5.0e+09 & 4.5e+09 & 5.3e+10 & 3.7e+09 & 3.1e+09 & 4.1e+10 & 4.6e+09 & 4.2e+09 & 5.2e+10 & 4.9e+09 & 4.3e+09\\
\enddata
\tablenotetext{a}{In area analyzed.  To scale to total M31 disk stellar mass, multiply these masses by 3.}
\label{free_total_mass}
\end{deluxetable*}

\begin{deluxetable*}{cccccccccc}
\tablecaption{Metallicity Distribution for the M31 Disk (Free Metallicity Fits).}
\tablehead{
\colhead{[Fe/H]$_{low}$} &
\colhead{[Fe/H]$_{high}$} &
\colhead{Padova Mass (M$_{\odot}$)} & 
\colhead{Padova Scaled (M$_{\odot}$)} &
\colhead{BaSTI Mass (M$_{\odot}$)} & 
\colhead{BaSTI Scaled (M$_{\odot}$)} &
\colhead{PARSEC Mass (M$_{\odot}$)} & 
\colhead{PARSEC Scaled (M$_{\odot}$)} &
\colhead{MIST Mass (M$_{\odot}$)} & 
\colhead{MIST Scaled (M$_{\odot}$)} 
}
\startdata
-2.4 & -2.2 & 6.0e+08 & 1.8e+09 & 6.0e+08 & 1.8e+09 & -0.0e+00 & -0.0e+00 & 1.1e+09 & 3.3e+09\\
-2.2 & -2.0 & 3.3e+08 & 1.0e+09 & 9.1e+07 & 2.7e+08 & 7.3e+08 & 2.2e+09 & 1.0e+09 & 3.1e+09\\
-2.0 & -1.8 & 3.8e+08 & 1.1e+09 & 2.3e+08 & 6.9e+08 & 4.5e+08 & 1.3e+09 & 1.2e+09 & 3.6e+09\\
-1.8 & -1.6 & 2.7e+08 & 8.0e+08 & 2.2e+08 & 6.6e+08 & 1.6e+08 & 4.8e+08 & 5.8e+08 & 1.7e+09\\
-1.6 & -1.4 & 4.9e+08 & 1.5e+09 & 2.2e+08 & 6.7e+08 & 2.9e+08 & 8.7e+08 & 7.6e+08 & 2.3e+09\\
-1.4 & -1.2 & 2.4e+08 & 7.3e+08 & 3.4e+08 & 1.0e+09 & 7.0e+08 & 2.1e+09 & 1.0e+09 & 3.0e+09\\
-1.2 & -1.0 & 6.0e+08 & 1.8e+09 & 8.0e+08 & 2.4e+09 & 7.5e+08 & 2.2e+09 & 1.8e+09 & 5.3e+09\\
-1.0 & -0.8 & 1.5e+09 & 4.4e+09 & 1.6e+09 & 4.8e+09 & 1.4e+09 & 4.1e+09 & 4.2e+09 & 1.3e+10\\
-0.8 & -0.6 & 2.5e+09 & 7.6e+09 & 4.5e+09 & 1.3e+10 & 2.3e+09 & 6.9e+09 & 2.9e+09 & 8.7e+09\\
-0.6 & -0.4 & 7.7e+09 & 2.3e+10 & 6.9e+09 & 2.1e+10 & 4.8e+09 & 1.4e+10 & 7.0e+09 & 2.1e+10\\
-0.4 & -0.2 & 8.4e+09 & 2.5e+10 & 1.0e+10 & 3.0e+10 & 7.9e+09 & 2.4e+10 & 8.3e+09 & 2.5e+10\\
-0.2 & 0.0 & 7.1e+09 & 2.1e+10 & 7.8e+09 & 2.3e+10 & 5.5e+09 & 1.7e+10 & 7.0e+09 & 2.1e+10\\
0.0 & 0.2 & 1.8e+10 & 5.5e+10 & 2.0e+10 & 6.0e+10 & 1.6e+10 & 4.9e+10 & 1.5e+10 & 4.4e+10\\
\enddata
\label{free_total_metallicity}
\end{deluxetable*}

\begin{deluxetable}{ccccccccccc}
\tablecaption{Metallicity Distribution in Radial Bins (Free Metallicity Fits).}
\tablehead{
\colhead{Radial Range (kpc)} &
\colhead{[Fe/H]$_{low}$} &
\colhead{[Fe/H]$_{high}$} &
\colhead{Padova Mass (M$_{\odot}$)} & 
\colhead{Padova Scaled (M$_{\odot}$)} &
\colhead{BaSTI Mass (M$_{\odot}$)} & 
\colhead{BaSTI Scaled (M$_{\odot}$)} &
\colhead{PARSEC Mass (M$_{\odot}$)} & 
\colhead{PARSEC Scaled (M$_{\odot}$)} &
\colhead{MIST Mass (M$_{\odot}$)} & 
\colhead{MIST Scaled (M$_{\odot}$)} 
}
\startdata
$<$5 & -2.4 & -2.2 & 2.4e+08 & 3.8e+08 & 1.5e+08 & 2.3e+08 & -0.0e+00 & -0.0e+00 & 1.6e+08 & 2.5e+08\\
$<$5 & -2.2 & -2.0 & 6.7e+07 & 1.1e+08 & 1.9e+07 & 3.1e+07 & 2.3e+08 & 4.8e+08 & 7.4e+08 & 1.2e+09\\
$<$5 & -2.0 & -1.8 & 1.2e+08 & 1.9e+08 & 3.6e+07 & 5.7e+07 & 1.2e+08 & 2.6e+08 & 4.4e+08 & 6.8e+08\\
$<$5 & -1.8 & -1.6 & 7.8e+07 & 1.2e+08 & 1.1e+08 & 1.8e+08 & 2.7e+07 & 5.6e+07 & 3.4e+08 & 5.3e+08\\
$<$5 & -1.6 & -1.4 & 7.9e+07 & 1.3e+08 & 3.9e+07 & 6.2e+07 & 7.3e+07 & 1.5e+08 & 3.6e+08 & 5.5e+08\\
$<$5 & -1.4 & -1.2 & 1.2e+08 & 1.9e+08 & 1.6e+08 & 2.6e+08 & 2.3e+08 & 4.9e+08 & 6.8e+08 & 1.1e+09\\
$<$5 & -1.2 & -1.0 & 2.8e+08 & 4.4e+08 & 2.8e+08 & 4.5e+08 & 2.5e+08 & 5.2e+08 & 9.0e+08 & 1.4e+09\\
$<$5 & -1.0 & -0.8 & 9.4e+08 & 1.5e+09 & 5.9e+08 & 9.4e+08 & 5.8e+08 & 1.2e+09 & 2.5e+09 & 3.8e+09\\
$<$5 & -0.8 & -0.6 & 2.0e+09 & 3.1e+09 & 2.9e+09 & 4.6e+09 & 1.1e+09 & 2.4e+09 & 9.1e+08 & 1.4e+09\\
$<$5 & -0.6 & -0.4 & 4.2e+09 & 6.7e+09 & 2.9e+09 & 4.6e+09 & 2.1e+09 & 4.4e+09 & 2.4e+09 & 3.7e+09\\
$<$5 & -0.4 & -0.2 & 3.5e+09 & 5.5e+09 & 3.5e+09 & 5.5e+09 & 2.4e+09 & 5.2e+09 & 3.2e+09 & 4.9e+09\\
$<$5 & -0.2 & 0.0 & 2.8e+09 & 4.5e+09 & 2.8e+09 & 4.4e+09 & 2.4e+09 & 5.1e+09 & 3.7e+09 & 5.7e+09\\
$<$5 & 0.0 & 0.2 & 1.1e+10 & 1.7e+10 & 1.2e+10 & 1.9e+10 & 9.3e+09 & 2.0e+10 & 9.6e+09 & 1.5e+10\\
5$-$12 & -2.4 & -2.2 & 2.4e+08 & 9.8e+08 & 2.4e+08 & 7.9e+08 & -0.0e+00 & -0.0e+00 & 6.3e+08 & 2.3e+09\\
5$-$12 & -2.2 & -2.0 & 2.0e+08 & 8.0e+08 & 4.6e+07 & 1.5e+08 & 2.5e+08 & 1.1e+09 & 2.4e+08 & 8.9e+08\\
5$-$12 & -2.0 & -1.8 & 2.2e+08 & 8.9e+08 & 9.0e+07 & 3.0e+08 & 2.6e+08 & 1.1e+09 & 6.3e+08 & 2.3e+09\\
5$-$12 & -1.8 & -1.6 & 1.5e+08 & 6.0e+08 & 8.2e+07 & 2.7e+08 & 9.8e+07 & 4.2e+08 & 1.9e+08 & 7.0e+08\\
5$-$12 & -1.6 & -1.4 & 3.3e+08 & 1.3e+09 & 1.1e+08 & 3.7e+08 & 1.7e+08 & 7.3e+08 & 3.2e+08 & 1.2e+09\\
5$-$12 & -1.4 & -1.2 & 1.0e+08 & 4.2e+08 & 1.2e+08 & 4.1e+08 & 3.9e+08 & 1.7e+09 & 2.2e+08 & 8.1e+08\\
5$-$12 & -1.2 & -1.0 & 2.6e+08 & 1.1e+09 & 3.4e+08 & 1.1e+09 & 3.8e+08 & 1.6e+09 & 6.4e+08 & 2.3e+09\\
5$-$12 & -1.0 & -0.8 & 3.4e+08 & 1.4e+09 & 7.1e+08 & 2.4e+09 & 6.3e+08 & 2.7e+09 & 1.4e+09 & 4.9e+09\\
5$-$12 & -0.8 & -0.6 & 4.3e+08 & 1.8e+09 & 1.2e+09 & 4.0e+09 & 8.2e+08 & 3.5e+09 & 1.2e+09 & 4.2e+09\\
5$-$12 & -0.6 & -0.4 & 2.5e+09 & 1.0e+10 & 2.7e+09 & 8.9e+09 & 1.7e+09 & 7.2e+09 & 3.0e+09 & 1.1e+10\\
5$-$12 & -0.4 & -0.2 & 3.3e+09 & 1.3e+10 & 4.8e+09 & 1.6e+10 & 3.4e+09 & 1.5e+10 & 3.9e+09 & 1.4e+10\\
5$-$12 & -0.2 & 0.0 & 2.8e+09 & 1.1e+10 & 3.6e+09 & 1.2e+10 & 2.2e+09 & 9.4e+09 & 2.8e+09 & 1.0e+10\\
5$-$12 & 0.0 & 0.2 & 6.4e+09 & 2.6e+10 & 7.0e+09 & 2.3e+10 & 6.0e+09 & 2.6e+10 & 4.3e+09 & 1.6e+10\\
$>$12 & -2.4 & -2.2 & 1.2e+08 & 8.2e+08 & 2.2e+08 & 1.2e+09 & -0.0e+00 & -0.0e+00 & 3.2e+08 & 1.9e+09\\
$>$12 & -2.2 & -2.0 & 7.0e+07 & 4.8e+08 & 2.6e+07 & 1.5e+08 & 2.5e+08 & 1.6e+09 & 5.1e+07 & 3.1e+08\\
$>$12 & -2.0 & -1.8 & 4.4e+07 & 3.0e+08 & 1.0e+08 & 5.8e+08 & 5.8e+07 & 3.7e+08 & 1.5e+08 & 9.1e+08\\
$>$12 & -1.8 & -1.6 & 4.0e+07 & 2.7e+08 & 2.5e+07 & 1.4e+08 & 3.5e+07 & 2.3e+08 & 4.7e+07 & 2.9e+08\\
$>$12 & -1.6 & -1.4 & 7.6e+07 & 5.2e+08 & 7.3e+07 & 4.1e+08 & 4.9e+07 & 3.2e+08 & 8.0e+07 & 4.8e+08\\
$>$12 & -1.4 & -1.2 & 2.1e+07 & 1.5e+08 & 5.7e+07 & 3.2e+08 & 8.6e+07 & 5.5e+08 & 1.0e+08 & 6.2e+08\\
$>$12 & -1.2 & -1.0 & 6.4e+07 & 4.4e+08 & 1.8e+08 & 1.0e+09 & 1.2e+08 & 7.9e+08 & 2.4e+08 & 1.5e+09\\
$>$12 & -1.0 & -0.8 & 1.7e+08 & 1.2e+09 & 2.9e+08 & 1.6e+09 & 1.7e+08 & 1.1e+09 & 4.1e+08 & 2.5e+09\\
$>$12 & -0.8 & -0.6 & 1.5e+08 & 1.0e+09 & 3.9e+08 & 2.2e+09 & 3.5e+08 & 2.3e+09 & 8.3e+08 & 5.1e+09\\
$>$12 & -0.6 & -0.4 & 1.0e+09 & 6.9e+09 & 1.4e+09 & 7.7e+09 & 9.8e+08 & 6.4e+09 & 1.7e+09 & 1.0e+10\\
$>$12 & -0.4 & -0.2 & 1.6e+09 & 1.1e+10 & 1.7e+09 & 9.8e+09 & 2.0e+09 & 1.3e+10 & 1.3e+09 & 7.7e+09\\
$>$12 & -0.2 & 0.0 & 1.5e+09 & 1.0e+10 & 1.4e+09 & 8.1e+09 & 9.5e+08 & 6.1e+09 & 5.8e+08 & 3.6e+09\\
$>$12 & 0.0 & 0.2 & 9.7e+08 & 6.7e+09 & 1.2e+09 & 6.8e+09 & 1.1e+09 & 7.0e+09 & 7.9e+08 & 4.8e+09\\
\enddata
\label{free_radial_metallicity}
\end{deluxetable}

\end{turnpage}

\clearpage

\begin{figure*}
\begin{center}
\includegraphics[width=3.1in]{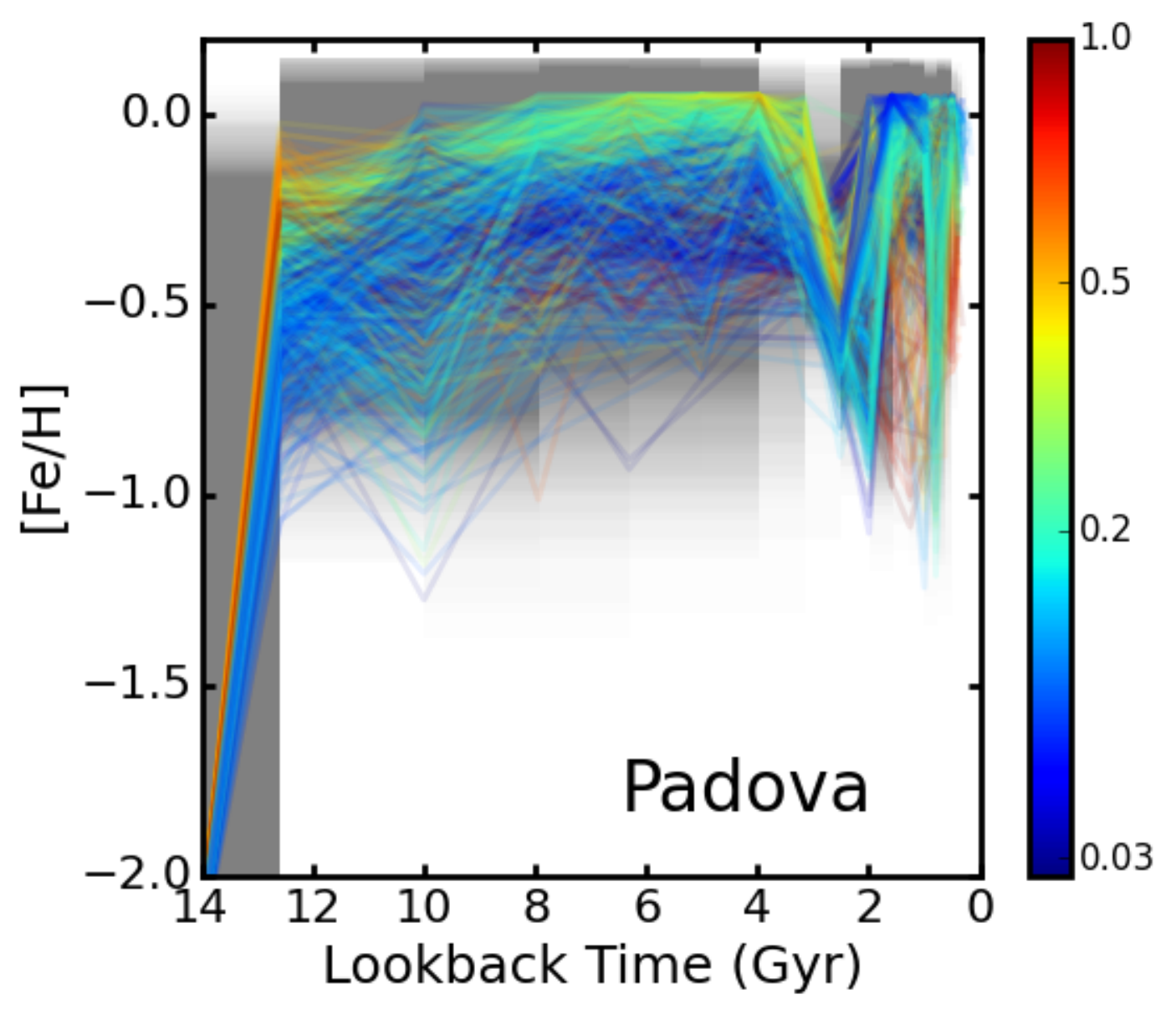}
\includegraphics[width=3.1in]{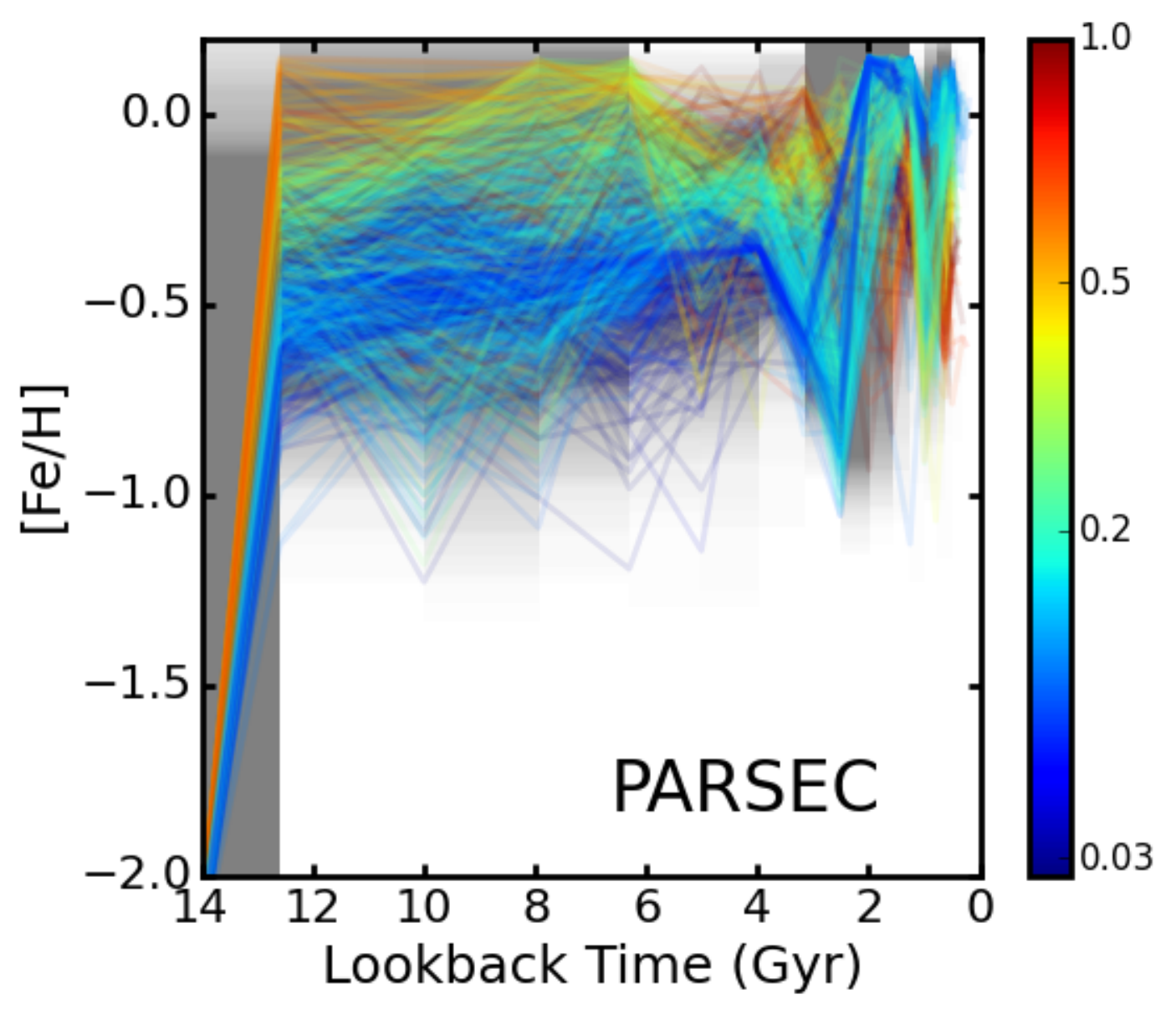}
\includegraphics[width=3.1in]{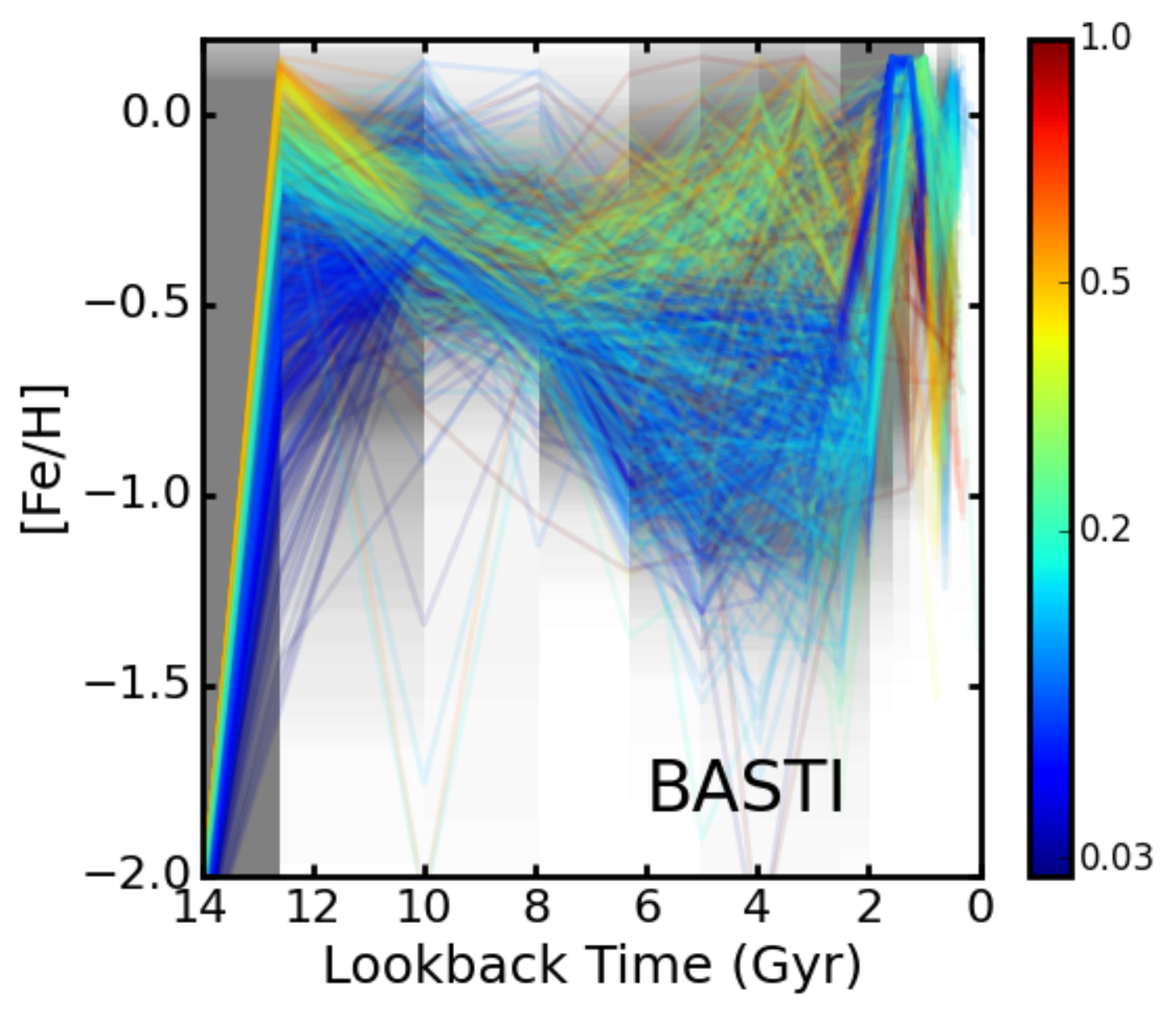}
\includegraphics[width=3.1in]{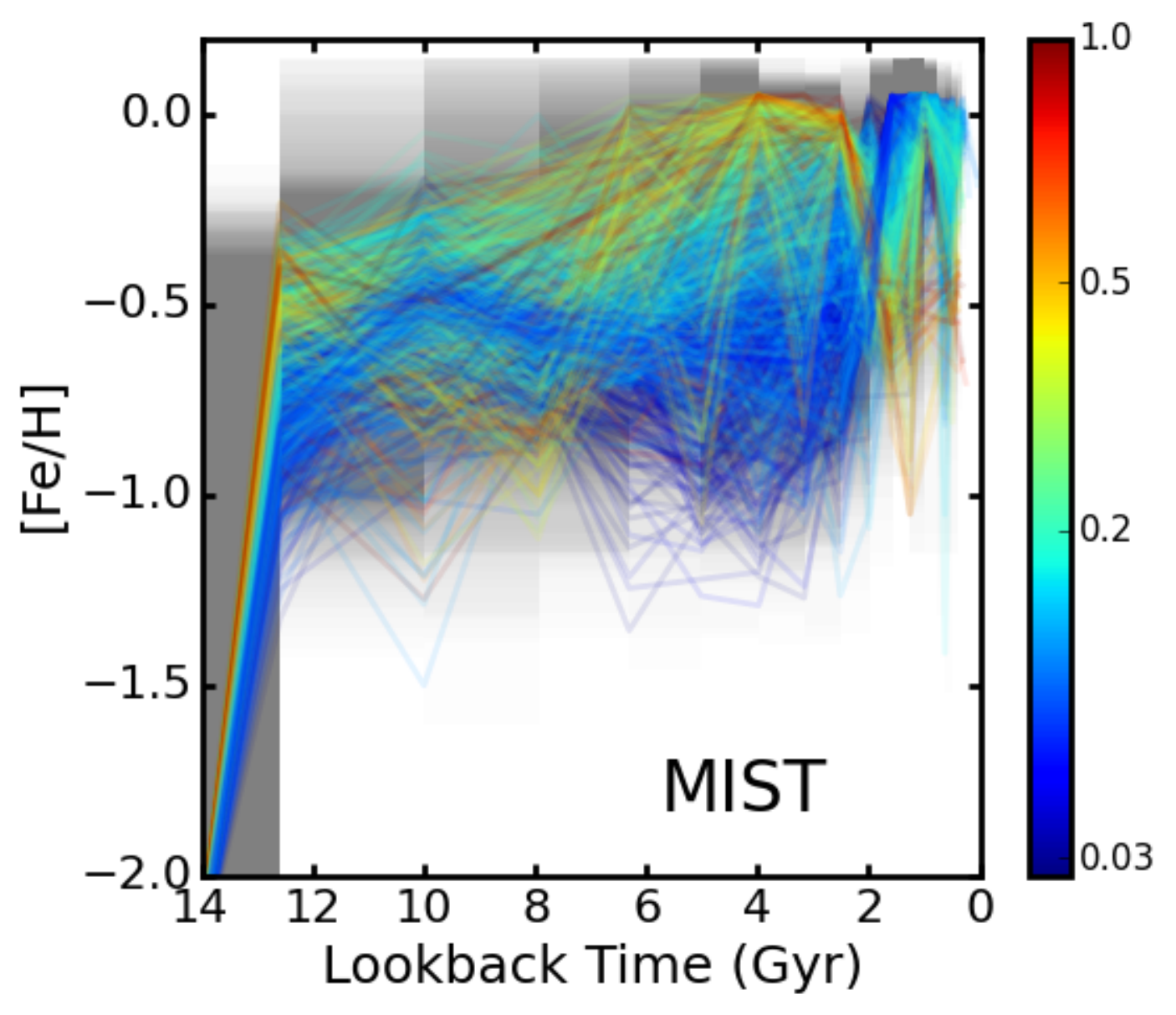}
\end{center}
\caption{Metallicity as a function of age for our fits of all regions. Panels are labeled with the models used to obtain the fits.  Lines are color-coded by their values in Figure~\ref{stellar_density_map}.}
\label{free_metallicity}
\end{figure*} 

\begin{figure*}
\begin{center}
\includegraphics[width=3.1in]{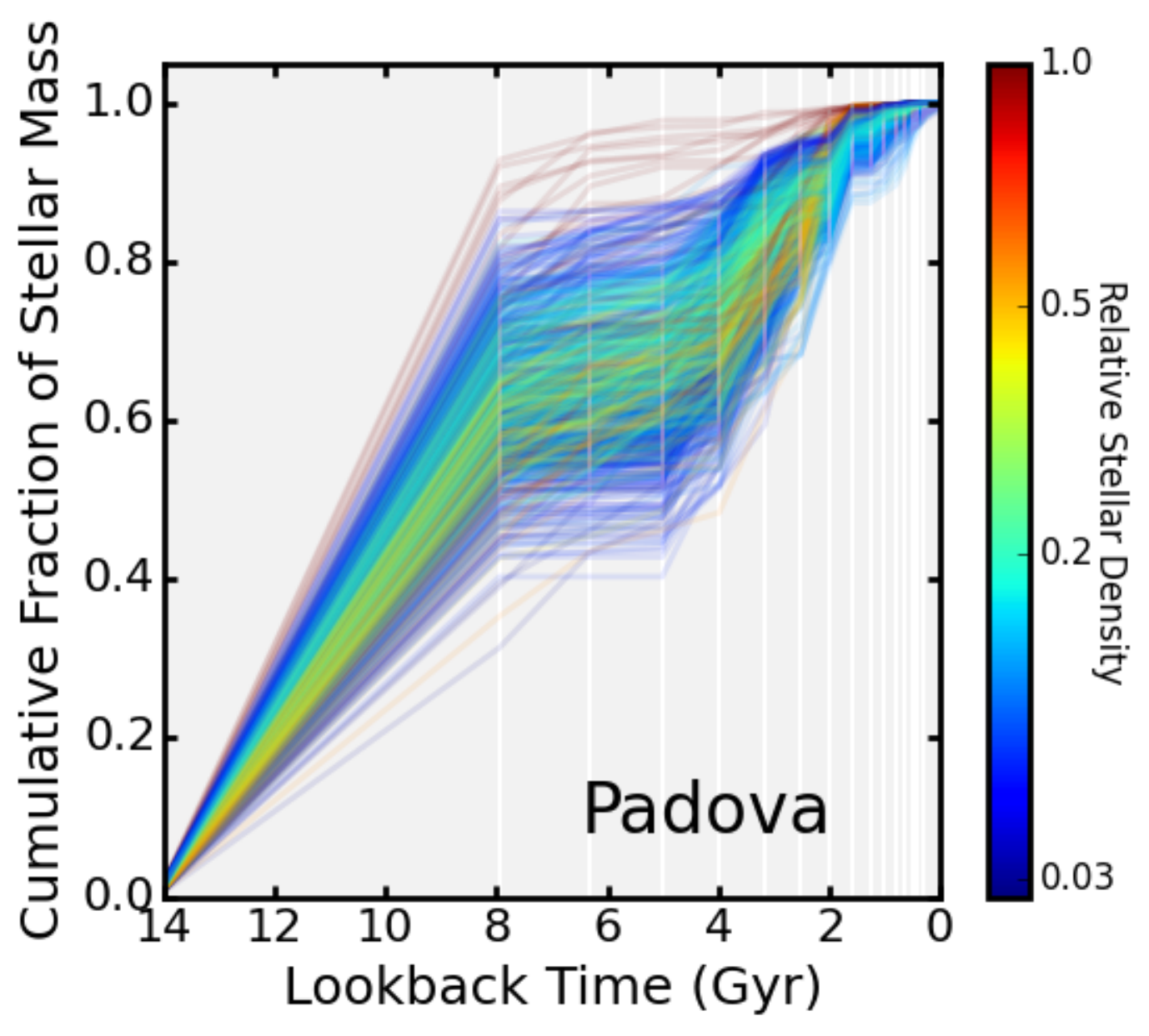}
\includegraphics[width=3.1in]{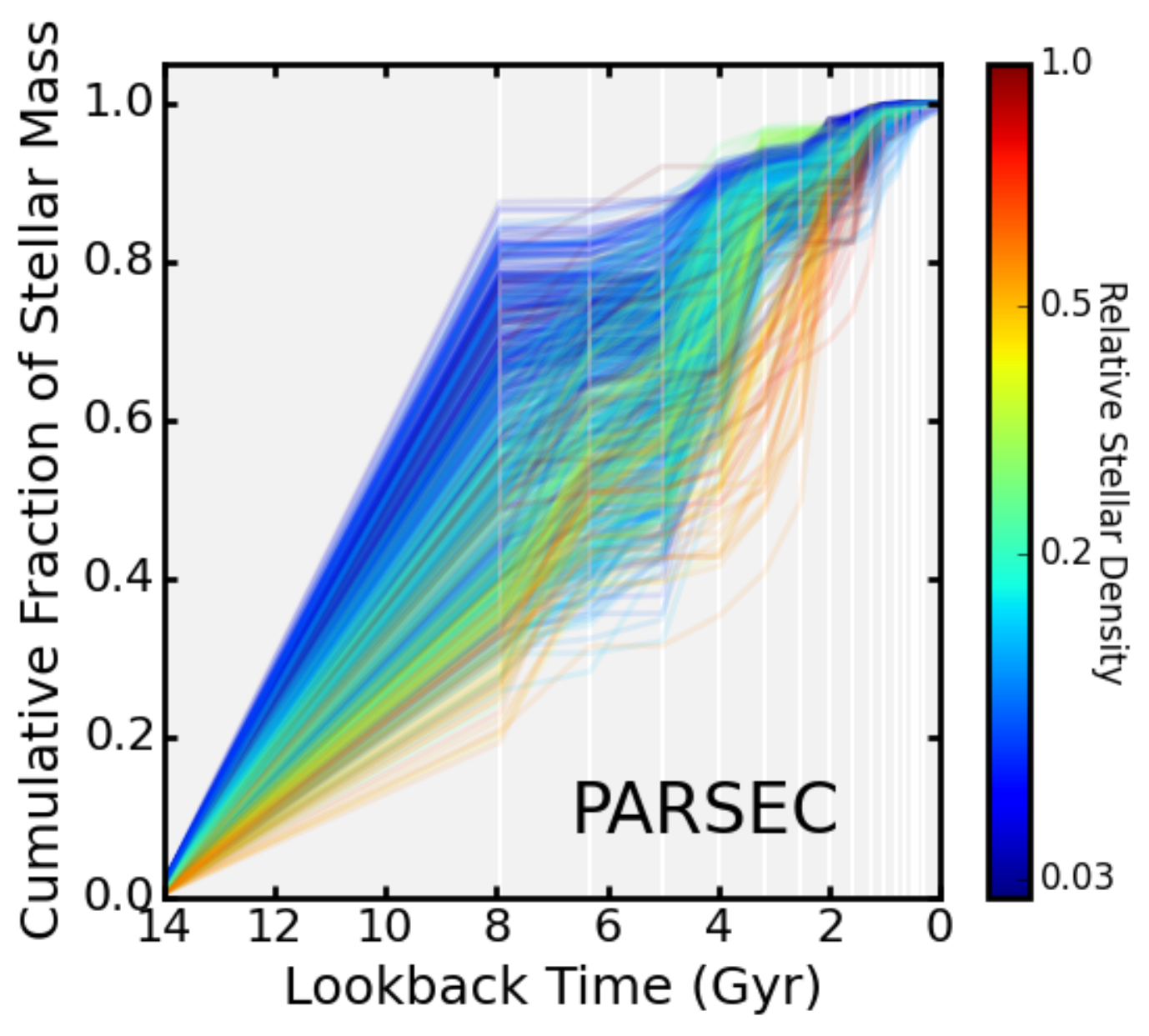}
\includegraphics[width=3.1in]{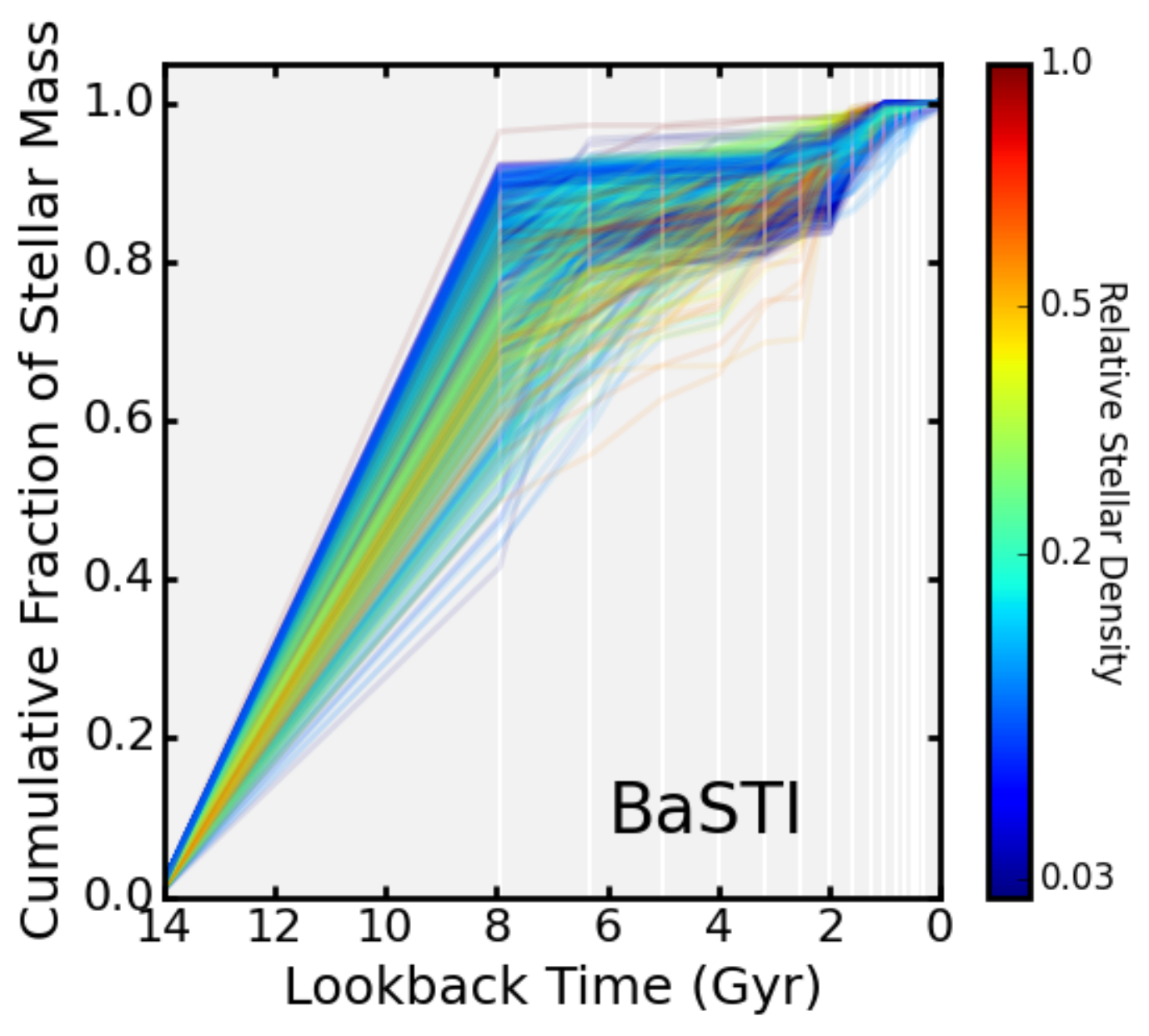}
\includegraphics[width=3.1in]{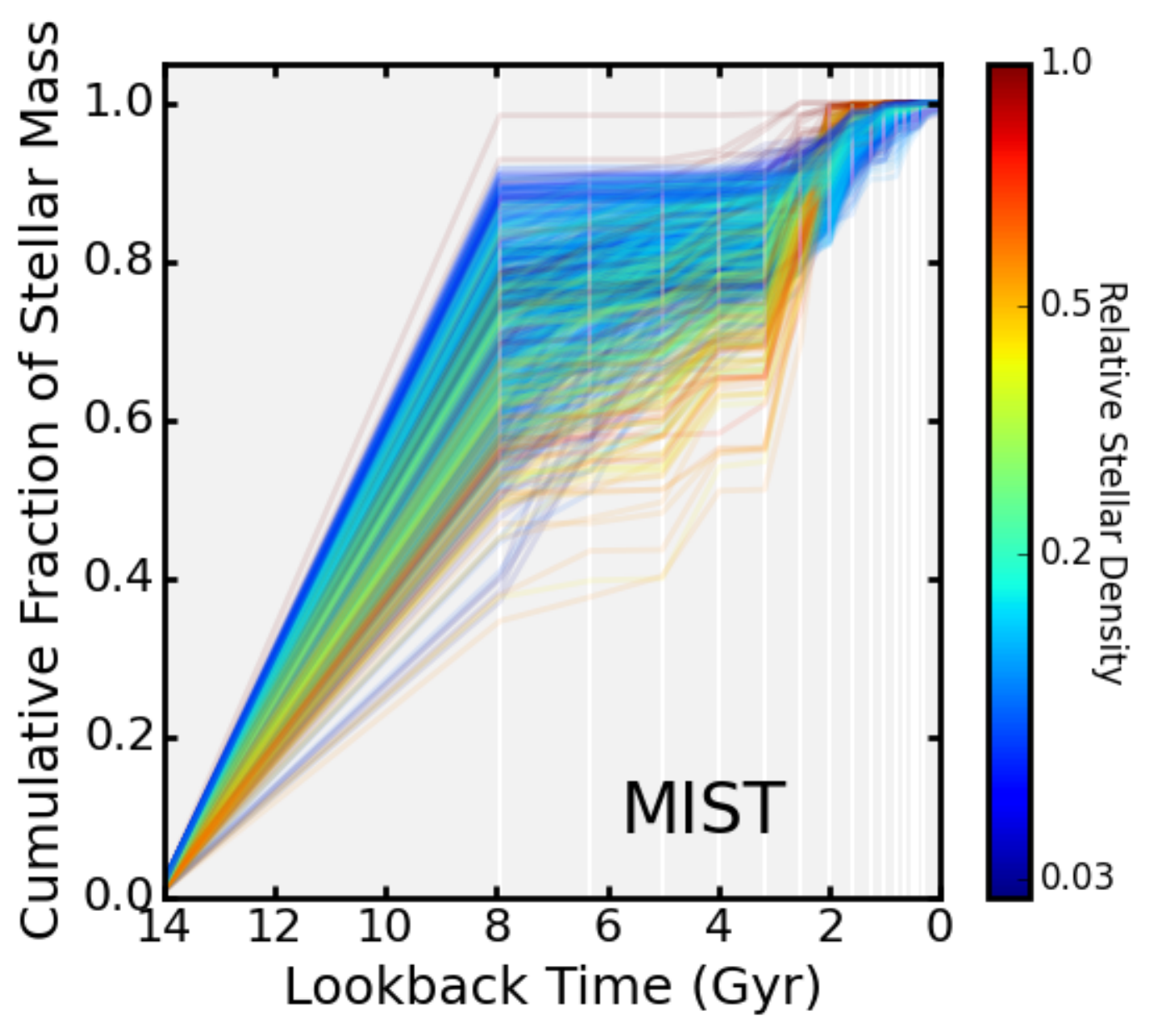}
\end{center}
\caption{Cumulative SFH vs. age for different stellar evolution model options. Each panel shows the best-fit cumulative age distribution for all 826 of our sub-regions, color-coded by the stellar density shown in Figure~\ref{stellar_density_map}.  Panels are labeled by with the names of the models used.  \label{free_sfhs}}

\end{figure*}

\begin{figure*}
\begin{center}
\includegraphics[width=3.1in]{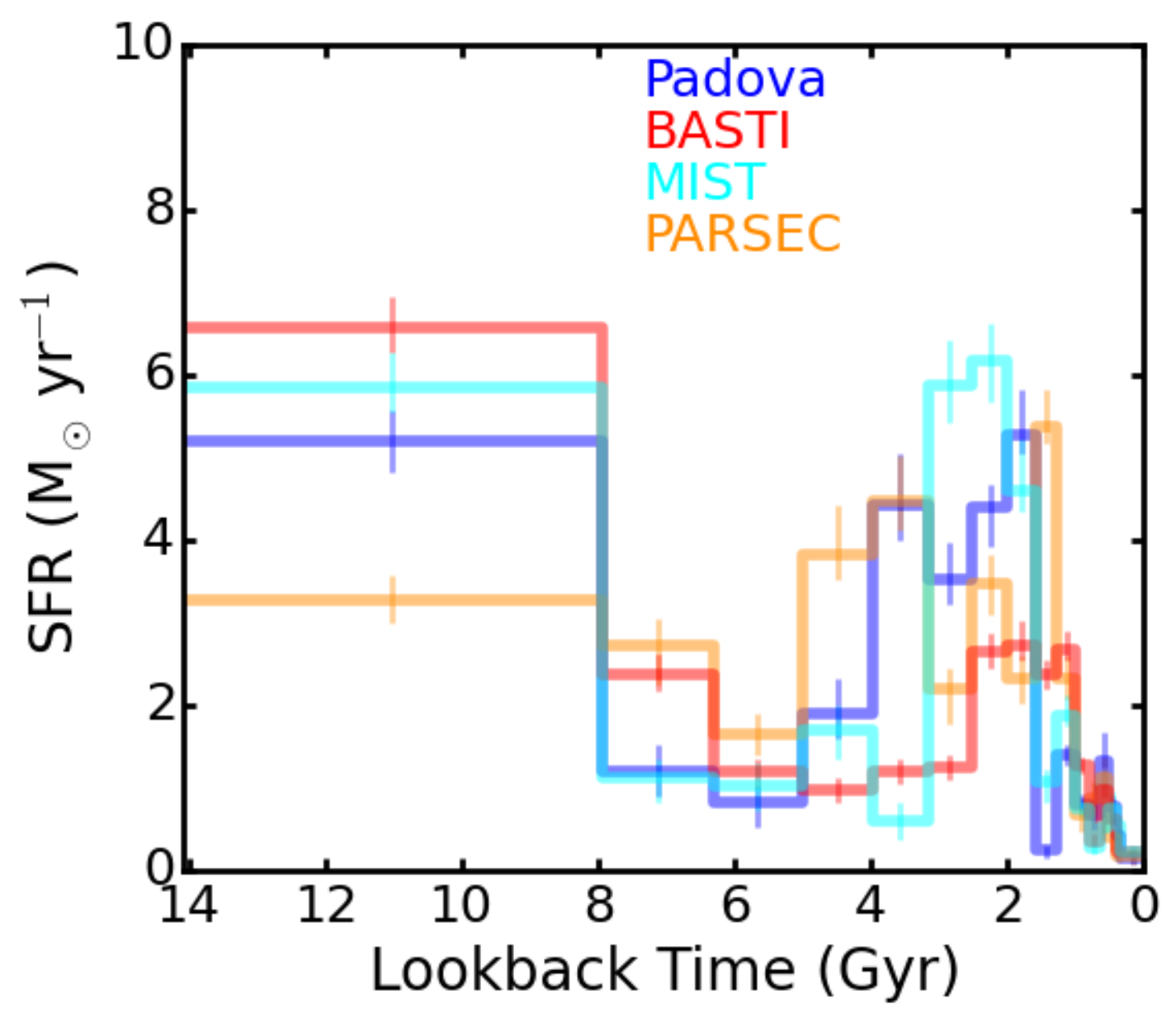}
\includegraphics[width=3.1in]{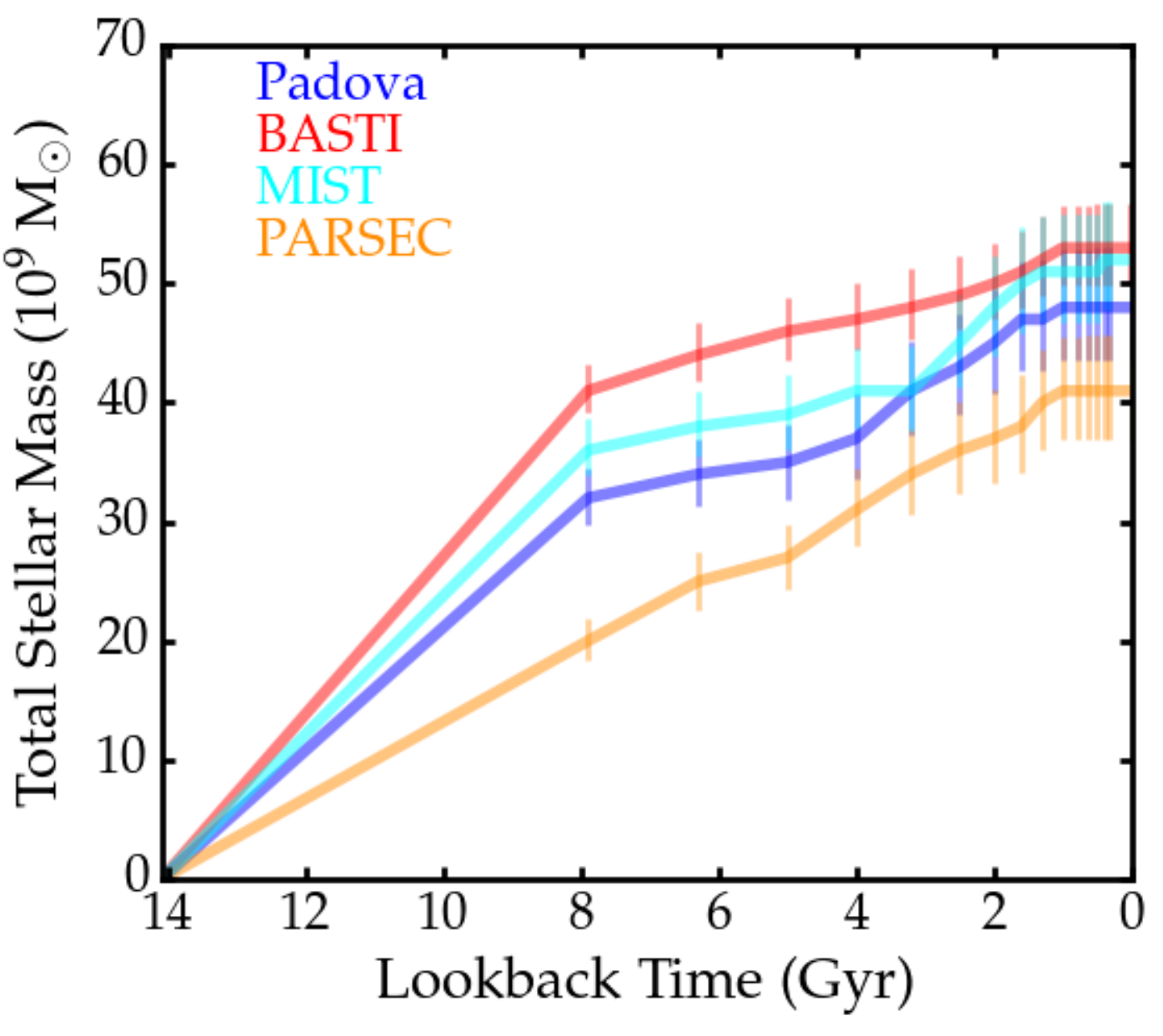}
\end{center}
\caption{Total star formation rate vs. age for the entire PHAT
  footprint, as measured using 3 different model sets imposing a
  common simple enrichment scenario throughout the survey.  The colors
  show the results for each of the three models sets fitted.   {\it Right:} Total
  accumulated mass as a function of lookback time, directly calculated
  from our fits.\label{free_total_sfh}}
\end{figure*}

\begin{figure*}
\begin{center}
\includegraphics[width=5.1in]{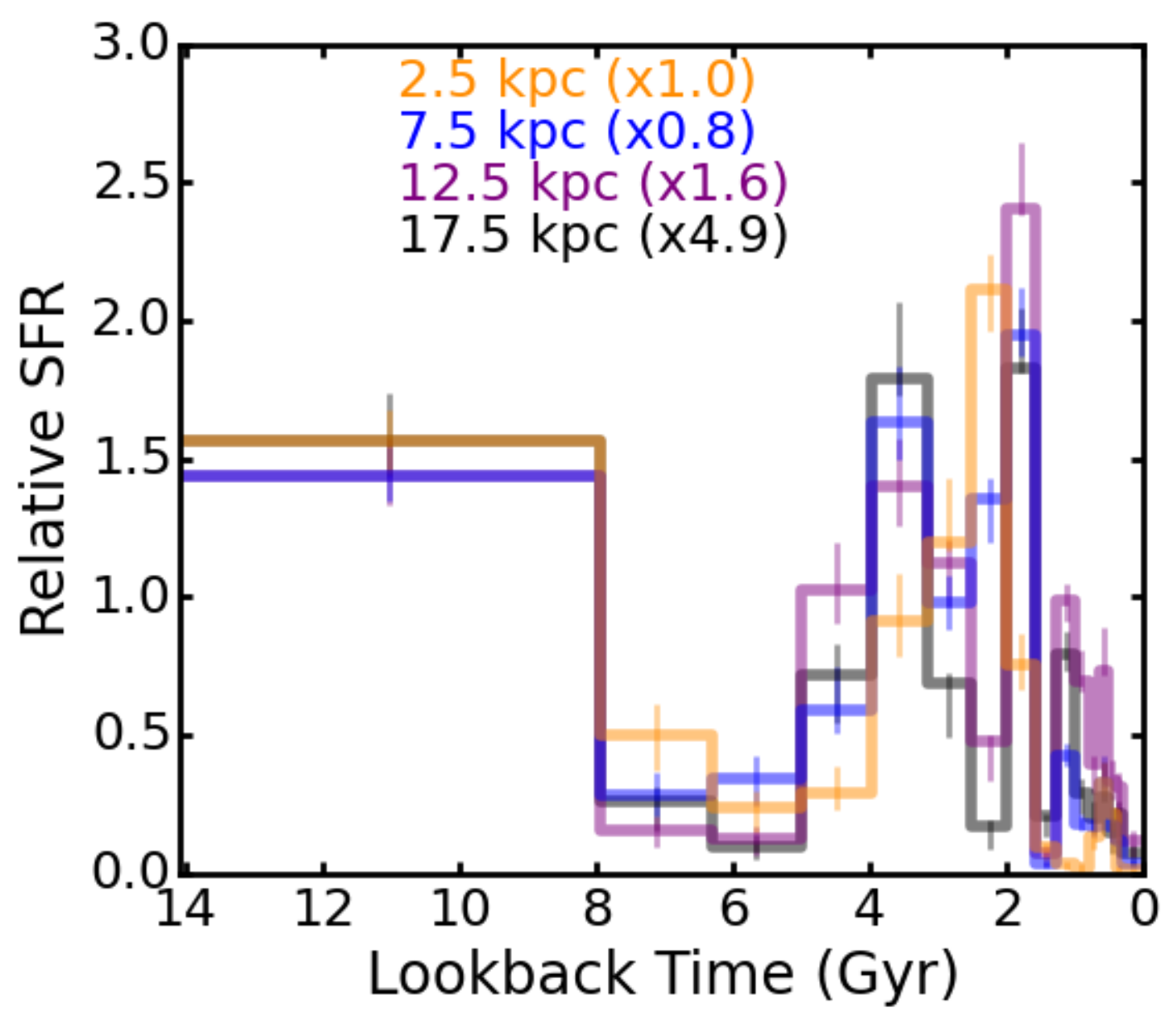}
\end{center}
\caption{Total relative star formation rate vs. age for 4 radial bins, color-coded by radius, normalized to have the same mean rate of 1 $M_{\odot}$~yr$^{-1}$. These are the results from the Padova model fits. \label{free_radial_sfhs}}
\end{figure*}


\begin{figure*}
\begin{center}
\includegraphics[width=3.1in]{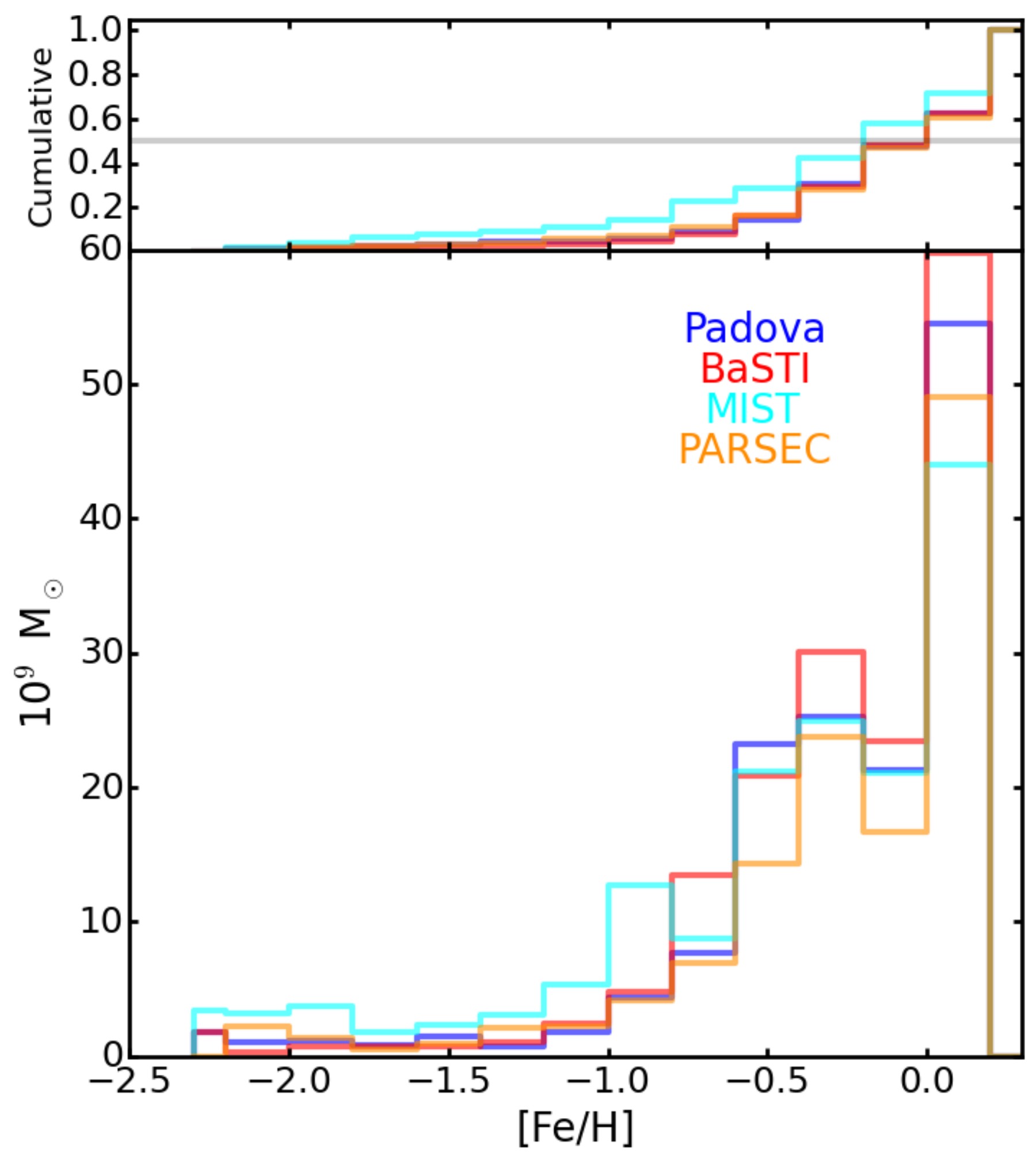}
\includegraphics[width=3.1in]{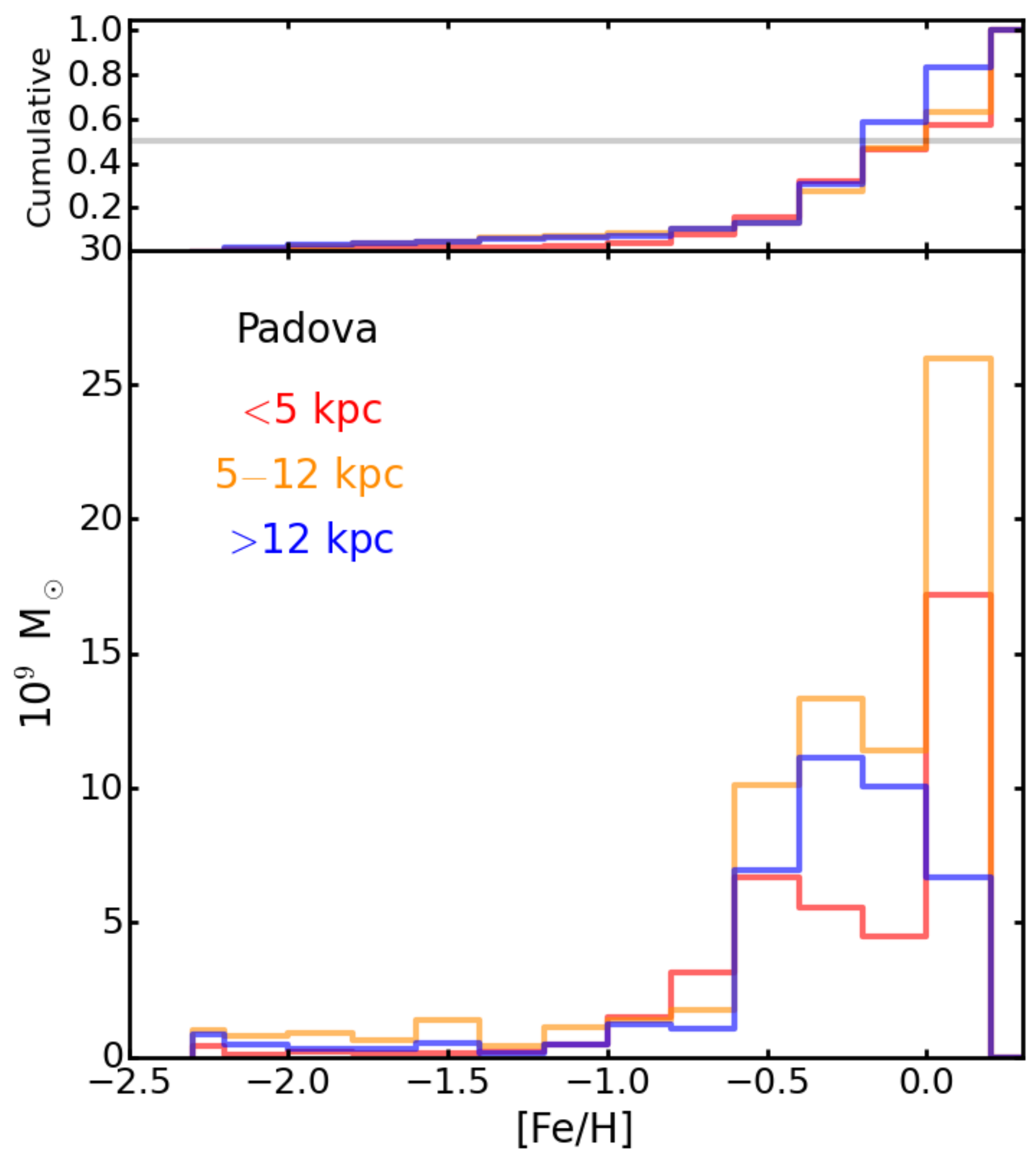}
\end{center}
\caption{Metallicity histograms from our free metallicity fits for the entire M31 disk, extrapolated
  from the entire area measured.  The top panel shows the cumulative fraction of the distribution as a function of metallicity. {\it Left:} The resulting distribution for all 4
  model sets is shown.  {\it Right:} Metallicity histograms for 3 radial bins from the Padova fits, correcting for area coverage.}
\label{free_metallicity_hist}
\end{figure*}

\begin{figure*}
\begin{center}
\includegraphics[width=3.1in]{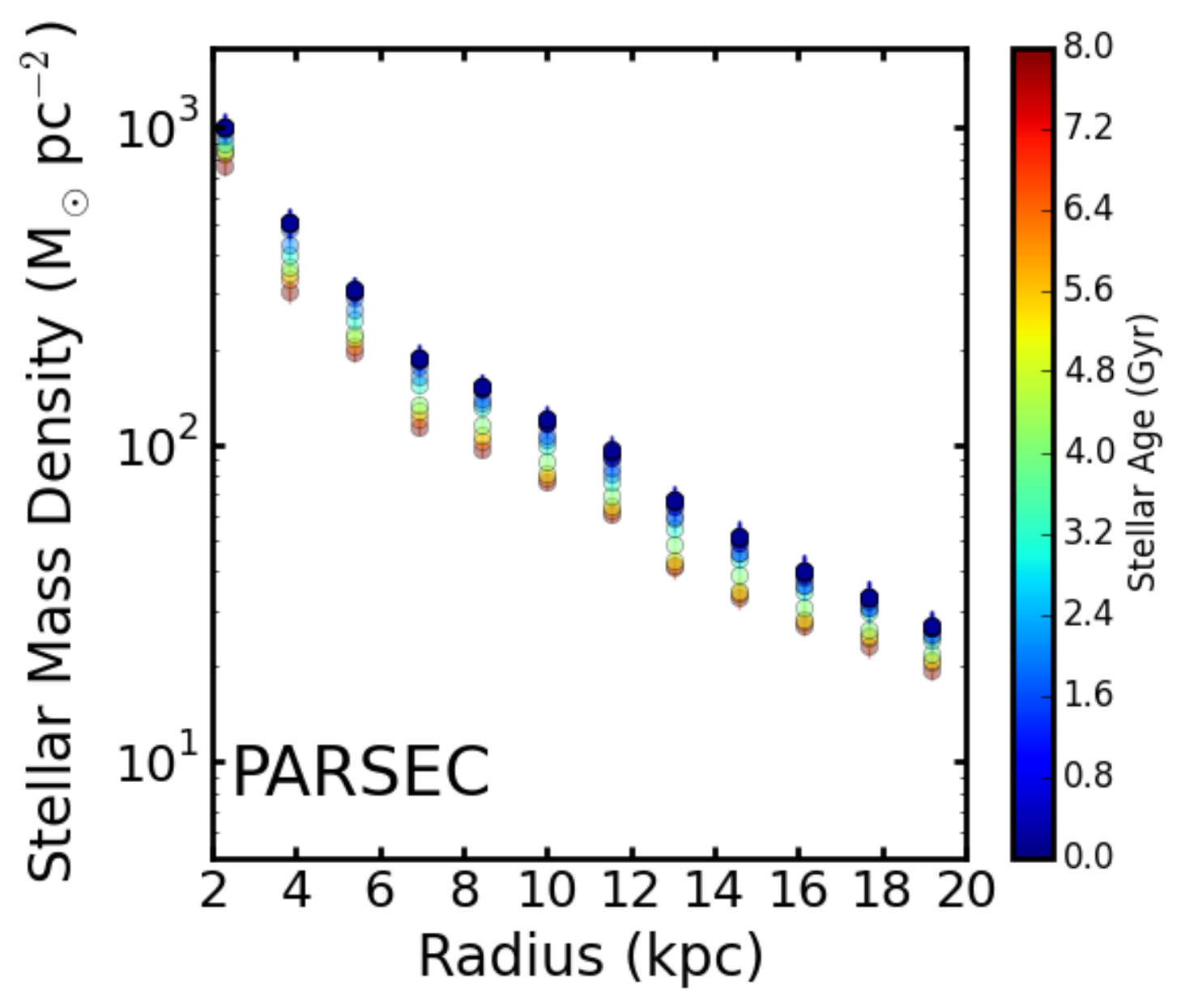}
\includegraphics[width=3.1in]{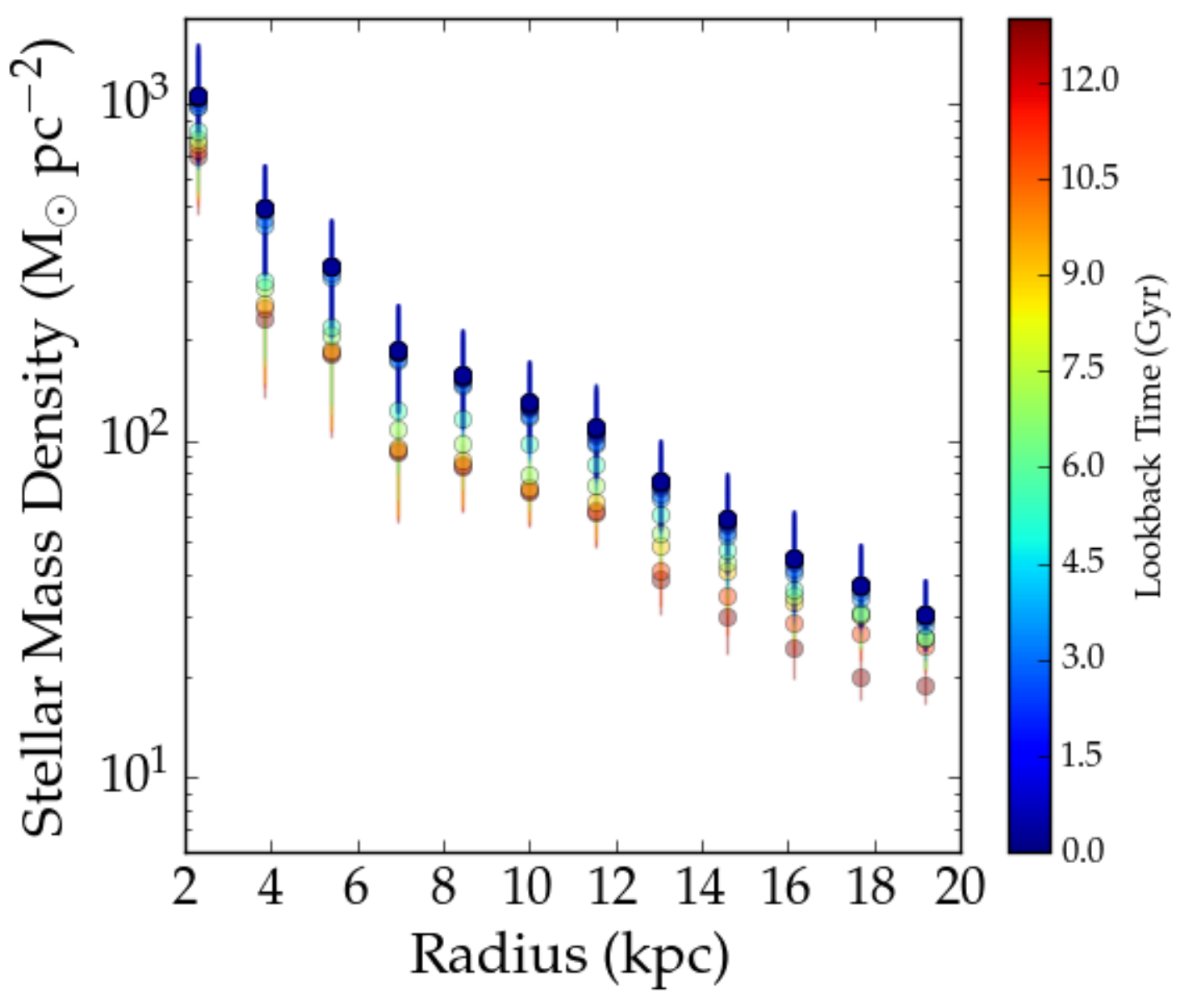}
\includegraphics[width=3.1in]{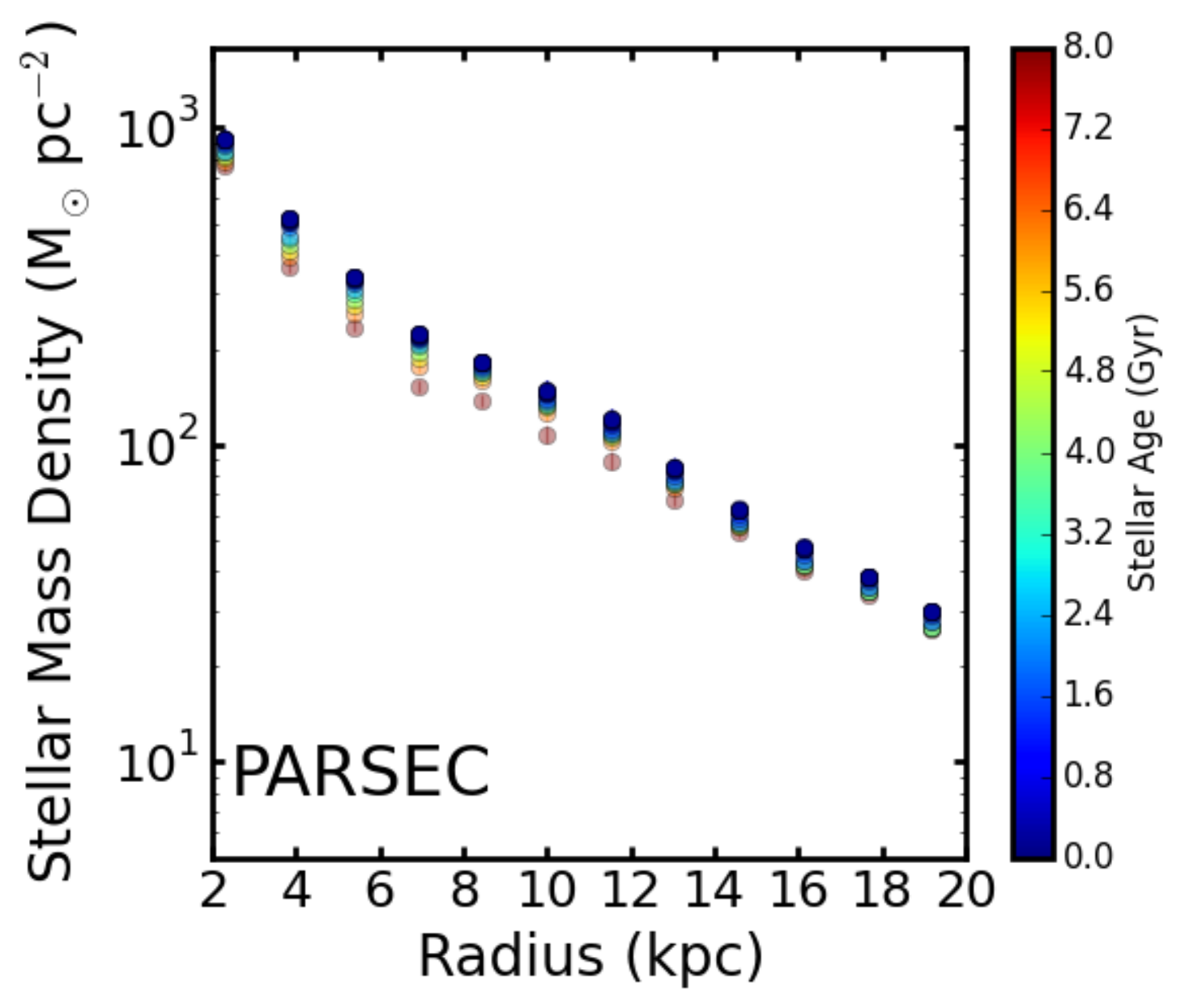}
\includegraphics[width=3.1in]{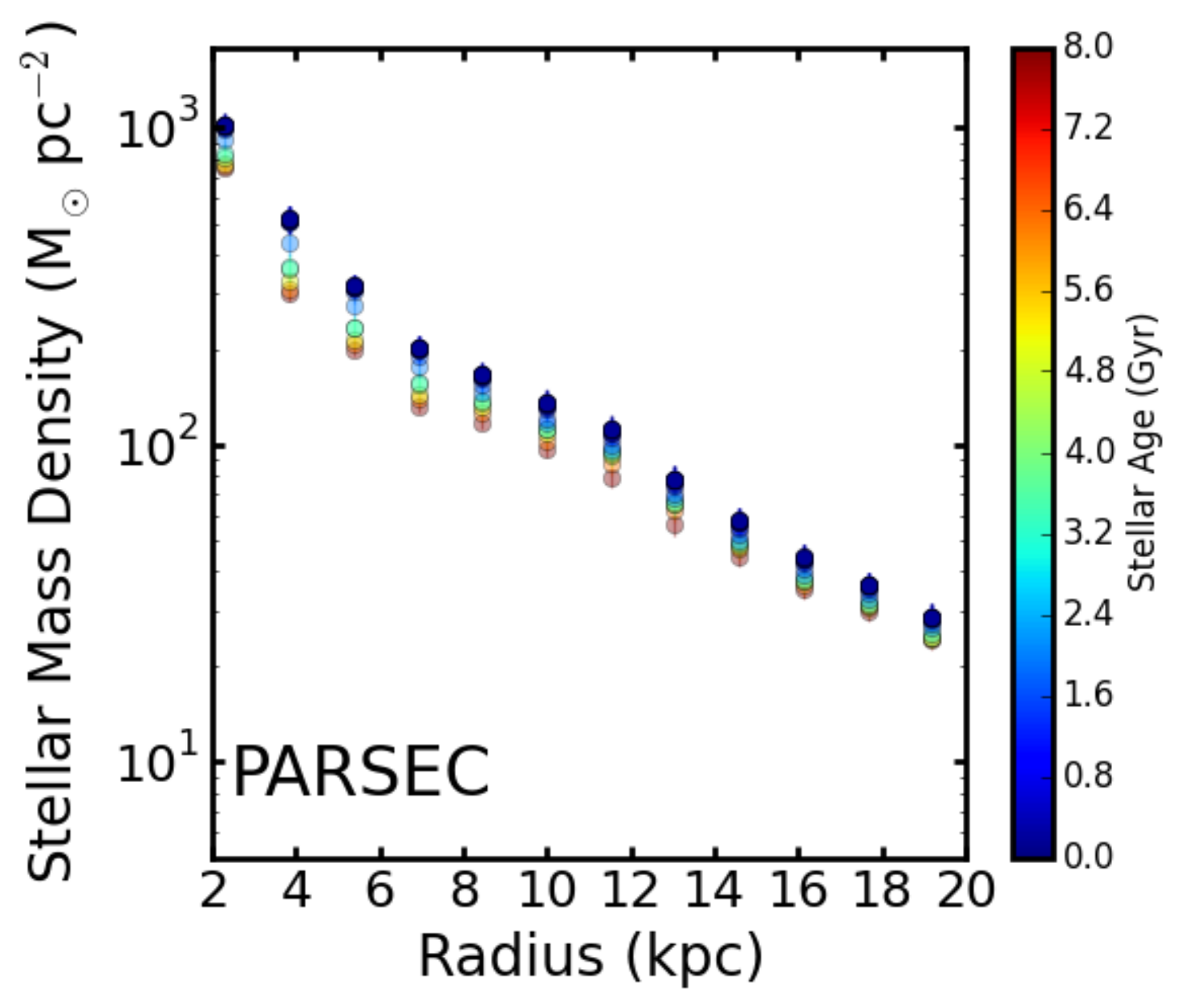}
\end{center}
\caption{Cumulative stellar mass surface density formed as a function
  of radius at several ages (lookback times) for the Padova SFHs shown in
  Figure~\ref{free_sfhs}.  Ratios of the results from the Padova SFHs
  to those from other model sets are provided for comparison.   Panels are labeled with the appropriate
  model set names.\label{free_density_vs_radius}}

\end{figure*}




\begin{figure*}
\begin{center}
\includegraphics[width=3.1in]{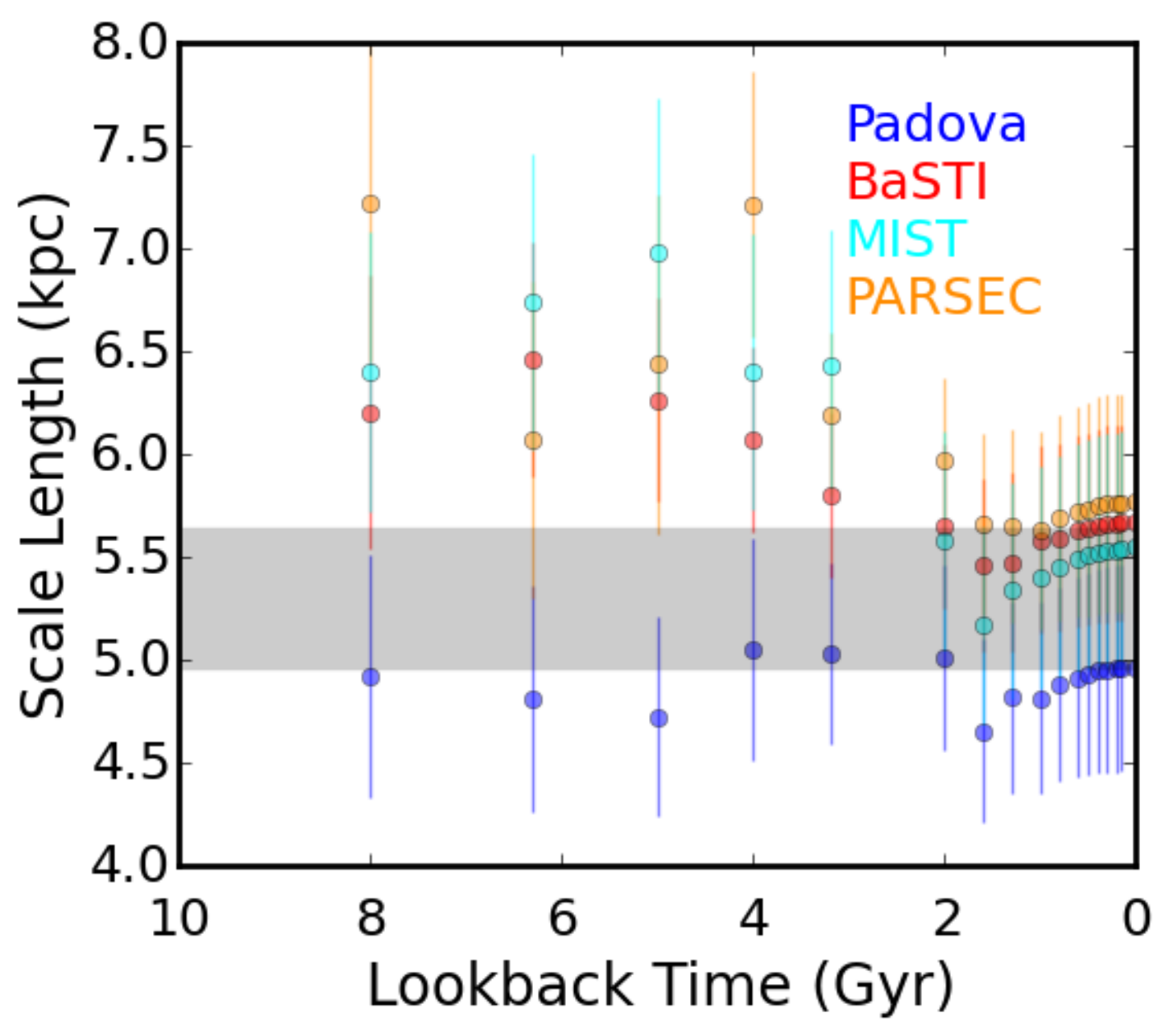}
\includegraphics[width=3.1in]{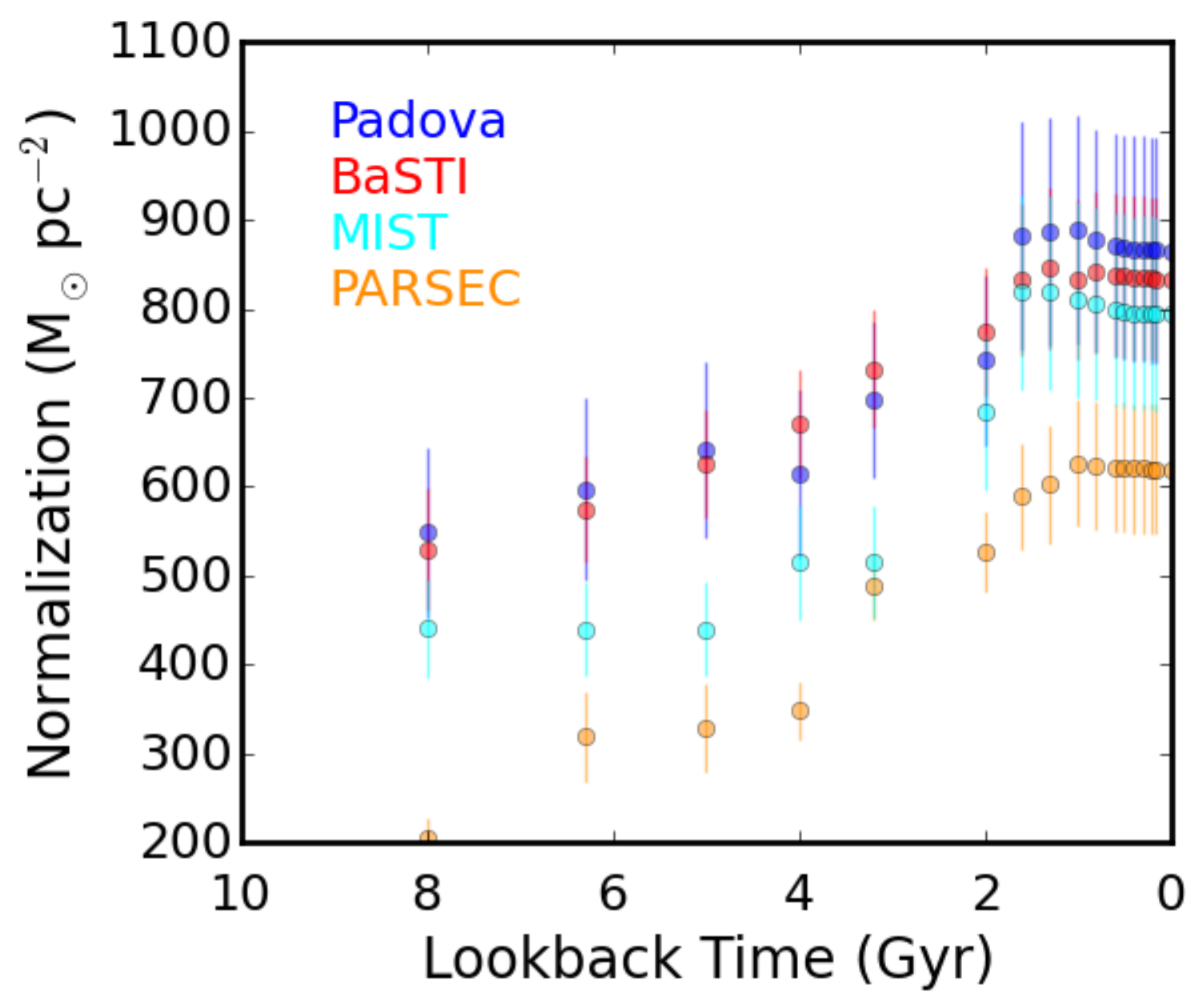}
\end{center}
\caption{Exponential fits to the disk profile of M31 as a function of
  time using the free metallicity fits from the scipy function {\tt
    curve\_fit}. {\it Left:} Disk scale length
  as a function of lookback time. {\it Right:} Normalization of the
  exponential disk as a function of lookback time.}
\label{free_exp_fits}
\end{figure*}

\begin{figure*}
\begin{center}
\includegraphics[height=2.2in]{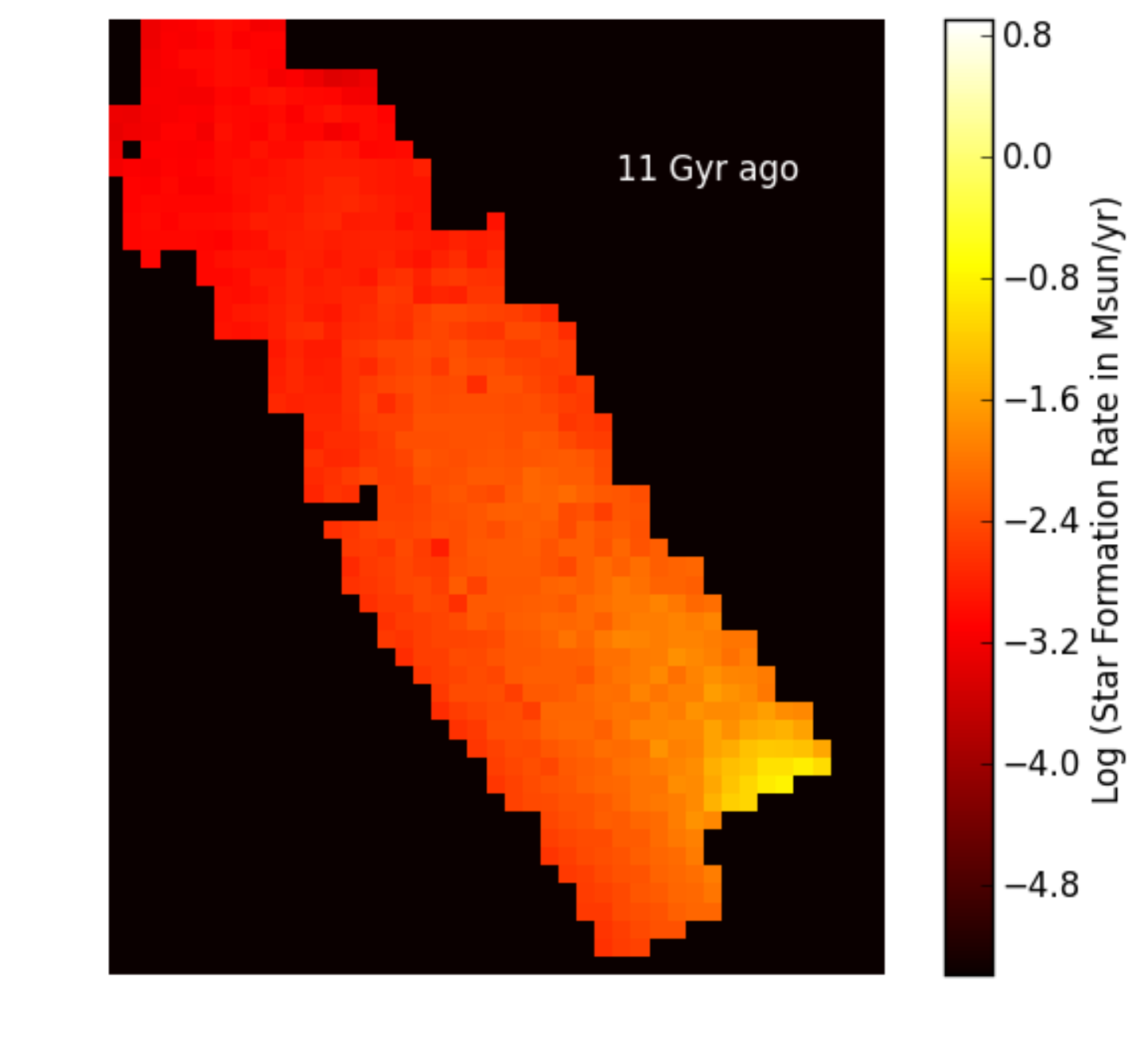}
\includegraphics[height=2.2in]{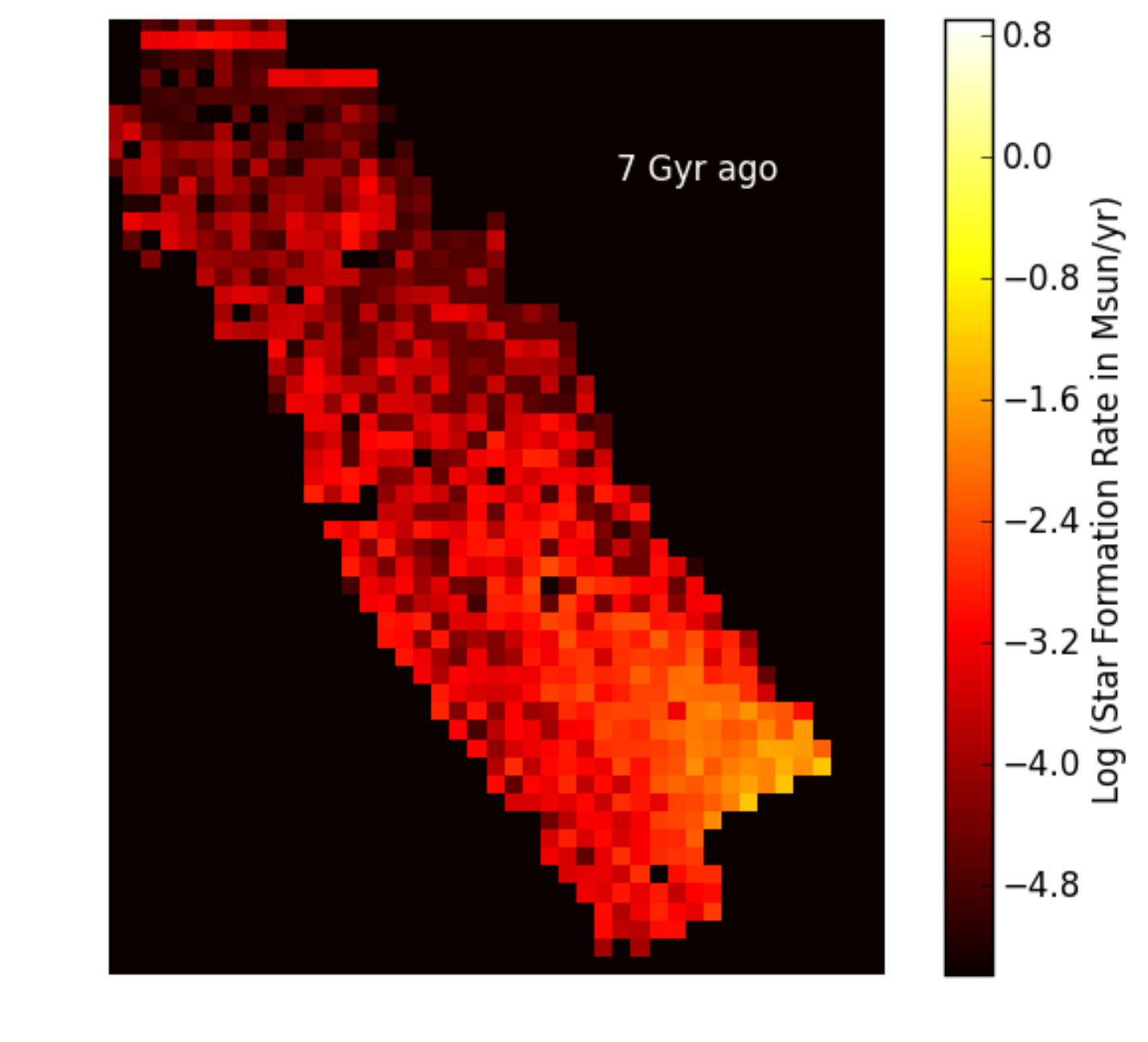}
\includegraphics[height=2.2in]{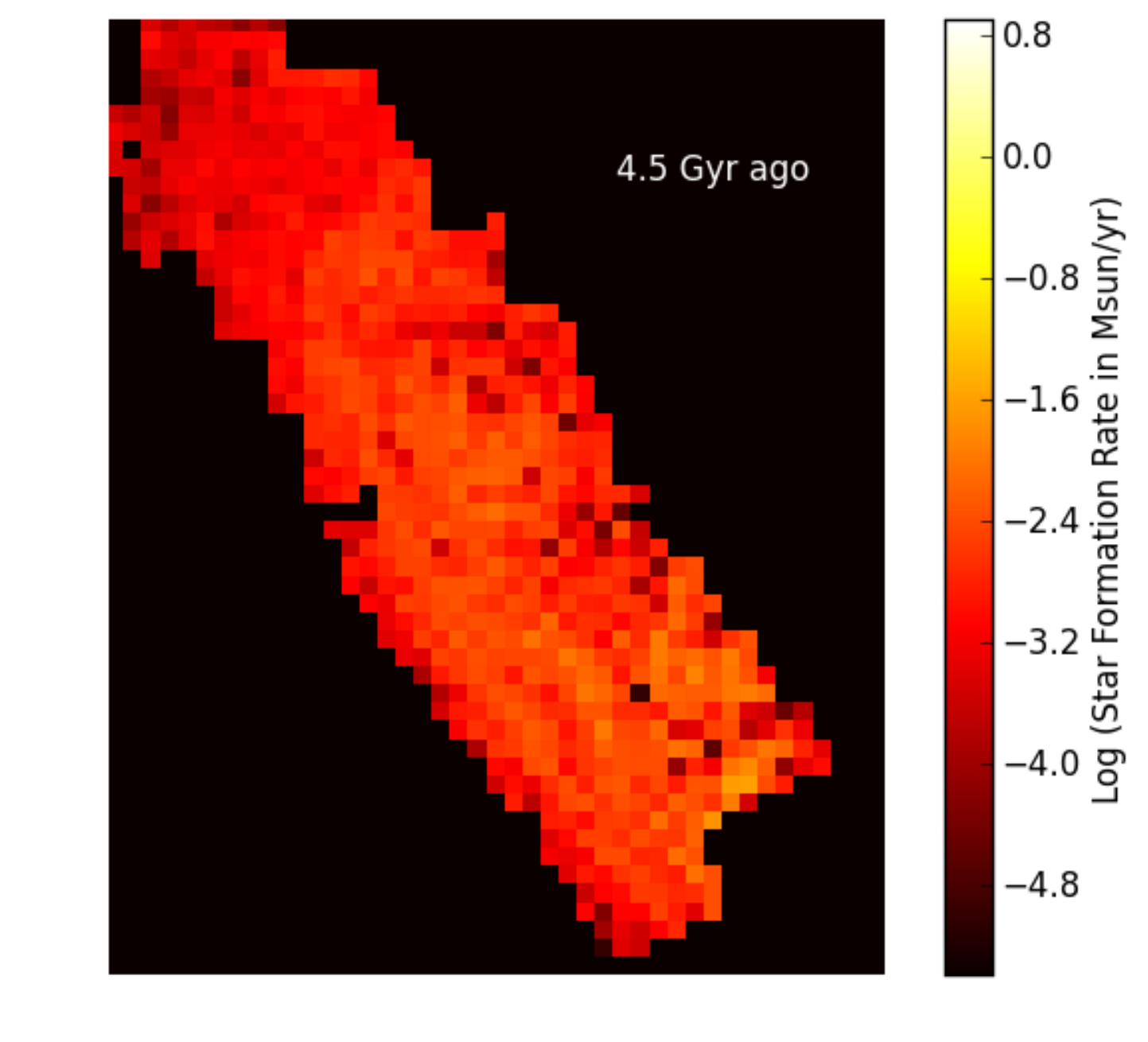}
\includegraphics[height=2.2in]{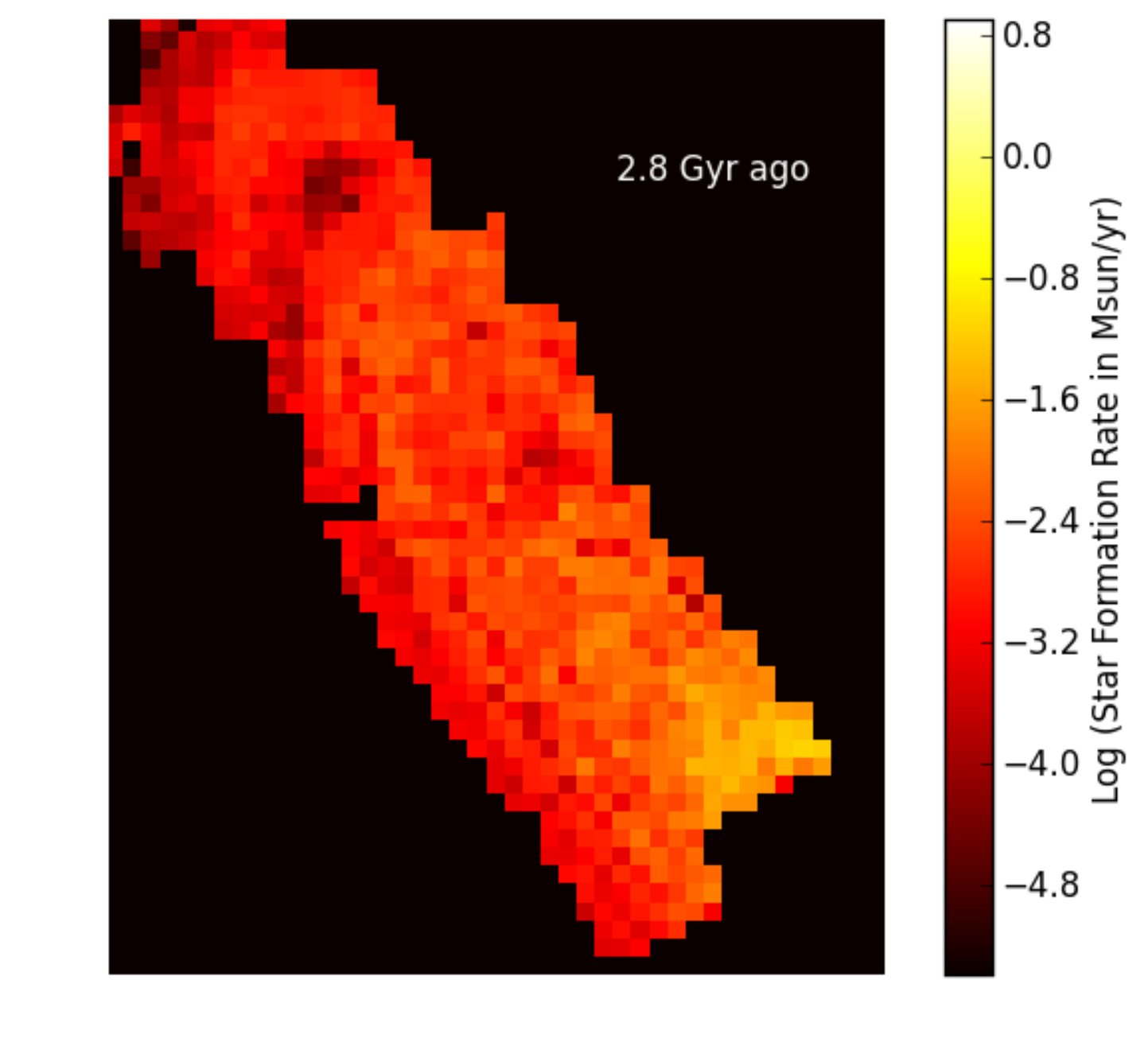}
\includegraphics[height=2.2in]{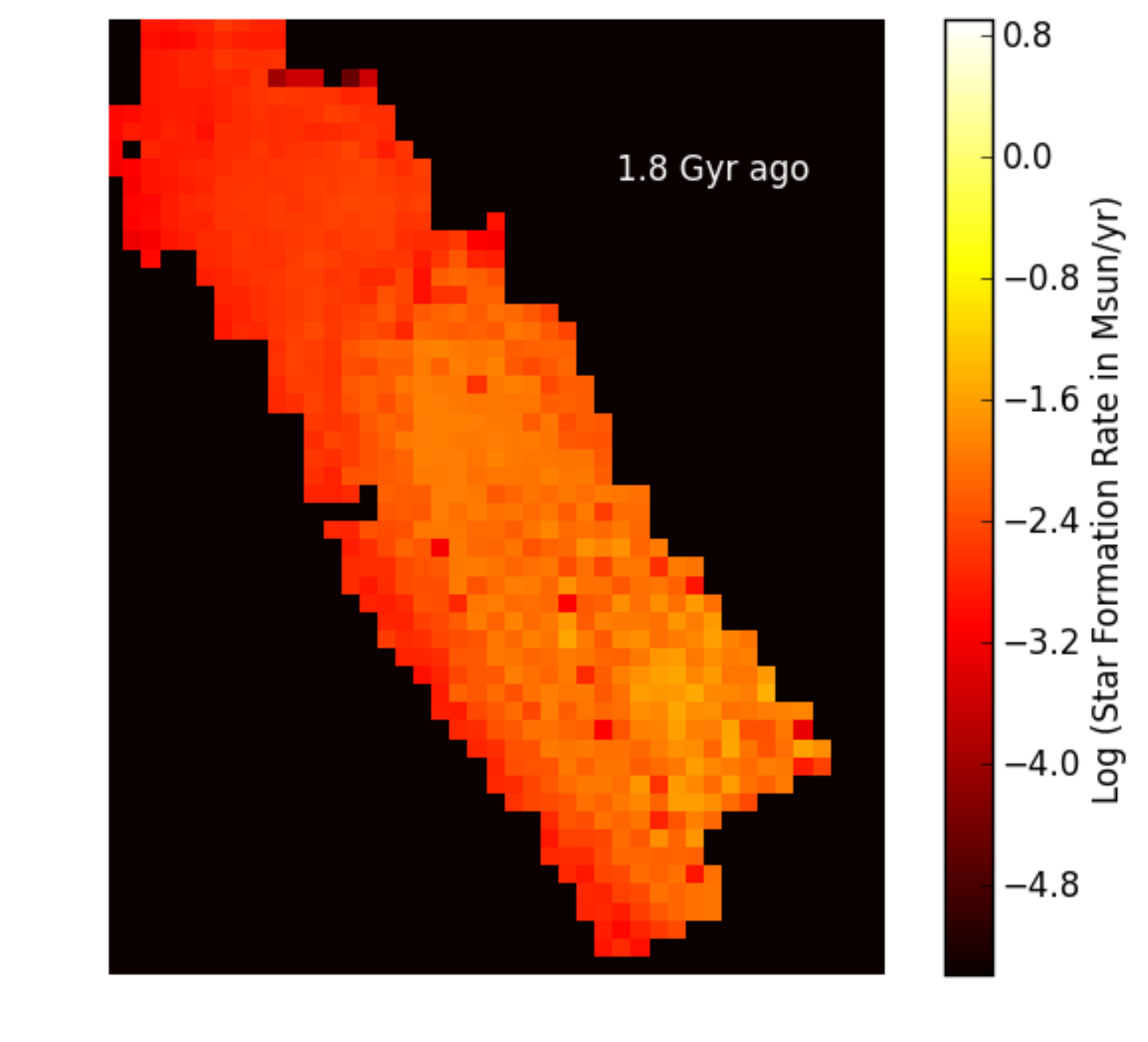}
\includegraphics[height=2.2in]{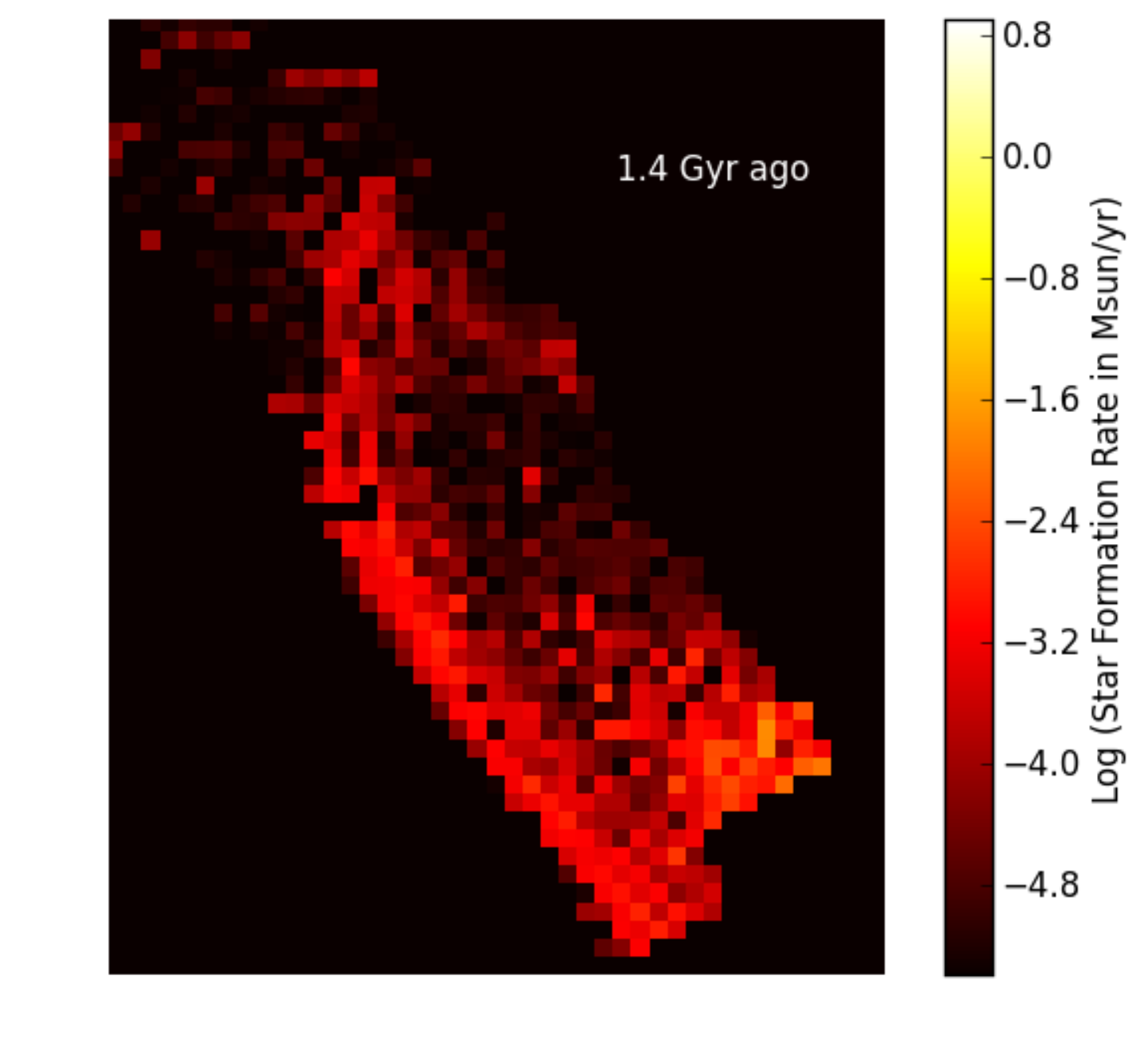}
\includegraphics[height=2.2in]{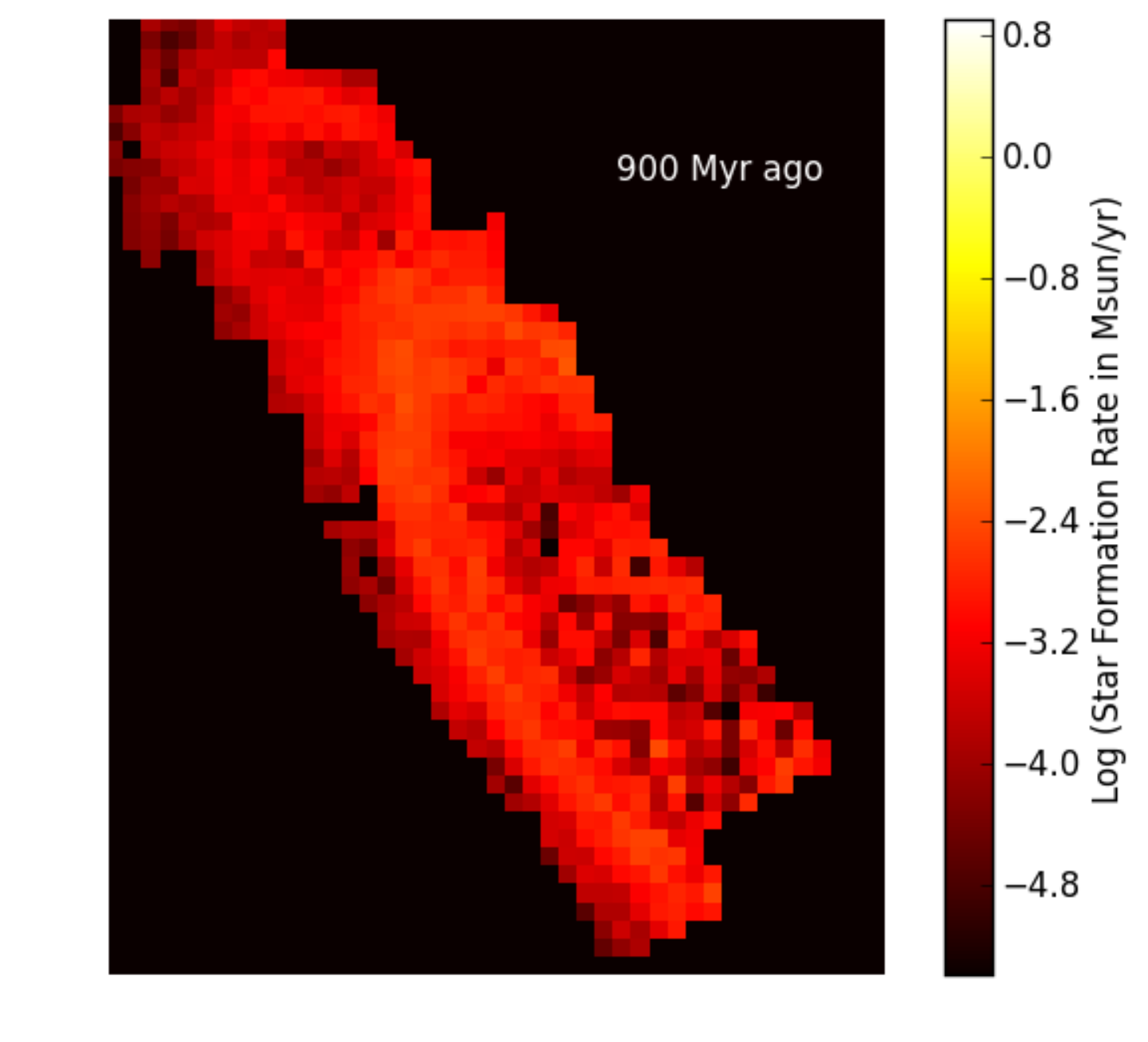}
\includegraphics[height=2.2in]{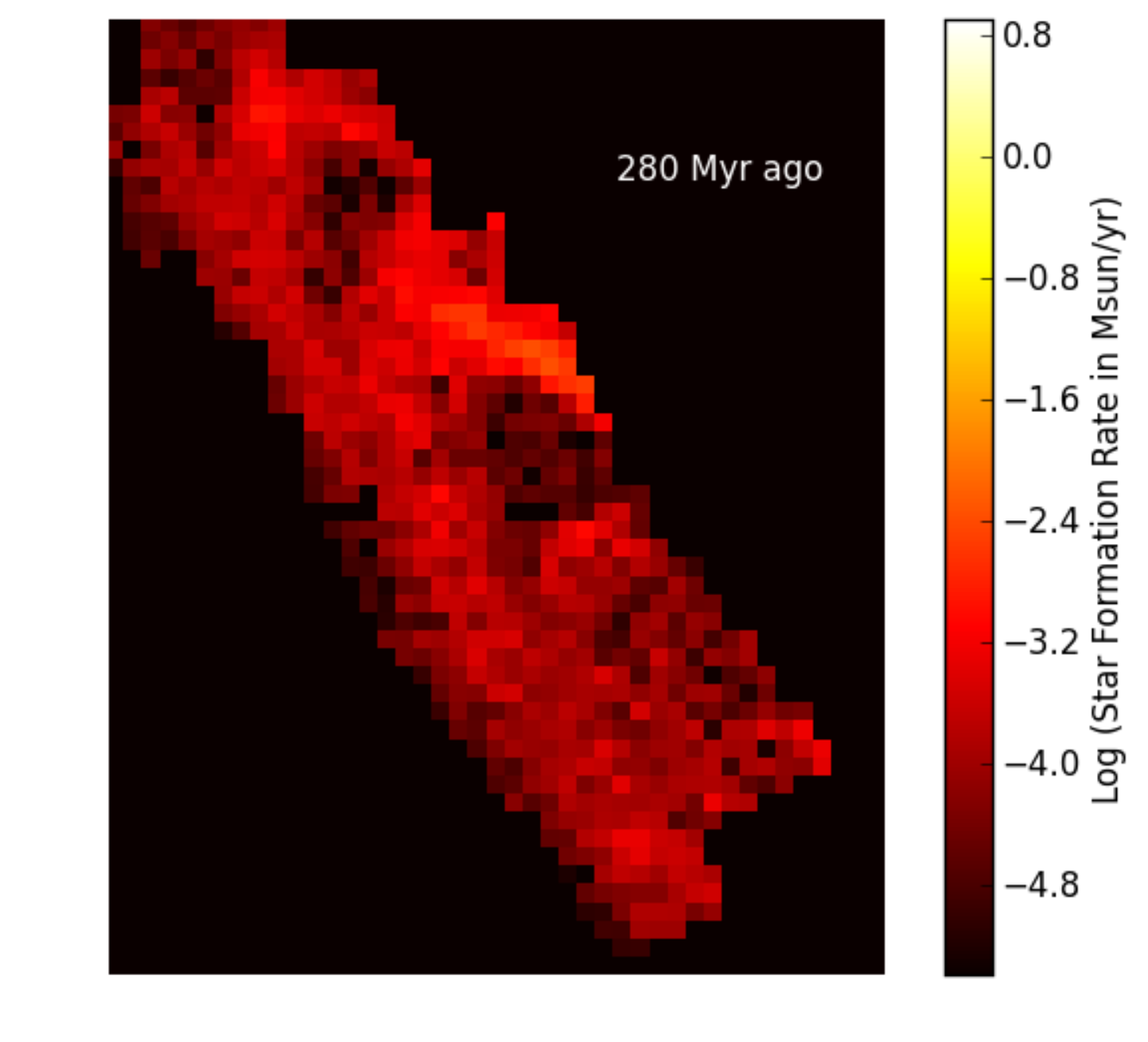}
\end{center}
\caption{Subsample of SFH maps from the star formation movie of the PHAT survey.}
\label{free_sfr_maps}
\end{figure*}

\begin{figure*}
\begin{center}
\includegraphics[width=6.2in]{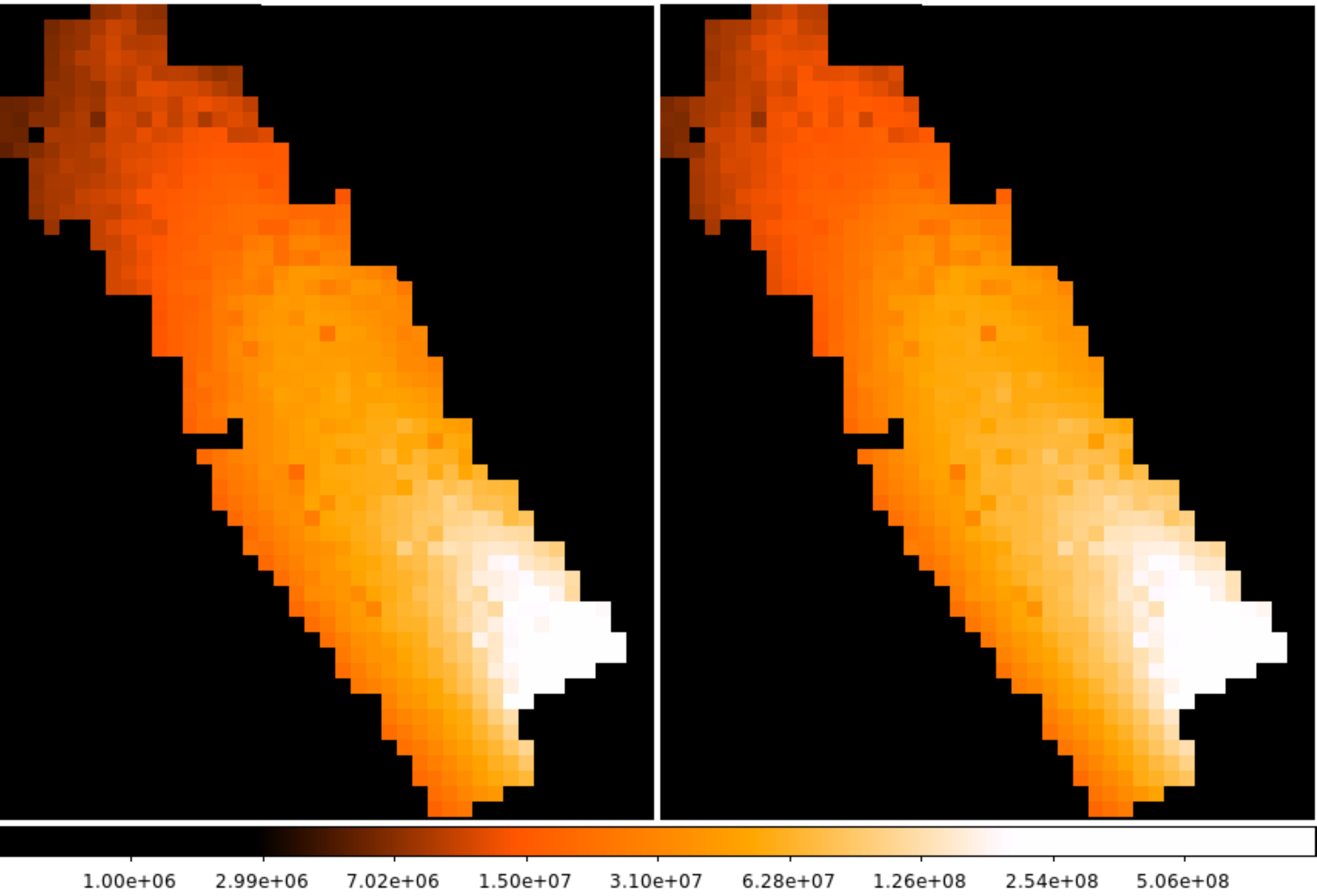}
\includegraphics[width=6.2in]{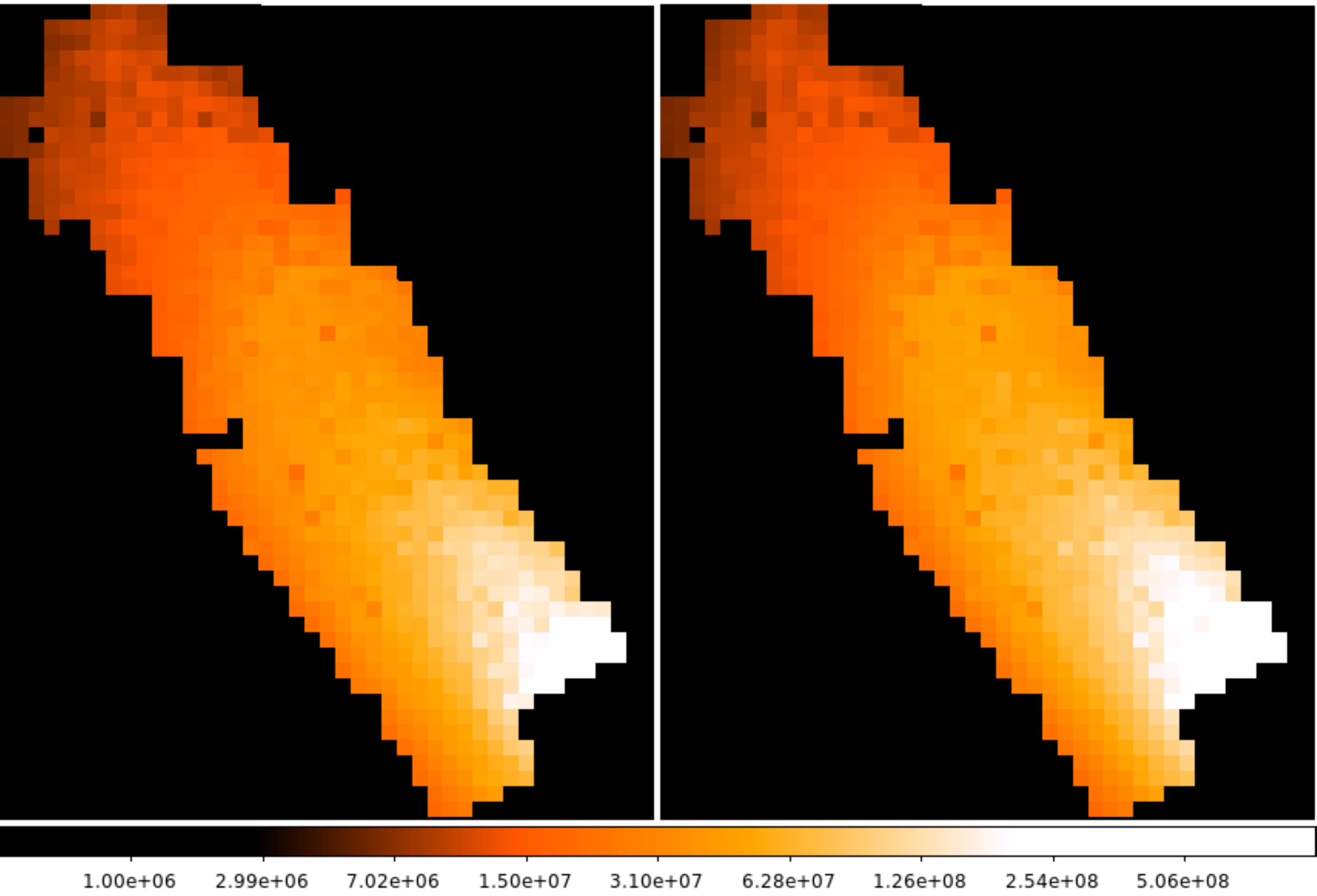}
\end{center}
\caption{Total stellar mass as a function of position in the PHAT
  survey when simple enrichment is assumed. {\it Left:} Total stellar mass in each 83$''{\times}83''$
  region as measured using the Padova models. {\it Right:} Same as
  {\it Left}, but as measured using the BaSTI
  models. \label{free_mass_maps}}
\end{figure*}

\begin{figure*}
\begin{center}
\includegraphics[width=6.2in]{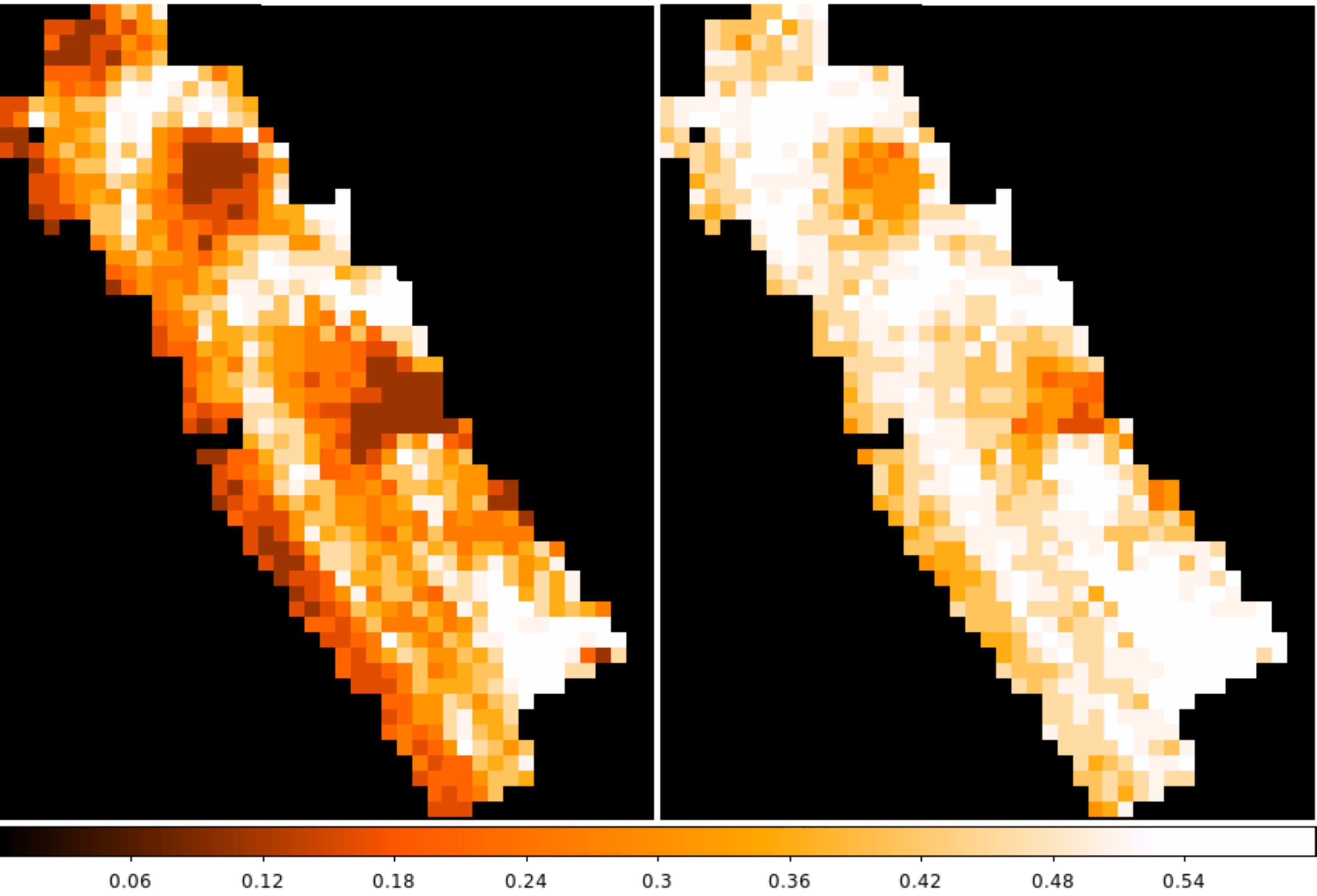}
\includegraphics[width=6.2in]{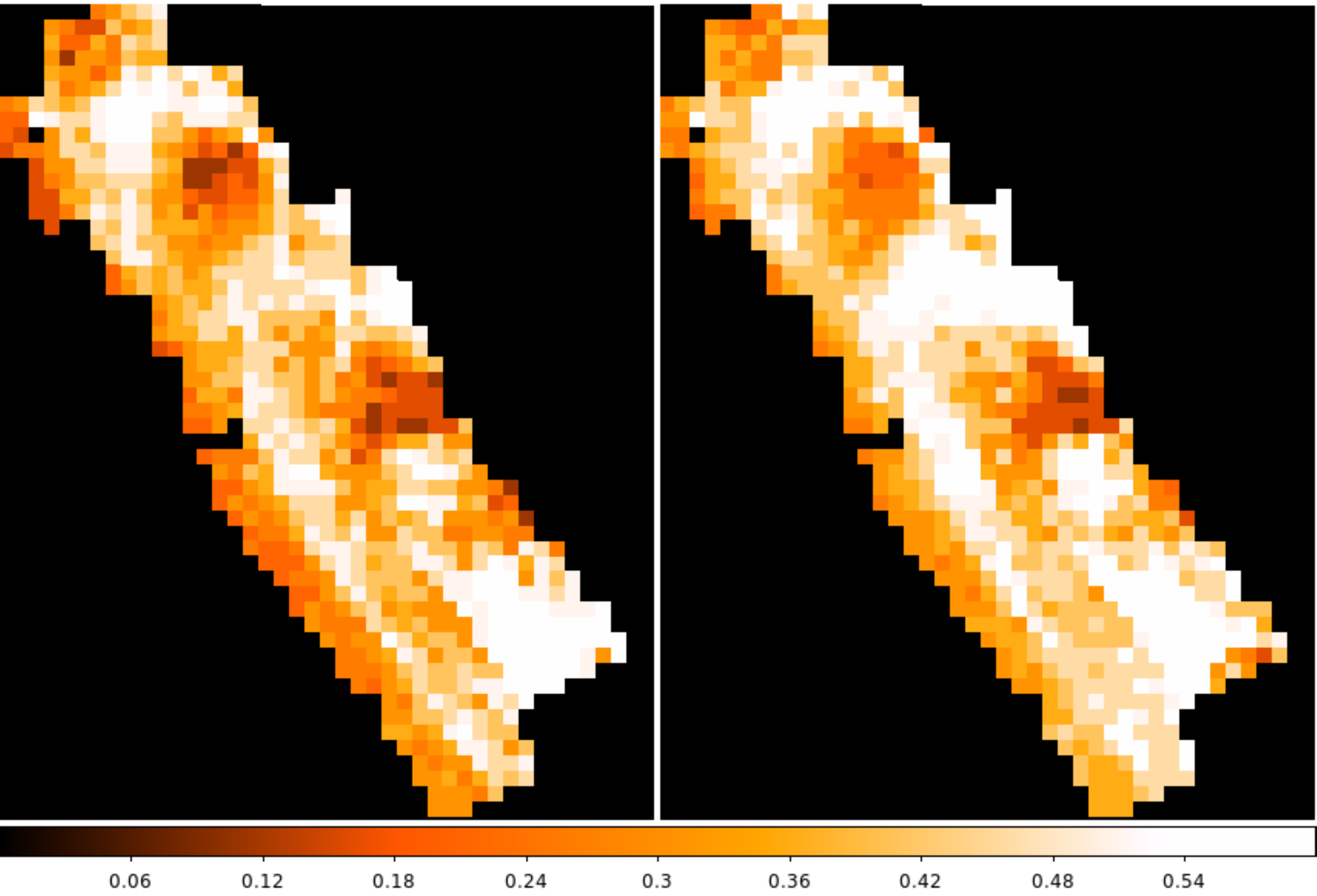}
\end{center}
\caption{Maps of the best-fitting foreground $A_{\rm VFG}$ values for the PHAT survey with free metallicity.  {\it Left:} Foreground $A_{\rm VFG}$ from the Padova fits. {\it Right:} Foreground $A_{\rm VFG}$ from the BaSTI fits. \label{free_av_maps}}

\end{figure*}

\begin{turnpage}

\begin{figure*}
\begin{center}
\includegraphics[width=3.0in]{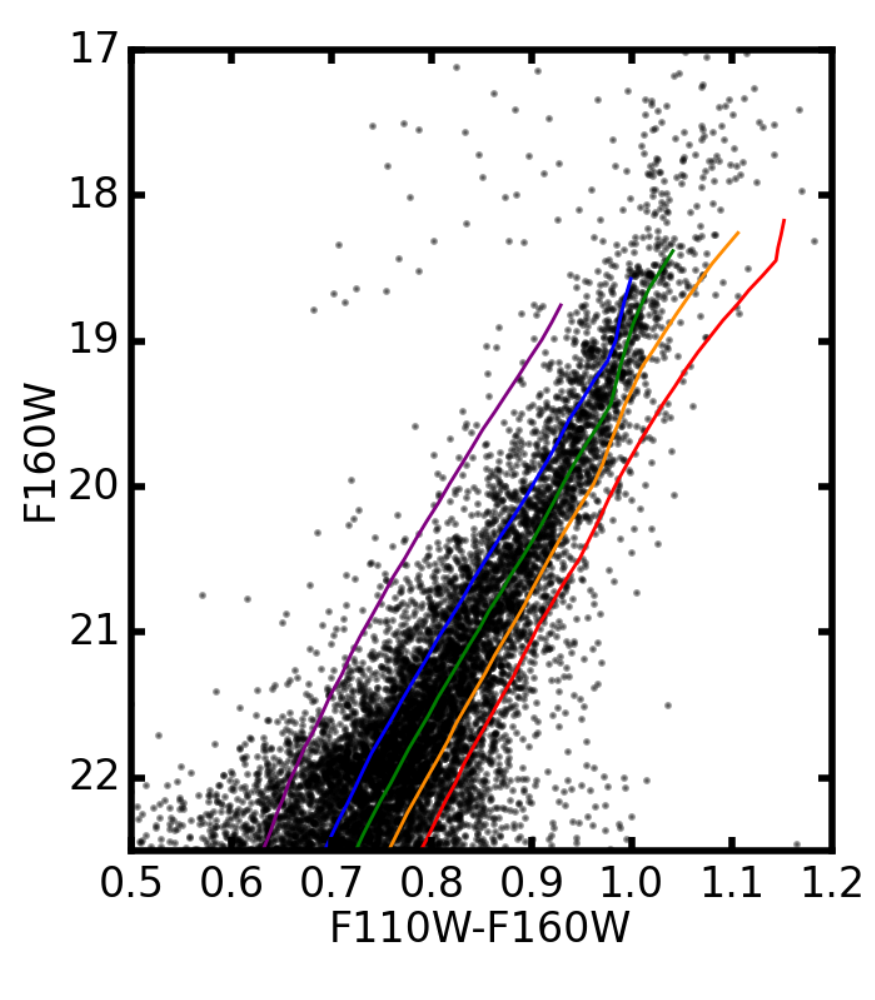}
\includegraphics[width=3.0in]{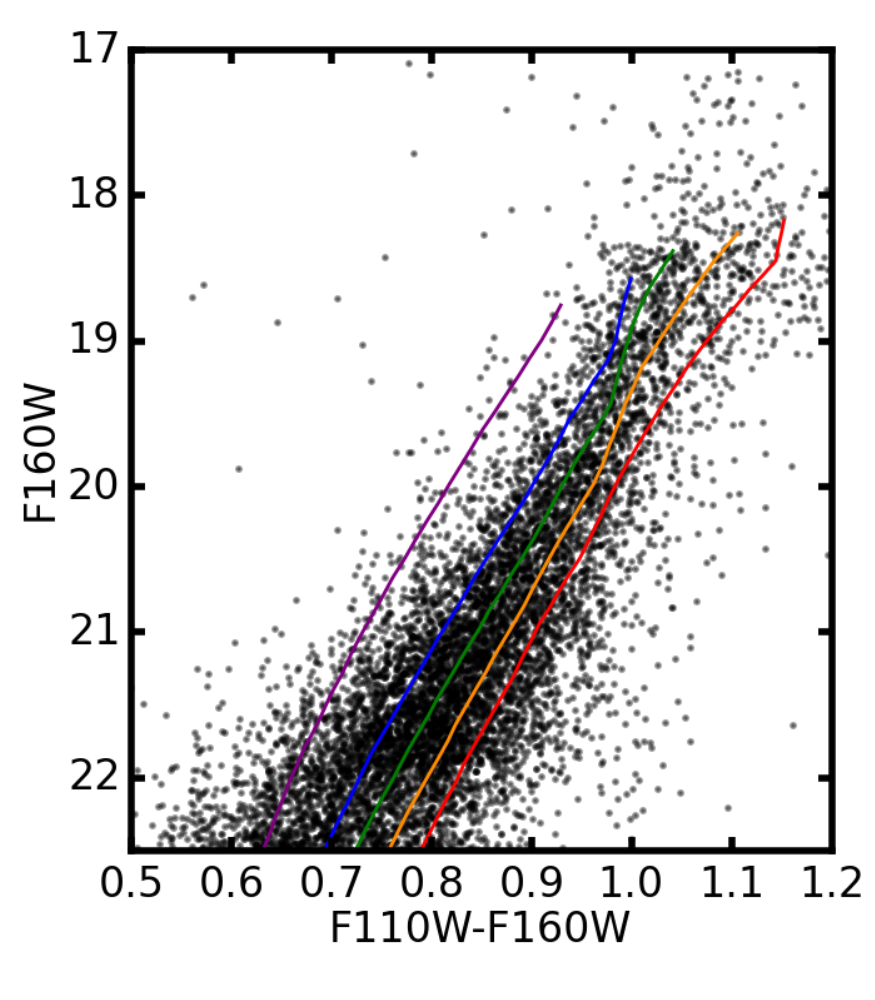}
\includegraphics[width=3.0in]{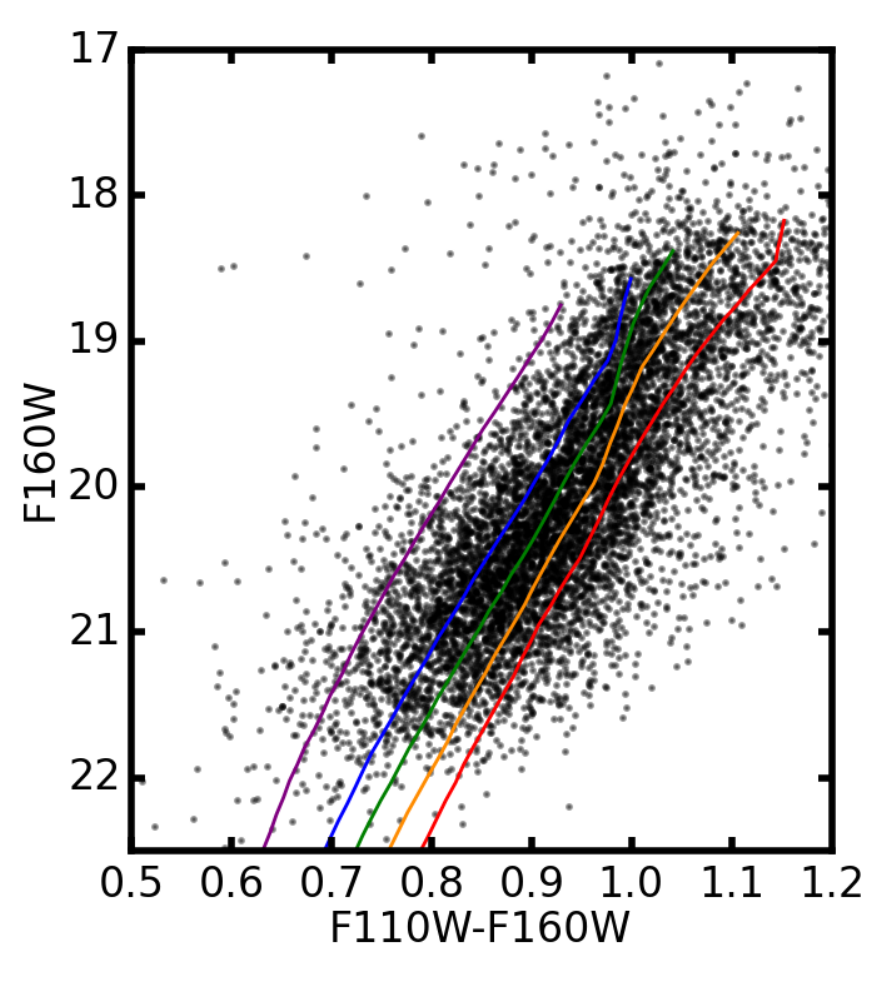}
\end{center}
\caption{NIR CMDs of the RGB taken from low (left), medium (middle), and high (right) stellar density regions with very low dust content according to the \citet{dalcanton2015} maps.  Overplotted are Padova 12 Gyr isochrones at a range of metallicities moved to the distance and foreground extinction toward M31.  From purple to red the [Fe/H] values are -1.2, -0.8, -0.5, -0.3, and 0.0.  There are very few stars with metallicities below -1.0.\label{isochrones}}

\end{figure*}
\end{turnpage}

\end{document}